\documentclass[11pt]{article}
\usepackage{floatrow}
\usepackage{float}
\usepackage{flafter}
\DeclareFloatFont{tiny}{\tiny}
\usepackage{jcappub} 
 \usepackage{graphicx}
\usepackage{epstopdf}
\usepackage{tabu}
\usepackage{overpic}
\usepackage{multirow}
\usepackage{paralist}
\usepackage{rviewport}

\usepackage{amssymb}
\usepackage{amsmath}
\usepackage{wrapfig}
\usepackage{array}
\usepackage{lineno}
\usepackage{booktabs,multirow,hhline}
\usepackage{comment}
\usepackage{lscape}

\newcommand{\beq}{\begin{equation}}

\newcommand{\CRT}{{\it CRT}\ }

\newcommand{\eeq}{\end{equation}}

\newcommand{\be}{\begin{equation}}
\newcommand{\ee}{\end{equation}}

\newcommand{\aap}{"Astro. \& Astrophys."}
\newcommand{\apj}{"Astrophys. J."}
\newcommand{\apjl}{"Astrophys. J. Lett."}

\newcommand{\prd}{"Phys. Rev. D"}

\def\og{\leavevmode\raise.3ex\hbox{$\scriptscriptstyle\langle\!\langle$~}}
\def\fg{\leavevmode\raise.3ex\hbox{~$\!\scriptscriptstyle\,\rangle\!\rangle$}}

\newcommand{\dg}{$^\circ$}

\begin{document}

\title{Deflections of UHECRs in the Galactic magnetic field}
\author[a,1]{Glennys R. Farrar\note{Corresponding author}}
\author[b]{and Michael S. Sutherland}
\affiliation[a]{Center for Cosmology and Particle Physics,
Department of Physics, New York University, NY, NY 10003, USA}
\affiliation[b]{Department of Physics \& Center for Cosmology and AstroParticle Physics, The Ohio State University, Columbus, OH}

\abstract{
We report results of the first comprehensive, high resolution study of the deflections of UHECRs using a realistic model of the Galactic magnetic field and extending to sufficiently low rigidities, $R \equiv E/Z \geq 10^{18}$V, to describe Fe in the UHE energy range above $\sim$55 EeV, or the mixed composition reported by the Pierre Auger Collaboration at the energy of the observed dipole anisotropy. 
We use the Jansson-Farrar (JF12) model, which has both a large scale coherent and a structured random component, and determine deflections for fifteen rigidities from $10^{18.0}$ to $10^{20.0}$ V, for two different coherence length choices for the random component, $L_{coh} = 30$ pc and 100 pc.  We also check the sensitivity of UHECR deflections to the particular realization of the random field, for 3 rigidity values.  For each rigidity and field model studied, the UHECR arrival direction distribution is determined for an arbitrary source direction, by inverting the trajectories of $> 5\times10^{7}$ isotropically-distributed anti-CRs of the given rigidity, which we backtrack using the code \textit{CRT}.  We present skyplots and tables characterizing the arrival directions, for representative $1^o$ sources.  

Except at high rigidity, the pattern of multiple images is very complex and depends strongly on the coherence length and source direction.  For almost all sources, average deflections grow rapidly as the rigidity falls below 10 EV and deflections commonly are greater than 90$^\circ$.   Magnification and demagnification can be strong at almost all rigidities, and varies significantly with source direction.  Much of the extragalactic celestial sphere, behind and south of the Galactic Center, cannot be seen in UHECRs below 10 EV.   
The pattern of deflections obtained with the coherent field alone can be significantly different than with the complete field model, even as regards the position of the centroid, for some source directions and rigidities.  Multiple images, sometimes very widely separated, are common for small coherence length, especially for low rigidity.} 
\maketitle

\flushbottom

\section{Introduction}
\label{intro}
\begin{figure}[t!]
\tiny{
\centering
\vspace{+0.05in}
\includegraphics[width= \textwidth]{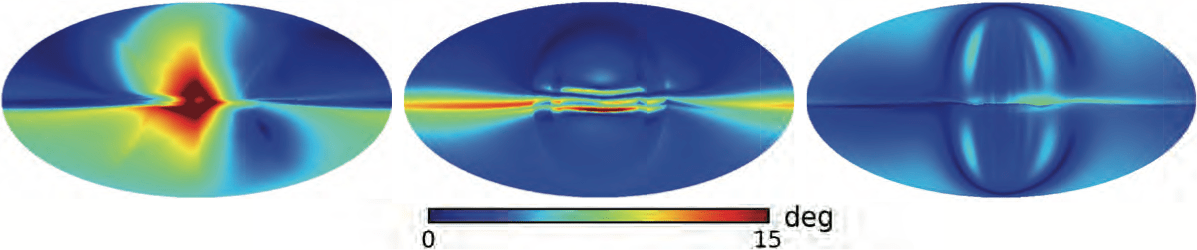}
\caption{The magnitude of the deflection of a 60 EeV proton, displayed by arrival direction, in (left to right) the Jansson and Farrar (JF12), Sun and Reich (SR10),  and Pshirkov, Tinyakov and Kronberg (PTK11) models of the large-scale GMF;  this is Fig. 11 of \cite{jf12a}.}\label{magdefscompare}
\vspace{+0.1in}
}
\end{figure}

Our understanding of magnetic fields in the Milky Way and of the global structure of the Galactic magnetic field (GMF) has developed over many decades, with a new generation of more sophisticated and quantitatively-constrained models emerging in the past few years, some of the most recent being due to Jansson and Farrar~\cite{jf12a,jf12b} (JF12), Sun and Reich (SR10, based on \cite{sun+08}) and Pshirkov, Tinyakov and Kronberg~\cite{pshirkov+11}  (PTK11).   
In spite of the progress, there are substantial differences between models, as can be seen in the differences in their predictions for deflections of UHECRs in the large-scale GMF shown in Fig. \ref{magdefscompare}.  

The most important differences between these GMF models are that i) only the JF12 model allows for a coherent poloidal component, designated the ``X" field, ii) only JF12 and SR10 are constrained with polarized synchrotron emission as well as Rotation Measure data and iii) only JF12 has a model of the random and ``striated-random" components\cite{jf12b}.  

\begin{figure}[t!]
\begin{overpic}[width=0.45\linewidth]{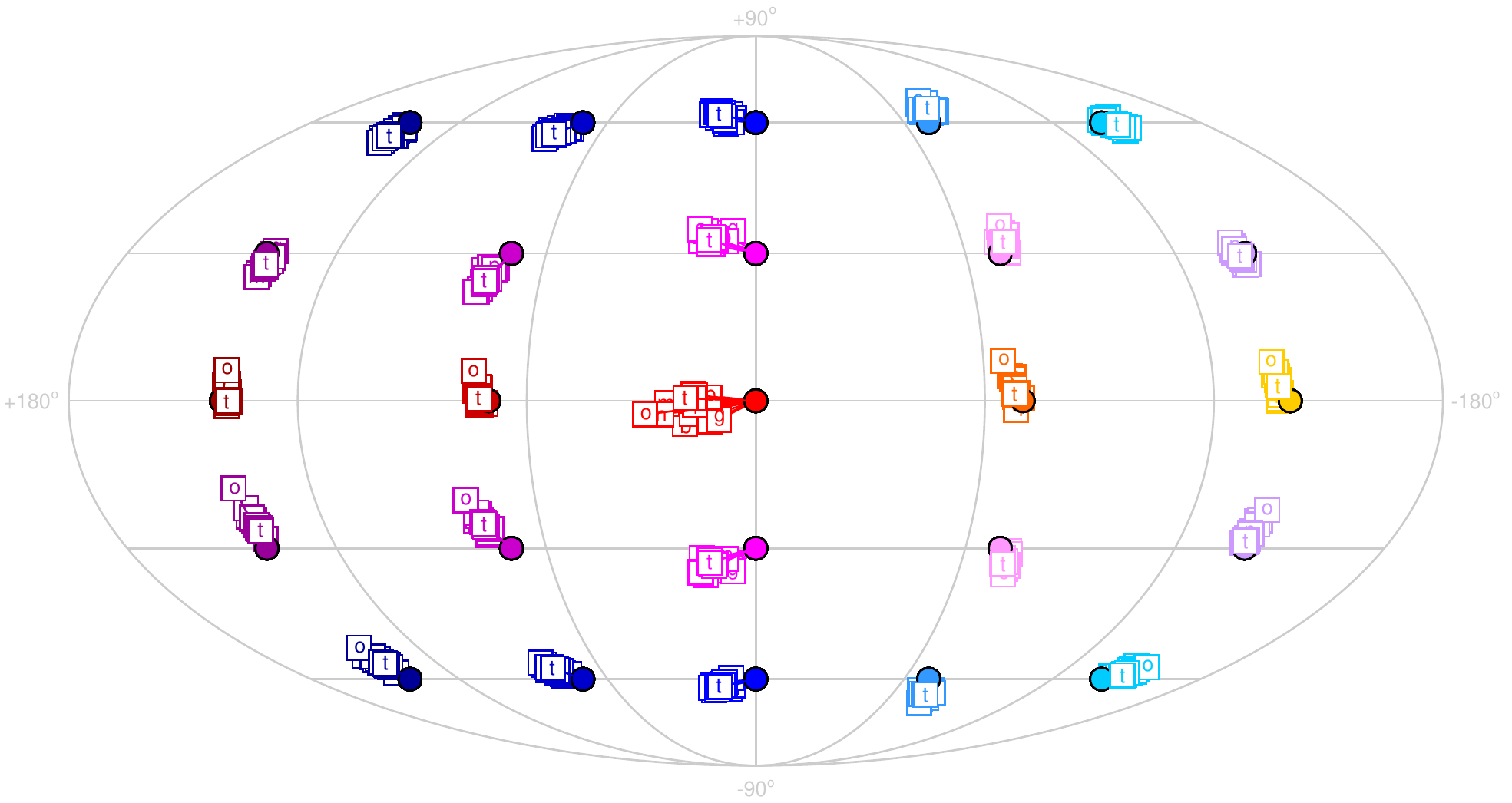}
\put (2,48) {\scriptsize $R = 60$~EV}
\end{overpic}
\begin{overpic}[width=0.45\linewidth]{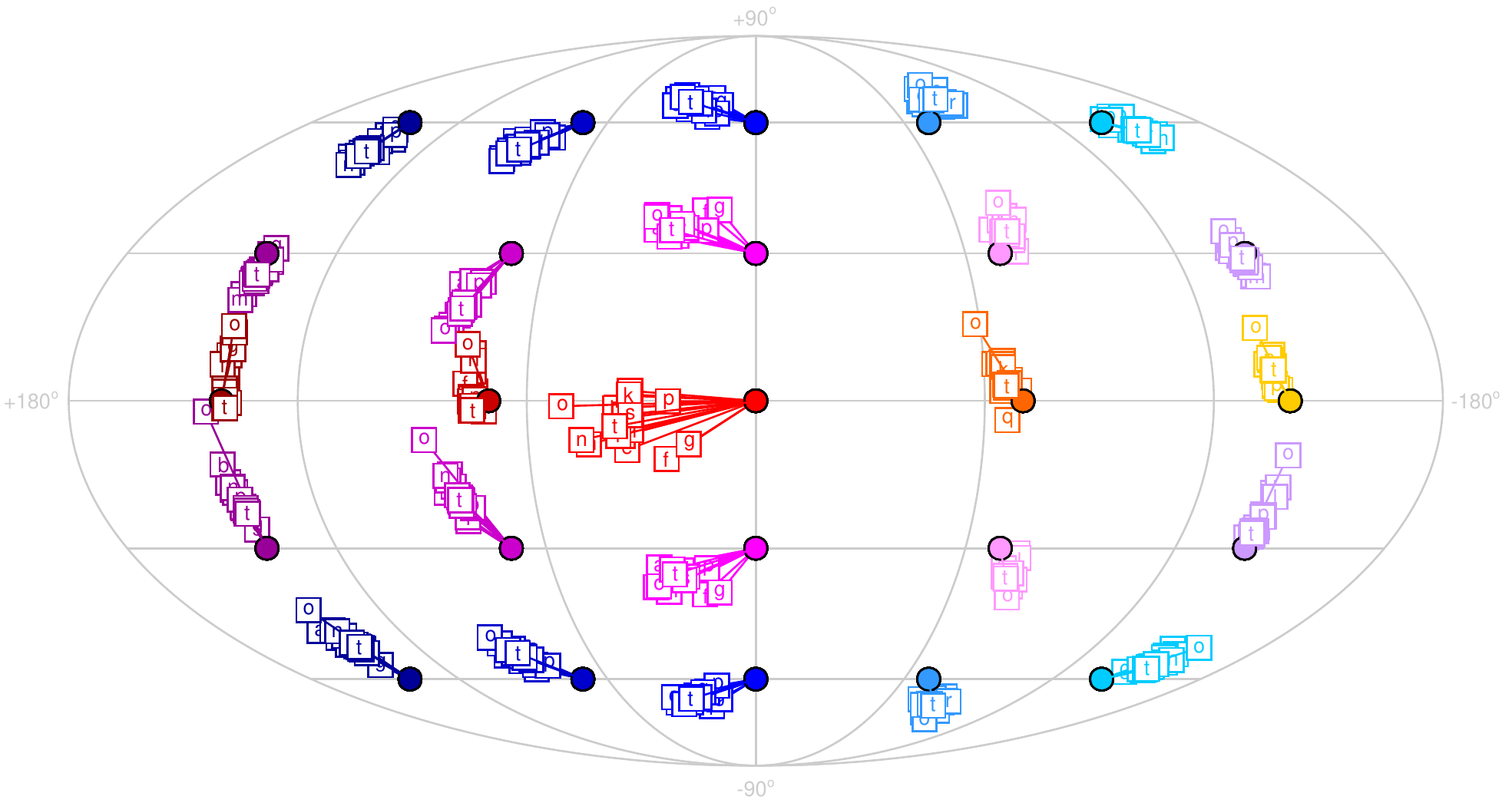}
\put (80,48) {\scriptsize $R = 30$~EV}
\end{overpic}\\
\begin{overpic}[width=0.45\linewidth]{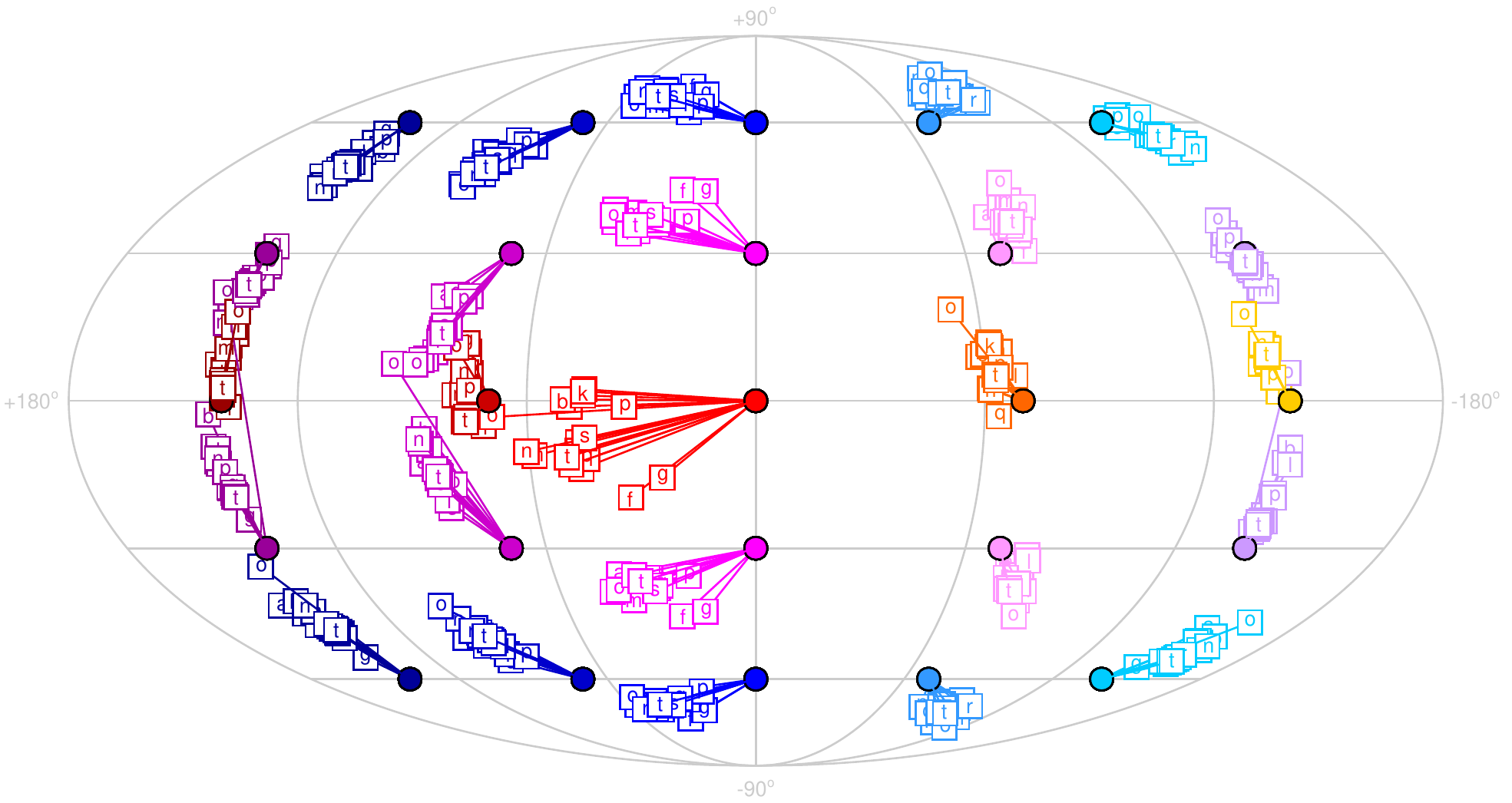}
\put (2,2) {\scriptsize $R = 20$~EV}
\end{overpic}
\begin{overpic}[width=0.45\linewidth]{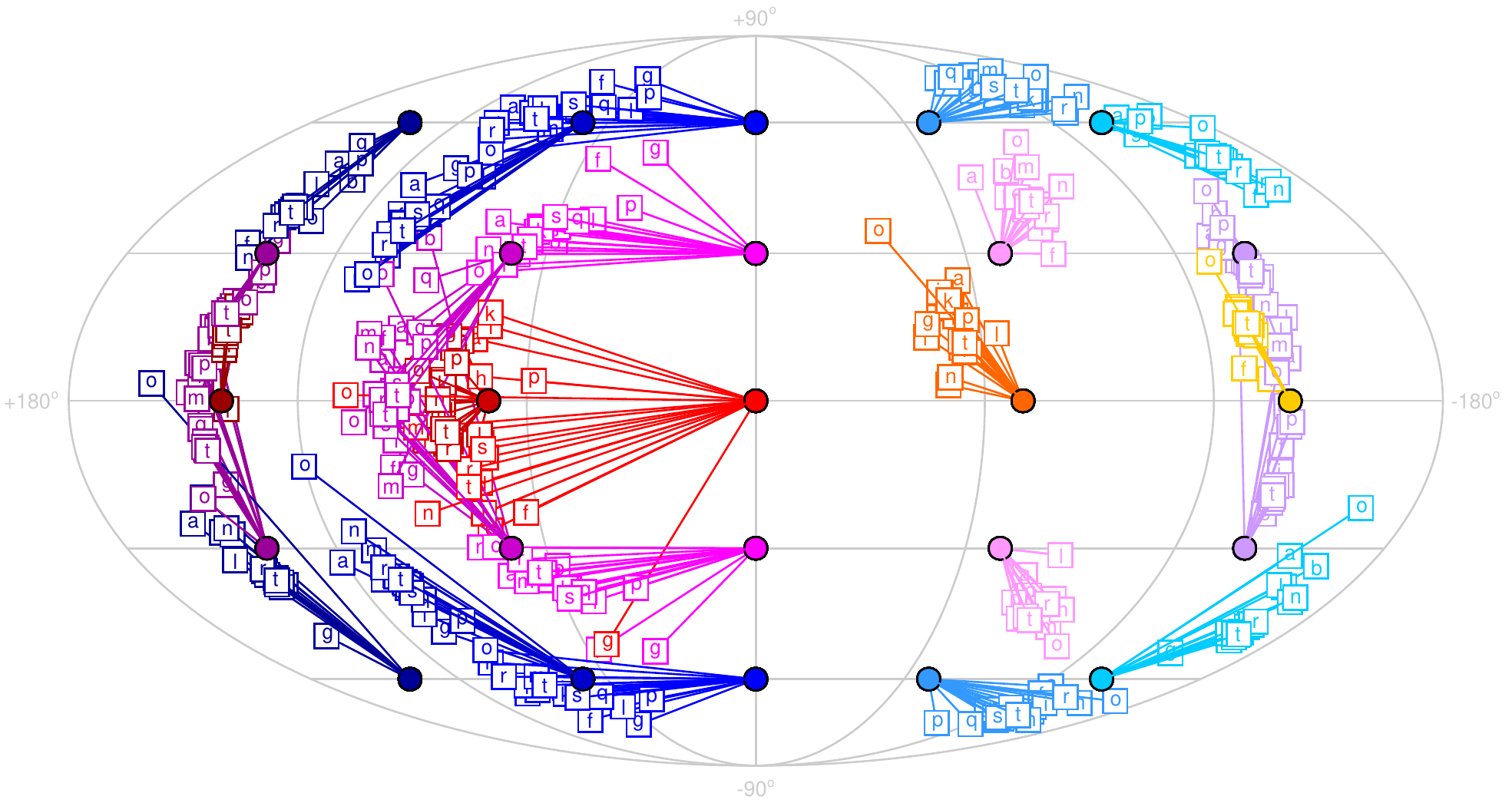}
\put (80,2) {\scriptsize $R = 10$~EV}
\end{overpic}
\caption[backtracking]{Source directions for selected CR arrival directions, corresponding to the different GMF model variants listed in the Table  of \cite{ufICRC17}, reproduced from  \cite{ufICRC17}; for details see \cite{ufICRC17}.  The sky maps are in Galactic coordinates and the particle rigidities
are indicated in the corners of each panel.}\label{fig:UFICRC}
\end{figure}

The recent study by Unger and Farrar \cite{ufICRC17}, explores how UHECR deflections are impacted by changes to the functional form of the coherent JF12 field model, use of newer polarized synchrotron data products, and use of different models of the cosmic ray and thermal electron distribution.   Fig. \ref{fig:UFICRC}, reproduced from  \cite{ufICRC17}, shows the deflections (due to the coherent field only), for selected CR arrival directions, in 19 different model variants.  One sees that model-to-model differences in the deflections can be large, so much work is needed to reduce the current GMF coherent model uncertainties.   

Our purpose here is primarily to explore the importance of the random field, at rigidities low enough to be applicable to observations \footnote{Rigidity, energy divided by charge, is the only important parameter for cosmic ray deflections.
Rigidity is properly measured in units of V but when only magnetic deflections and not energy losses are of concern, V and eV may be used interchangeably since knowing the deflection of a proton of energy E in eV specifies the deflection of any CR with the same value of E/Z.}.  This is the first work to quantitatively investigate the impact of the random magnetic field on UHECR deflections, with a realistic model of the random field and for multiple coherence lengths and field realizations. 

Measurements of the Pierre Auger Observatory \cite{augerXmaxMeas14,augerXmaxInterp14} indicate that the composition moves from being dominantly protons and/or He in the EeV region to having average charge $Z\approx 7$ at $E \sim 30$ EeV \footnote{The composition measurements are consistent between the Pierre Auger and TA collaborations \cite{augerTA_wgComp15}, but Auger's much greater statistics enables it to make geometric cuts to obtain an essentially uniform acceptance.  However, for both observatories, inferences on composition from the data depend on extrapolating models of UHE hadron physics which introduces a significant uncertainty.}.  Thus we analyze rigidities down to 1 EV. 
For each rigidity and field model studied, the UHECR arrival direction distribution is determined for an arbitrary source direction, by inverting the trajectories of $> 5\times10^{7}$ isotropically-distributed anti-CRs of the given rigidity.  This very high number of trajectories is needed, in order to have sufficiently high statistics to resolve structure in the arrival direction patterns and to obtain an accurate measurement of demagnification factors and their spatial and spectral variation.

A more detailed discussion of the arrival directions of UHECRs originating from Cen A was given in \cite{kfs14}.
That paper explored the variations due to individual realizations of the random field, but it focused on the coherent field and investigated only a limited number of rigidities propagated through the random field.
Here we have simulated with a higher resolution, for a finer sampling of rigidities propagated through various forms of the random field, and extended the analysis to the entire sky.

We take the JF12 model as a semi-realistic magnetic field model to investigate whether the random field significantly modifies deflections compared to those in the regular field alone.  The Jansson and Farrar model (JF12) gives a much better fit to the rotation measure and synchrotron emission data than other models of the coherent field \cite{jf12a}, and is the only model fitting the random magnetic field \cite{jf12b}, so we restrict ourselves to JF12 for this study.  A comparison of deflections without a random field, in the the PTK11 and JF12 coherent-only GMF models, is given in \cite{Erdmann+16}\footnote{Ref. \cite{Erdmann+16} also reports (Sec. 3.5) on the impact of different realizations of a random field on deflections, but at lower resolution and without sufficient details about their procedure for us to be able to comment on the differences between their results and ours.}.

Although no current model can be trusted in detail, JF12 as noted above fits the largest range of observations and should be good enough to draw a number of qualitative conclusions about deflections of the UHECRs.  We note however that the random field in the disk is particularly uncertain, due to the wide variation in the synchrotron emission from the Galactic disk deduced by WMAP and Planck, depending on which analysis procedure is used (\cite{ufICRC17} and Unger and Farrar, in preparation).  In addition, uncertainties in $n_{\rm cre}$ (cosmic-ray electron density) and potential correlations between CR electron density and the random field strength can have strong impacts on the interpretation of the synchrotron emission data \cite{fCRAS14,ufICRC17}. 

We assess the validity of the ``uniform Gaussian smearing" assumption commonly made in the literature to account for the impact of the random field.   We also explore the conditions required in a realistic field model, to observe multiple imaging and magnification.    As we shall see, the consequences of GMF deflection can be dramatic and will need to be taken into account in any careful future effort towards source identification.  

In the present paper we focus exclusively on Galactic deflections.   Extragalactic deflections are generally expected to be much smaller than those arising from the GMF, but for a recent study of UHECR deflections in the maximally-strong EGMF consistent with Planck limits, see \cite{Batista+EGMF&CRdefs16}.
The typical cosmic ray deflection magnitude in a turbulent extragalactic magnetic field, in the limit of many small deflections, can be estimated to be \cite{waxmanME,fjfg12}:
\begin{equation}
\label{EGdefs}
\delta\theta_{EG}\approx0.15^\circ \left(\frac{D}{3.8\ \mbox{Mpc}}\frac{\lambda_{EG}}{100\ \mbox{kpc}}\right)^{\frac{1}{2}}\left(\frac{B_{EG}}{1\ \mbox{nG}}\right)\left(\frac{Z}{E_{100}}\right),
\end{equation}
where $D$ is the source distance, $B_{EG}$ and $\lambda_{EG}$ are the r.m.s. strength and coherence length of the extragalactic magnetic field, $Z$ is the charge of the UHECR in units of the proton charge, and $E_{100}$ is the UHECR energy in units of 100 EeV.   
A 100 EeV proton originating from Cen A would thus be deflected by only $\approx0.15^{\circ}$ in a turbulent extragalactic field of 1 nG RMS strength and 100 kpc coherence length; this deflection is small compared to that from the regular and random Galactic field of JF12.
Ultimately, source studies should aspire to modeling possible deflections in the local extragalactic magnetic field as well.  

Thanks to the GZK horizon, UHECRs come from a limited source distance up to $\approx 200$ Mpc for protons and Fe, and smaller for intermediate masses.   This may assist in the problem of source ID with complex magnetic imaging and mixed composition, especially when event-by-event composition information improves with the AugerPrime upgrade \cite{AugerPrime} commencing in 2018.

\section{UHECR deflections}
Since UHECR energy losses in the Galaxy are negligible, Liouville's theorem guarantees that if the illumination from external sources is isotropic, the observed UHECR sky must be isotropic, separately for each rigidity.
Sometimes it is incorrectly argued from this that there can be no magnification of sources, multiple imaging, etc., contrary to the expectation from lensing.
The resolution of this seeming paradox is simply that while the observed sky looks isotropic if the galaxy is illuminated isotropically, that does not mean that the observed sky is illuminated by every source direction.
Sources in some directions can make no contribution whatsoever, or have a greatly reduced flux relative to their intrinsic emission.

\begin{figure}[t]
\vspace{-0.15in}
\centering
\includegraphics[width=\textwidth]{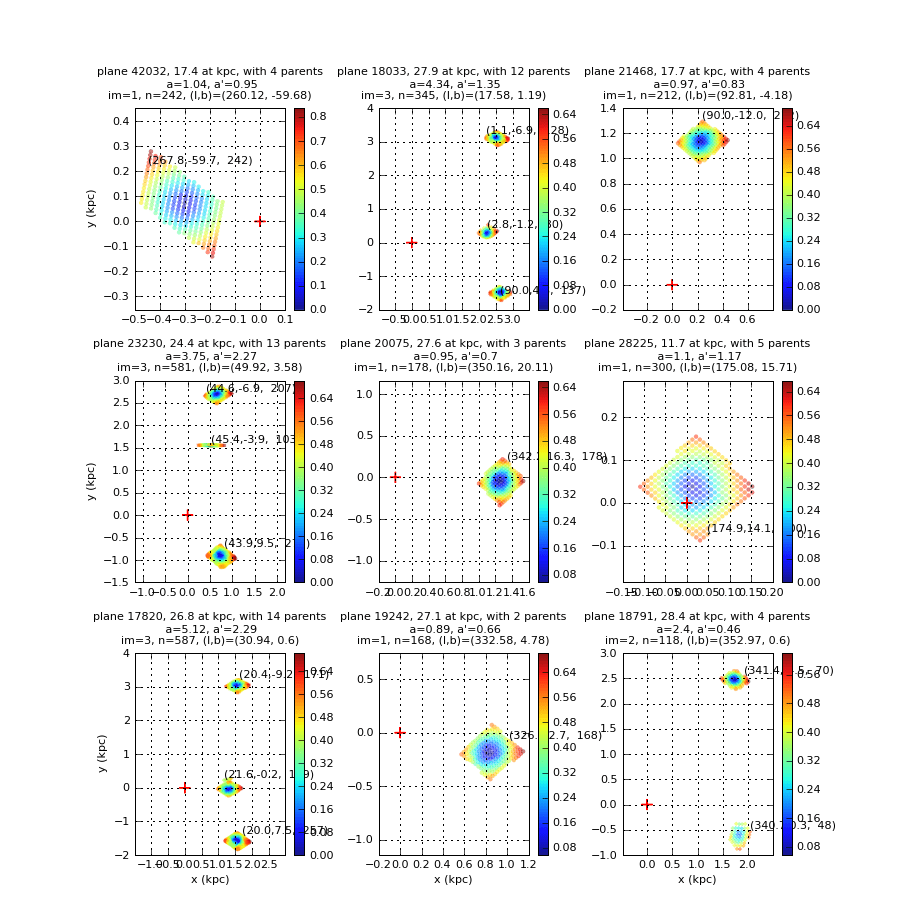}
\vspace{-0.25in}
\caption{The arrival plane regions contributing to detected 60 EV CRs in the JF12 coherent field for a sources located at various Galactic coodinates (shown in the title of each panel).
Photons would arrive at (0,0) which is indicated by a red ``+''.
The colorbar shows the angular distance of the UHECR from the center of its image.  Courtesy R. Jansson.}
\label{arrivalplane}
\end{figure}

A simple thought experiment illustrates this paradox resolution.
UHECRs from a particular distant source arrive to the Galaxy as a uniform, parallel beam.
Define an ``arrival plane'' transverse to the direction to the source, at sufficient distance that the GMF has not yet produced significant deflection.
The subsequent trajectory of a UHECR in the GMF will depend on the CR's rigidity and where it intersects the arrival plane.
Most locations on the arrival plane do not illuminate the Earth at all -- those UHECRs' trajectories simply miss Earth altogether.
It can happen that, for a sufficiently low rigidity or in a complex enough magnetic field, more than one region of the transverse plane illuminates the Earth.
Fig. \ref{arrivalplane}, courtesy R. Jansson, shows this phenomenon for some illustrative source directions by showing the position on the arrival plane from a given source of CRs which hit the detector; their diamond-shapes reflect the pixelization of the arrival direction binning.
It is striking that some source directions produce multiple images, while other directions only produce one image and yet others (not shown) produce none.
It is common that the angular distances and directions of the images from the source direction vary.

It is intuitively plausible, and required by the conservation of specific intensity along a CR trajectory (flux per steradian, c.f., \cite{rybickiLightman}), that CRs in each of the non-overlapping patches follow distinct trajectories, such that in the limit of infinite angular resolution they arrive in non-overlapping images.
As the rigidity decreases, the complexity of the trajectories and the number of images increases; within a given detector resolution, images may merge.
The sizes of the various regions in the arrival plane which illuminate the detector, are in general not equal.
Since the flux is uniform in the arrival plane, the relative numbers of events in the various detected images are proportional to the area in the arrival plane which produces them, or equivalently, to the angular size of the given arrival-direction image.
This means that the total flux from a given source can be magnified or demagnified relative to the total flux that would be received from the source in the absence of the GMF.
The magnification factor is the flux from a standard source in the given direction, summed over its images, relative to the flux in the absence of the GMF.
As we shall see, the GMF has a highly-non-trivial impact on UHECR observations.

The conservation of CR intensity (flux per steradian) between the arrival plane and the detector enables one to replace the very computationally expensive forward-tracking from a given source direction, by all-sky backtracking to make high-resolution all-sky deflection maps.  This is accomplished by following the anti-particle, which simply means changing the sign of the particle charge and reversing its incoming velocity vector.
Our backtracking simulations are performed with the numerical tool \CRT \cite{crt10}, running on the NASA Pleiades supercomputer cluster.
\CRT uses adaptive Runge-Kutta integration methods to determine the trajectory of a charged particle through a magnetic field according to the relativistic Lorentz force.  
The most recent version of this code includes the JF12 model as well as a method for generating and simulating Kolmogorov turbulent fields.

We take as the field model a superposition of the analytic JF12 coherent large scale regular field plus a small scale turbulent field whose RMS field strength is given by the JF12 random field model.
A striated field component is not included in this study. 
We use several different realizations of the turbulent field to understand the extent of variations in predicted UHECR arrival distributions coming from details of the random field.
Each turbulent realization is implemented by creating a Kolmogorov random field with the desired $\lambda_{\rm max}$ on a lattice whose RMS field strength is $1 \, \mu$G, as discussed in  \cite{kfs14}.
During the \CRT simulation, this unit box is repeated throughout the simulation volume and locally re-scaled by the position-dependent JF12 random field strength during each step.
The resulting random field is no longer perfectly divergence-free; it can be ``cleaned" by standard methods but, as expected and verified, doing so makes negligible difference in UHECR trajectories because the length scale of the variation of $B_{\rm rand}$, the rms strength of the random field, is much greater than the maximum coherence length of the turbulent field.
It should be emphasized that this treatment of the random field with a fixed, explicit realization of the random field that all UHECRs  in the simulation see, is the only correct approach.  The much easier and computationally faster method used in many early studies, in which the random component of the GMF was produced ``on-the-fly'' for each CR, gives misleading results by effectively averaging over an ensemble of random fields, thus smoothing and broadening the images.  

The Larmor radius of a 1 EV CR in a 1 $\mu$G field is 1.1 kpc.  We take the minimum lattice spacing in the unit box to be 5 pc.  This assures that the small-scale structure in the field is adequately described for CR rigidities of interest here, even in the regions with largest field strengths, and that there is range of scales even for the 30 pc coherence length cases.   Several different methods of computing the basic unit box with the Kolmogorov spectrum were investigated in \cite{kfs14}, as was the sensitivity to reducing the JF12 disk random field strengths by a factor of 3, as argued in \cite{fCRAS14} may be the consequence of using a more realistic, modern model of $n_{\rm cre}$ than the GALPROP model used by JF12, or one of the newer models of Galactic synchrotron emission discussed in \cite{ufICRC17}.  

The turbulent random component is described by multiple realizations of the random field described by a Kolmogorov power spectrum.
The first realization, termed KRF6, has a coherence length $L_{\rm coh}$ of 100 pc, a typical maximum size of supernova remnants\cite{Gaensler:1995, Haverkorn:2008} and a widely adopted value in the literature\footnote{Throughout, we quote the coherence length of a random field configuration based on the relation between maximum wavelength of the Fourier modes used to generate it given in \cite{harari+CohLength02}: $ L_{\rm coh} \approx \lambda_{\rm max} / 5$.}.
A second realization, termed KRF10, has a smaller coherence length of 30 pc.
This was motivated by Beck and Wielebinski \cite{beckWielebinski13}, who estimated the magnetic turbulence (coherence) length from a combination of typical beam depolarisation and Faraday dispersion in galaxies to be $\approx 30$ pc;  a value of $\approx 50$ pc was estimated for M51 in \cite{fletcher+cohLen11}.
The two values of coherence length used here thus bracket the expected deflection properties of the Galactic magnetic field for UHECR; note that in general the coherence length should be expected to vary with position within the galaxy and in the Galactic plane may be considerably smaller \cite{Haverkorn:2008}.  
A third realization, termed KRF11, is complementary to KRF10 in that it was generated using the same parameters but used a different random seed value.

We backtrack cosmic rays from the Earth with isotropically distributed initial directions calculated using the HEALPix\footnote{http://healpix.sourceforge.net} software package.
For resolution index 11, this pixelates the sky with more than $5\times10^{7}$ pixels each covering $8\times10^{-4}$ deg$^{2}$.
One particle is backtracked with a fixed rigidity $R$ from the center of each pixel.
The Larmor radius of a relativistic charged particle in a magnetic field $B$ depends only on its rigidity, so the use of rigidity allows a single simulation to represent a variety of primary particle compositions with appropriately scaled total primary energies.
An event with rigidity $R$ could be a proton with energy $R$ or an iron nucleus with energy 26$R$.
For the KRF6 and KRF10 realizations, simulated rigidities include $10^{18.0}$, $10^{18.1}$, $10^{18.2}$, $10^{18.3}$, $10^{18.4}$, $10^{18.5}$, $10^{18.6}$, $10^{18.8}$, $10^{19.0}$, $10^{19.2}$, $10^{19.4}$, $10^{19.5}$, $10^{19.6}$, $10^{19.8}$, and $10^{20.0}$ V.
In the case of KRF11, only $10^{18.5}$, $10^{19.0}$, and $10^{19.5}$ V are simulated due to time constraints on Pleiades.

\section{Magnification Effects of the GMF}

It was pointed out by Harari, Mollerach, and Roulet in a series of papers beginning in the 1990s, that the GMF acts as a lens whose properties change with rigidity, in principle giving rise to multiple images and magnification; c.f., \cite{hmrLensing00}.   Simulations were performed by Giacinti, Semikoz and collaborators \cite{giacinti+HiZturbGMF11} using hypothetical GMF models of the day, to investigate the phenomenon.  They included a turbulent component for the field, taking it to be a non-structured Kolmogorov random field whose rms field strength $\sim e^{-z/z_0} e^{-r/r_0}$, and allowing for different parameters.  They considered a single rigidity, R = 2.3 EV, appropriate to 60 EeV iron, and used a much lower resolution than we do:   $10^5$ rather than $> 5 \times 10^7$ isotropically distributed back-tracked events for each rigidity and field configuration studied.   

\begin{figure}[H]
\hspace{-0.3in}
\centering
\begin{minipage}[b]{0.48 \textwidth}
\includegraphics[width=1. \textwidth]{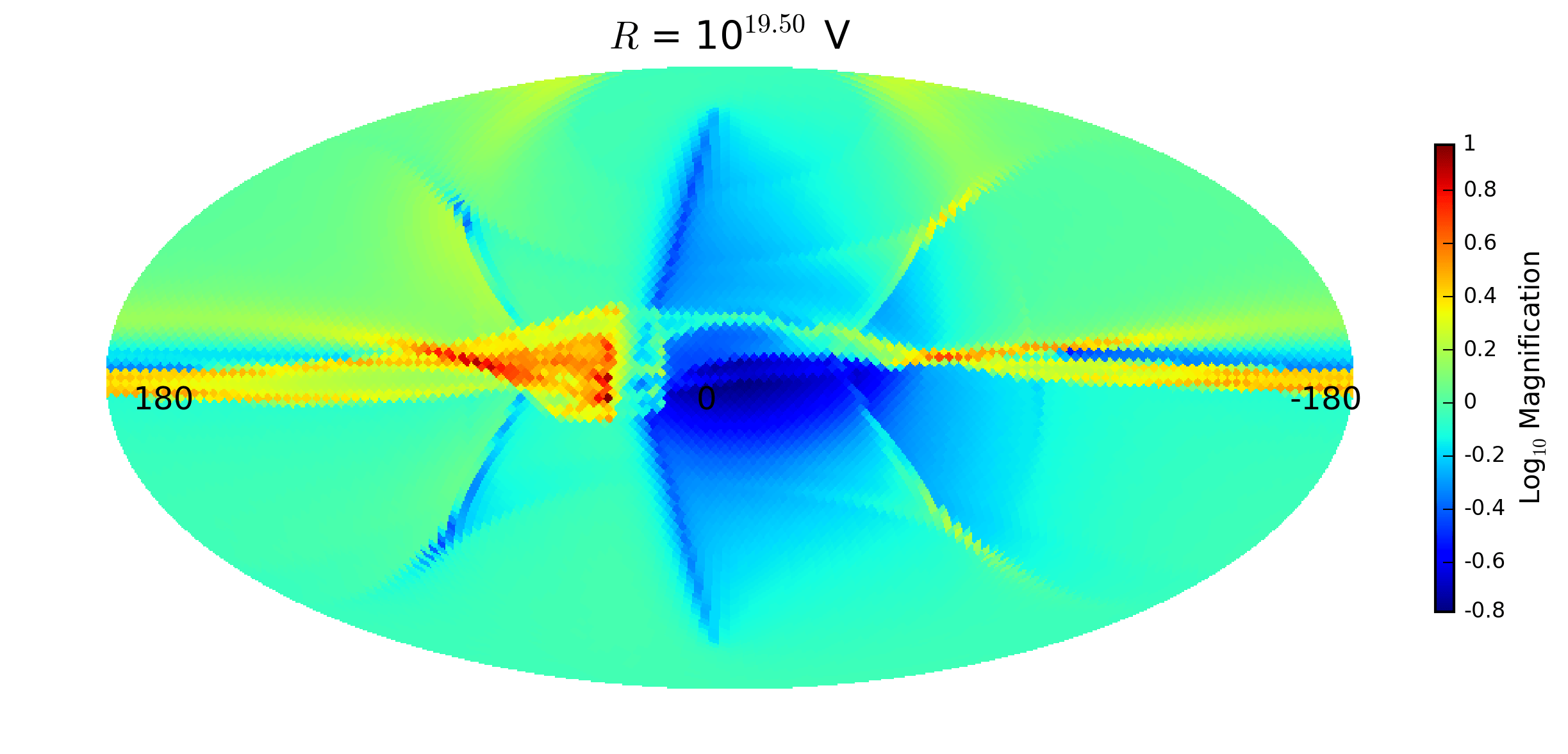}
\end{minipage}
\begin{minipage}[b]{0.48 \textwidth}
\includegraphics[width=1. \textwidth]{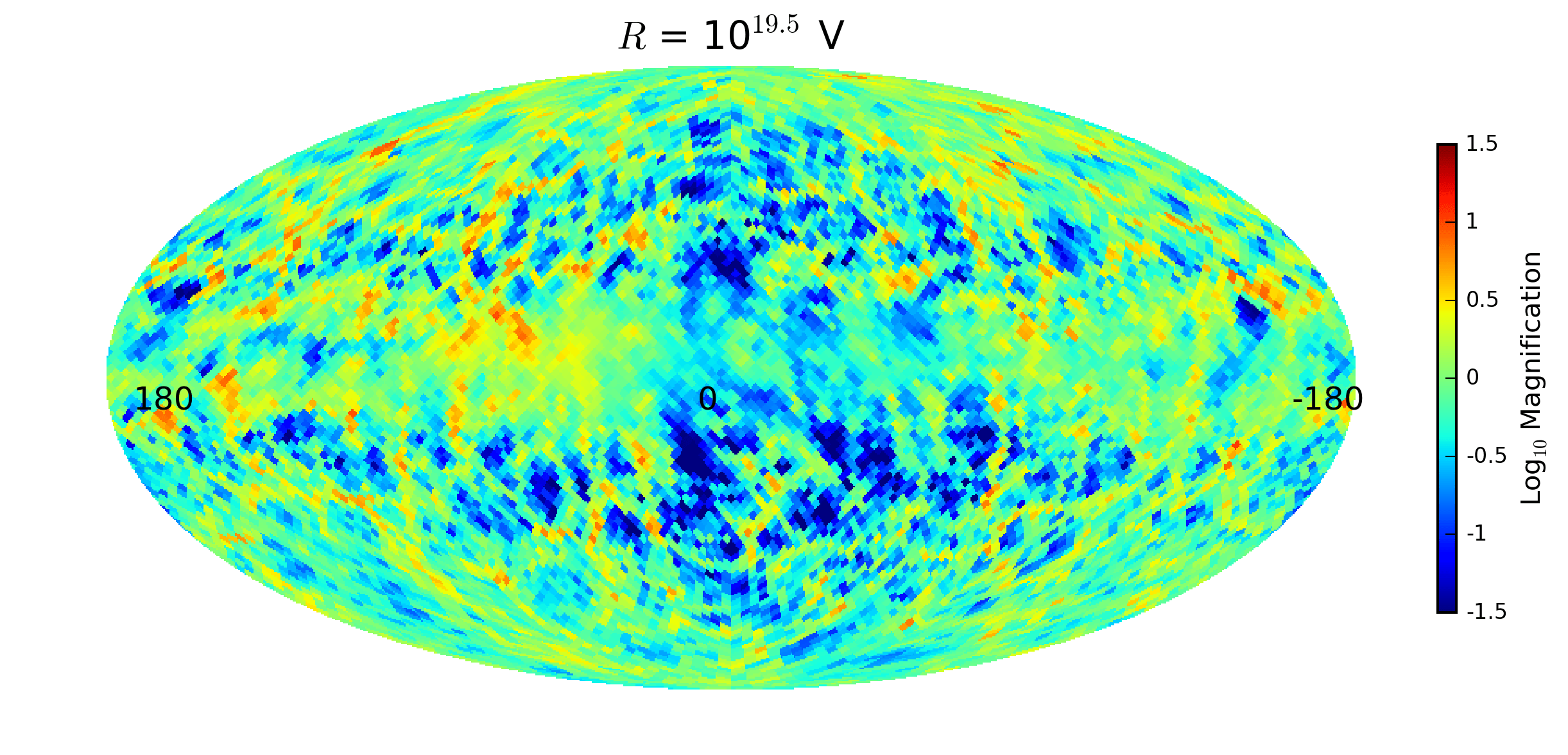}
\end{minipage}
\begin{minipage}[b]{0.48 \textwidth}
\includegraphics[width=1. \textwidth]{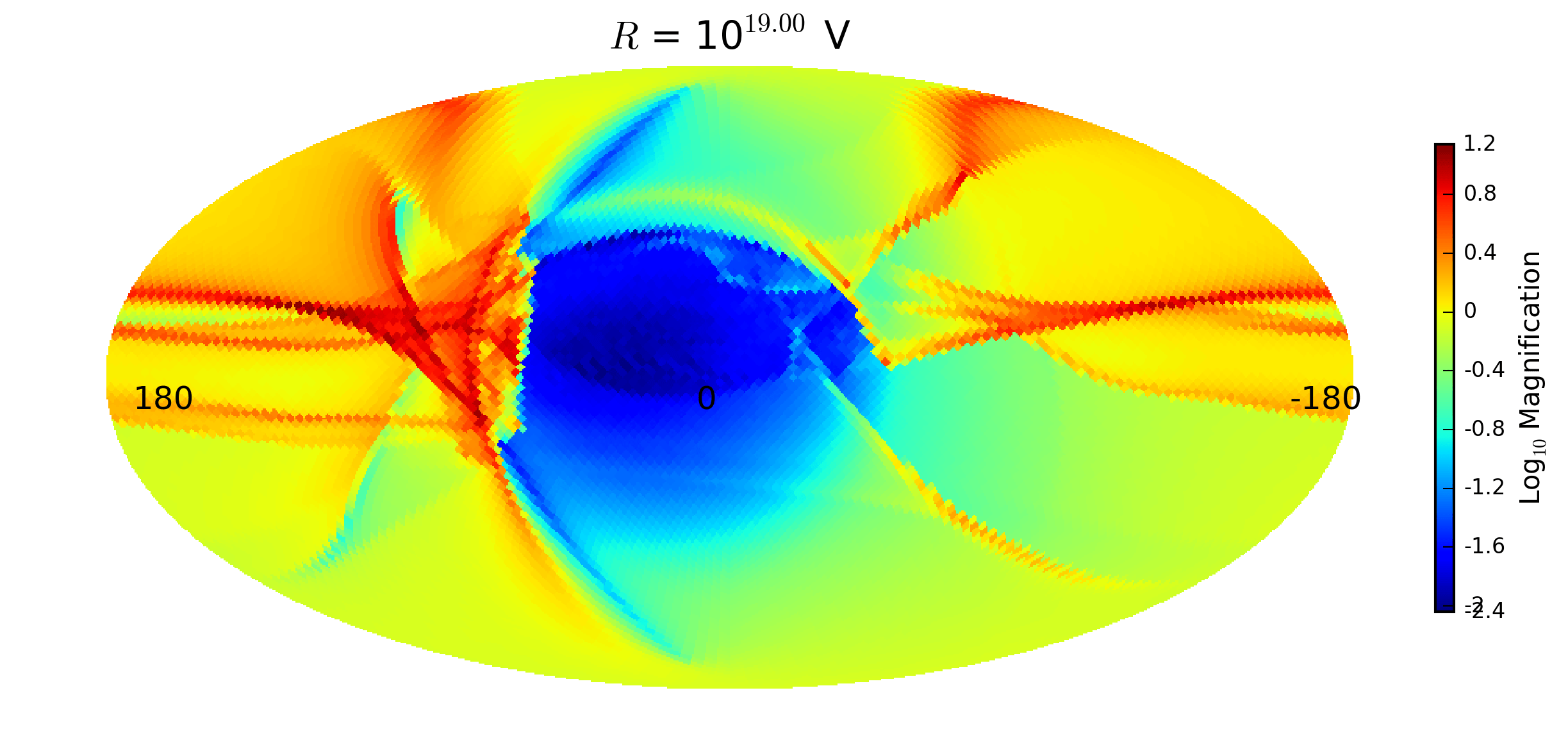}
\end{minipage}
\begin{minipage}[b]{0.48 \textwidth}
\includegraphics[width=1. \textwidth]{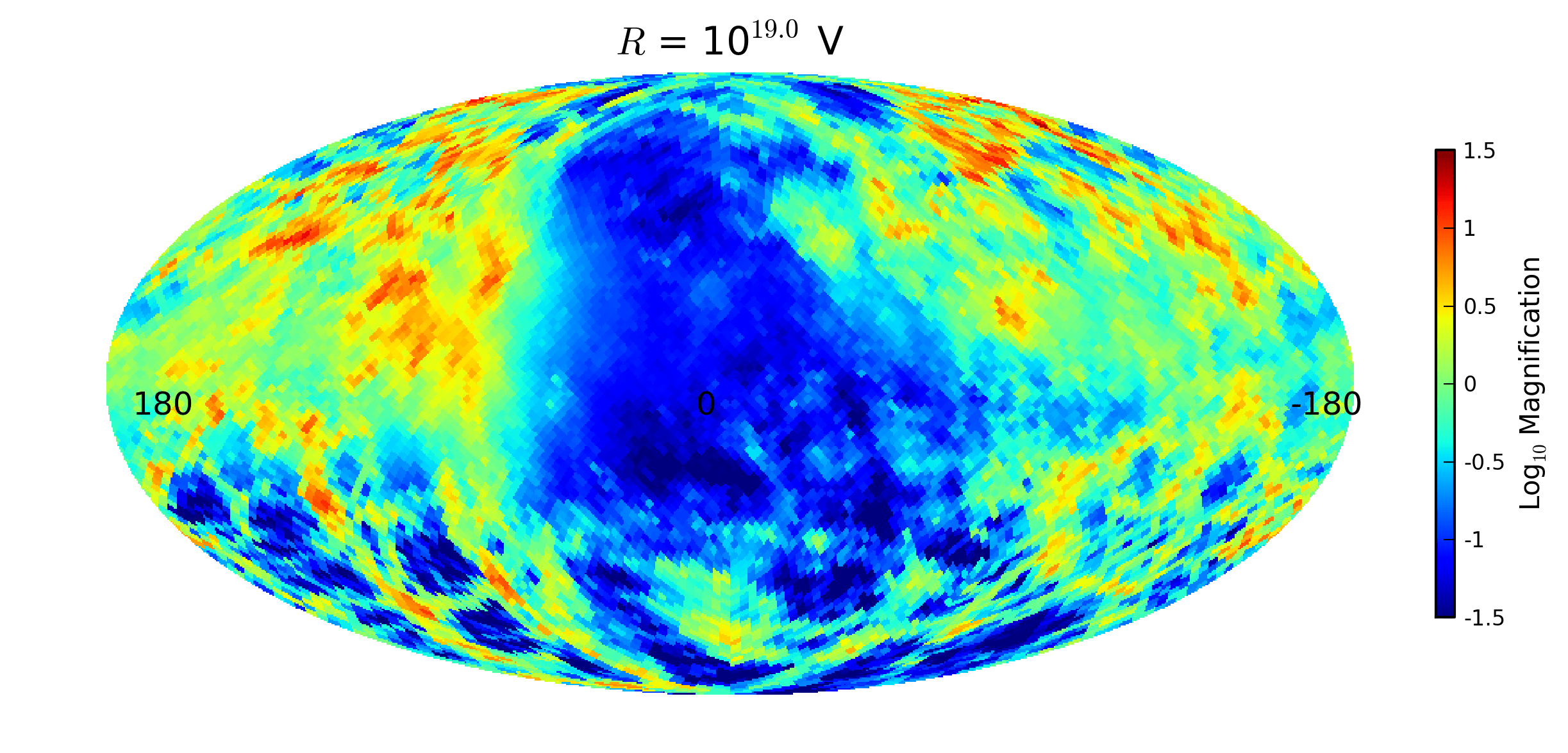}
\end{minipage}
\begin{minipage}[b]{0.48 \textwidth}
\includegraphics[width=1. \textwidth]{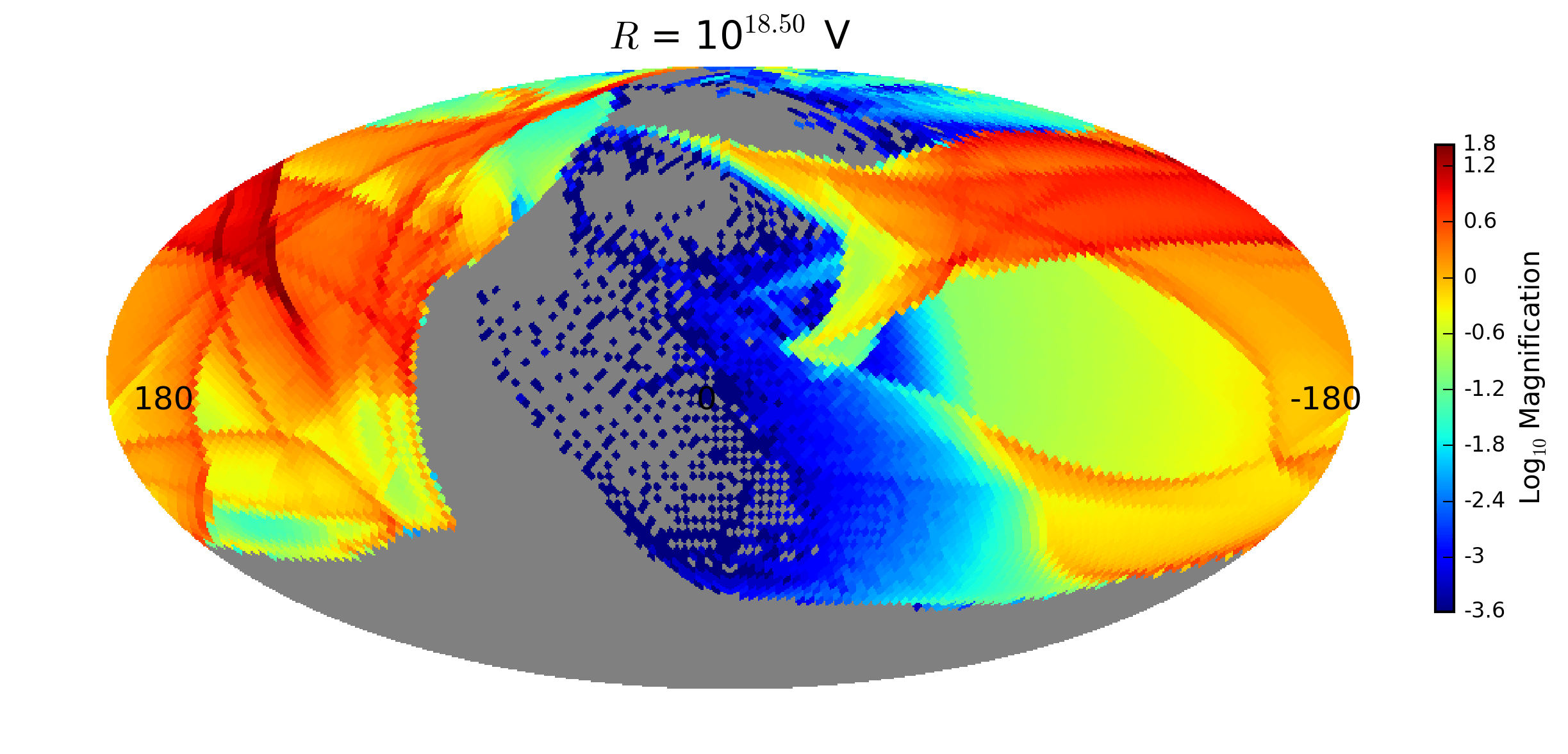}
\end{minipage}
\begin{minipage}[b]{0.48 \textwidth}
\includegraphics[width=1. \textwidth]{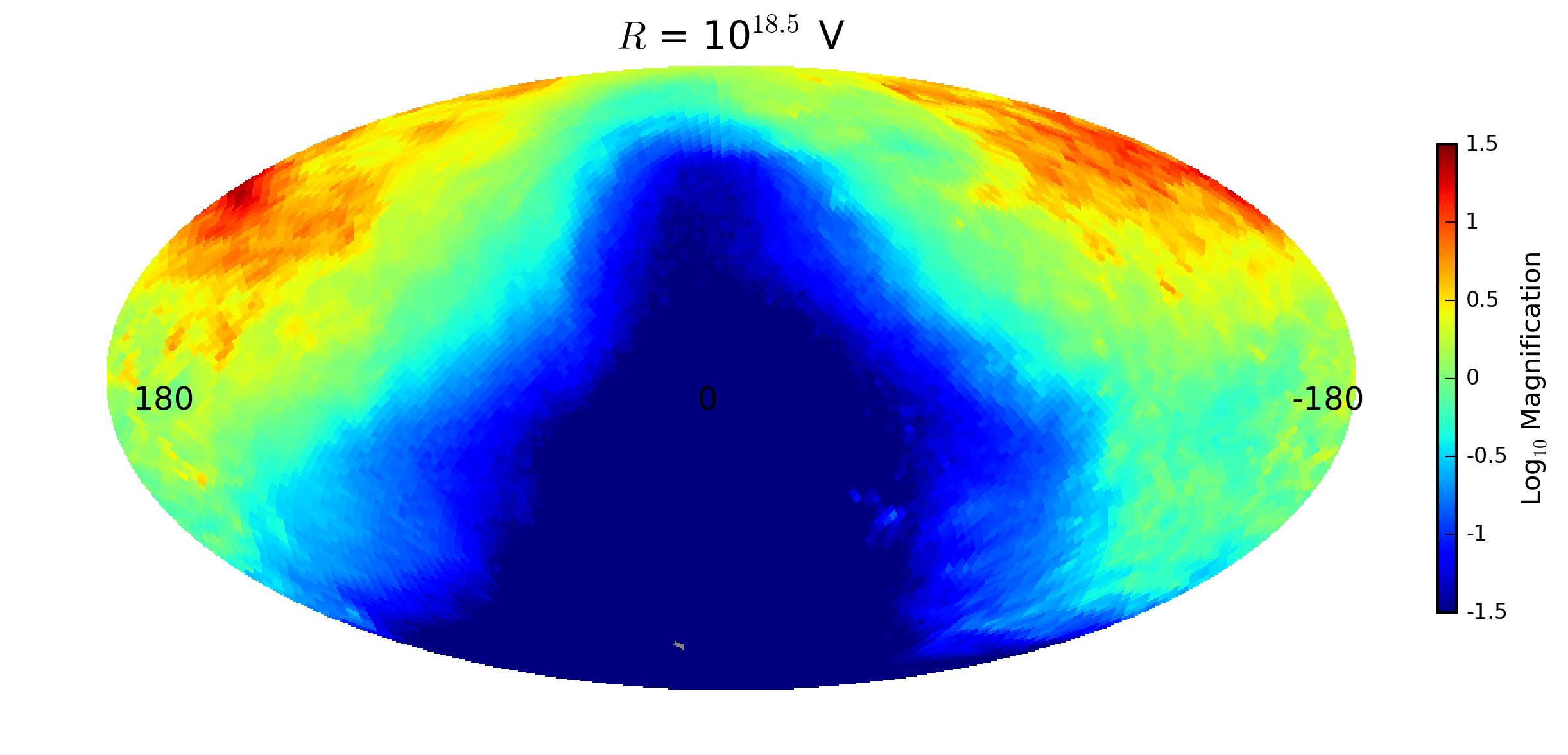}
\end{minipage}
\vspace{-0.1in}
\caption{Magnification maps 
for log($R$/V)=19.5, 19.0, and 18.5 from top to bottom.  The magnification is defined to be the flux from a standard source in the given direction, summed over its images, relative to the flux in the absence of the GMF.
The left column shows the simulation results with only the coherent JF12 component and the right column shows the coherent plus the KRF6 turbulent realization with $L_{\rm coh}=100$ pc.
The colorbar shows the log10 factor of the magnification of each pixel.  The linear features in the coherent-field-only maps on the left result from the abrupt transitions in coherent field strength at the edges of spiral arms in the simple JF12 modeling, which is naturally smoothed and made more physical by including the turbulent component.  One sees that gross features such as the demagnification of sources toward the Galactic center and magnification of those toward the outer, Northern galaxy, are governed by the coherent field.} 
\label{plt:magmaps_coh-krf10}
\vspace{-0.1in}
\end{figure}

 We thus present here the first thorough examination of the magnification phenomenon in a magnetic field applicable for the Milky Way and at sufficiently high resolution to resolve fine structure.  
Figure \ref{plt:magmaps_coh-krf10} shows the all-sky magnification factors for representative rigidities for the JF12 coherent field only (left) and for the total of coherent field plus KRF6 realization of the random field (right).  It demonstrates the importance of including random fields, to have a realistic description of GMF deflections.  As could be expected, inclusion of the random field removes the unphysical artifacts due to sharp boundaries between regions in the simple JF12 parameterization.   The gross features, e.g., the demagnification of sources toward the Galactic center and magnification of those toward the outer, Northern galaxy, are governed by the coherent field.

It is striking that, at lower rigidities, we become blind to sources in a large portion of the extragalactic universe.  
Another important feature of these maps are the high magnification regions towards the northern galactic anti-center, which at high rigidity are isolated in regions of small angular size, but merge and become smoother at lower rigidity.  
It is also noteworthy that at these low rigidities, the random field has a qualitatively important impact.
The presence of the turbulent field acts to smear high magnification regions, but not to such an extent as to mask the effects of the coherent field.

\begin{figure}[t]
\hspace{-0.3in}
\centering
\begin{minipage}[b]{0.48 \textwidth}
\includegraphics[width=1. \textwidth]{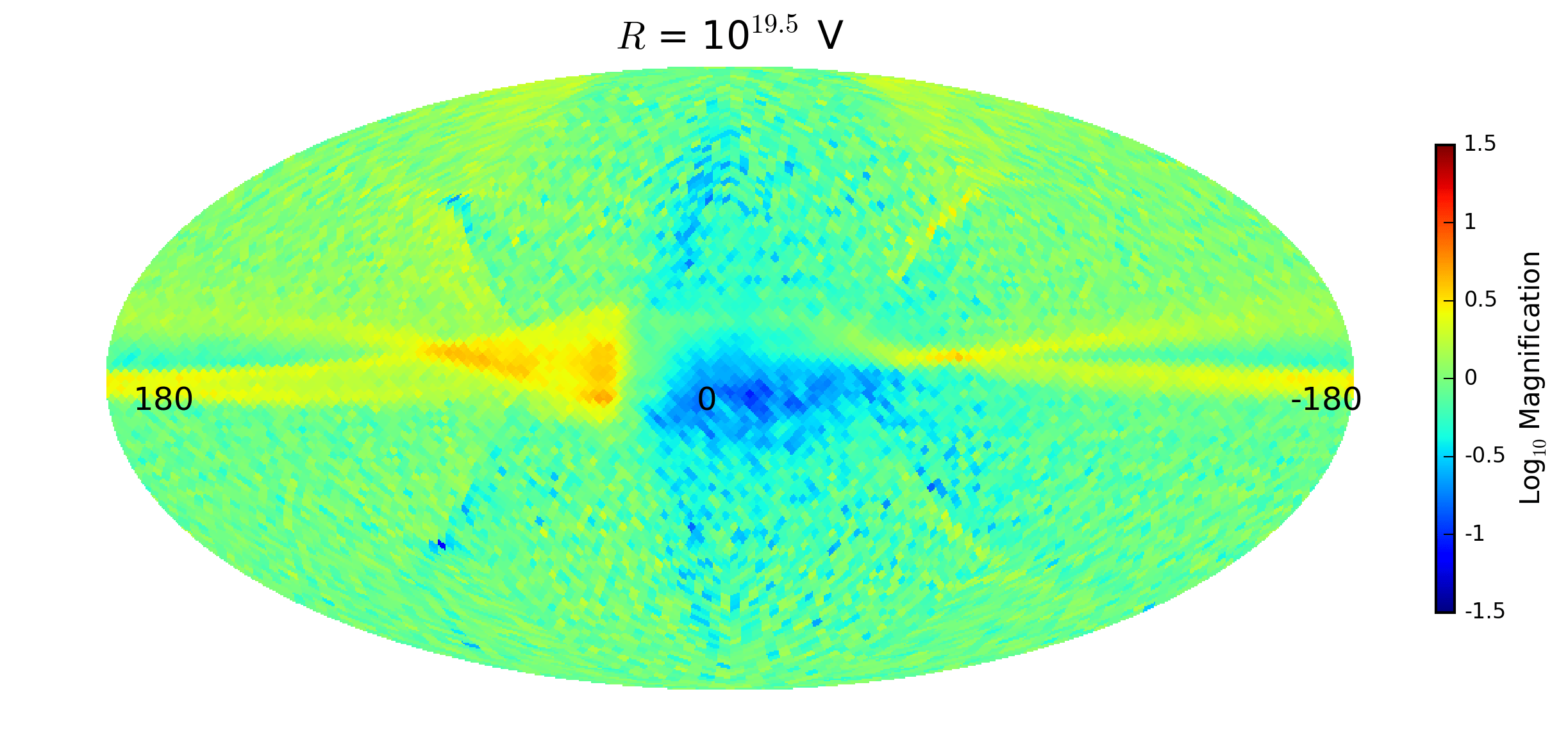}
\end{minipage}
\begin{minipage}[b]{0.48 \textwidth}
\includegraphics[width=1. \textwidth]{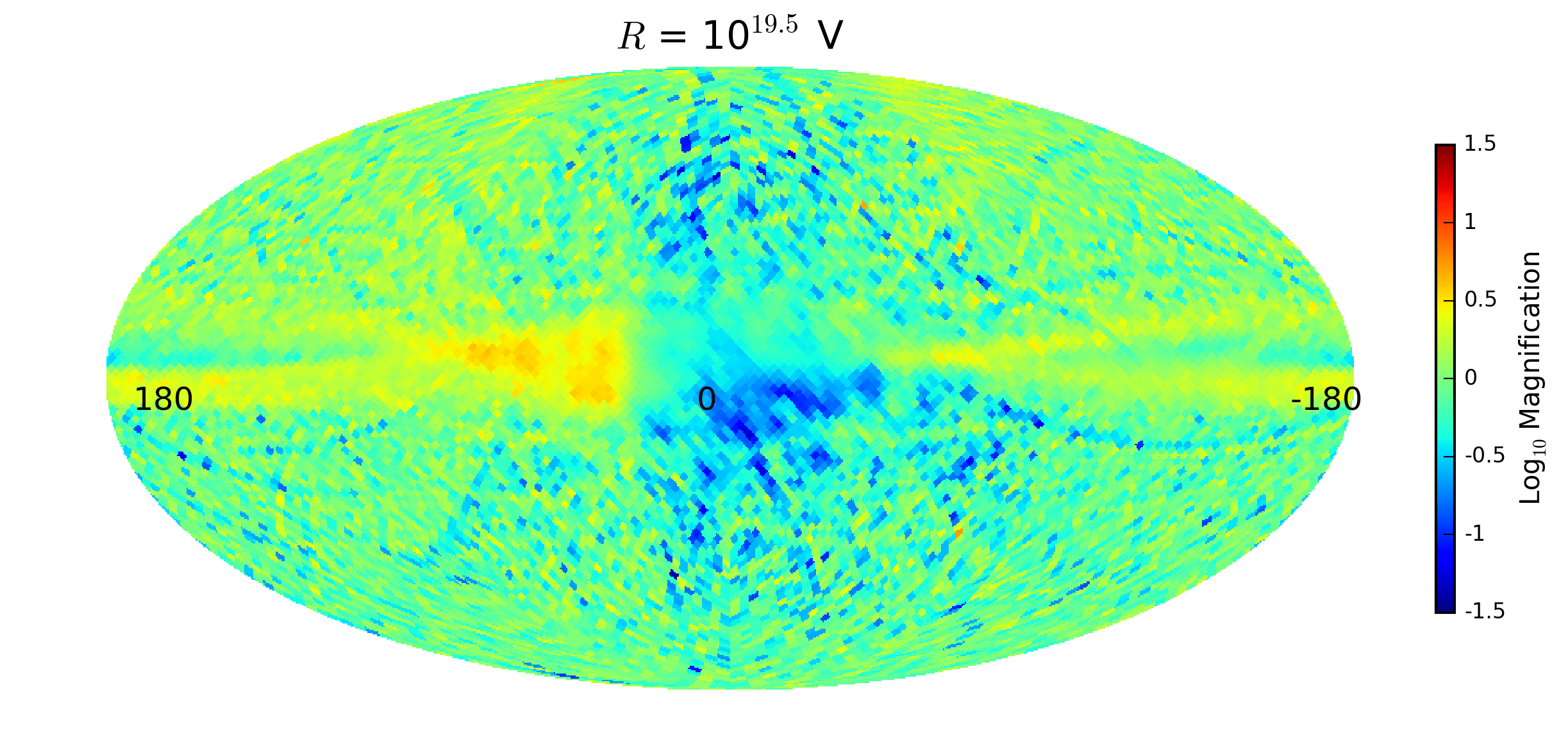}
\end{minipage}
\begin{minipage}[b]{0.48 \textwidth}
\includegraphics[width=1. \textwidth]{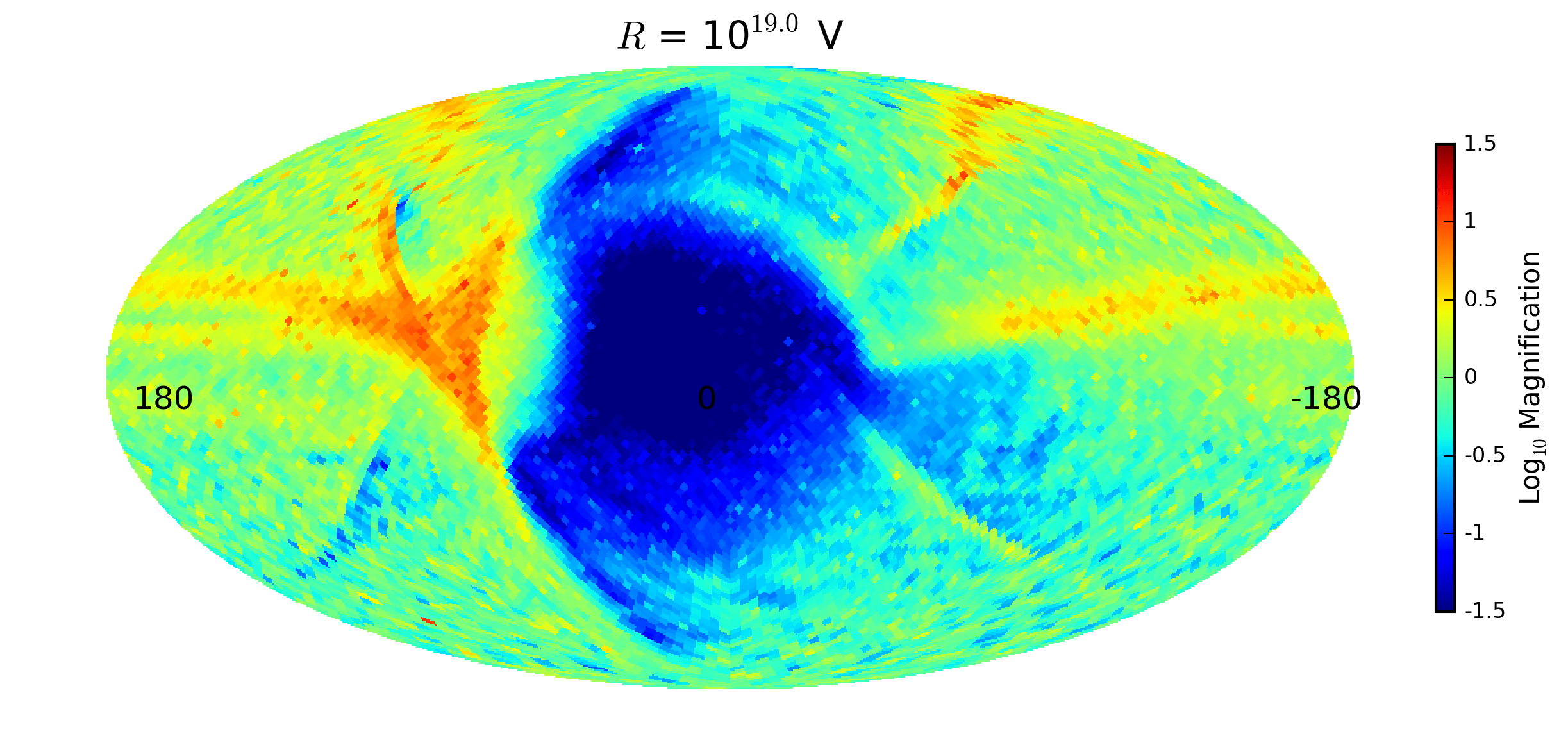}
\end{minipage}
\begin{minipage}[b]{0.48 \textwidth}
\includegraphics[width=1. \textwidth]{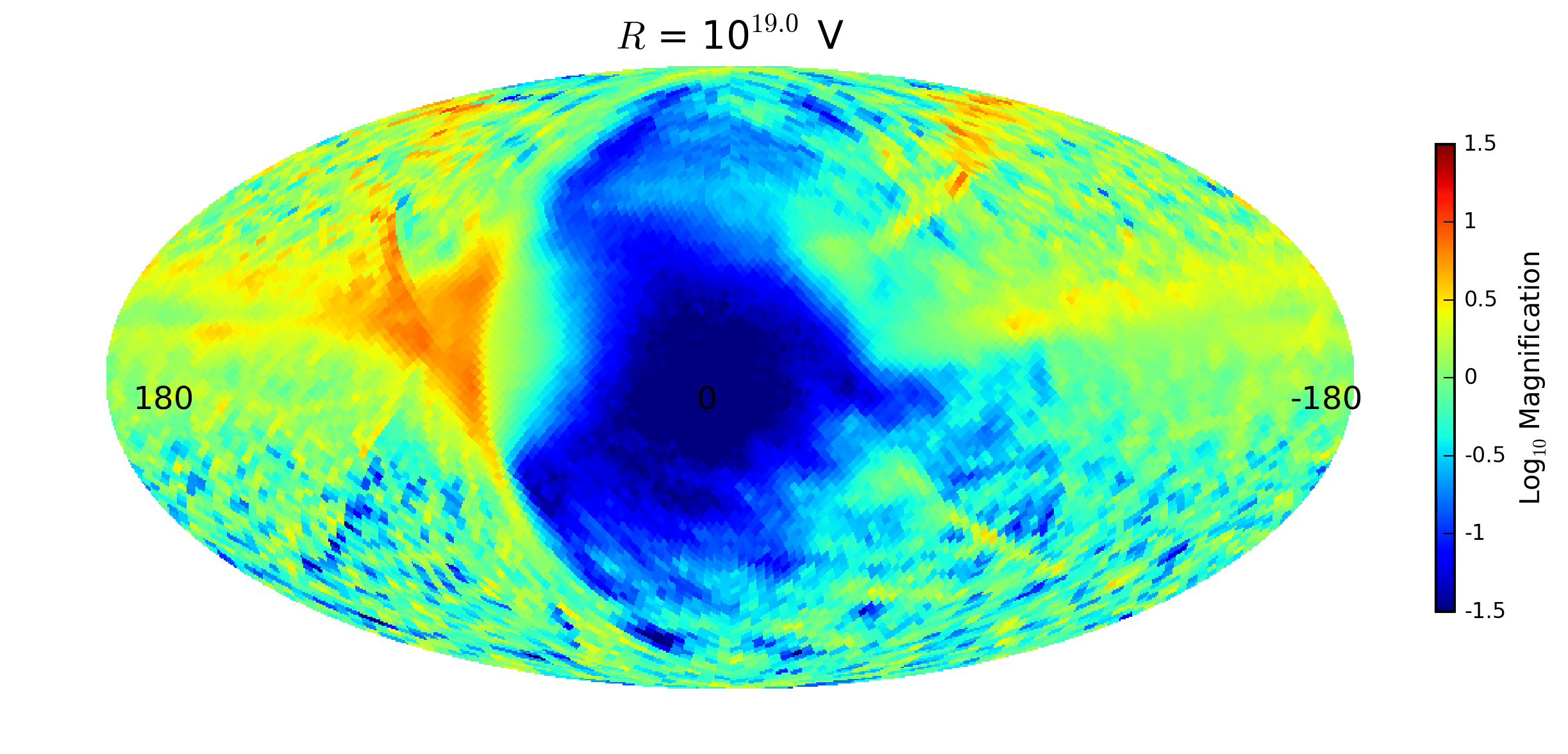}
\end{minipage}
\begin{minipage}[b]{0.48 \textwidth}
\includegraphics[width=1. \textwidth]{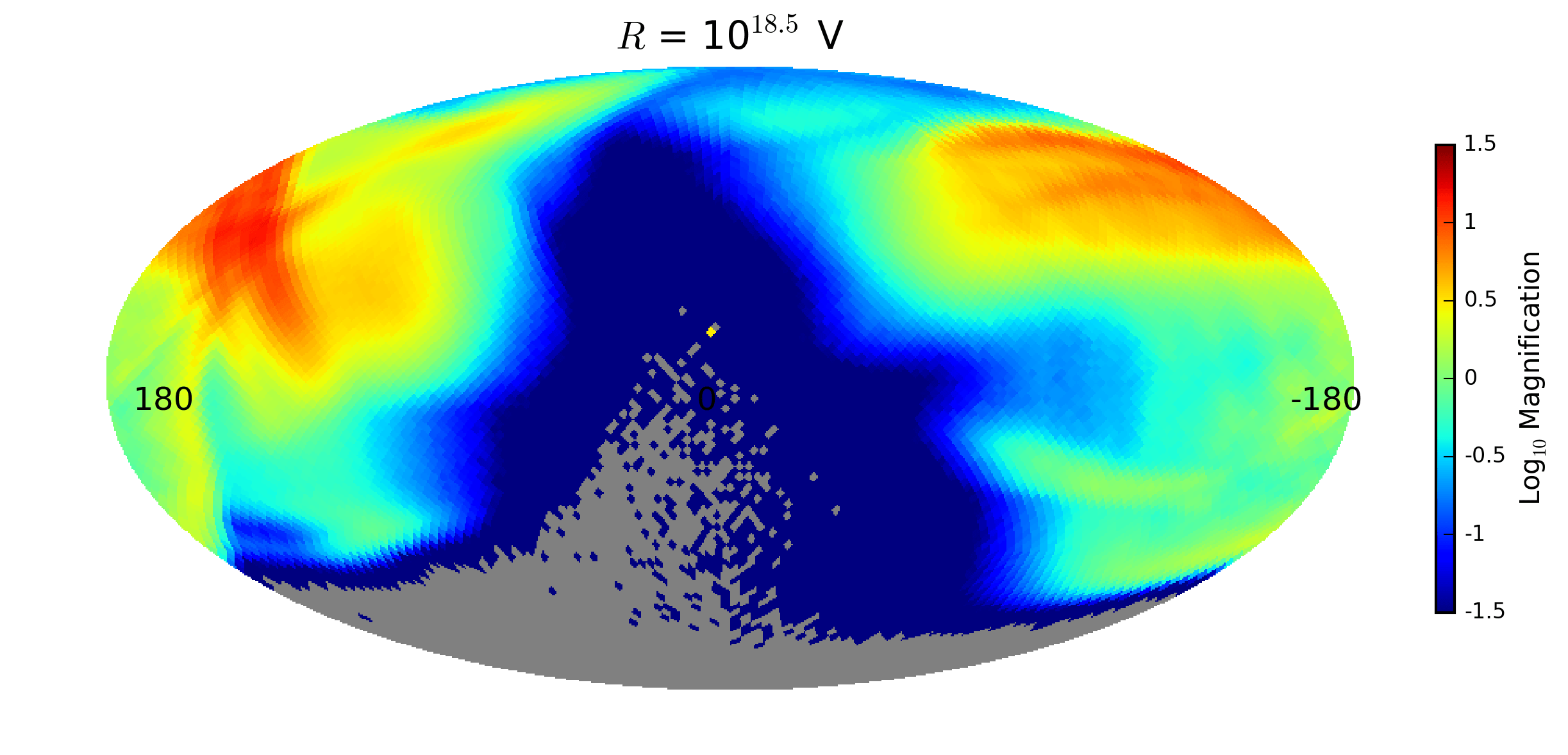}
\end{minipage}
\begin{minipage}[b]{0.48 \textwidth}
\includegraphics[width=1. \textwidth]{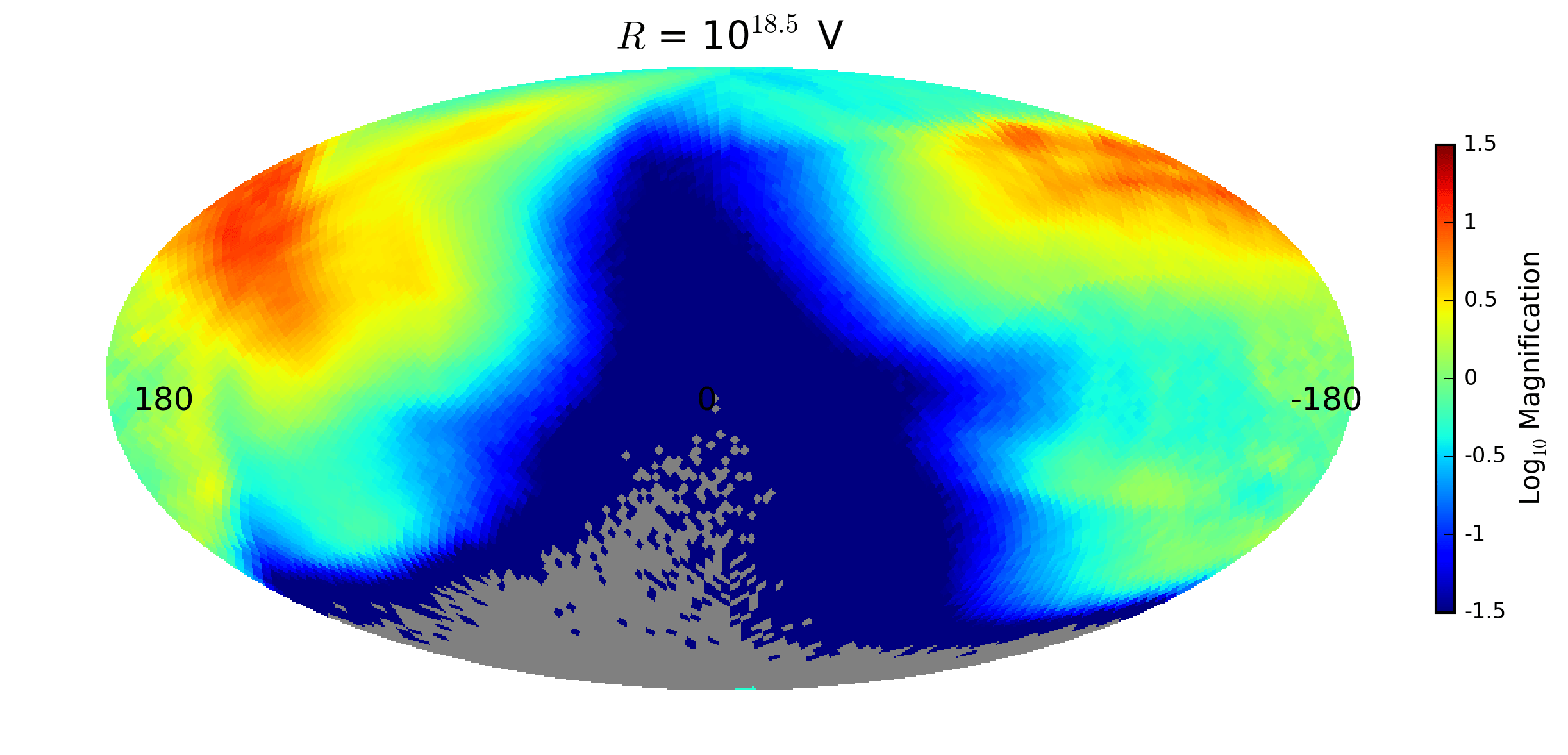}
\end{minipage}
\vspace{-0.1in}
\caption{Magnification maps for log($R$/V)=19.5, 19.0, and 18.5 from top to bottom.
Left column shows KRF10 and right column shows KRF11; both random fields have the same coherence length (30 pc) but are distinct realizations.
The colorbar shows the log10 factor of the magnification of each pixel.
Grey pixels indicate that no events backtracked to that region.}
\label{plt:magmaps_krf10-krf11}
\vspace{-0.1in}
\end{figure}

Figure \ref{plt:magmaps_krf10-krf11} shows analogous magnification plots for the KRF10 (left) and KRF11 (right) field realizations, both having $L_{coh} = 30$ pc.
The magnification effects are similar between the two different field realizations.  Gross features are preserved, which are driven largely by the coherent field and the coherence length of the random field.  Qualitatively, the magnification is insensitive to the particular construction of a realization.  Not surprisingly, the specific loci of high magnification regions at high rigidity are not identical between KRF10 and KRF11, so it will only be possible to predict statistical attributes of the magnification.
\newpage

\section{Arrival direction distributions as a function of source direction }
\label{results}
In order to communicate an impression of the arrival direction distributions of individual sources, we adopt two strategies: \\
$\bullet$ First, we consider 7 source directions towards galaxies discussed as plausible UHECR acceleration sites, as listed in Table \ref{tbl:srcpos}.  For each source direction, we display the rigidity dependence of the arrival directions in a sequence of separate skyplots, for two choices of coherence length.  The first 5 sources are the only radio galaxies satisfying the Hillas criterion for CR acceleration within $z<0.023$, as determined assuming equipartition \cite{vV+RadioGals12}.   M82 was proposed as the origin of the TA hotspot \cite{HeKusenko+TAhotspot16}\footnote{Several other sources in addition to M82 were also proposed in \cite{HeKusenko+TAhotspot16}, but they are in similar locations in the sky.} and M87 is a long-discussed candidate whose direction is not represented by the others. \\
$\bullet$ 
Second, we consider 14 source directions consisting of a uniform grid of 12 source directions defined from a low resolution HEALPix map as listed in Table \ref{tbl:hpxpos}, plus the north and south Galactic poles.  These grid directions yield lines-of-sight projected through a variety of Galactic magnetic environments.  
The skyplots for the grid directions, and Tables of general arrival distribution quantities for all the sources, are located in Appendices \ref{appdx:HEALPixSkyplots} and \ref{appdx:HEALPixTables}.

For 3 rigidities, we investigate how the arrival direction distributions change with the specific realization of the random field, for a 30 pc coherence length. These plots are shown in Appendix \ref{appdx:realizations}.

\begin{table}
\small
\begin{center}
\begin{tabular}[!t]{ c|ccc }
Name	&  Redshift		& L	& B 	\\
		&	($z$)		& [$^{\circ}$]	& [$^{\circ}$] \\
\hline
Centaurus A	& 0.0018	& 309.52	& 19.42 \\
UGC 1841		& 0.021	& 140.25	& -16.77 \\
NGC 1128		& 0.023	& 170.26	& -44.93 \\
NGC 4782		& 0.015	& 304.14	& 50.29 \\
CGCG 114-025	 & 0.017	& 46.97	& 10.54 \\
M82			& 0.00067	 &141.41	& 40.57\\
M87			& 0.00428	& 283.78   & 74.49    
\end{tabular}
\caption{Seven candidate UHECR accelerators, and their distances and Galactic coordinates.}
\label{tbl:srcpos}
\end{center}
\end{table}

\begin{table}[t]
\footnotesize
\begin{center}
  \begin{tabular}[!t]{ c|cc }
    Name	& Gal. long. & Gal. lat. \\
    		& [$^{\circ}$]	& [$^{\circ}$] \\
\hline
    HPX 1		& $45^{\circ}$	& $41.8103^{\circ}$ \\
    HPX 2		& $135^{\circ}$	& $41.8103^{\circ}$ \\
    HPX 3		& $-135^{\circ}$	& $41.8103^{\circ}$ \\
    HPX 4		& $-45^{\circ}$	& $41.8103^{\circ}$ \\
    HPX 5		& $0^{\circ}$	& $0^{\circ}$ \\
    HPX 6		& $90^{\circ}$	& $0^{\circ}$ \\
    HPX 7		& $180^{\circ}$	& $0^{\circ}$ \\
    HPX 8		& $-90^{\circ}$	& $0^{\circ}$ \\
    HPX 9		& $45^{\circ}$	& $-41.8103^{\circ}$ \\
    HPX 10		& $135^{\circ}$	& $-41.8103^{\circ}$ \\
    HPX 11		& $-135^{\circ}$	& $-41.8103^{\circ}$ \\
    HPX 12		& $-45^{\circ}$	& $-41.8103^{\circ}$ \\
    North Pole   	& $0^{\circ}$	& $90^{\circ}$ \\
    South Pole  	& $0^{\circ}$	& $-90^{\circ}$ \\
  \end{tabular}
\caption{Locations of representative source directions sampled from a resolution 0 HEALPix skymap.
Column 1 lists the source name and columns 2 and 3 give the sky position in Galactic coordinates.
The north and south Galactic poles are also included for completeness.}
\label{tbl:hpxpos}
\end{center}
\end{table}

These skyplots are made as follows.  For each all-sky backtracking simulation, we identify particles that backtrack to regions within a specified solid angle around a chosen source direction.
In order to have a large enough sample of events, we use a region of $1^{\circ}$ radius; this would contain about 4000 events in the absence of deflections, given our resolution level.
The choice of $1^\circ$ is also motivated on physical grounds:  for Cen A, a $1^{\circ}$ angular window encompasses the central region that could contain UHECR acceleration and emission sites, and most other sources are located at larger distances for which the extragalactic arrival direction smearing is also plausibly of degree scale, given the rigidities considered.  Each arriving CR is shown as a semi-transparent disk.

\begin{figure}[H]
\hspace{-0.3in}
\centering
\begin{minipage}[b]{0.48 \textwidth}
\includegraphics[width=1. \textwidth]{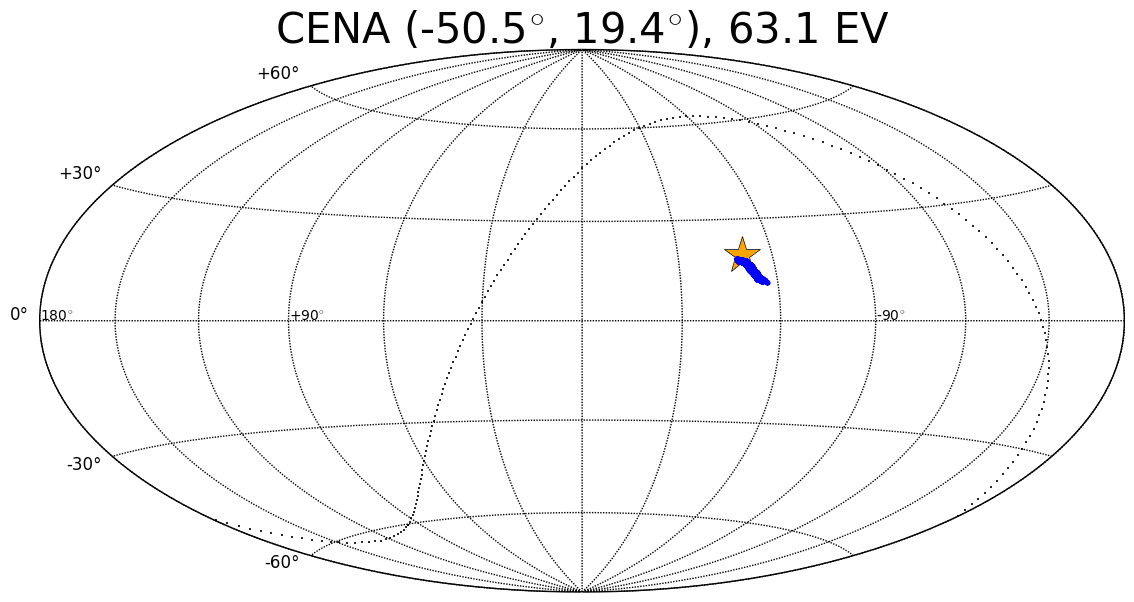}
\end{minipage}
\begin{minipage}[b]{0.48 \textwidth}
\includegraphics[width=1. \textwidth]{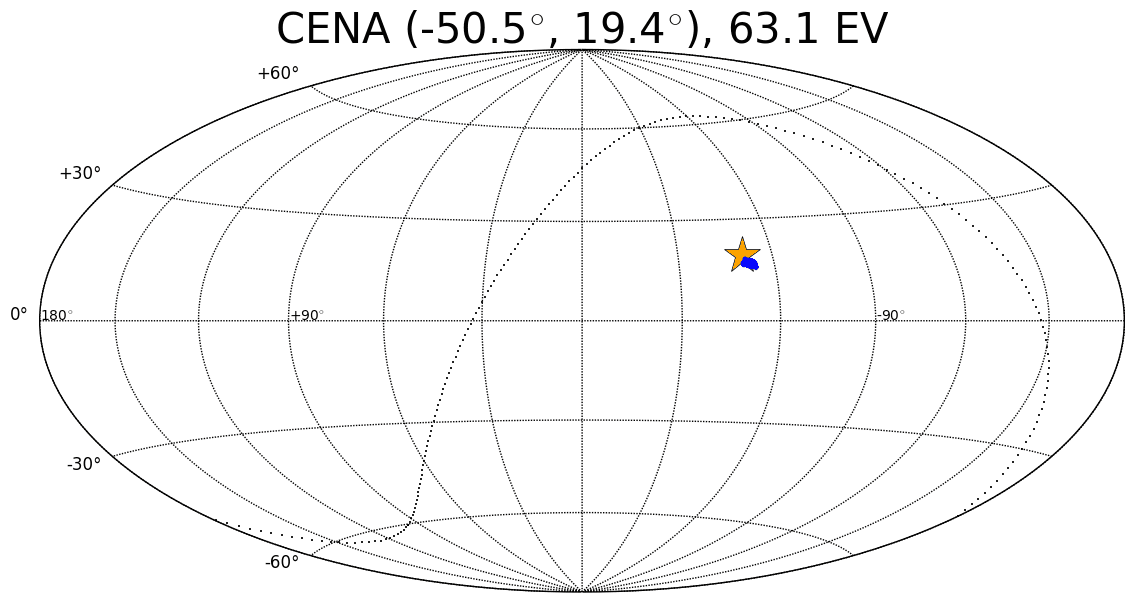}
\end{minipage}
\begin{minipage}[b]{0.48 \textwidth}
\includegraphics[width=1. \textwidth]{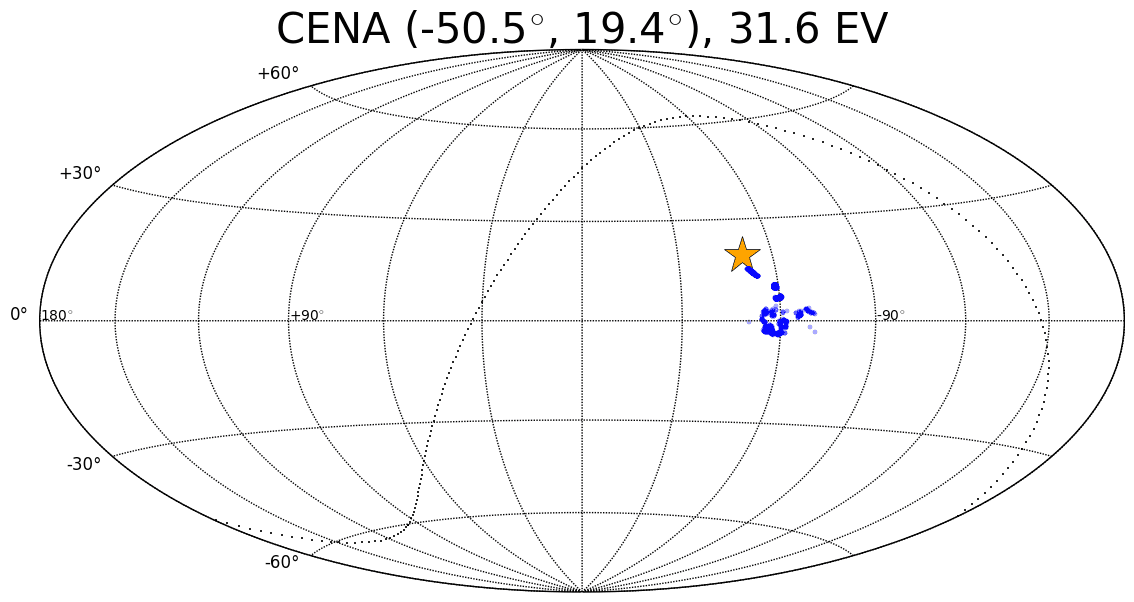}
\end{minipage}
\begin{minipage}[b]{0.48 \textwidth}
\includegraphics[width=1. \textwidth]{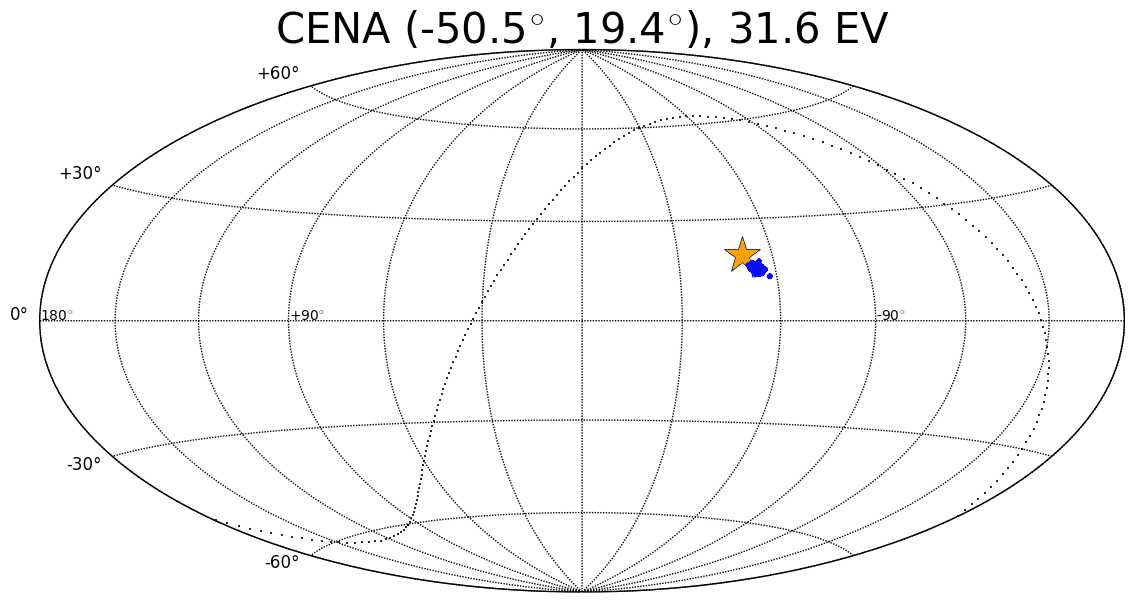}
\end{minipage}
\begin{minipage}[b]{0.48 \textwidth}
\includegraphics[width=1. \textwidth]{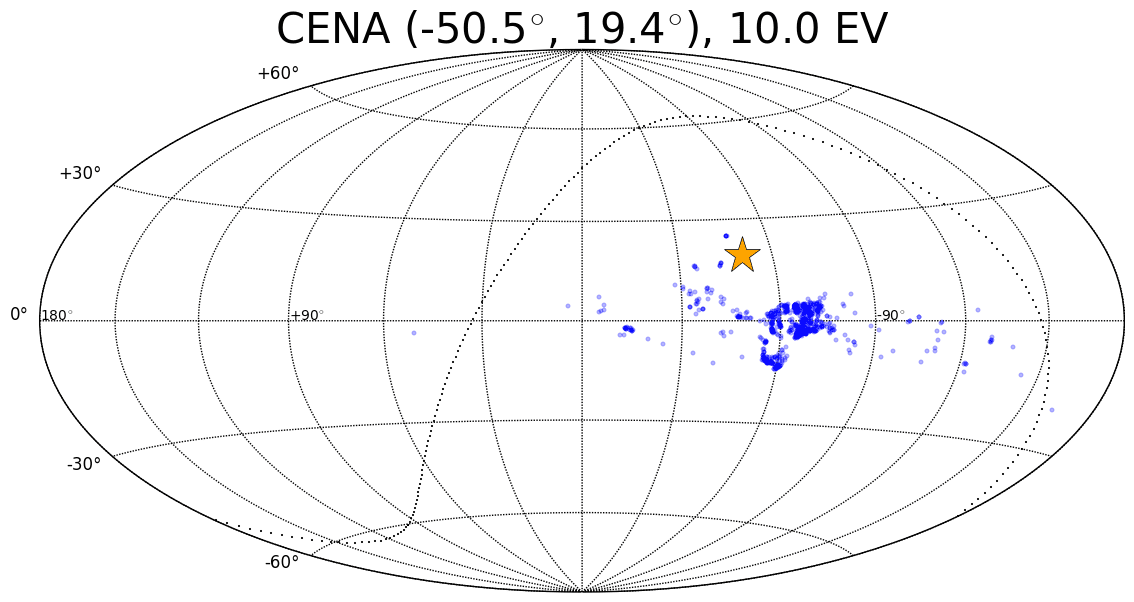}
\end{minipage}
\begin{minipage}[b]{0.48 \textwidth}
\includegraphics[width=1. \textwidth]{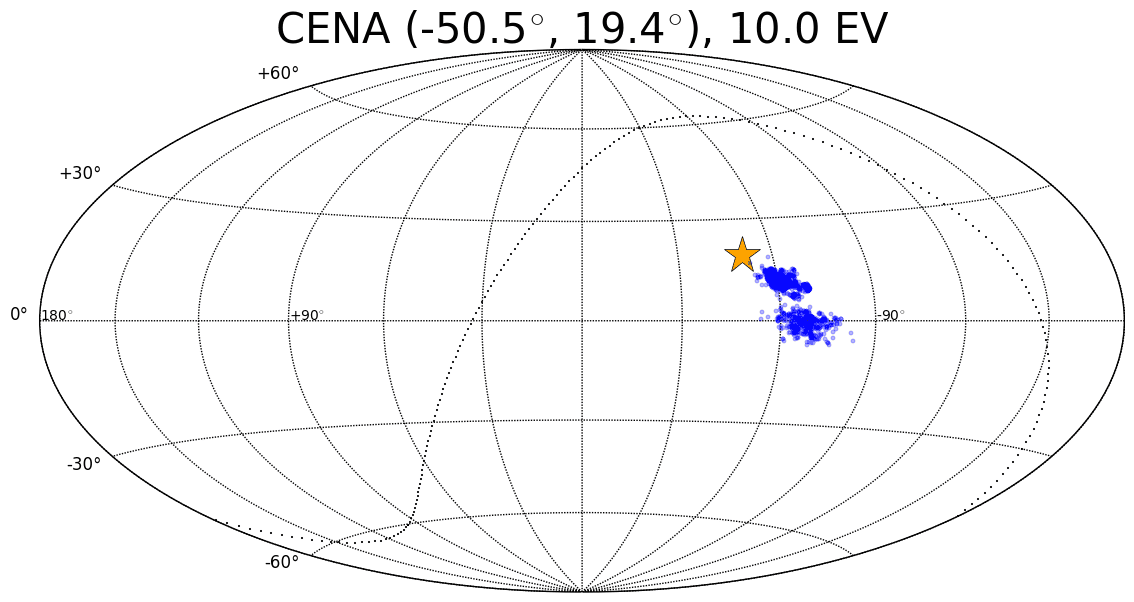}
\end{minipage}
\begin{minipage}[b]{0.48 \textwidth}
\includegraphics[width=1. \textwidth]{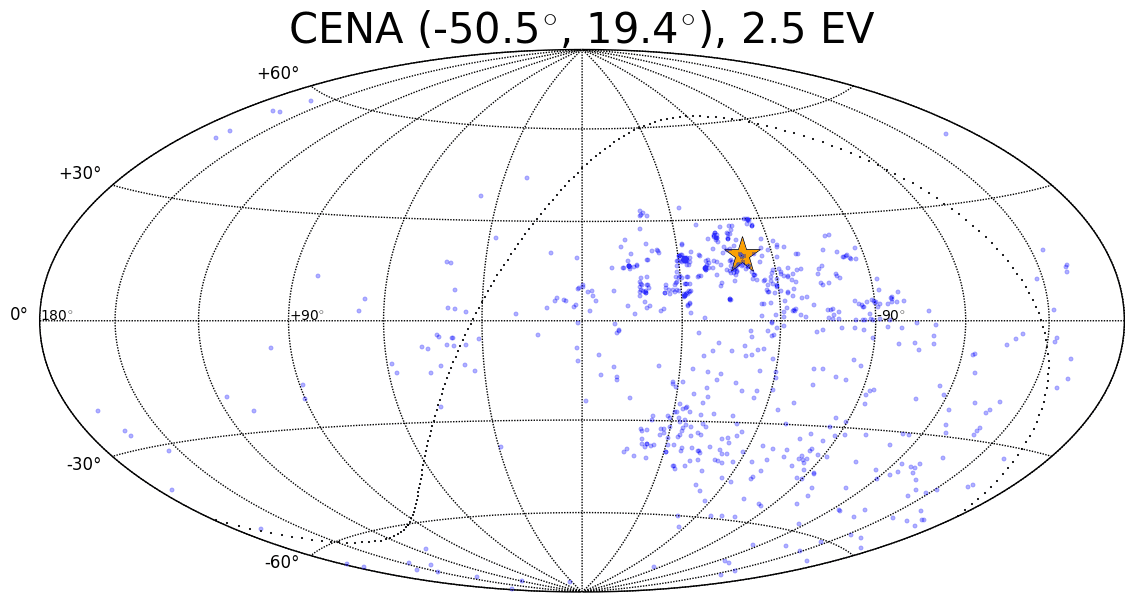}
\end{minipage}
\begin{minipage}[b]{0.48 \textwidth}
\includegraphics[width=1. \textwidth]{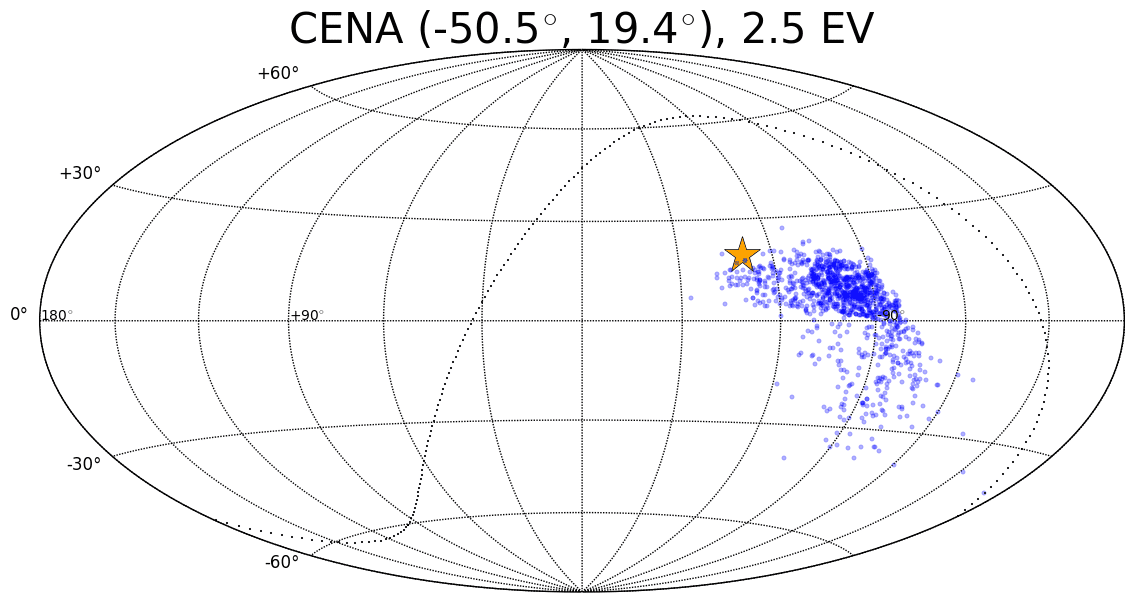}
\end{minipage}
\begin{minipage}[b]{0.48 \textwidth}
\includegraphics[width=1. \textwidth]{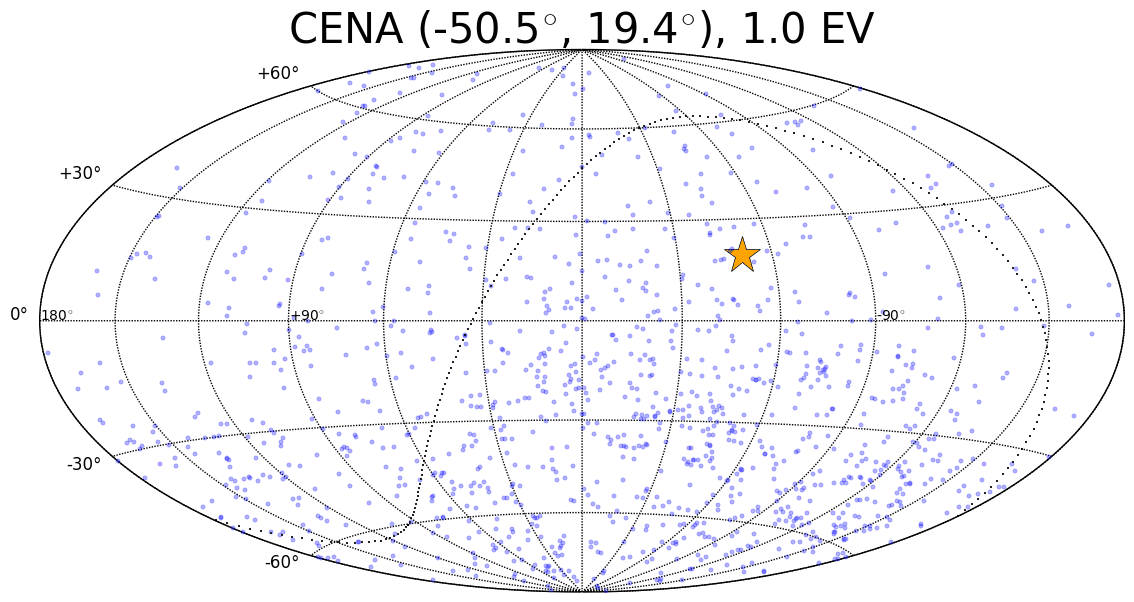}
\end{minipage}
\begin{minipage}[b]{0.48 \textwidth}
\includegraphics[width=1. \textwidth]{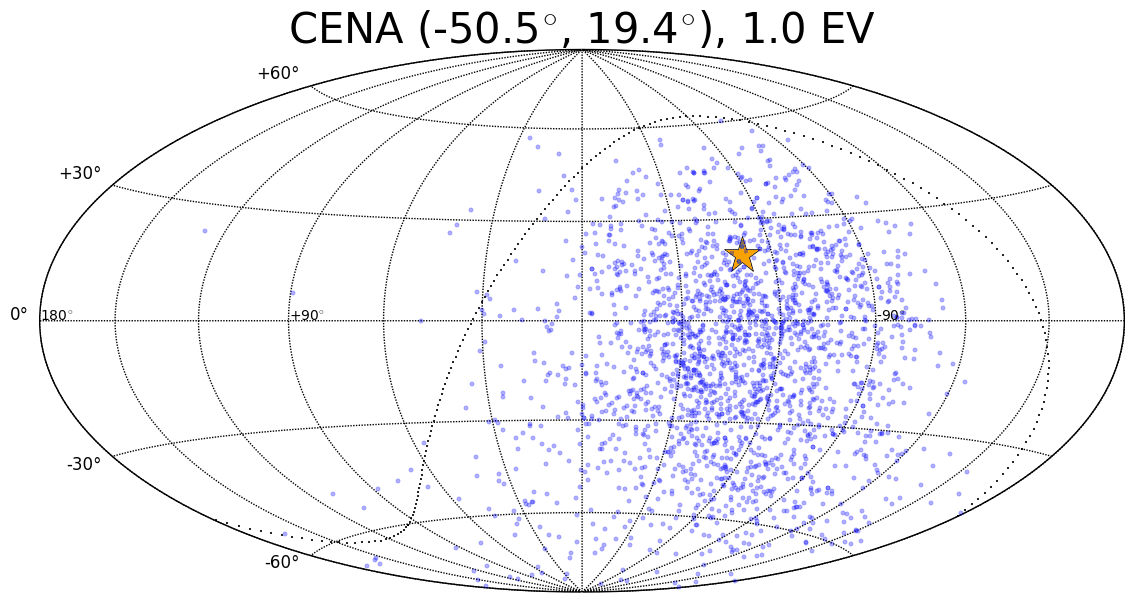}
\end{minipage}
\vspace{-0.3in}
\caption{Arrival direction distributions for selected rigidities (log($R$/V) = 18.0 - 19.8) for cosmic rays from Cen A (marked with green star) located at ($\ell$, $b$) = ($-50.5^{\circ}$, $19.4^{\circ}$).  The left column shows the KRF6 ($L_{coh}$=100 pc) realization and the right column shows the KRF10 ($L_{coh}$=30 pc) realization.   The sky map is in Galactic coordinates and the dotted line indicates decl. $\delta=0^{\circ}$.} 
\label{plt:cena}
\vspace{-0.1in}
\end{figure}

\begin{figure}[htb]
\hspace{-0.3in}
\centering
\begin{minipage}[b]{0.48 \textwidth}
\includegraphics[width=1. \textwidth]{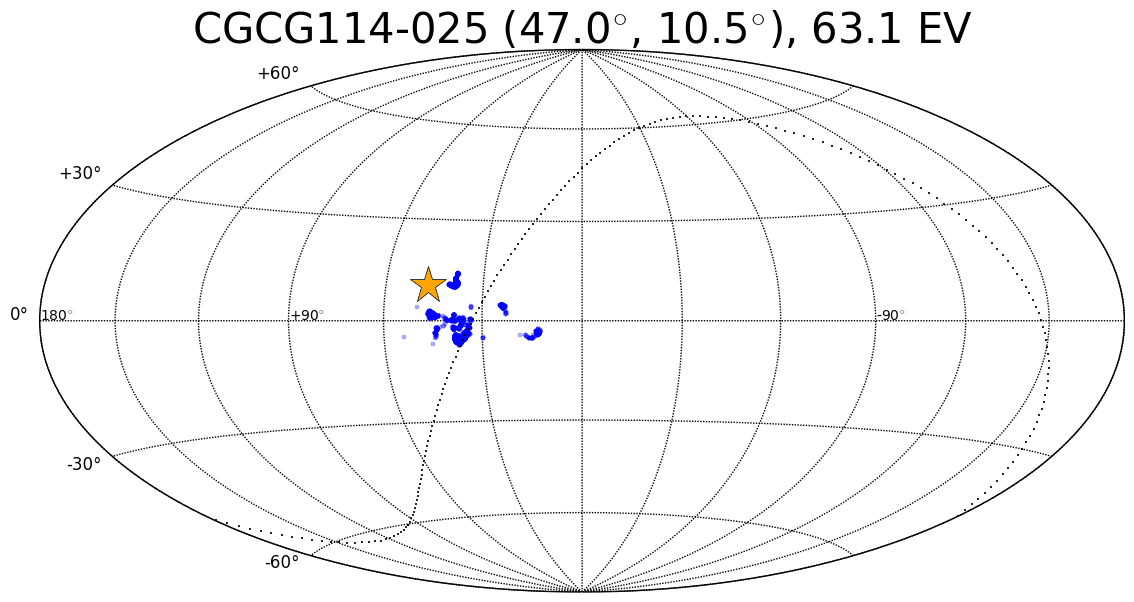}
\end{minipage}
\begin{minipage}[b]{0.48 \textwidth}
\includegraphics[width=1. \textwidth]{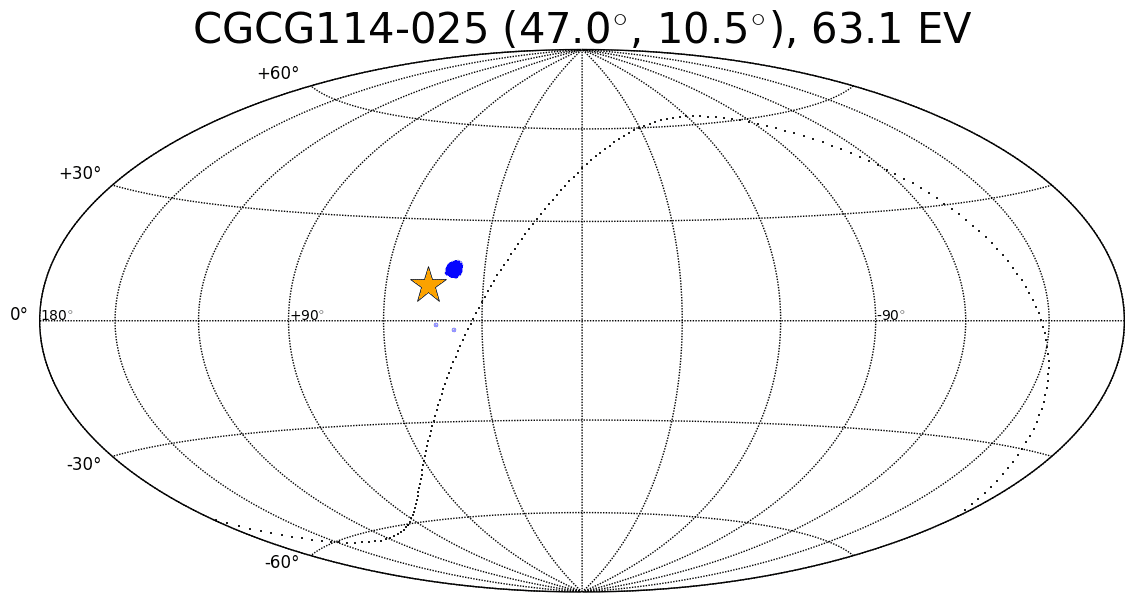}
\end{minipage}
\begin{minipage}[b]{0.48 \textwidth}
\includegraphics[width=1. \textwidth]{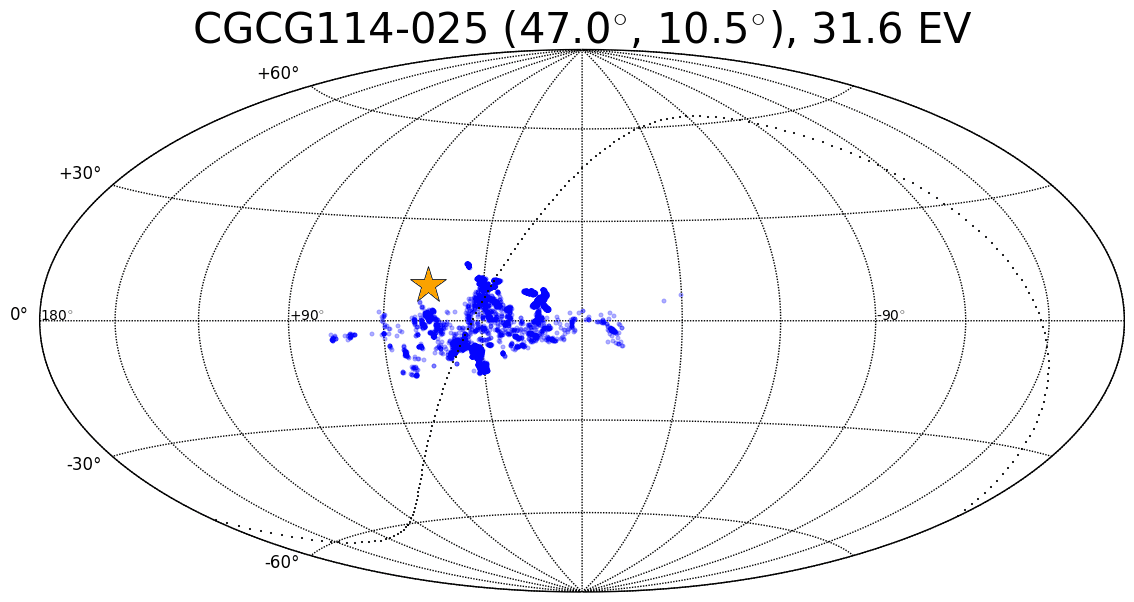}
\end{minipage}
\begin{minipage}[b]{0.48 \textwidth}
\includegraphics[width=1. \textwidth]{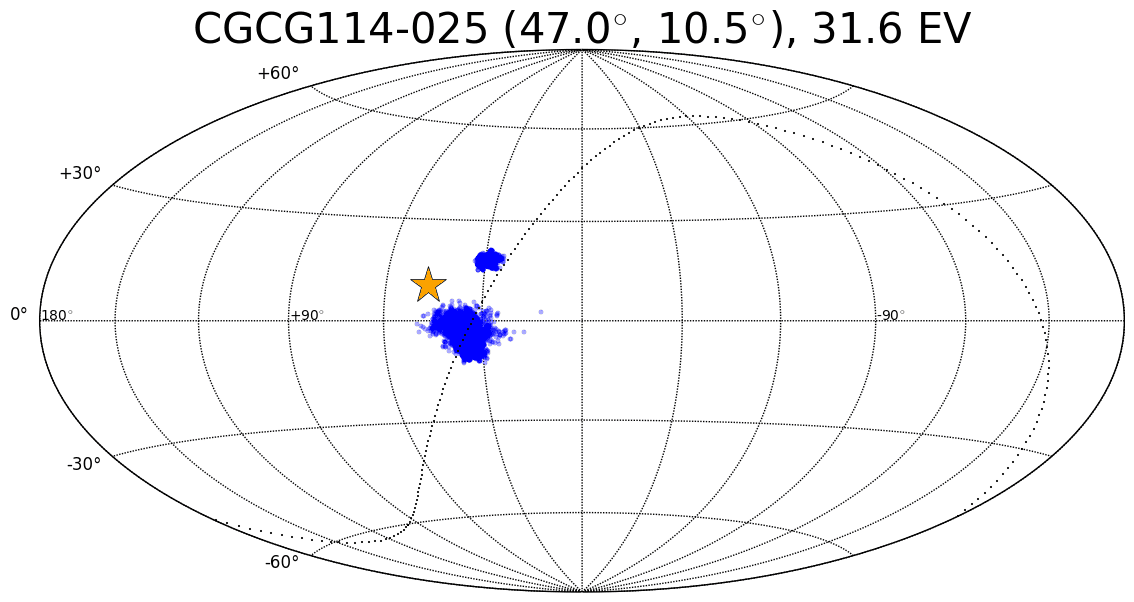}
\end{minipage}
\begin{minipage}[b]{0.48 \textwidth}
\includegraphics[width=1. \textwidth]{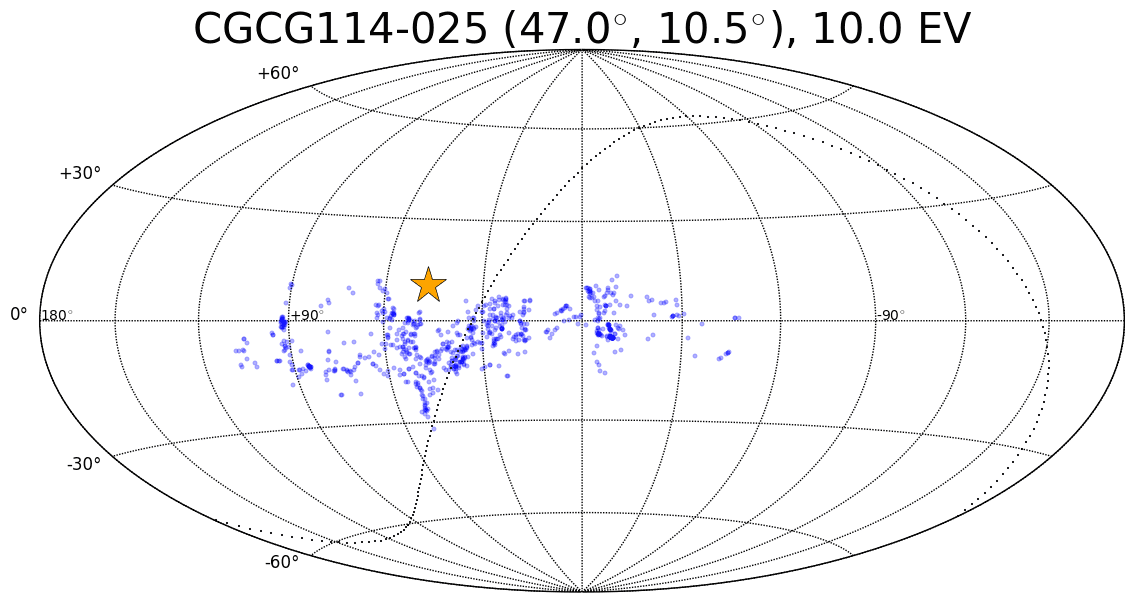}
\end{minipage}
\begin{minipage}[b]{0.48 \textwidth}
\includegraphics[width=1. \textwidth]{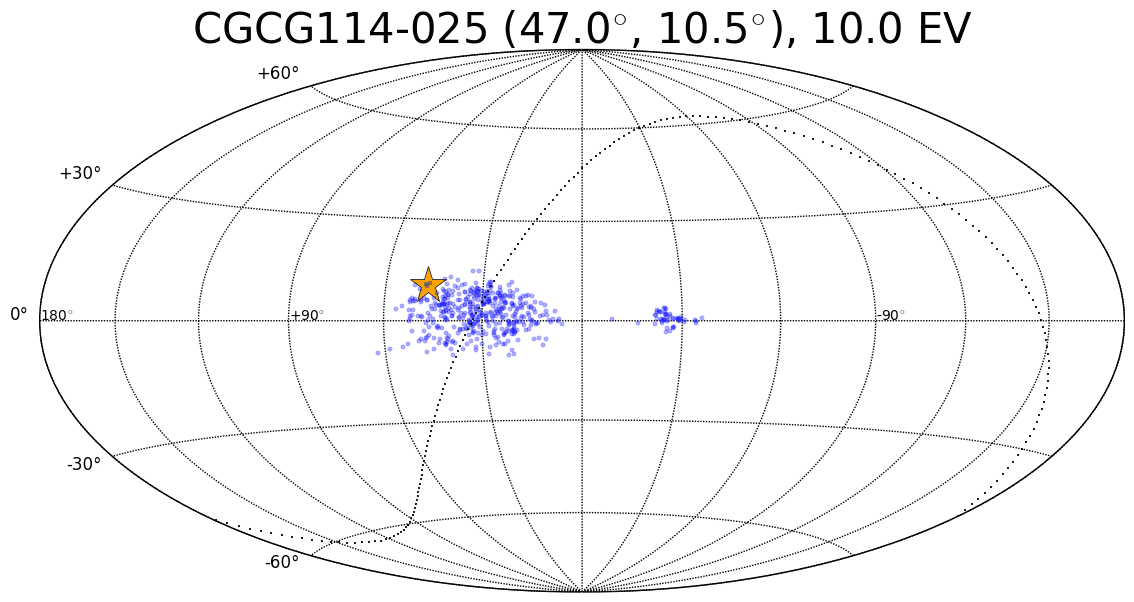}
\end{minipage}
\begin{minipage}[b]{0.48 \textwidth}
\includegraphics[width=1. \textwidth]{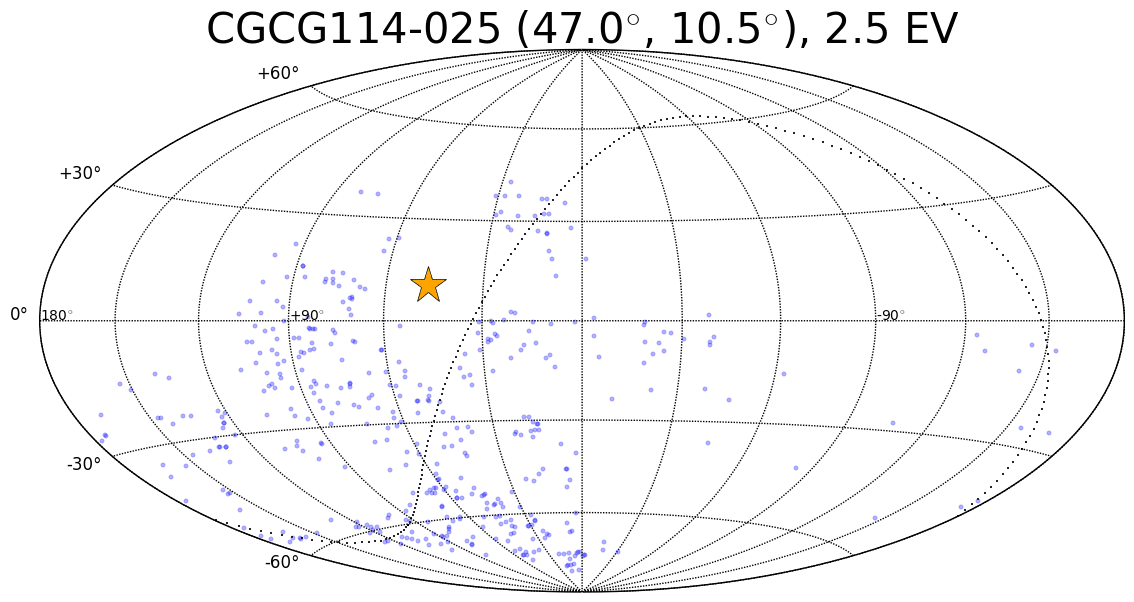}
\end{minipage}
\begin{minipage}[b]{0.48 \textwidth}
\includegraphics[width=1. \textwidth]{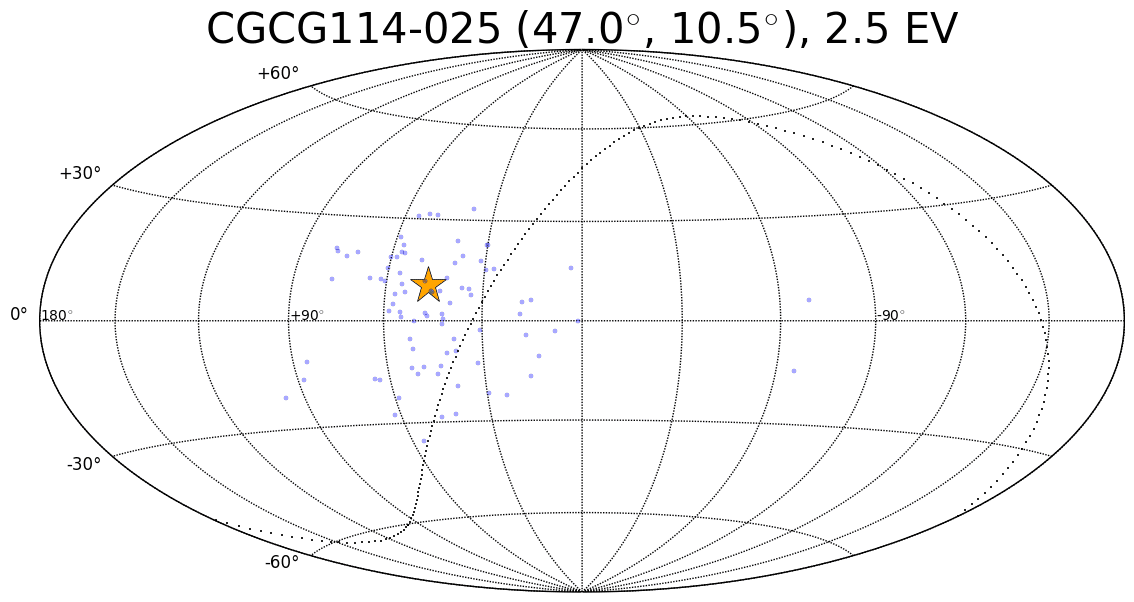}
\end{minipage}
\begin{minipage}[b]{0.48 \textwidth}
\includegraphics[width=1. \textwidth]{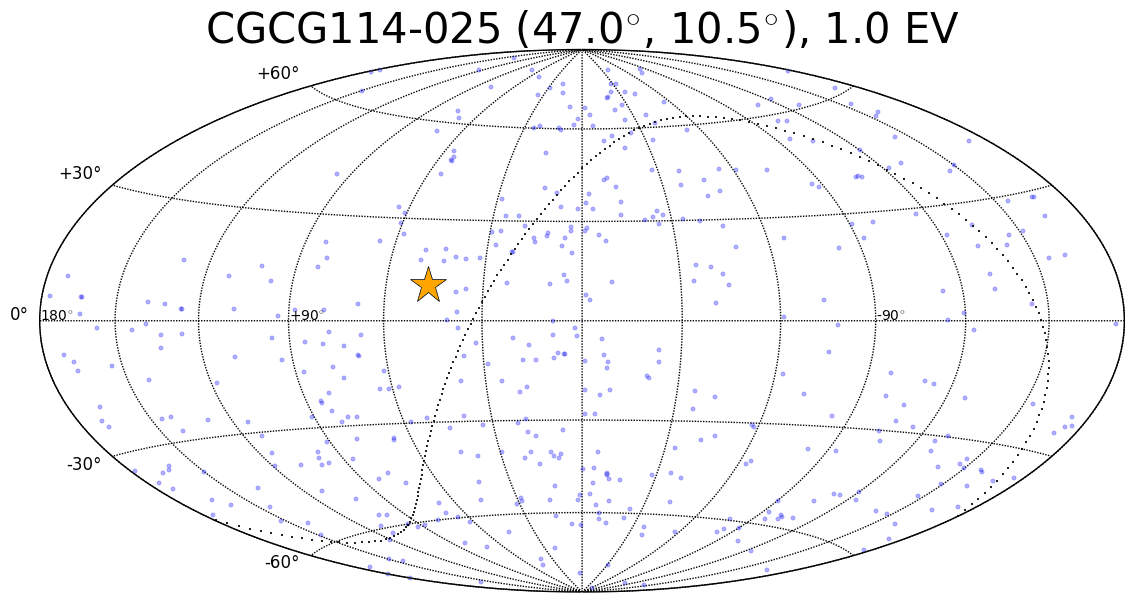}
\end{minipage}
\begin{minipage}[b]{0.48 \textwidth}
\includegraphics[width=1. \textwidth]{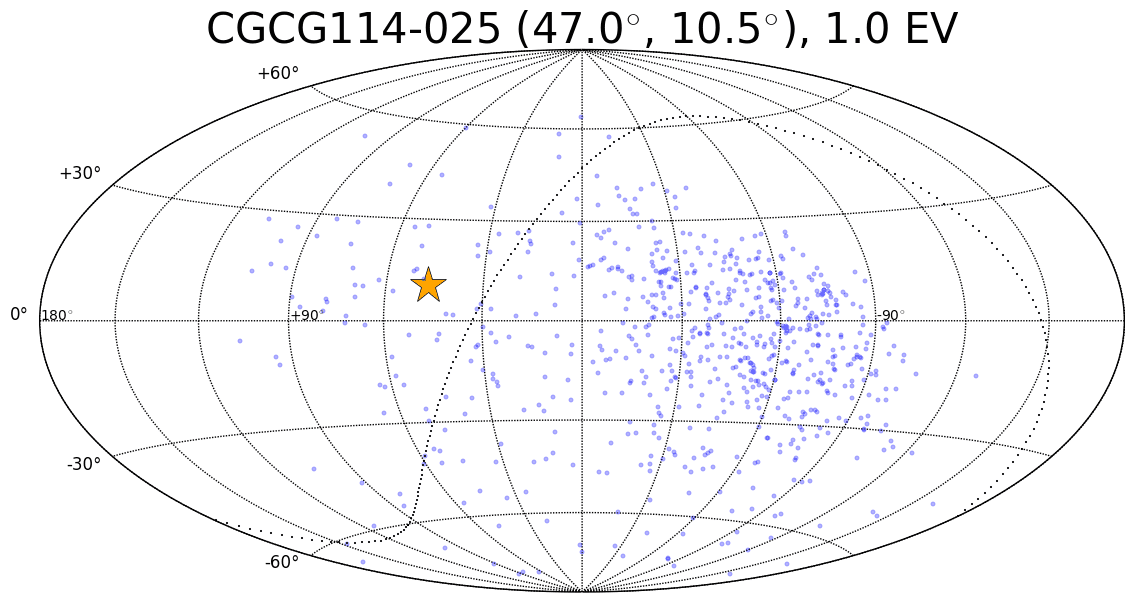}
\end{minipage}
\vspace{-0.3in}
\caption{As in Fig. \ref{plt:cena} but for CGCG114-025, located at ($\ell$, $b$) = ($46.97^{\circ}$, $10.54^{\circ}$).} 
\label{plt:cgcg114-025}
\vspace{-0.1in}
\end{figure}

\begin{figure}[htb]
\hspace{-0.3in}
\centering
\begin{minipage}[b]{0.48 \textwidth}
\includegraphics[width=1. \textwidth]{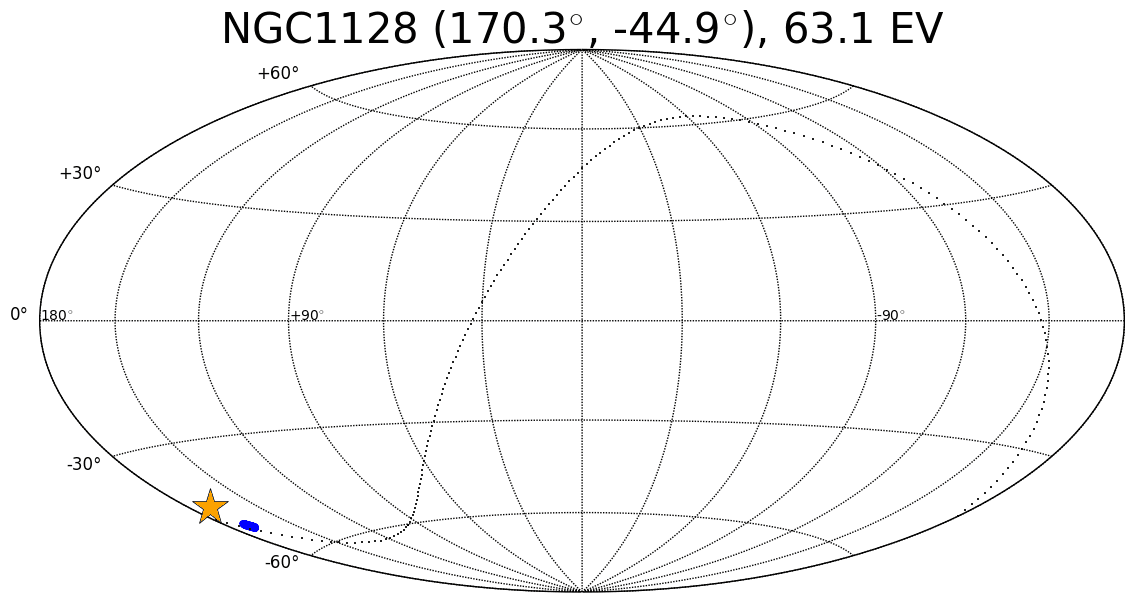}
\end{minipage}
\begin{minipage}[b]{0.48 \textwidth}
\includegraphics[width=1. \textwidth]{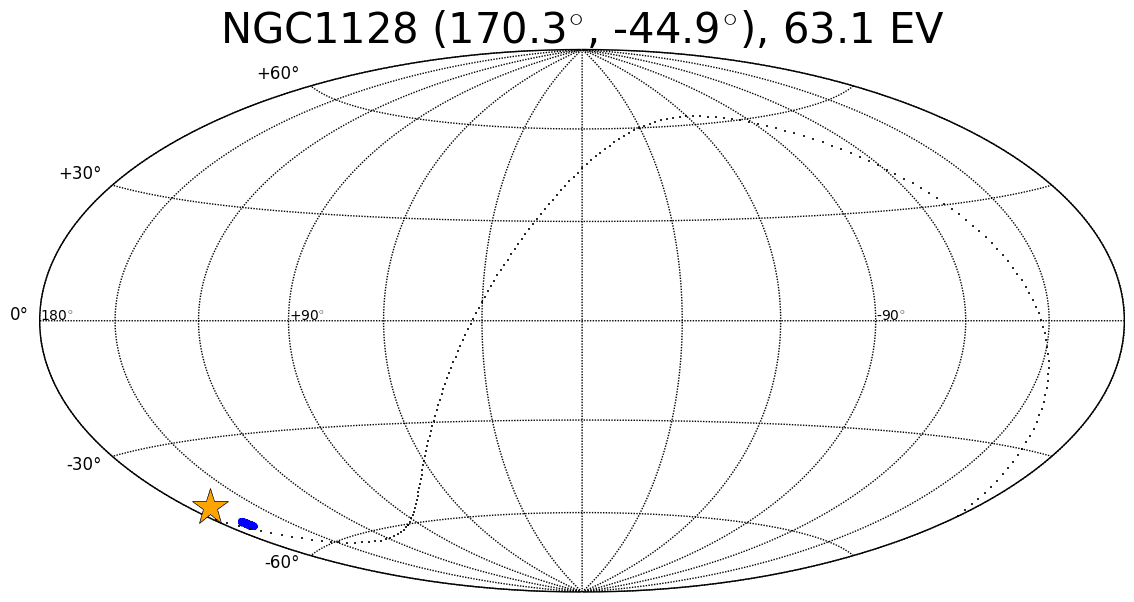}
\end{minipage}
\begin{minipage}[b]{0.48 \textwidth}
\includegraphics[width=1. \textwidth]{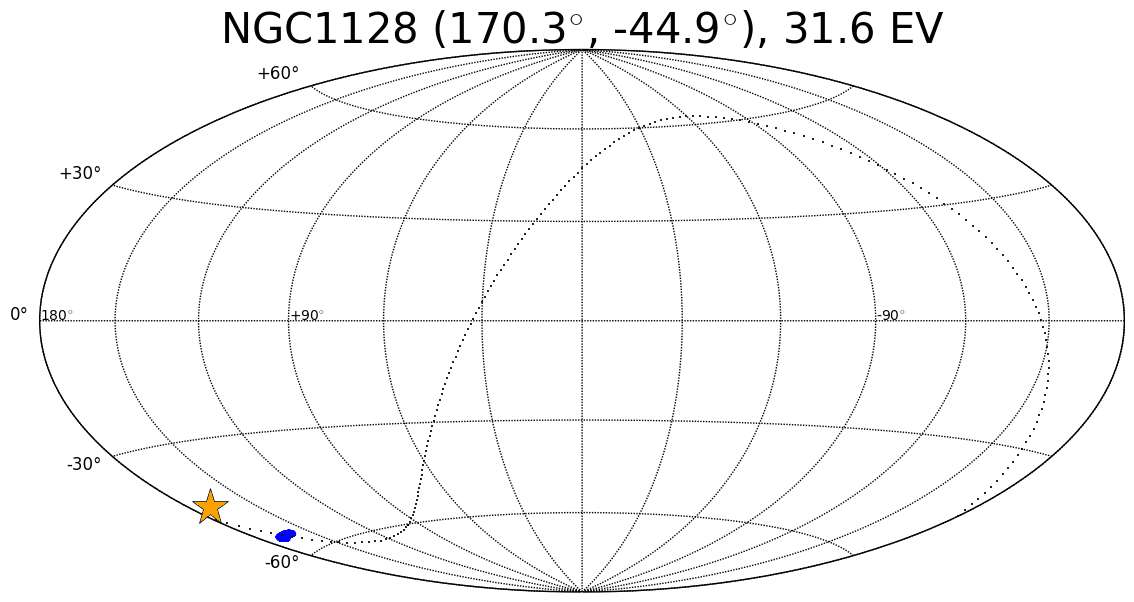}
\end{minipage}
\begin{minipage}[b]{0.48 \textwidth}
\includegraphics[width=1. \textwidth]{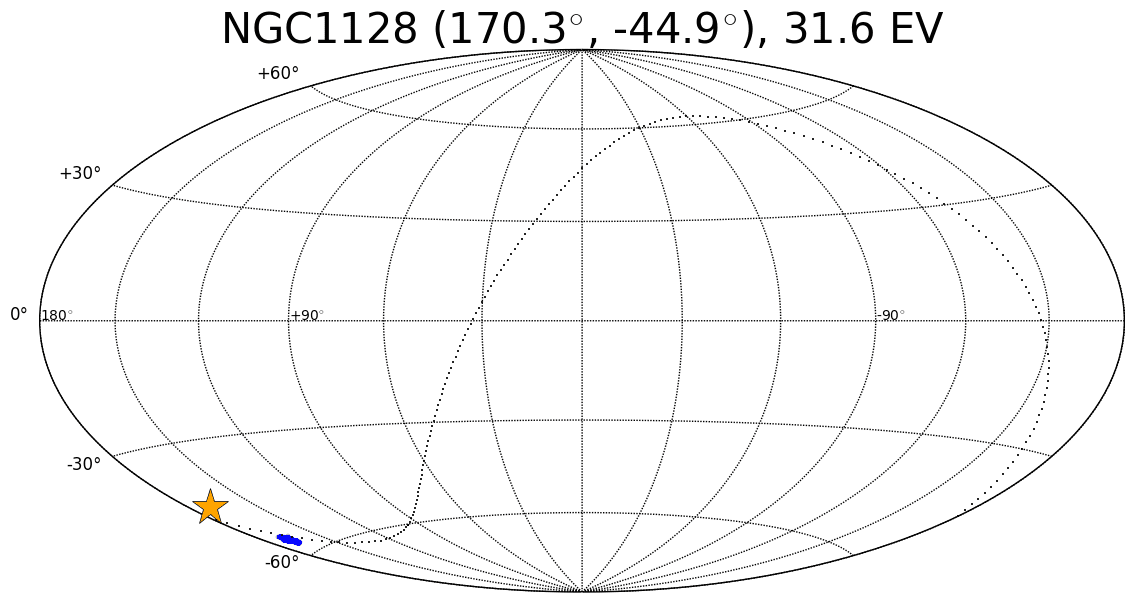}
\end{minipage}
\begin{minipage}[b]{0.48 \textwidth}
\includegraphics[width=1. \textwidth]{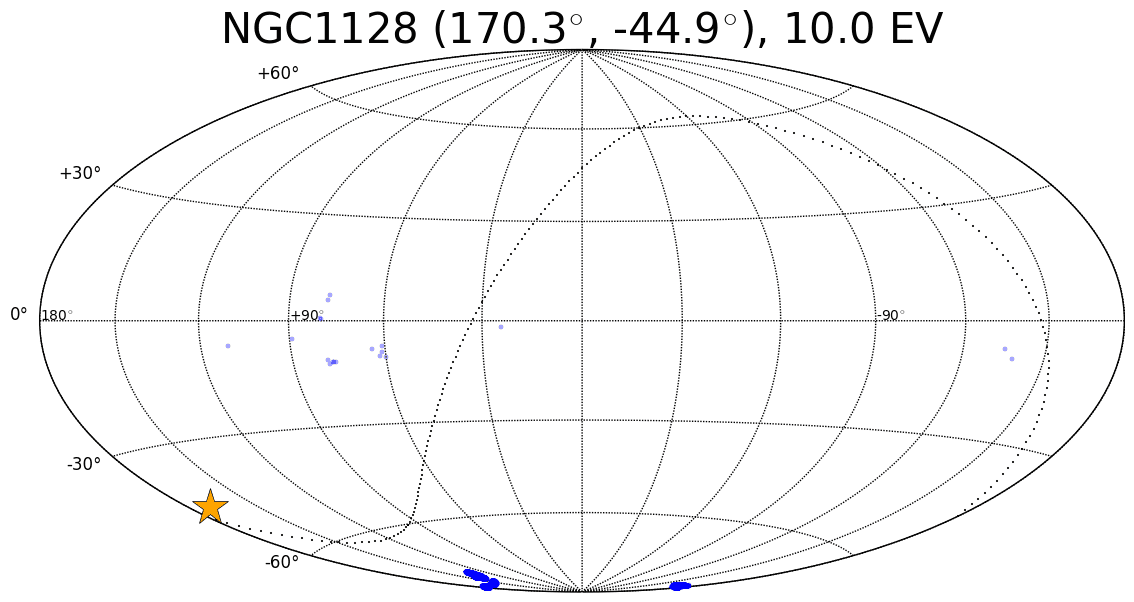}
\end{minipage}
\begin{minipage}[b]{0.48 \textwidth}
\includegraphics[width=1. \textwidth]{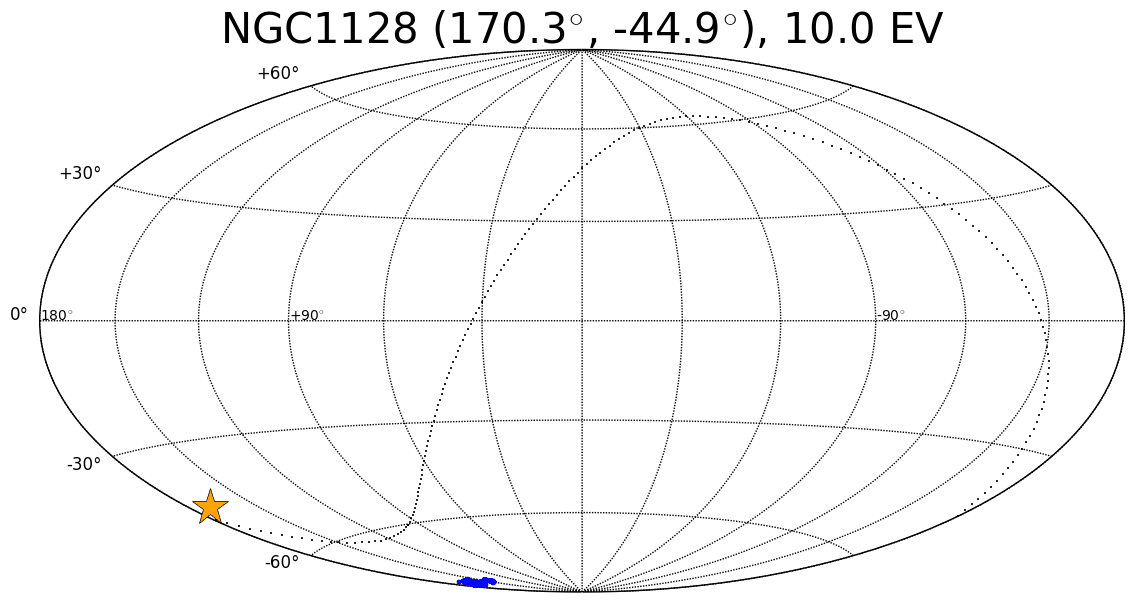}
\end{minipage}
\begin{minipage}[b]{0.48 \textwidth}
\includegraphics[width=1. \textwidth]{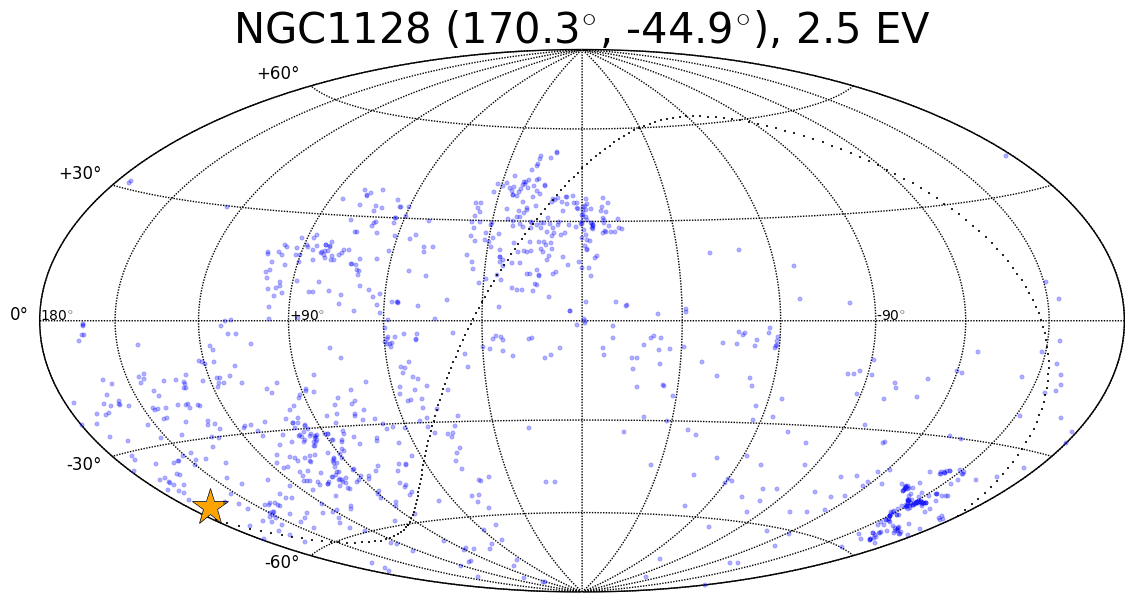}
\end{minipage}
\begin{minipage}[b]{0.48 \textwidth}
\includegraphics[width=1. \textwidth]{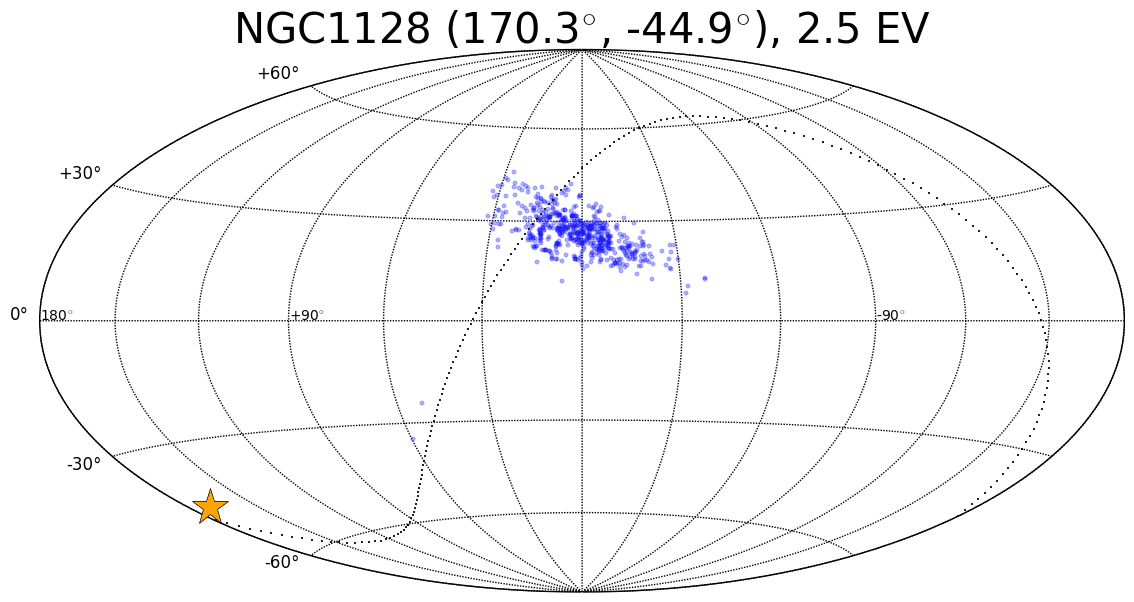}
\end{minipage}
\begin{minipage}[b]{0.48 \textwidth}
\includegraphics[width=1. \textwidth]{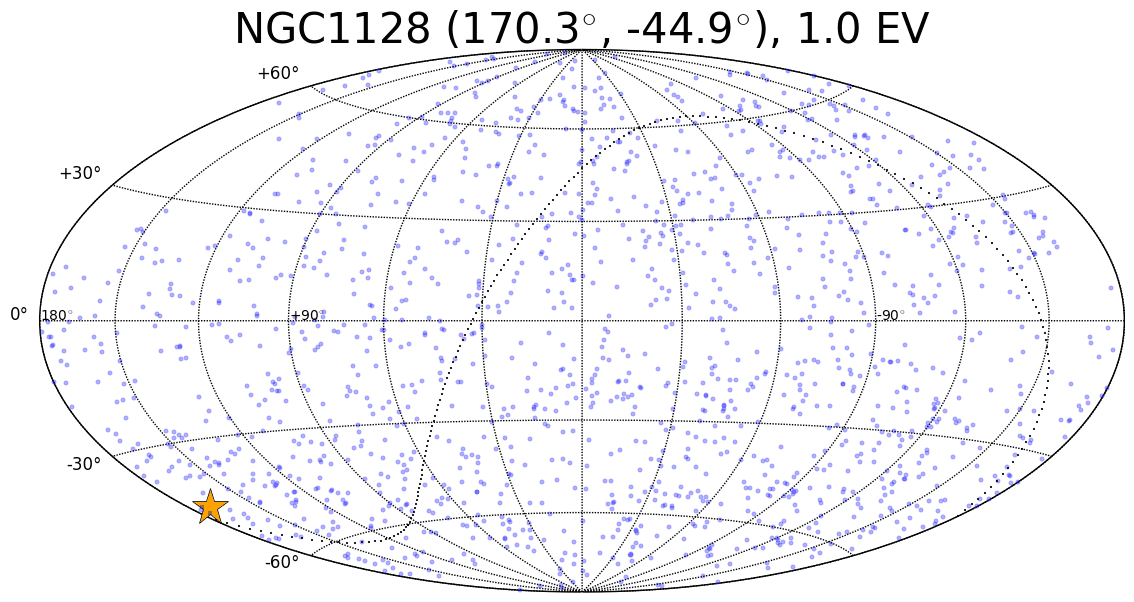}
\end{minipage}
\begin{minipage}[b]{0.48 \textwidth}
\includegraphics[width=1. \textwidth]{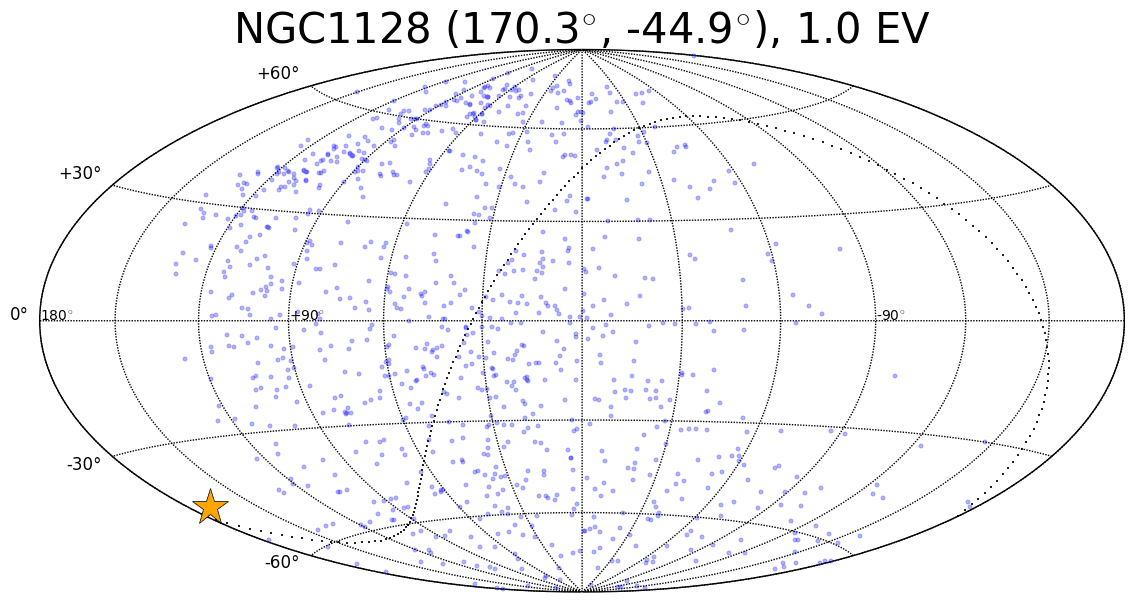}
\end{minipage}
\vspace{-0.3in}
\caption{As in Fig. \ref{plt:cena} but for NGC 1128, located at ($\ell$, $b$) = ($170.3^{\circ}$, $-44.9^{\circ}$).
} 
\label{plt:ngc1128}
\vspace{-0.1in}
\end{figure}

\begin{figure}[htb]
\hspace{-0.3in}
\centering
\begin{minipage}[b]{0.48 \textwidth}
\includegraphics[width=1. \textwidth]{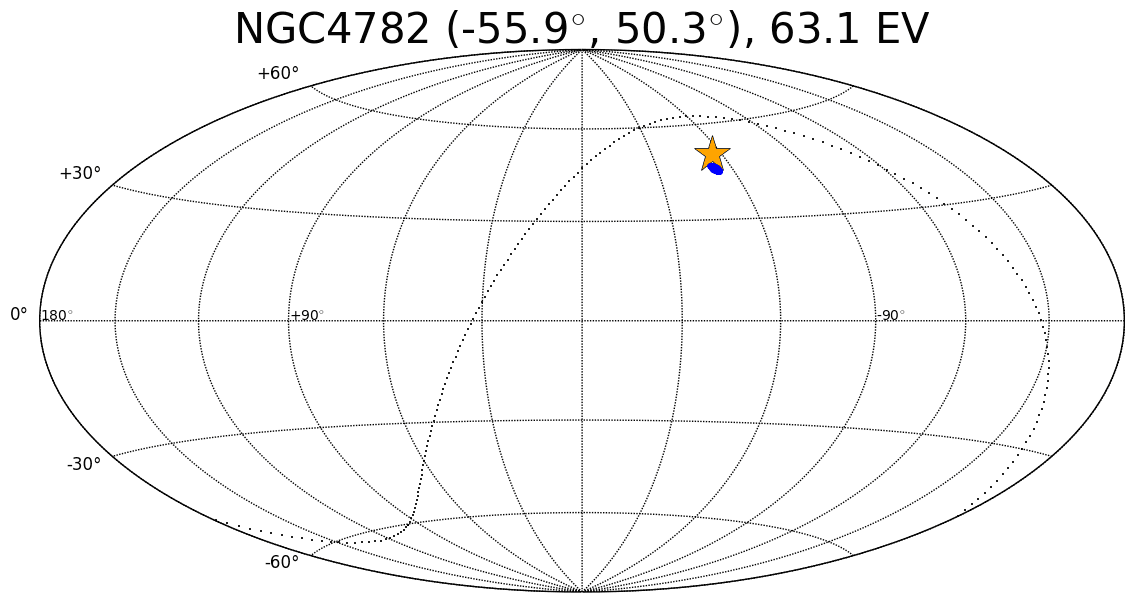}
\end{minipage}
\begin{minipage}[b]{0.48 \textwidth}
\includegraphics[width=1. \textwidth]{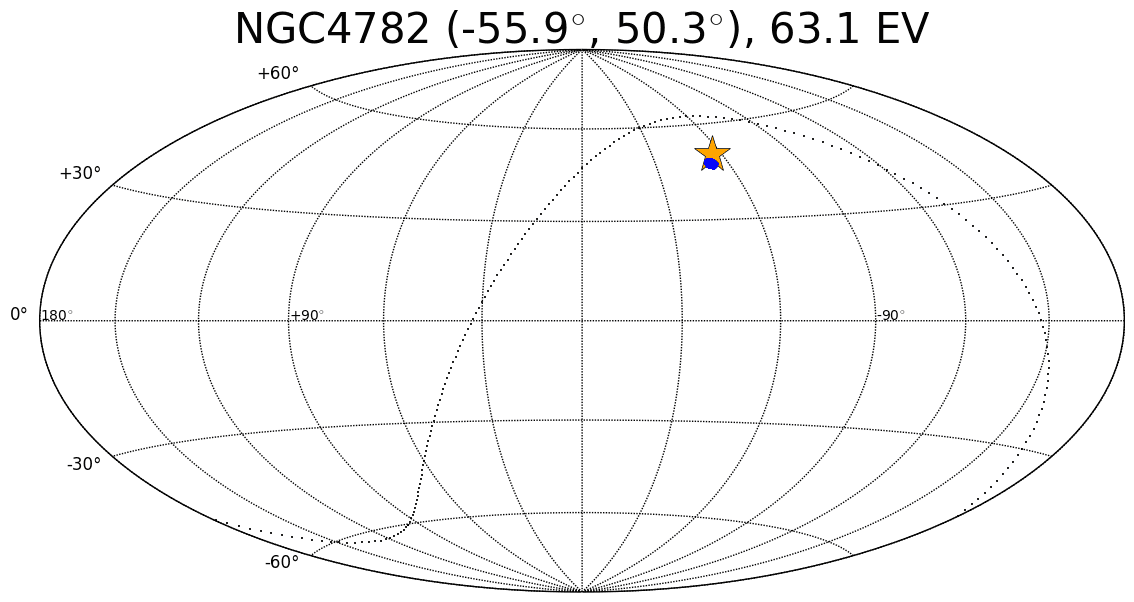}
\end{minipage}
\begin{minipage}[b]{0.48 \textwidth}
\includegraphics[width=1. \textwidth]{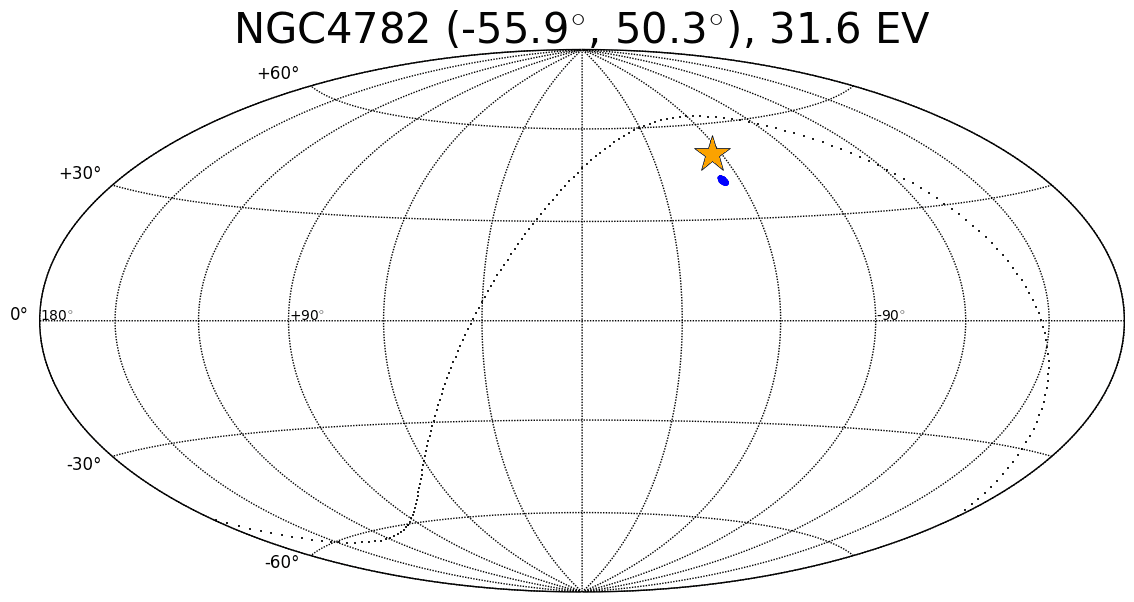}
\end{minipage}
\begin{minipage}[b]{0.48 \textwidth}
\includegraphics[width=1. \textwidth]{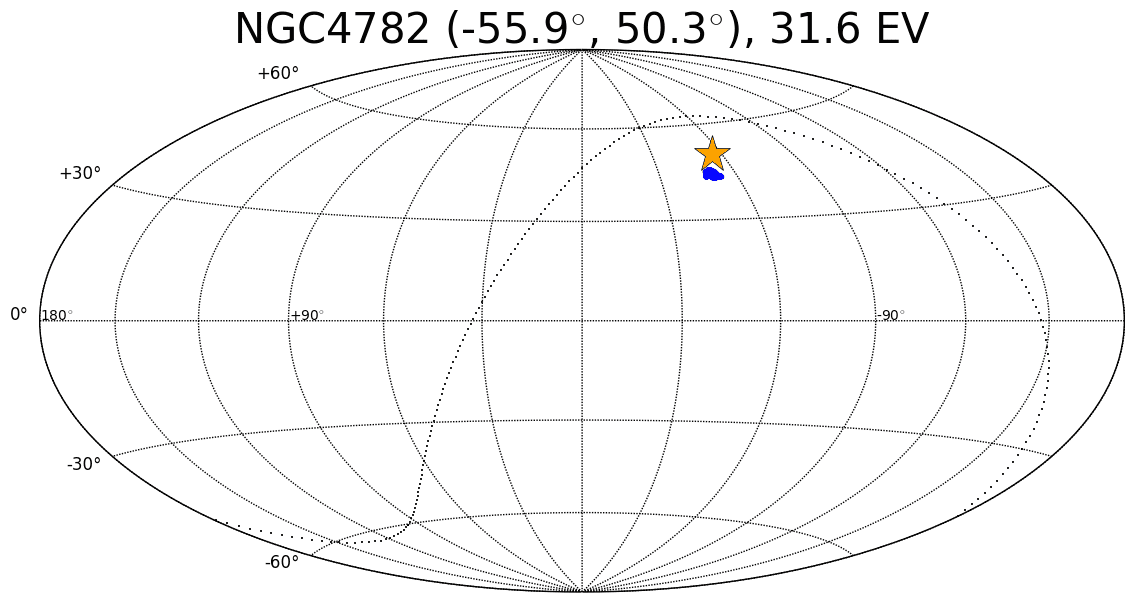}
\end{minipage}
\begin{minipage}[b]{0.48 \textwidth}
\includegraphics[width=1. \textwidth]{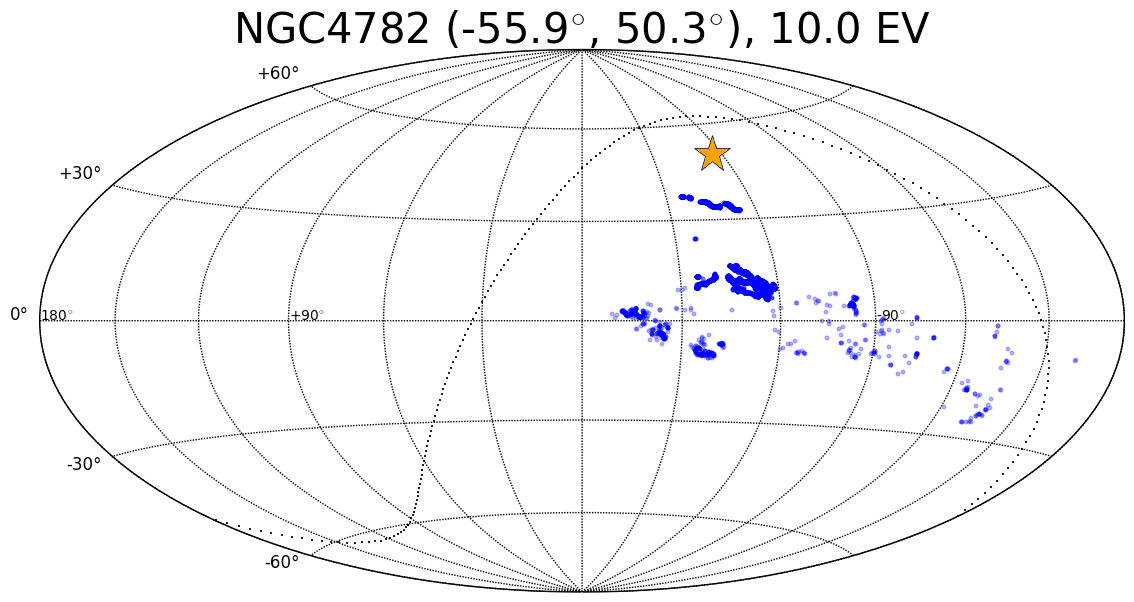}
\end{minipage}
\begin{minipage}[b]{0.48 \textwidth}
\includegraphics[width=1. \textwidth]{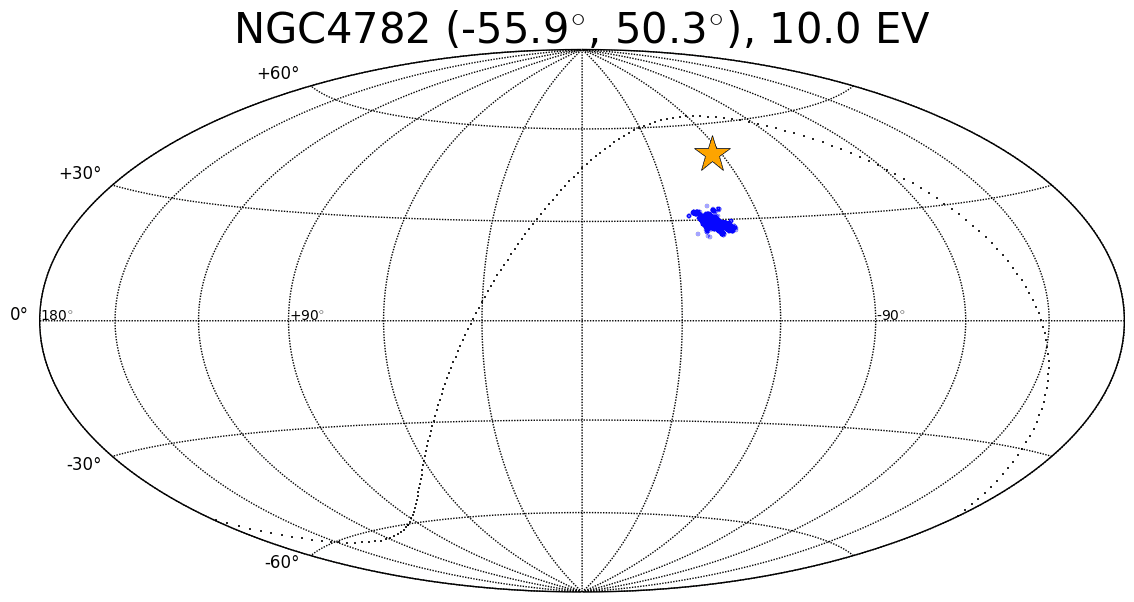}
\end{minipage}
\begin{minipage}[b]{0.48 \textwidth}
\includegraphics[width=1. \textwidth]{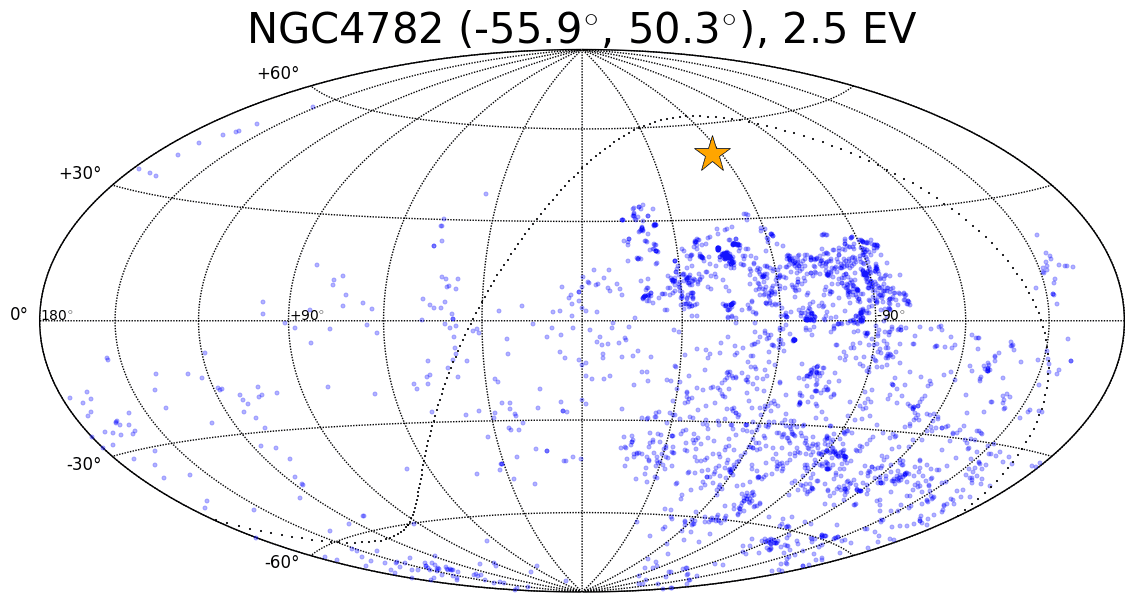}
\end{minipage}
\begin{minipage}[b]{0.48 \textwidth}
\includegraphics[width=1. \textwidth]{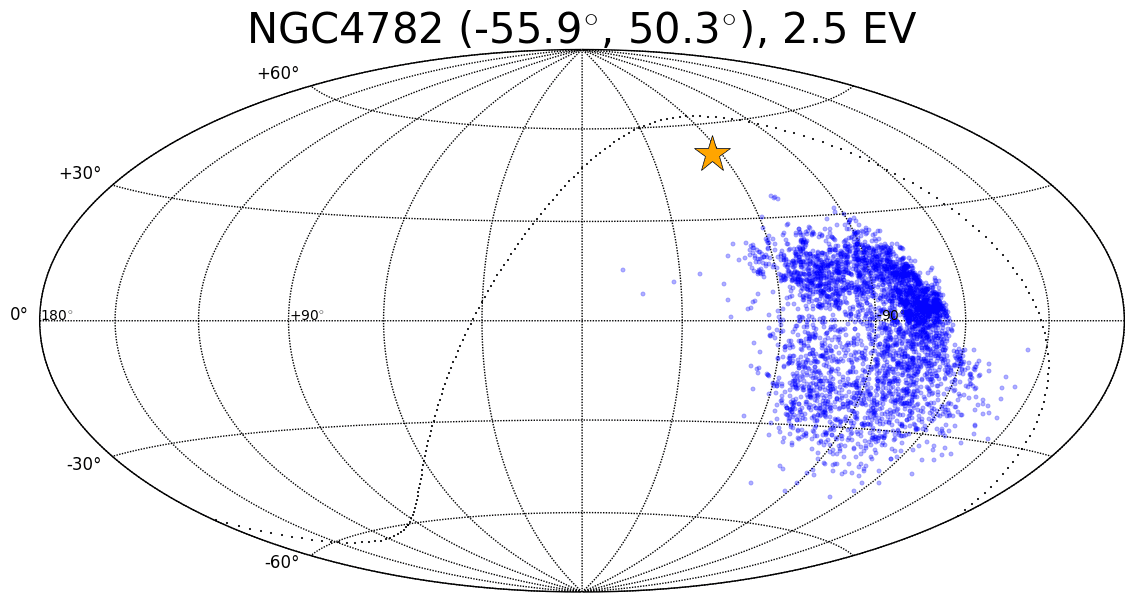}
\end{minipage}
\begin{minipage}[b]{0.48 \textwidth}
\includegraphics[width=1. \textwidth]{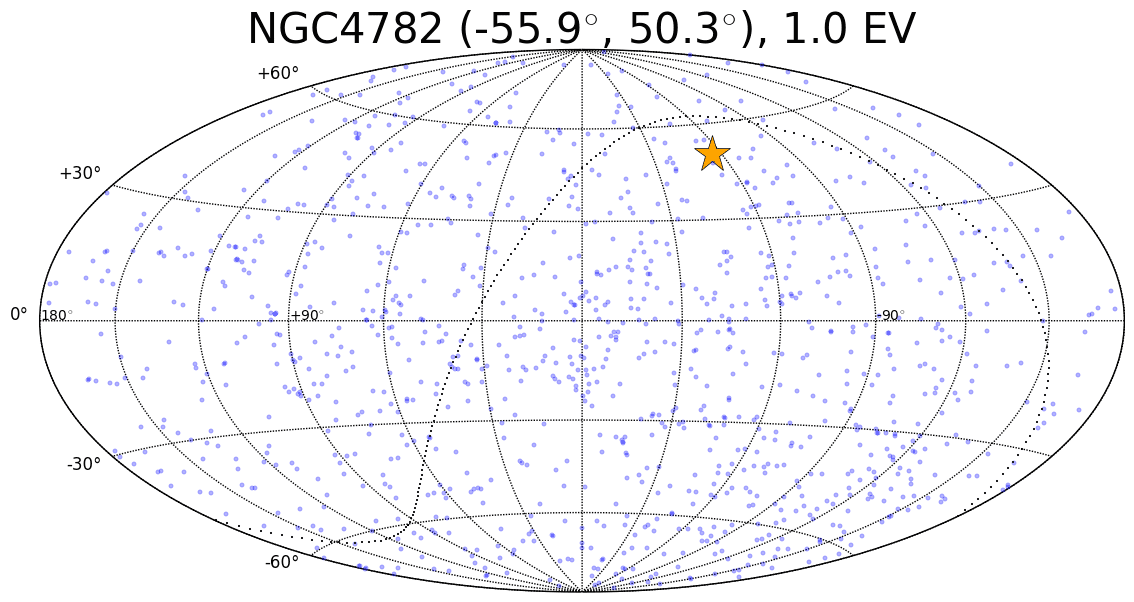}
\end{minipage}
\begin{minipage}[b]{0.48 \textwidth}
\includegraphics[width=1. \textwidth]{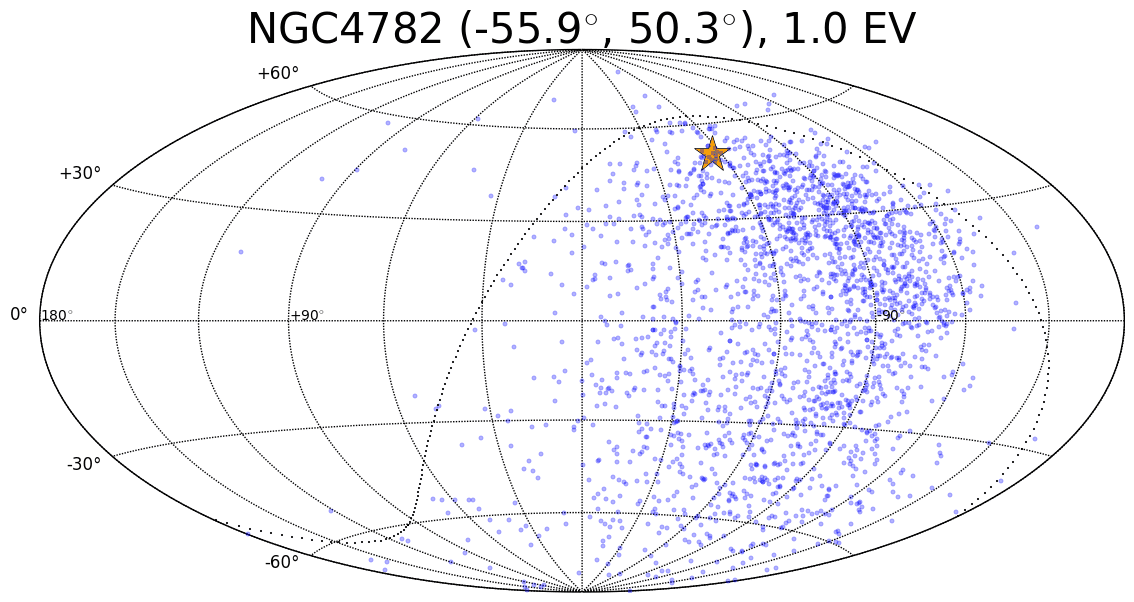}
\end{minipage}
\vspace{-0.3in}
\caption{As in Fig. \ref{plt:cena} but for NGC 4782 (marked with green star) located at ($\ell$, $b$) = ($-55.9^{\circ}$, $50.3^{\circ}$).} 
\label{plt:ngc4782}
\vspace{-0.1in}
\end{figure}

\begin{figure}[htb]
\hspace{-0.3in}
\centering
\begin{minipage}[b]{0.48 \textwidth}
\includegraphics[width=1. \textwidth]{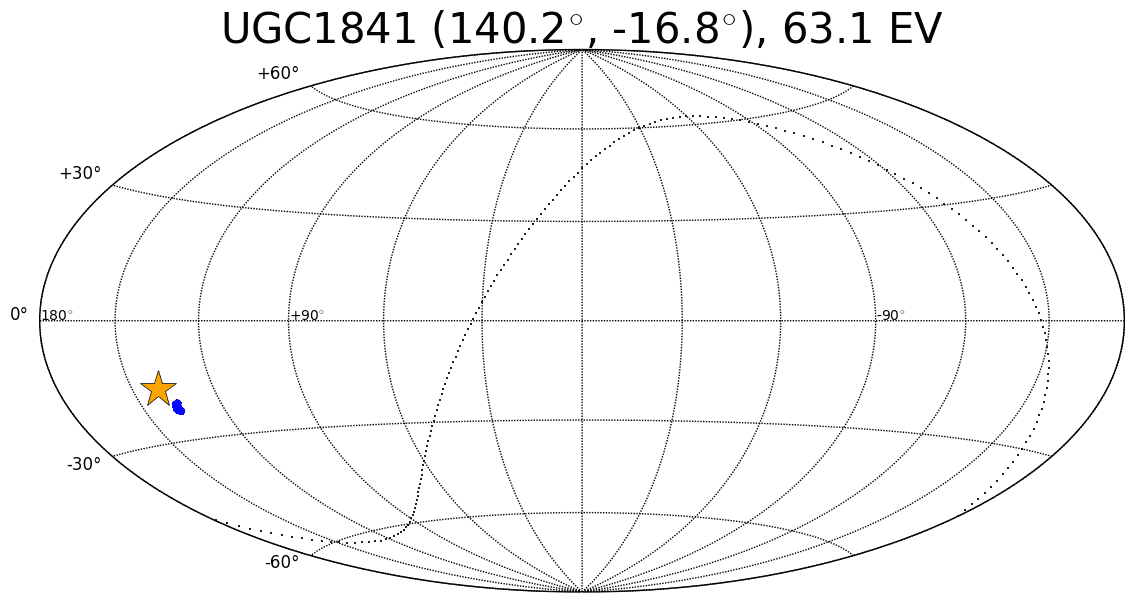}
\end{minipage}
\begin{minipage}[b]{0.48 \textwidth}
\includegraphics[width=1. \textwidth]{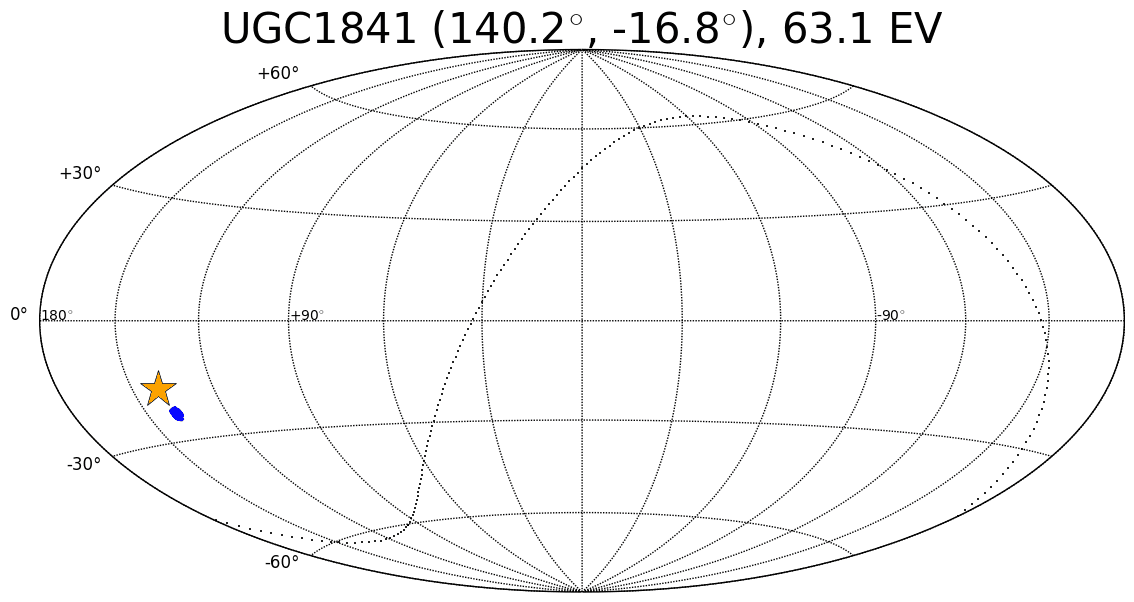}
\end{minipage}
\begin{minipage}[b]{0.48 \textwidth}
\includegraphics[width=1. \textwidth]{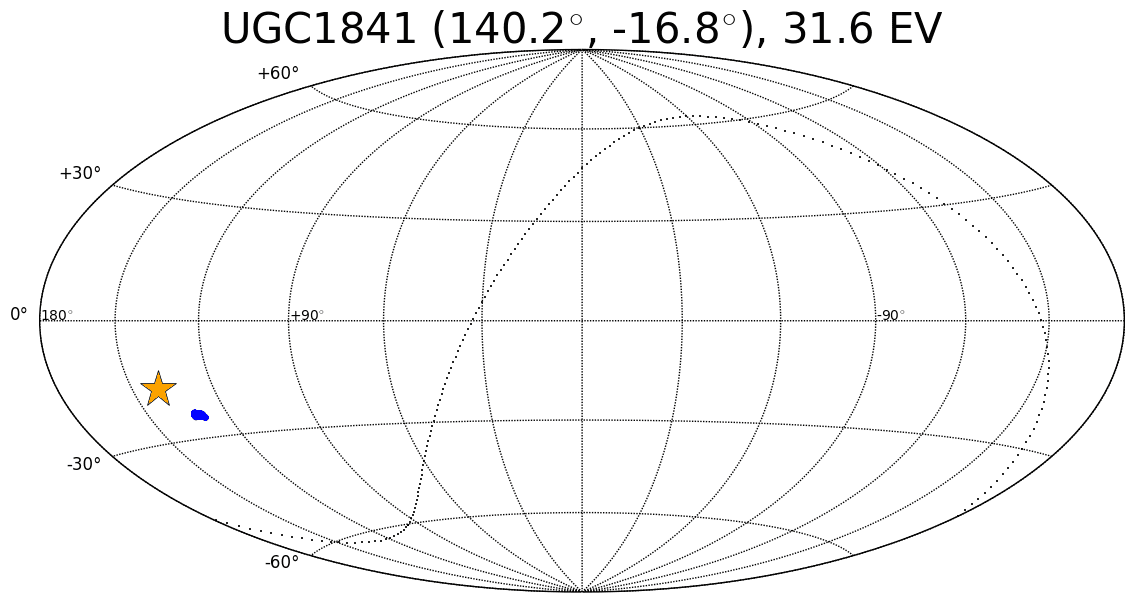}
\end{minipage}
\begin{minipage}[b]{0.48 \textwidth}
\includegraphics[width=1. \textwidth]{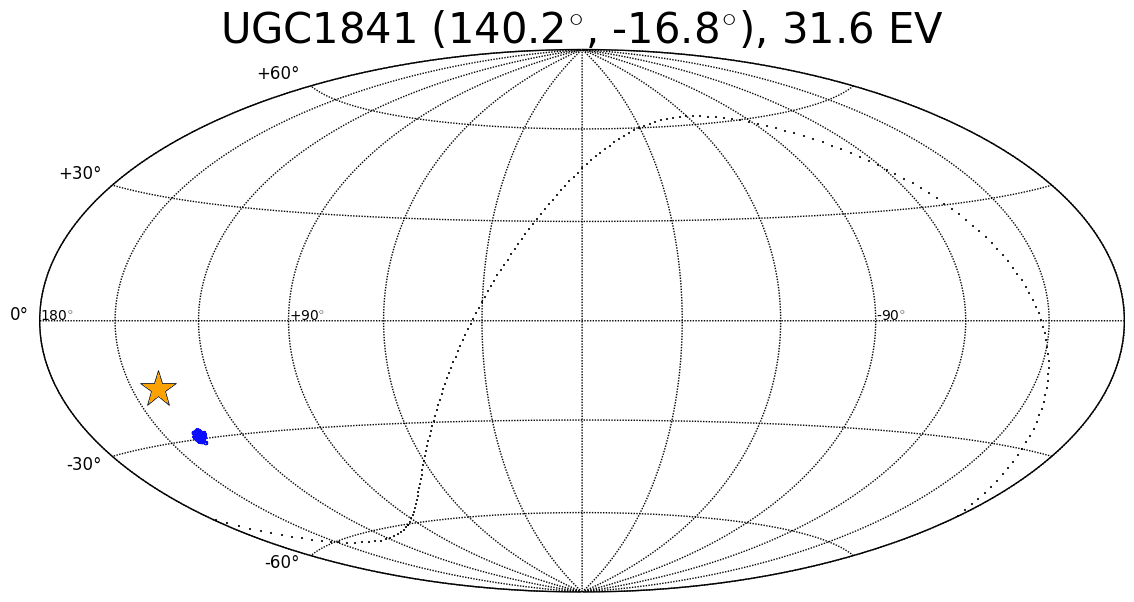}
\end{minipage}
\begin{minipage}[b]{0.48 \textwidth}
\includegraphics[width=1. \textwidth]{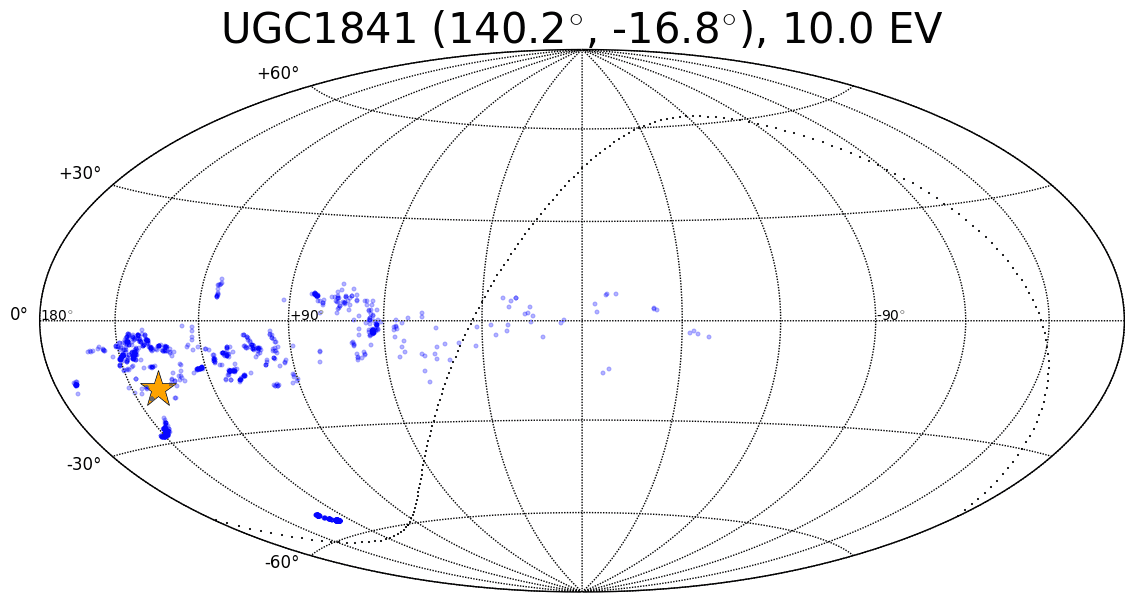}
\end{minipage}
\begin{minipage}[b]{0.48 \textwidth}
\includegraphics[width=1. \textwidth]{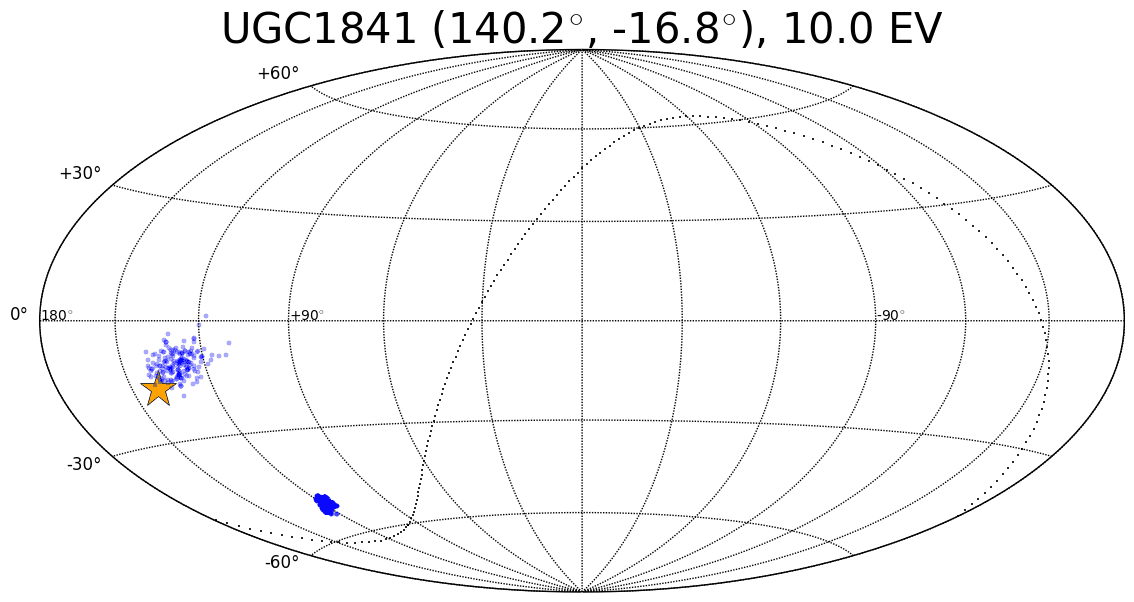}
\end{minipage}
\begin{minipage}[b]{0.48 \textwidth}
\includegraphics[width=1. \textwidth]{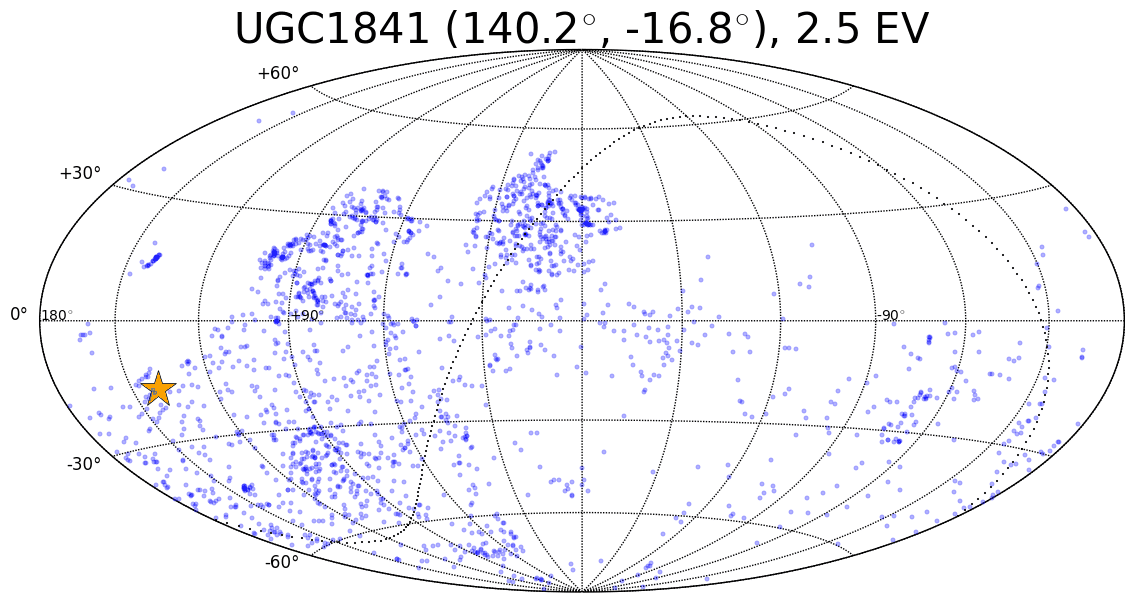}
\end{minipage}
\begin{minipage}[b]{0.48 \textwidth}
\includegraphics[width=1. \textwidth]{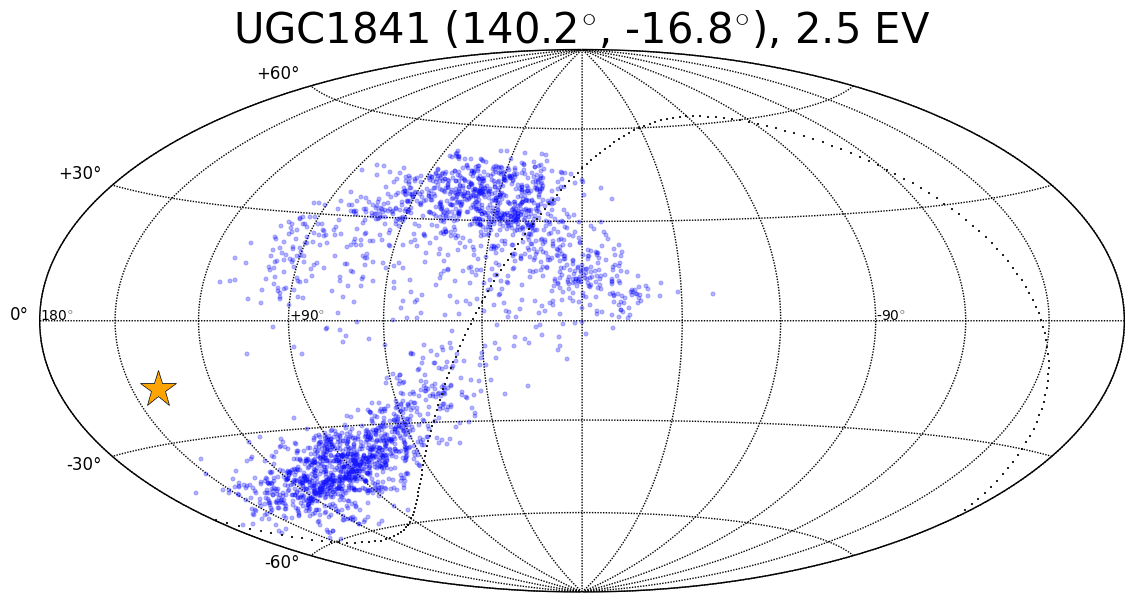}
\end{minipage}
\begin{minipage}[b]{0.48 \textwidth}
\includegraphics[width=1. \textwidth]{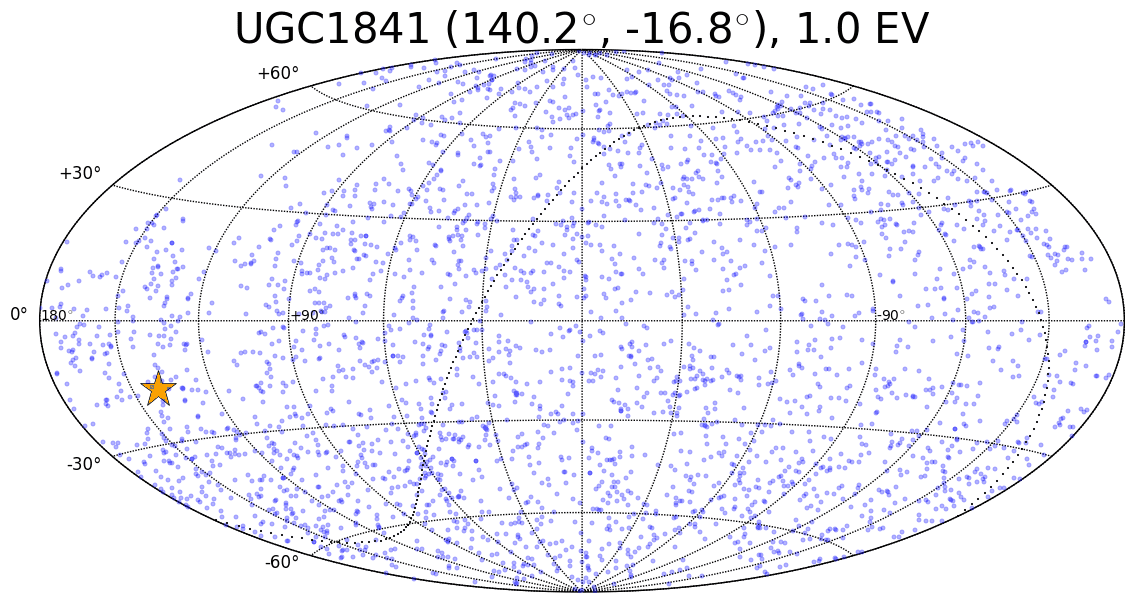}
\end{minipage}
\begin{minipage}[b]{0.48 \textwidth}
\includegraphics[width=1. \textwidth]{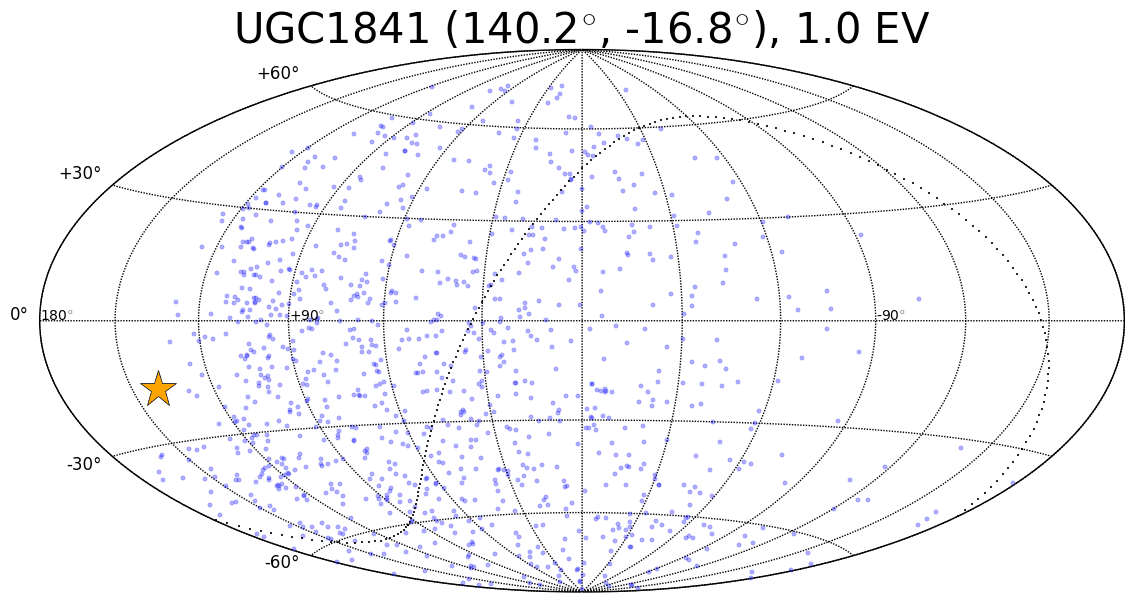}
\end{minipage}
\vspace{-0.3in}
\caption{As in Fig. \ref{plt:cena}, but for  UGC 1841 (marked with green star) located at ($\ell$, $b$) = ($140.2^{\circ}$, $-16.8^{\circ}$).} 
\label{plt:ugc1841}
\vspace{-0.1in}
\end{figure}

\begin{figure}[htb]
\hspace{-0.3in}
\centering
\begin{minipage}[b]{0.48 \textwidth}
\includegraphics[width=1. \textwidth]{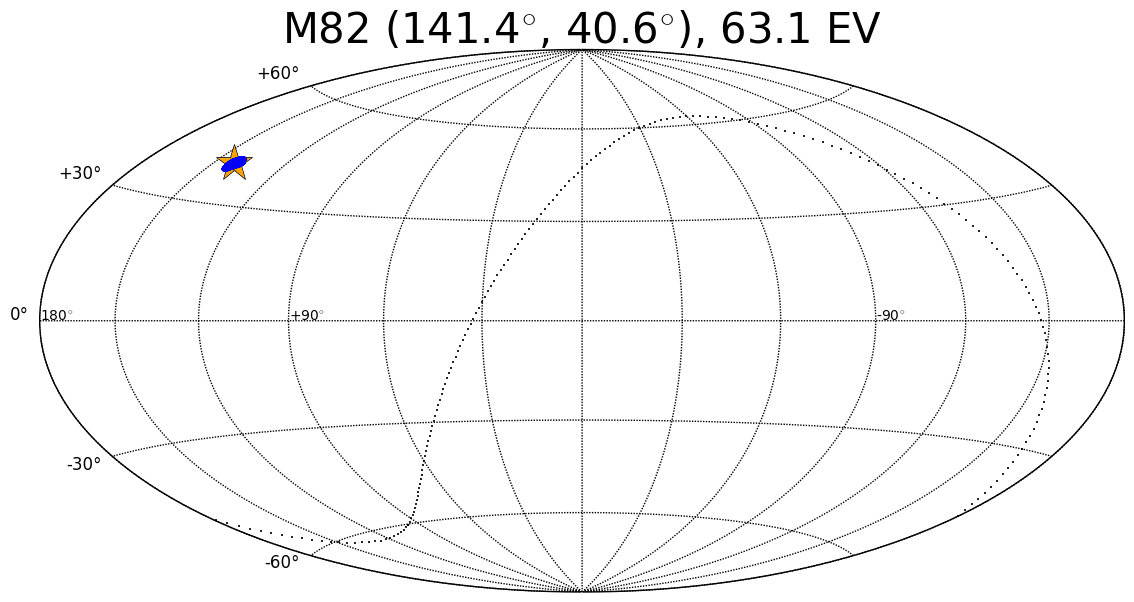}
\end{minipage}
\begin{minipage}[b]{0.48 \textwidth}
\includegraphics[width=1. \textwidth]{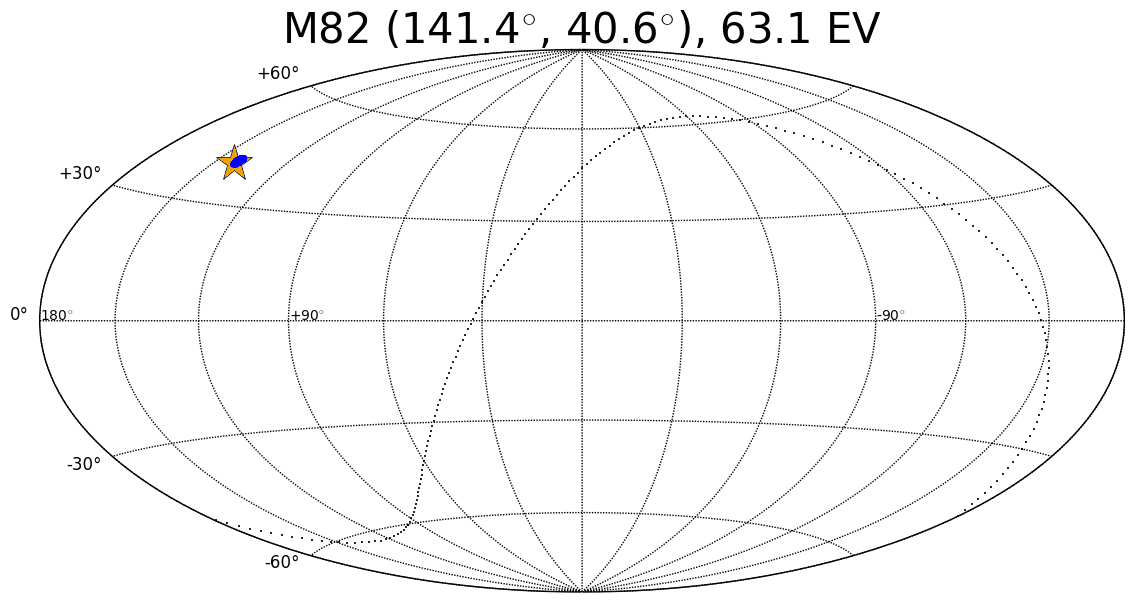}
\end{minipage}
\begin{minipage}[b]{0.48 \textwidth}
\includegraphics[width=1. \textwidth]{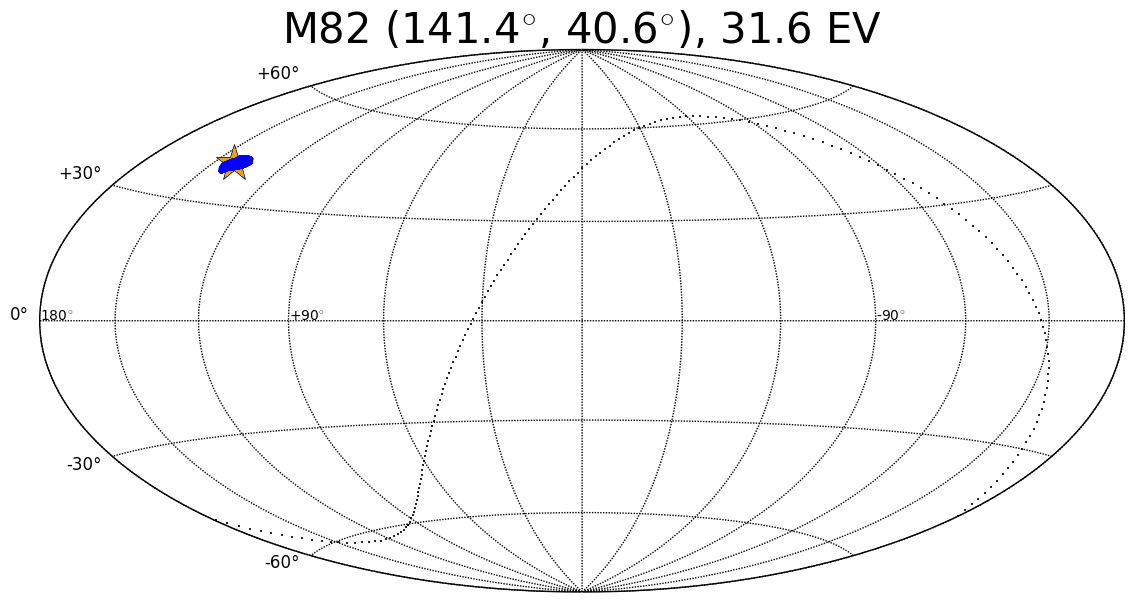}
\end{minipage}
\begin{minipage}[b]{0.48 \textwidth}
\includegraphics[width=1. \textwidth]{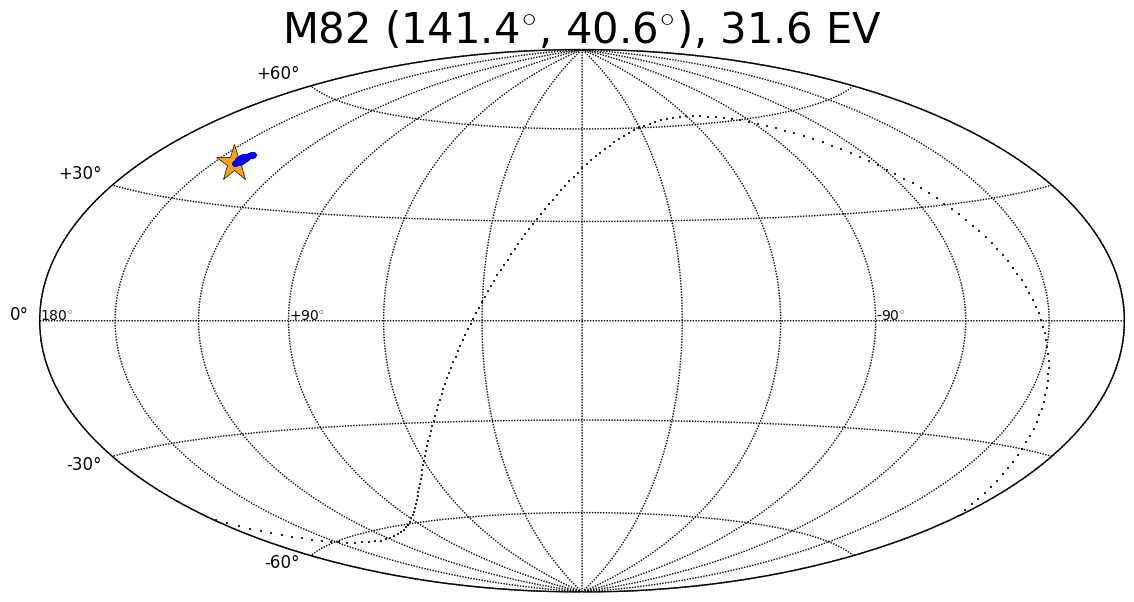}
\end{minipage}
\begin{minipage}[b]{0.48 \textwidth}
\includegraphics[width=1. \textwidth]{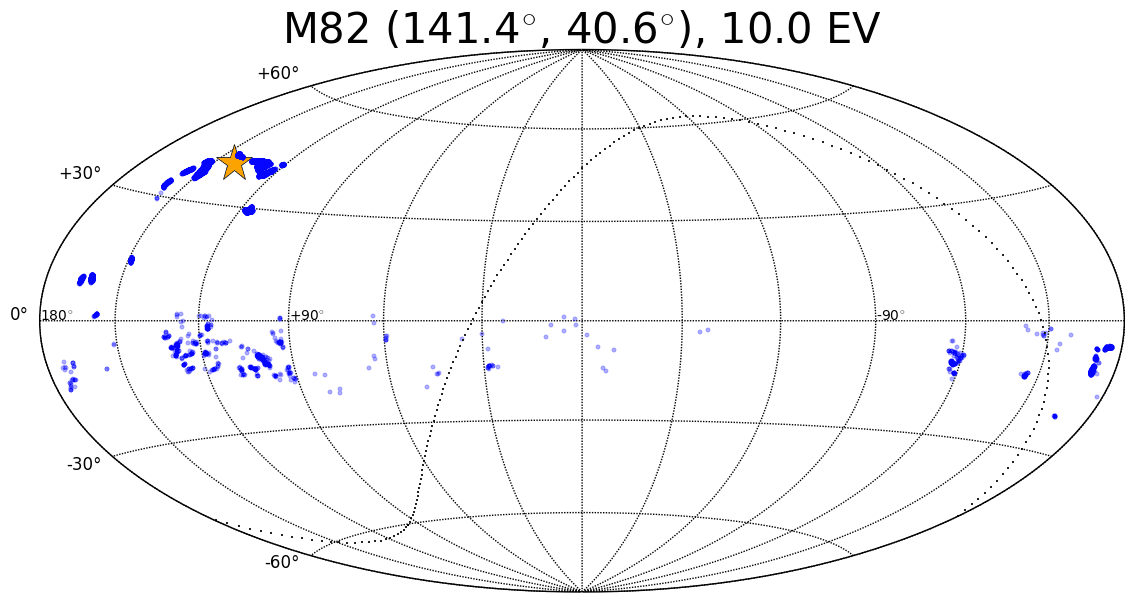}
\end{minipage}
\begin{minipage}[b]{0.48 \textwidth}
\includegraphics[width=1. \textwidth]{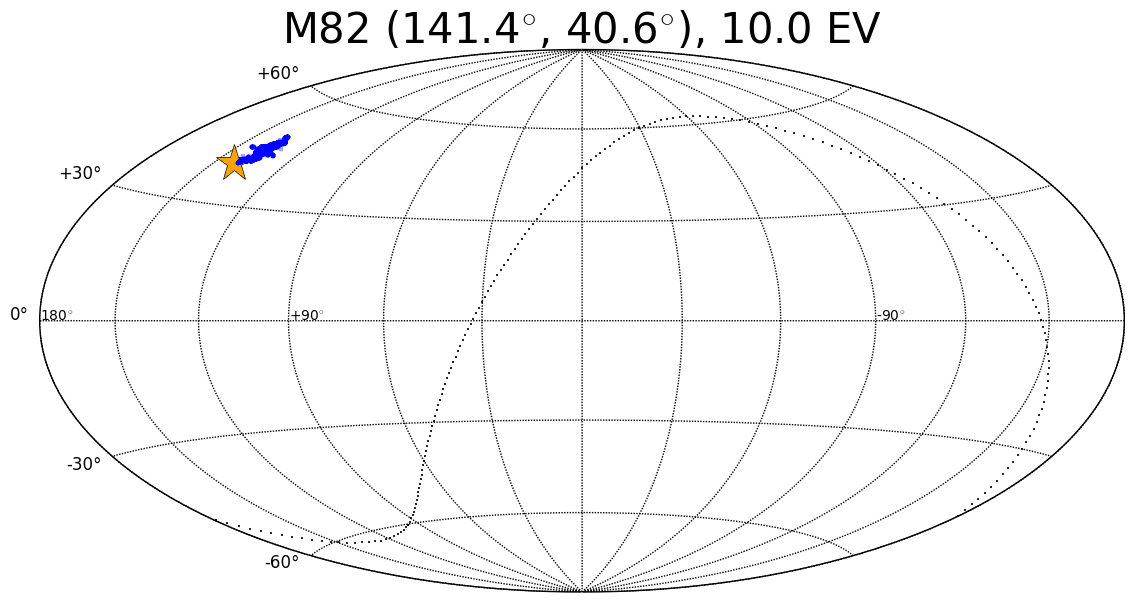}
\end{minipage}
\begin{minipage}[b]{0.48 \textwidth}
\includegraphics[width=1. \textwidth]{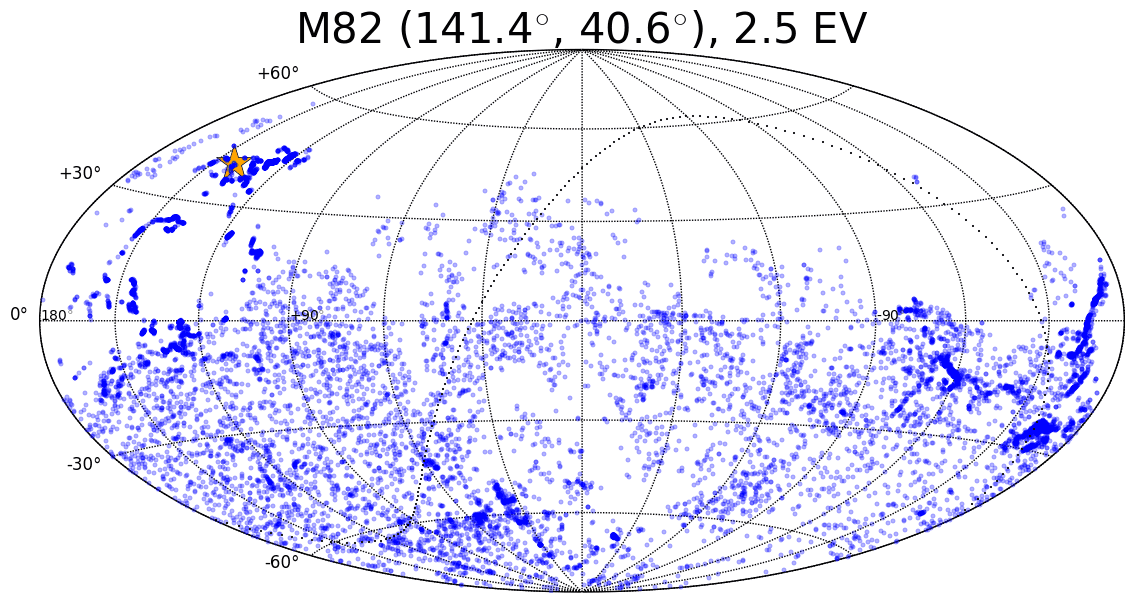}
\end{minipage}
\begin{minipage}[b]{0.48 \textwidth}
\includegraphics[width=1. \textwidth]{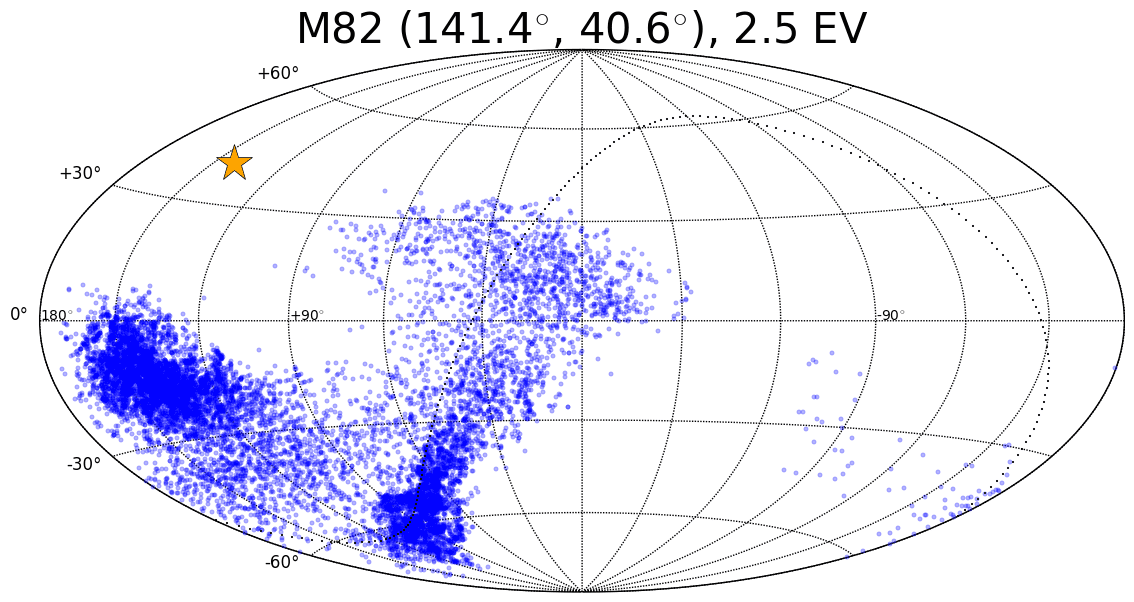}
\end{minipage}
\begin{minipage}[b]{0.48 \textwidth}
\includegraphics[width=1. \textwidth]{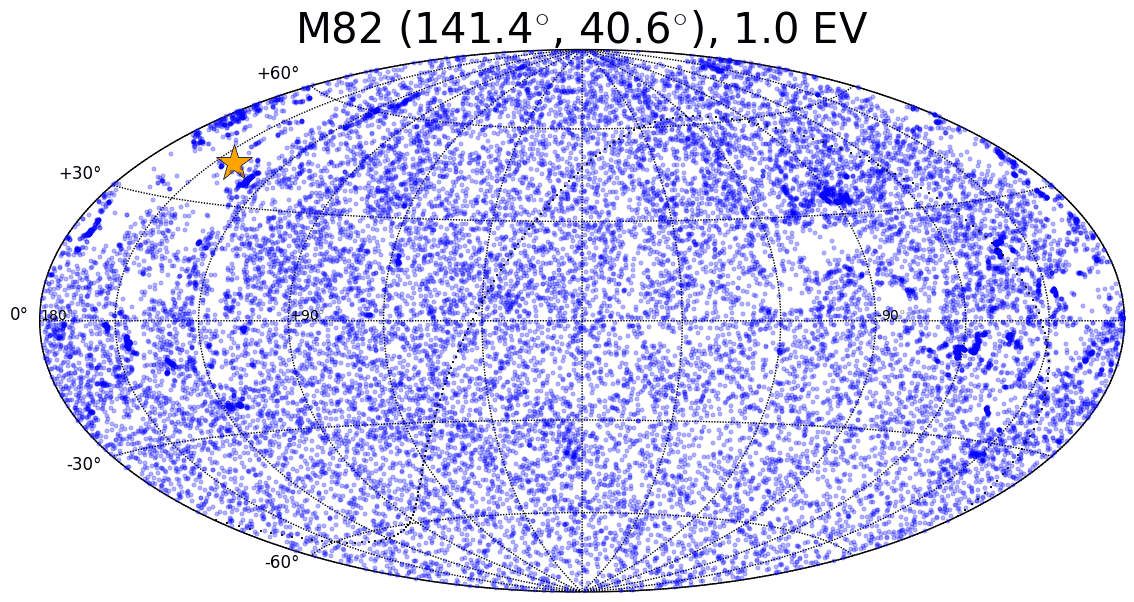}
\end{minipage}
\begin{minipage}[b]{0.48 \textwidth}
\includegraphics[width=1. \textwidth]{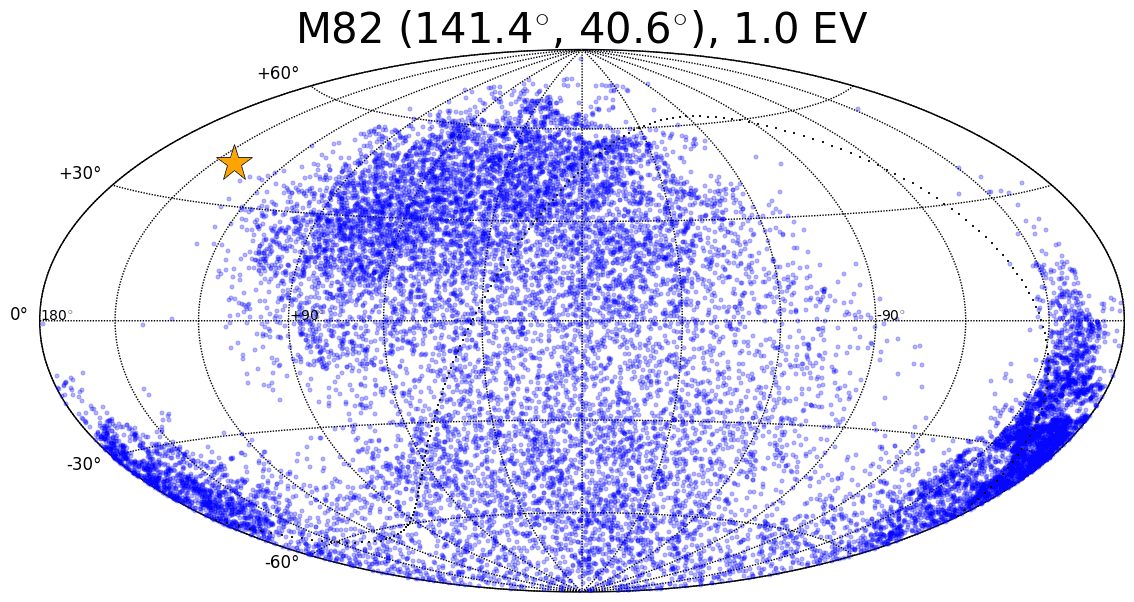}
\end{minipage}
\vspace{-0.3in}
\caption{As in Fig. \ref{plt:cena}, but for M82 (marked with green star) located at ($\ell$, $b$) = ($141.41^{\circ}$, $40.57^{\circ}$).} 
\label{plt:m82}
\vspace{-0.1in}
\end{figure}

\begin{figure}[H]
\hspace{-0.3in}
\centering
\begin{minipage}[b]{0.48 \textwidth}
\includegraphics[width=1. \textwidth]{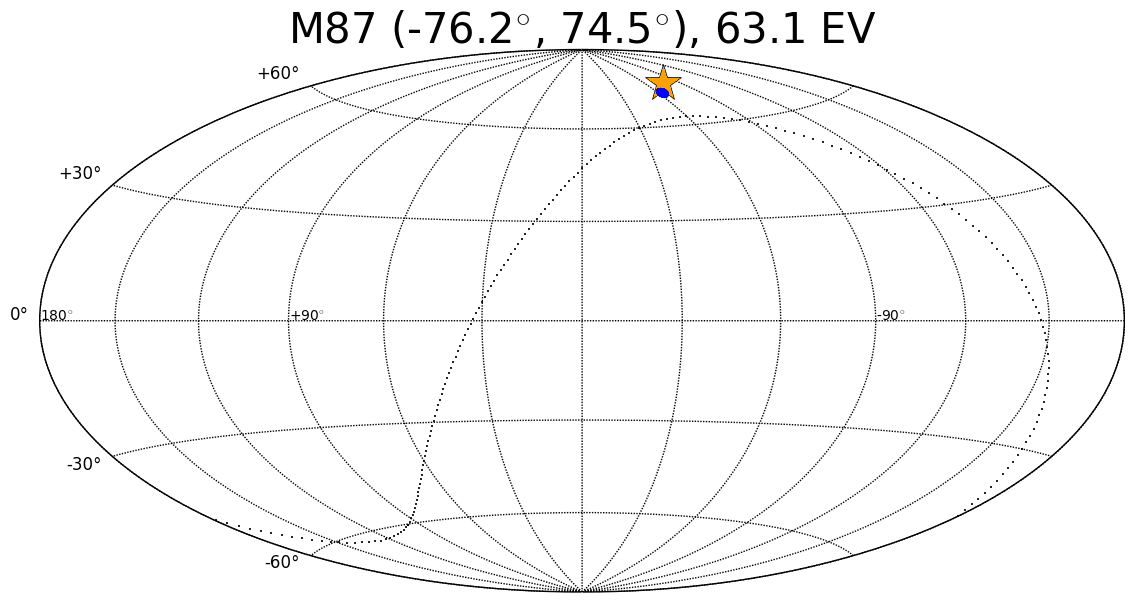}
\end{minipage}
\begin{minipage}[b]{0.48 \textwidth}
\includegraphics[width=1. \textwidth]{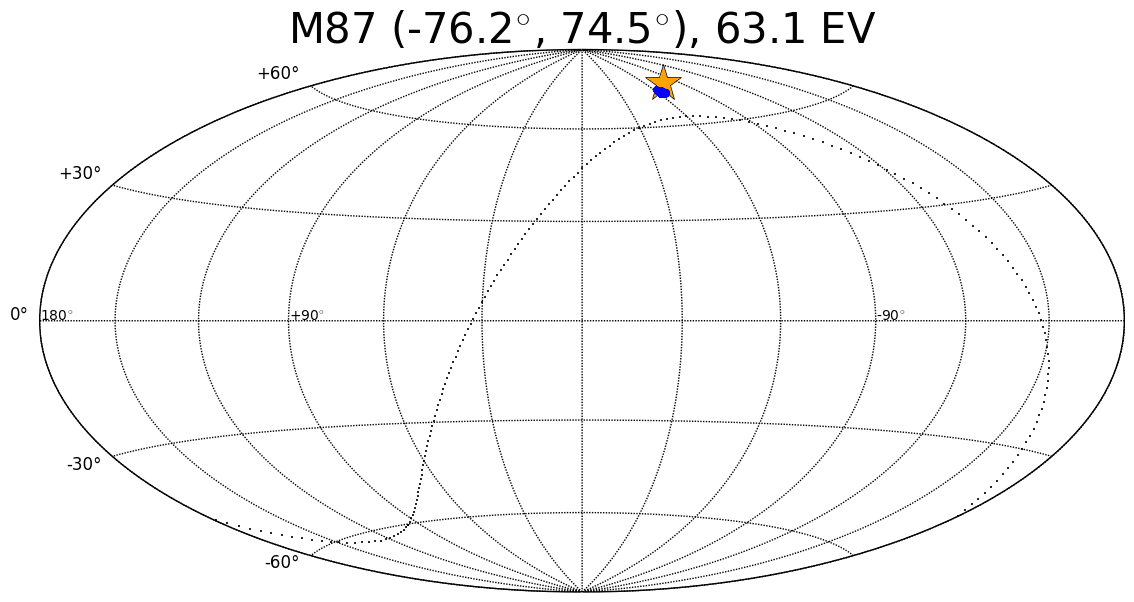}
\end{minipage}
\begin{minipage}[b]{0.48 \textwidth}
\includegraphics[width=1. \textwidth]{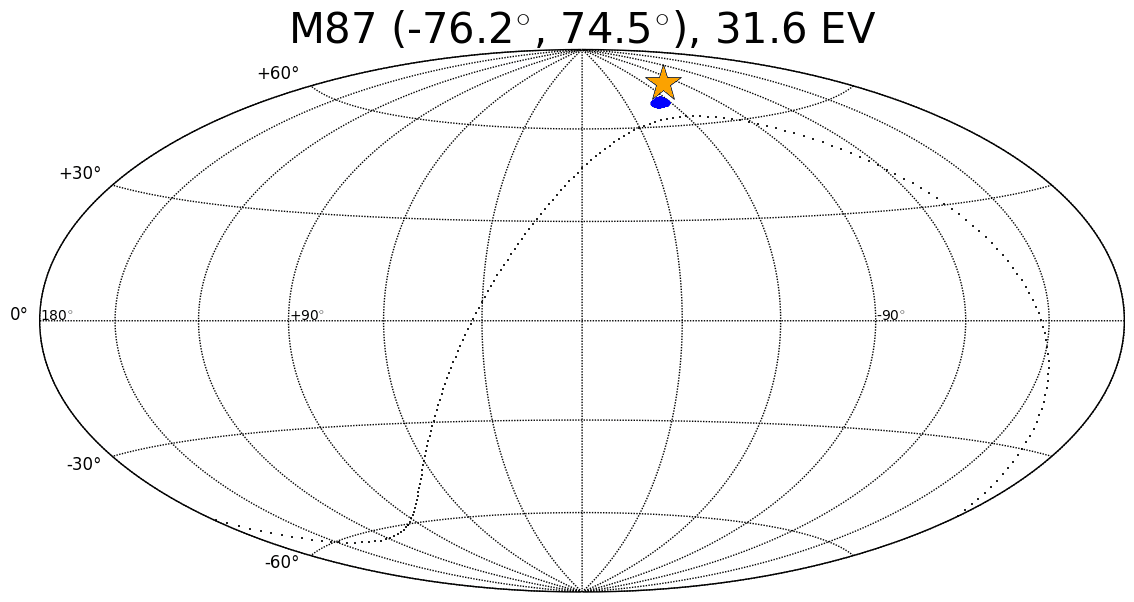}
\end{minipage}
\begin{minipage}[b]{0.48 \textwidth}
\includegraphics[width=1. \textwidth]{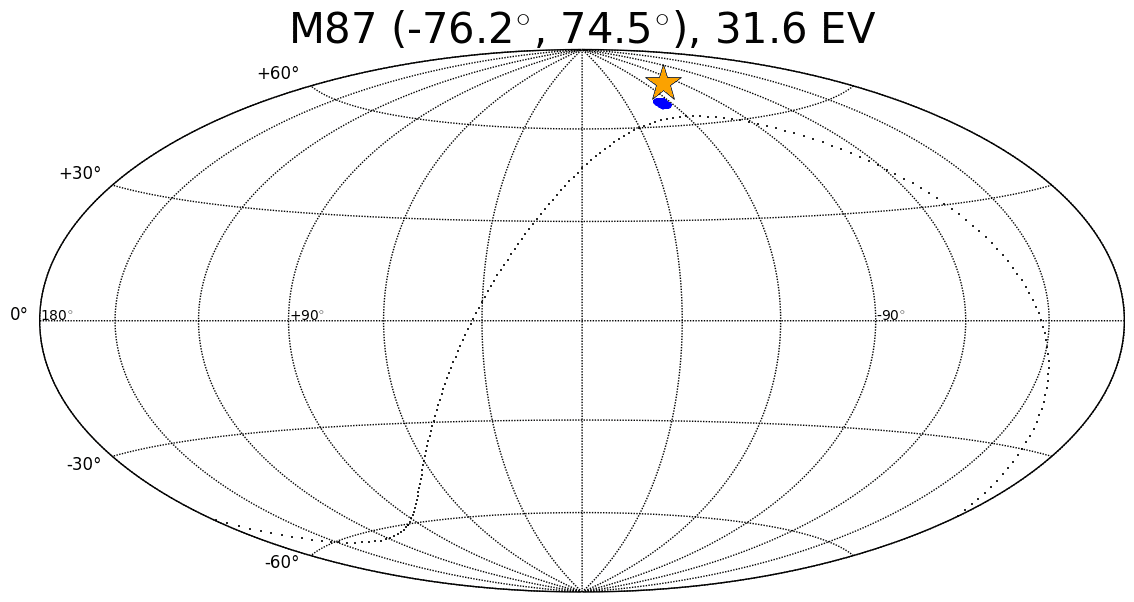}
\end{minipage}
\begin{minipage}[b]{0.48 \textwidth}
\includegraphics[width=1. \textwidth]{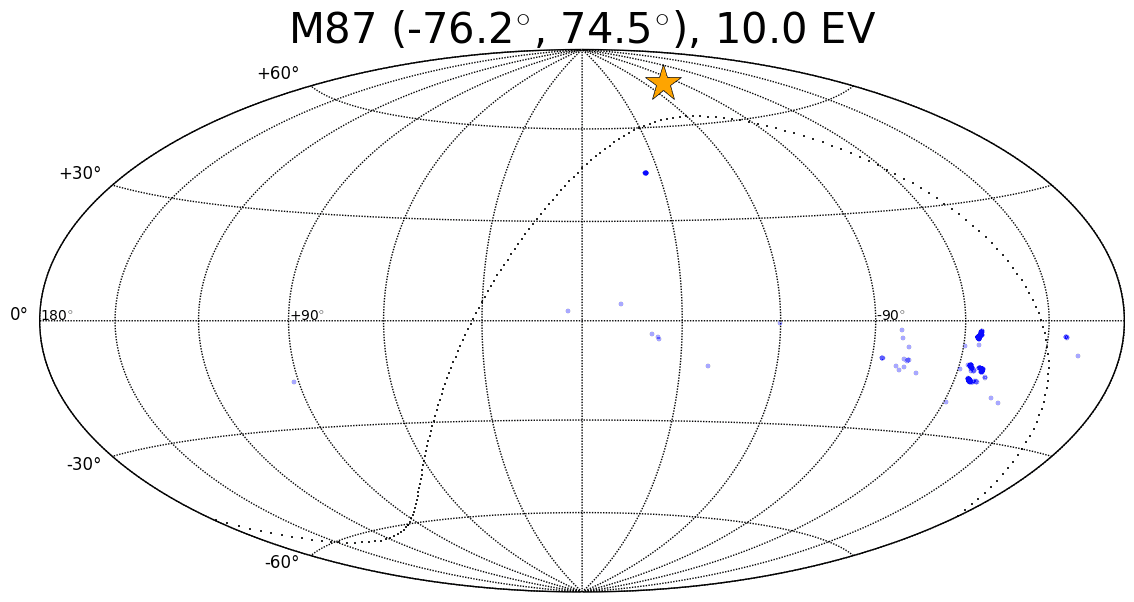}
\end{minipage}
\begin{minipage}[b]{0.48 \textwidth}
\includegraphics[width=1. \textwidth]{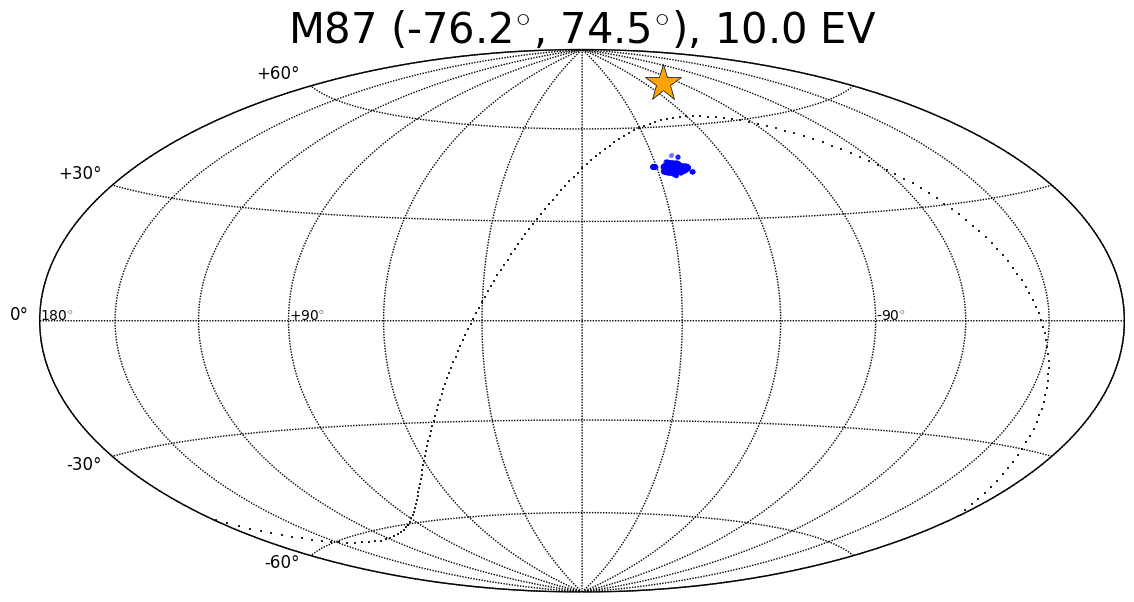}
\end{minipage}
\begin{minipage}[b]{0.48 \textwidth}
\includegraphics[width=1. \textwidth]{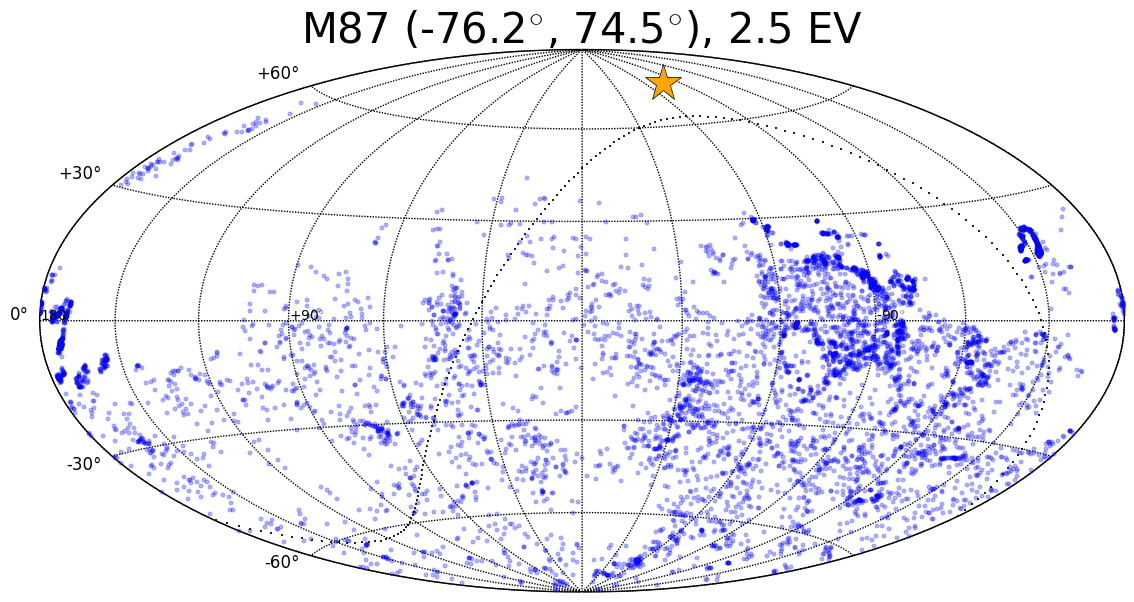}
\end{minipage}
\begin{minipage}[b]{0.48 \textwidth}
\includegraphics[width=1. \textwidth]{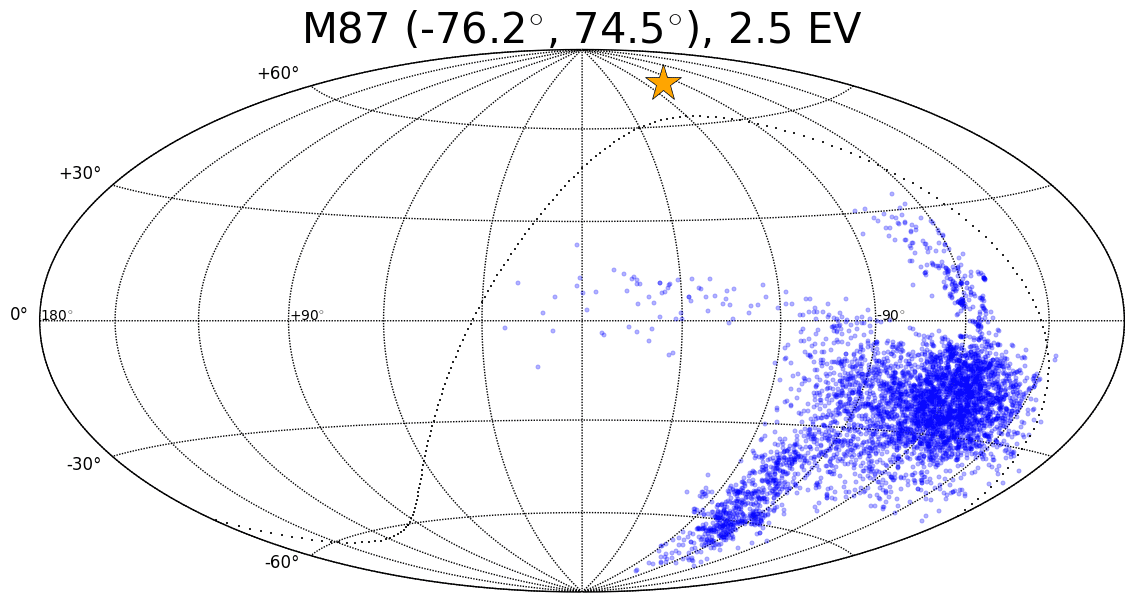}
\end{minipage}
\begin{minipage}[b]{0.48 \textwidth}
\includegraphics[width=1. \textwidth]{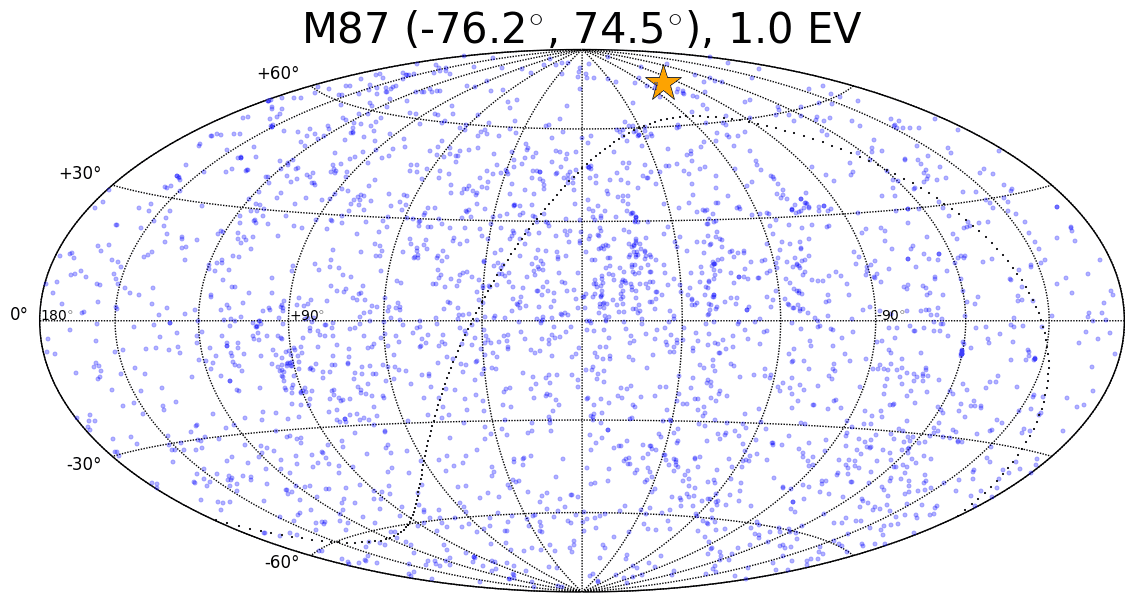}
\end{minipage}
\begin{minipage}[b]{0.48 \textwidth}
\includegraphics[width=1. \textwidth]{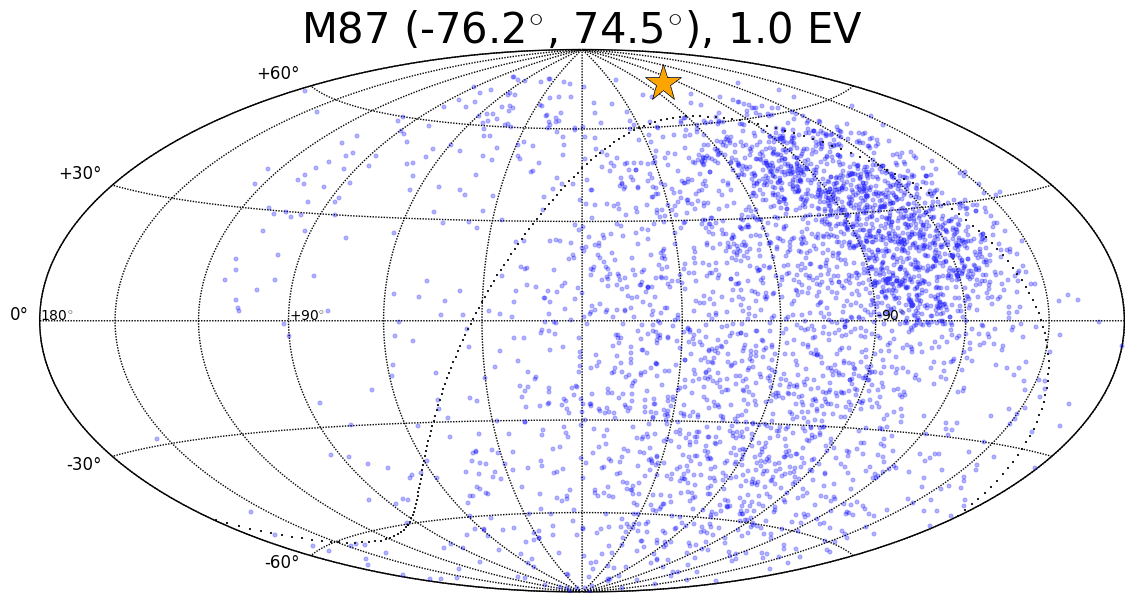}
\end{minipage}
\vspace{-0.3in}
\caption{As in Fig. \ref{plt:cena}, but for M87 (marked with green star) located at ($\ell$, $b$) = ($283.78^{\circ}$, $74.49^{\circ}$).} 
\label{plt:m87}
\vspace{-0.1in}
\end{figure}

Figures \ref{plt:cena}-\ref{plt:m87} show arrival direction distribution skymaps for various rigidities for each of the 7 radio galaxy source candidates listed in Table \ref{tbl:srcpos}.
Tables of general arrival distribution quantities for these sources are located in Appendix \ref{appdx:RadioTables}.
In all of these figures, the left column shows the KRF6 ($L_{coh}$=100 pc) realization and the right column shows the KRF10 ($L_{coh}$=30 pc) realization.
At the highest rigidities, all direction distributions lie near the source positions as expected.
In all cases, the distributions shift in position and increase in areal extent with lower rigidity.
The shifts vary in direction due to the different probed field geometries and magnitudes along the particle trajectory.
Although we do not include $10^{20}$ V maps in Figure \ref{plt:cgcg114-025}, for CGCG114 there are two distinct distribution regions even at this very high rigidity.  These regions are located at roughly ($45^{\circ}$, $-5^{\circ}$) and ($40^{\circ}$, $12^{\circ}$) covering only about a few square degrees.
The existence of imaging even at such high rigidities is likely due to the close proximity of the source direction to the Galactic plane where there are high field strengths and geometries allowing for magnification effects even at this rigidity.
More typically, significant dispersal of the arrival direction distribution across the sky starts at about $10^{18.6-8}$ V.  

Figures \ref{plt:hpx19o8}-\ref{plt:hpx_xtra_18o3_18o0} (Appendix \ref{appdx:HEALPixSkyplots}) show the arrival direction distributions for the regularly-spaced source directions listed in Table \ref{tbl:hpxpos} at the beginning of this section.  
At high rigidity, the distributions tend to lie near their sources, although shifts in the centroids are evident on the scale of a few degrees and increase with lower rigidity.  The size of the arrival direction distributions varies with the source position, even for identical rigidities.
The numbers of events observed at a given rigidity depends on latitude, with more being seen on average from sources near the Galactic plane; this is an example of magnification.  
The strong field within the disk appears to have a focusing and trapping effect, even at high rigidities.
At intermediate rigidities, the distributions begin to collect along the Galactic plane, particularly $10^{19}$ V and below.
Below about $10^{18.6}$ V, the distributions begin to encompass a significantly large fraction of the sky but tend to preferentially populate the southern Galactic sky.
Tables of general arrival distribution quantities for these sources are located in Appendix \ref{appdx:HEALPixTables}.

Figures \ref{plt:hpx19o5_krf10_krf11}-\ref{plt:hpx18o5_krf10_krf11} (Appendix \ref{appdx:realizations}) compare the arrival distributions for both $L_{coh} = 30$ pc realizations (KRF 10 and KRF11).
For all three rigidity values, the distributions for each source are similar, especially when compared to the corresponding $L_{coh} = 100$ pc distributions.
The basic properties (centroid, widths, number of peaks, etc.) are retained, although the outliers in the distributions vary due to the different random generation of the realizations.

\newpage
\newpage
\newpage
\newpage
\newpage
\newpage

\newpage

\newpage

\section{Discussion of deflection plots}

A key result illustrated by the skyplots of the previous section and the Appendices, is how variable the images of sources in different directions are, in spite of the original sources being identical, uniform, circular disks.  In particular, the images are rarely well-represented by simple azimuthally-symmetric distributions about their centroid, such as a Gaussian or top-hat.  

\begin{figure}[H]
\hspace{-0.3in}
\centering
\begin{minipage}[b]{0.32 \textwidth}
\includegraphics[width=1. \textwidth]{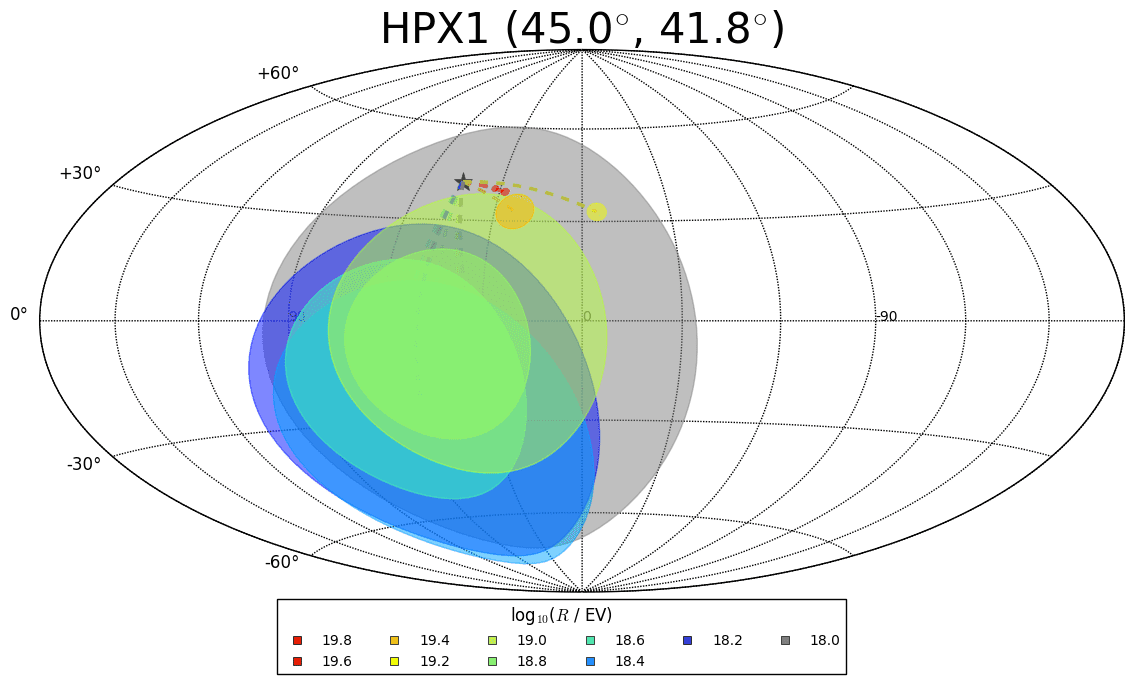}
\end{minipage}
\begin{minipage}[b]{0.32 \textwidth}
\includegraphics[width=1. \textwidth]{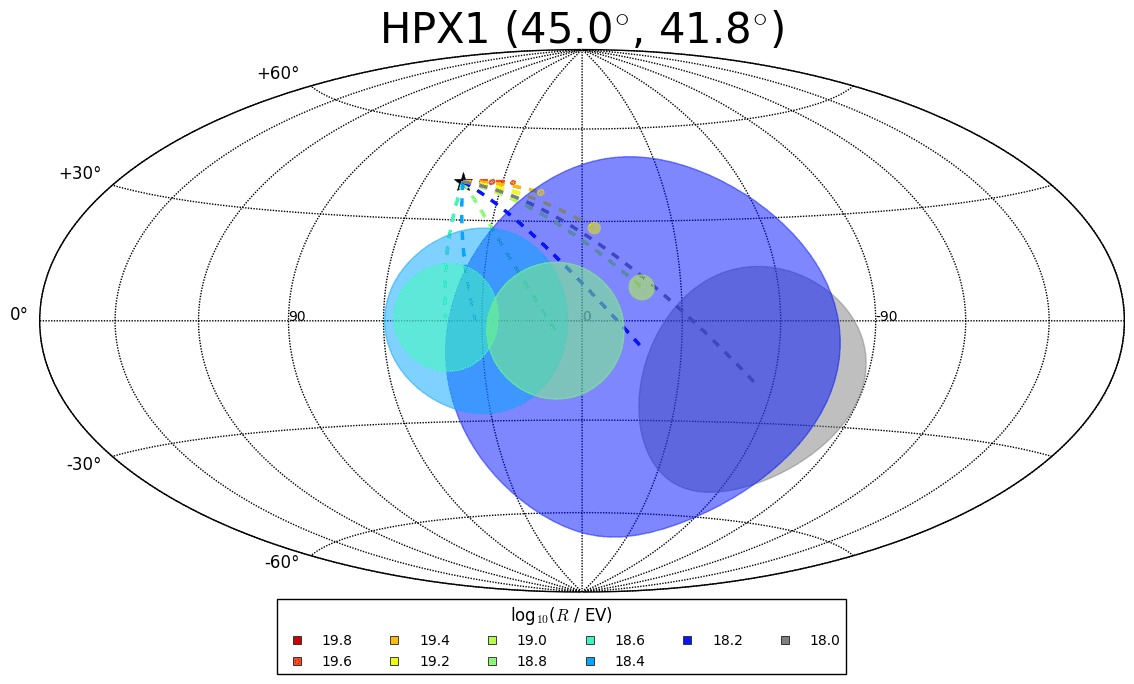}
\end{minipage}
\begin{minipage}[b]{0.32 \textwidth}
\includegraphics[width=1. \textwidth]{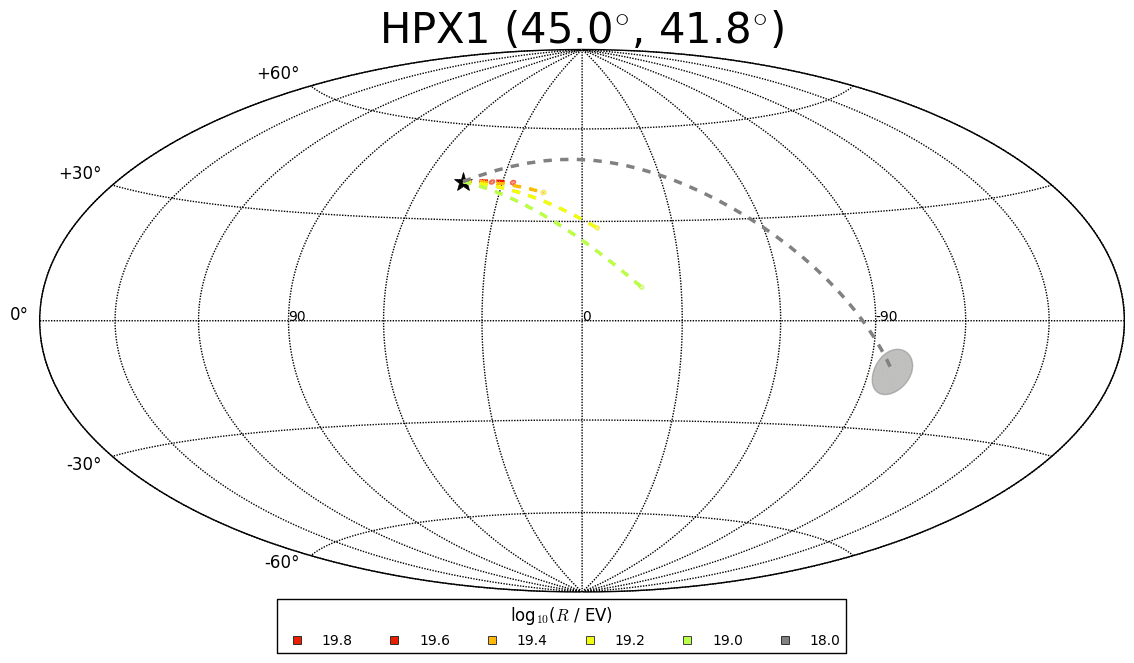}
\end{minipage}
\begin{minipage}[b]{0.32 \textwidth}
\includegraphics[width=1. \textwidth]{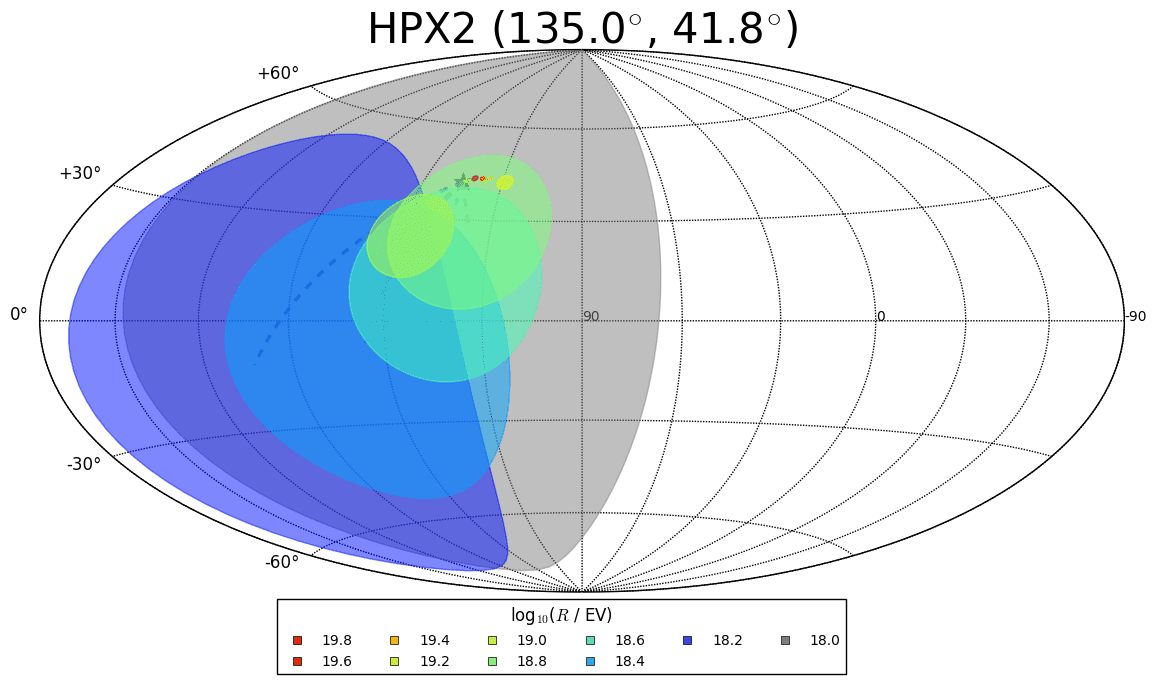}
\end{minipage}
\begin{minipage}[b]{0.32 \textwidth}
\includegraphics[width=1. \textwidth]{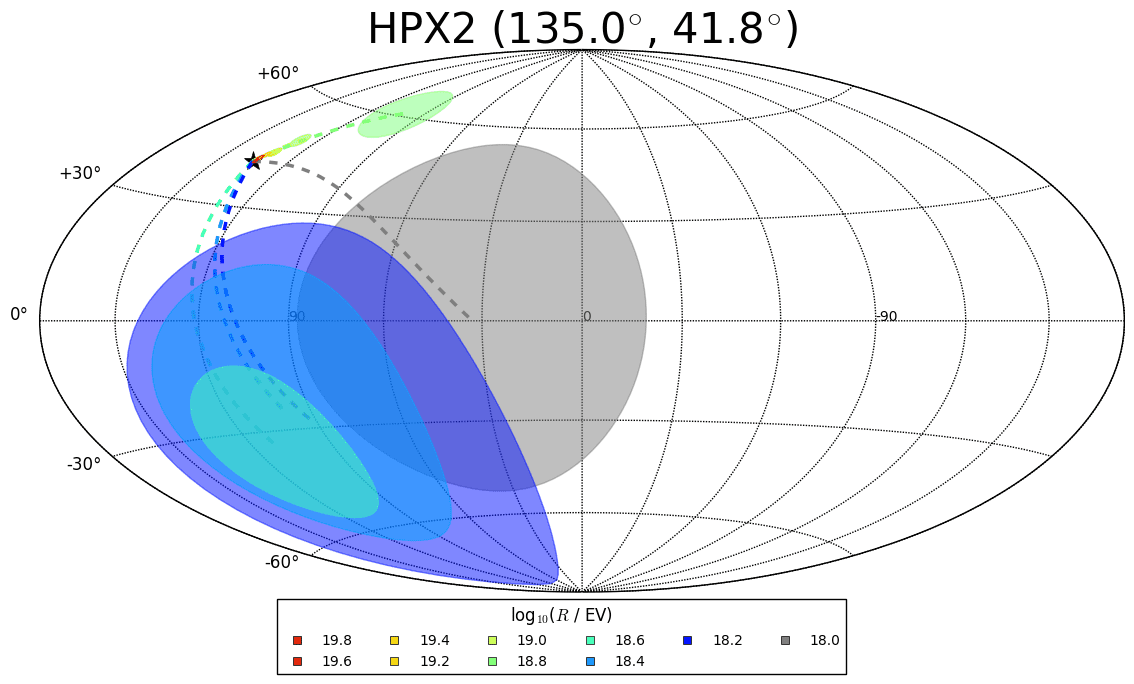}
\end{minipage}
\begin{minipage}[b]{0.32 \textwidth}
\includegraphics[width=1. \textwidth]{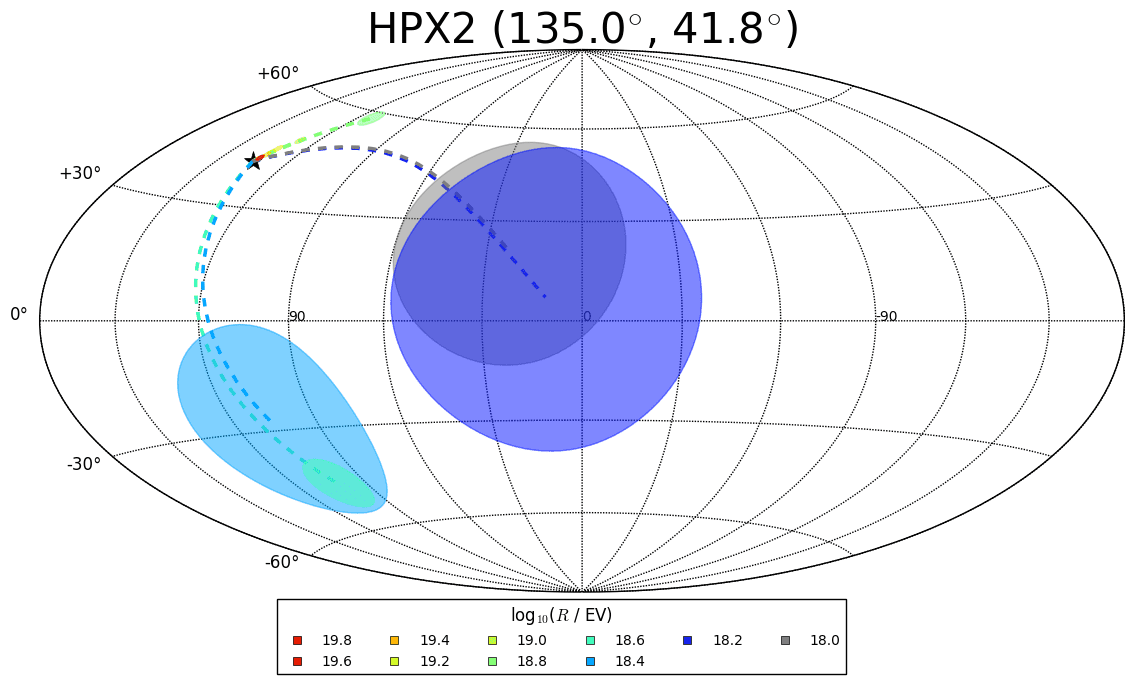}
\end{minipage}
\begin{minipage}[b]{0.32 \textwidth}
\includegraphics[width=1. \textwidth]{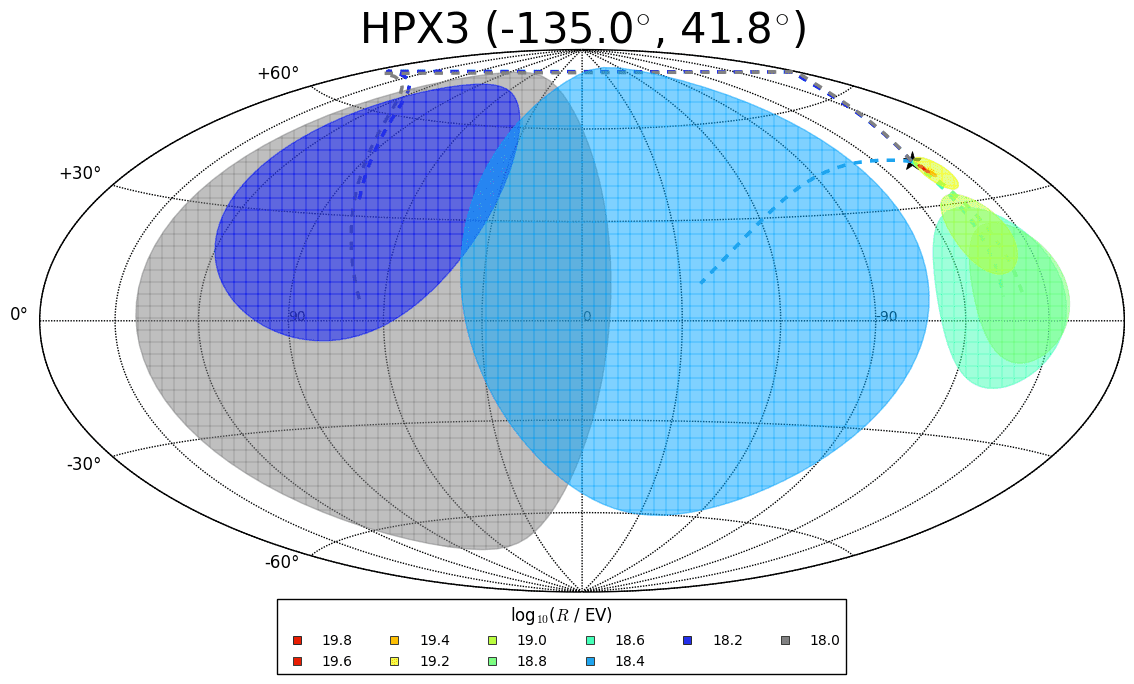}
\end{minipage}
\begin{minipage}[b]{0.32 \textwidth}
\includegraphics[width=1. \textwidth]{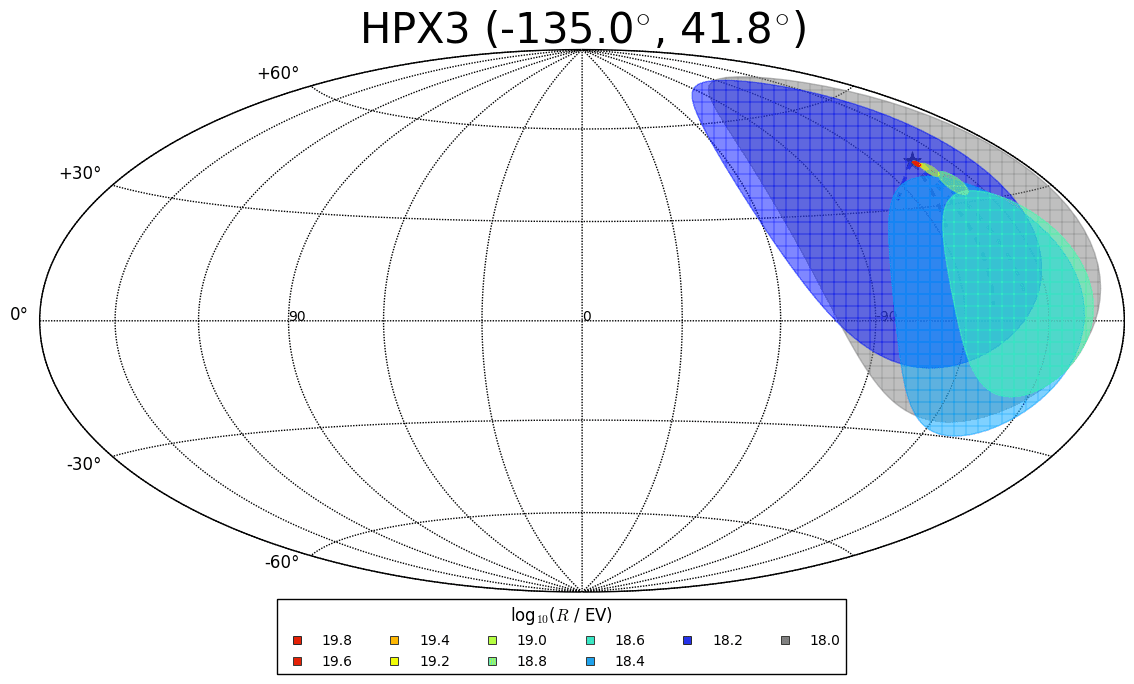}
\end{minipage}
\begin{minipage}[b]{0.32 \textwidth}
\includegraphics[width=1. \textwidth]{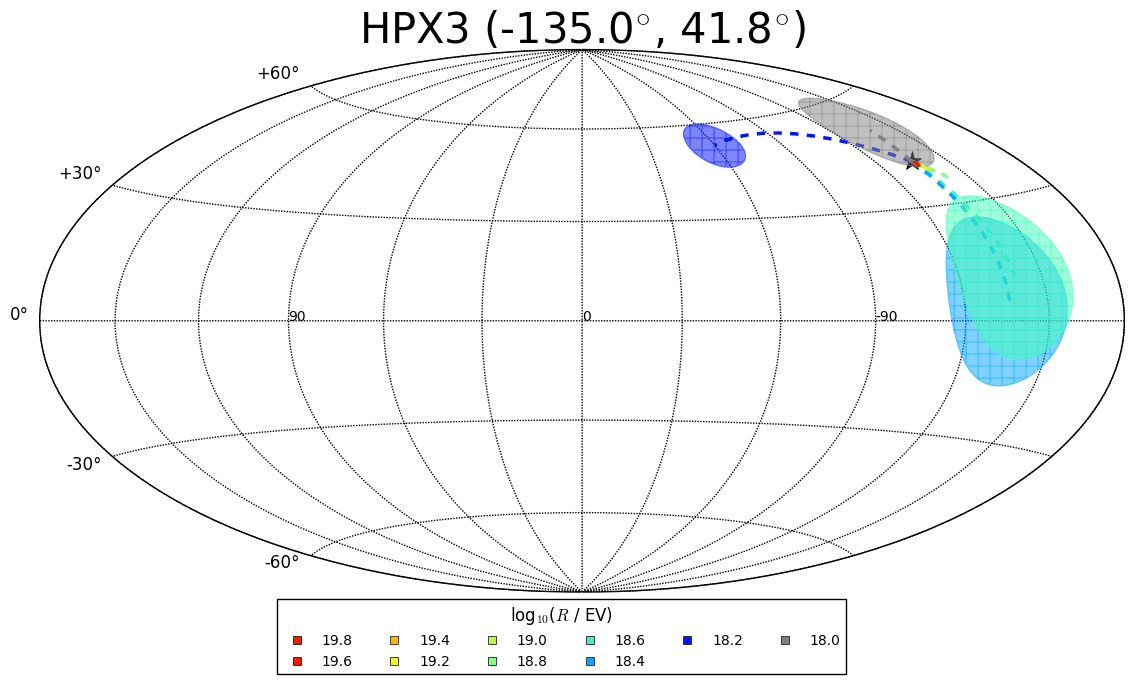}
\end{minipage}
\begin{minipage}[b]{0.32 \textwidth}
\includegraphics[width=1. \textwidth]{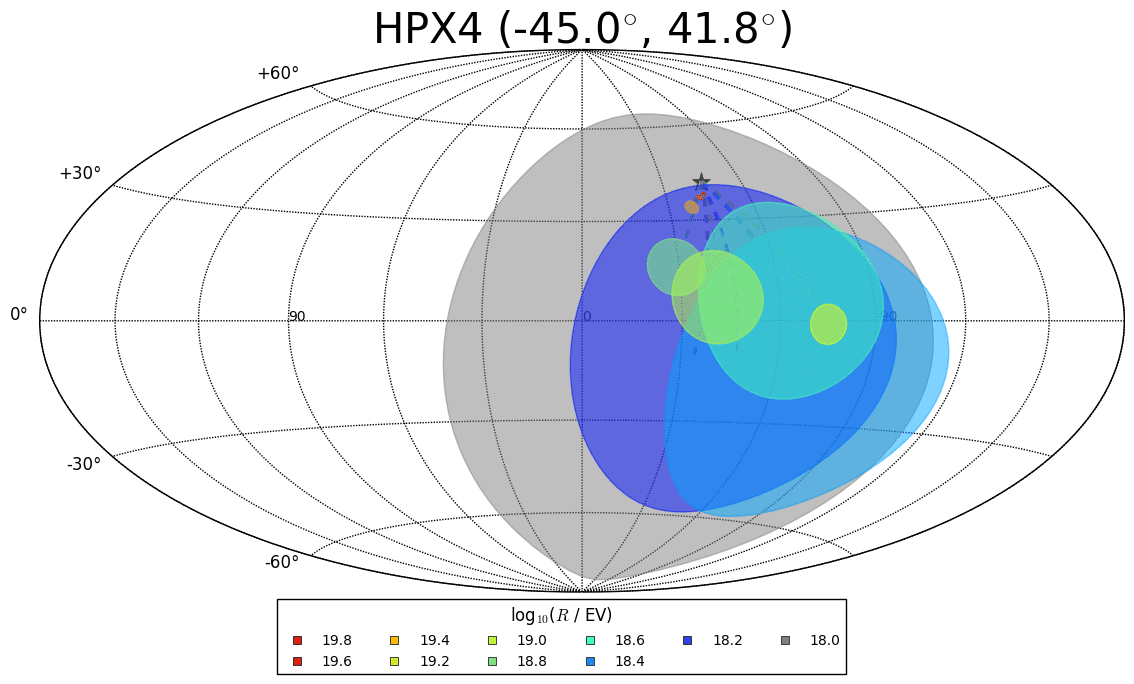}
\end{minipage}
\begin{minipage}[b]{0.32 \textwidth}
\includegraphics[width=1. \textwidth]{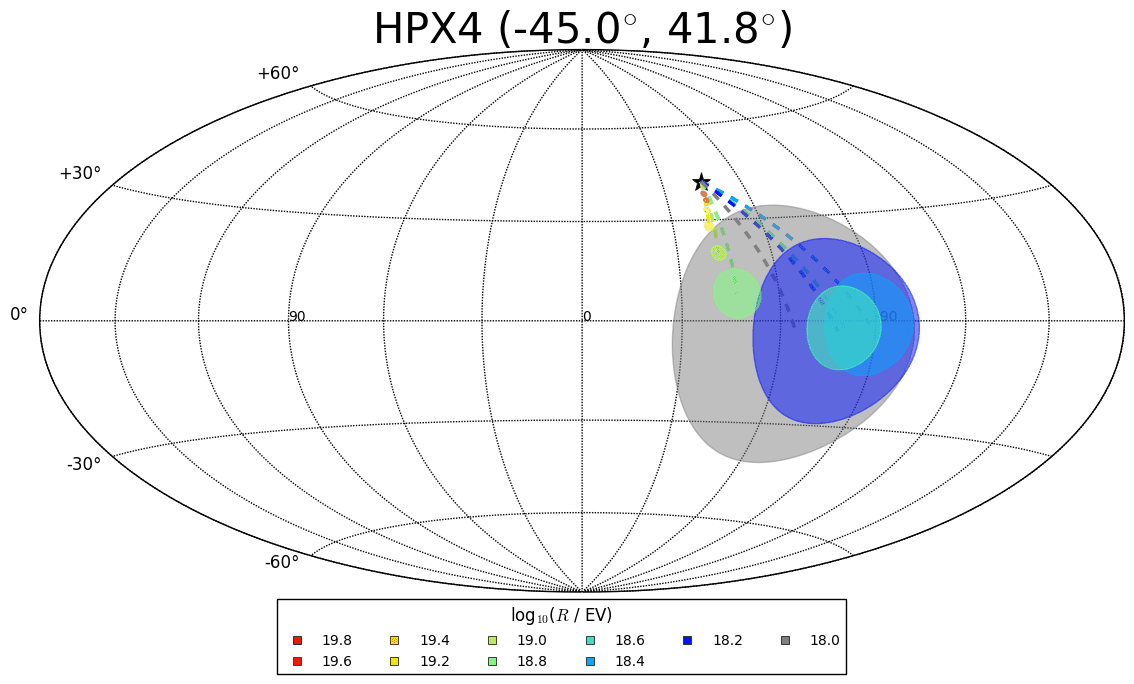}
\end{minipage}
\begin{minipage}[b]{0.32 \textwidth}
\includegraphics[width=1. \textwidth]{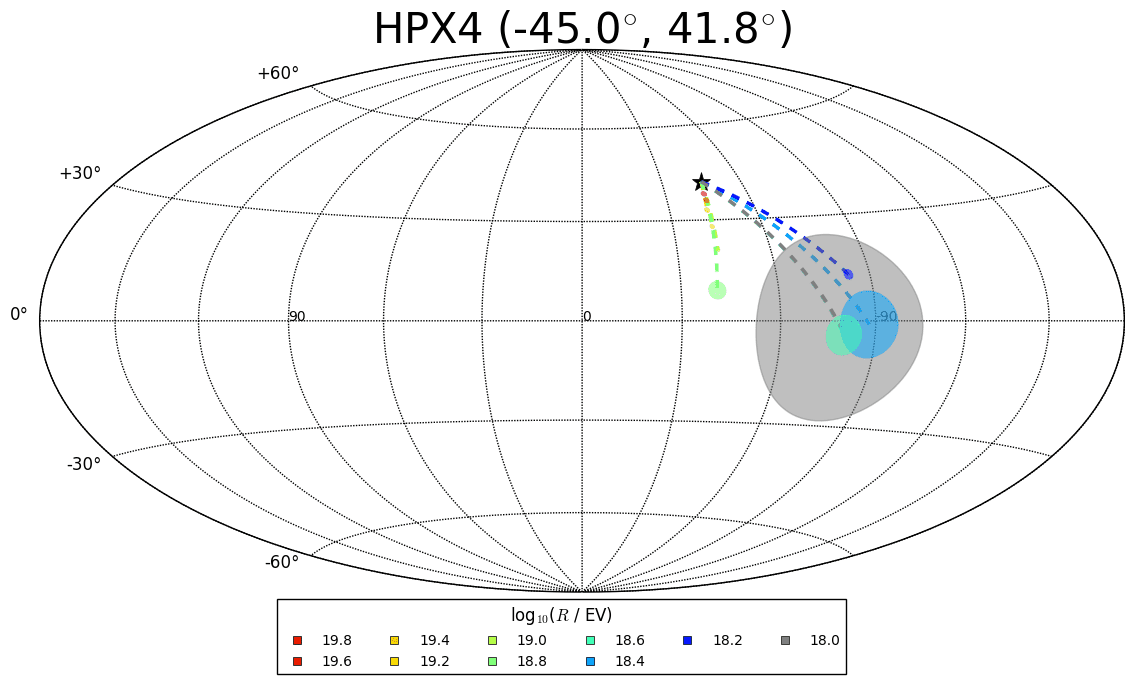}
\end{minipage}
\begin{minipage}[b]{0.32 \textwidth}
\includegraphics[width=1. \textwidth]{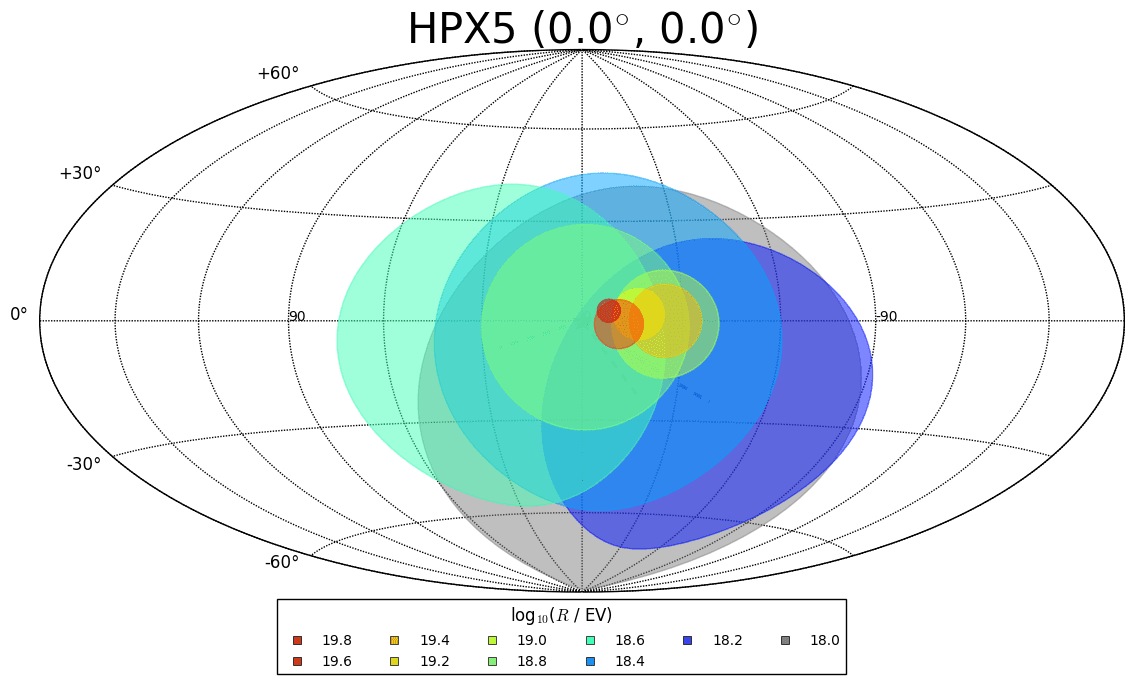}
\end{minipage}
\begin{minipage}[b]{0.32 \textwidth}
\includegraphics[width=1. \textwidth]{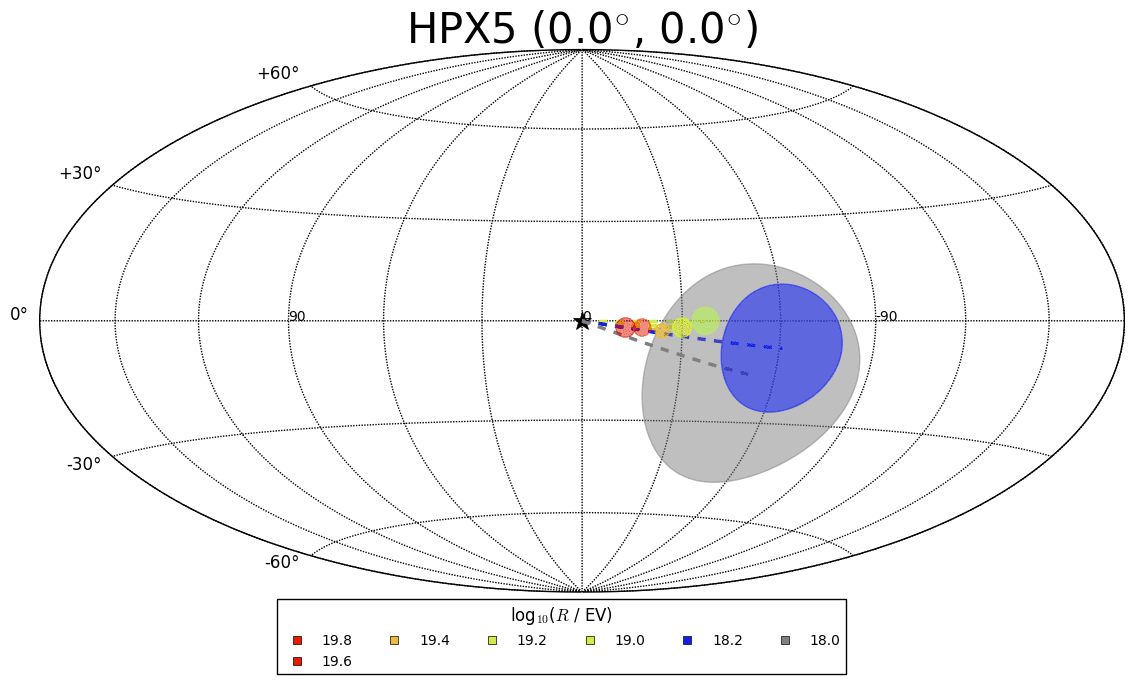}
\end{minipage}
\begin{minipage}[b]{0.32 \textwidth}
\includegraphics[width=1. \textwidth]{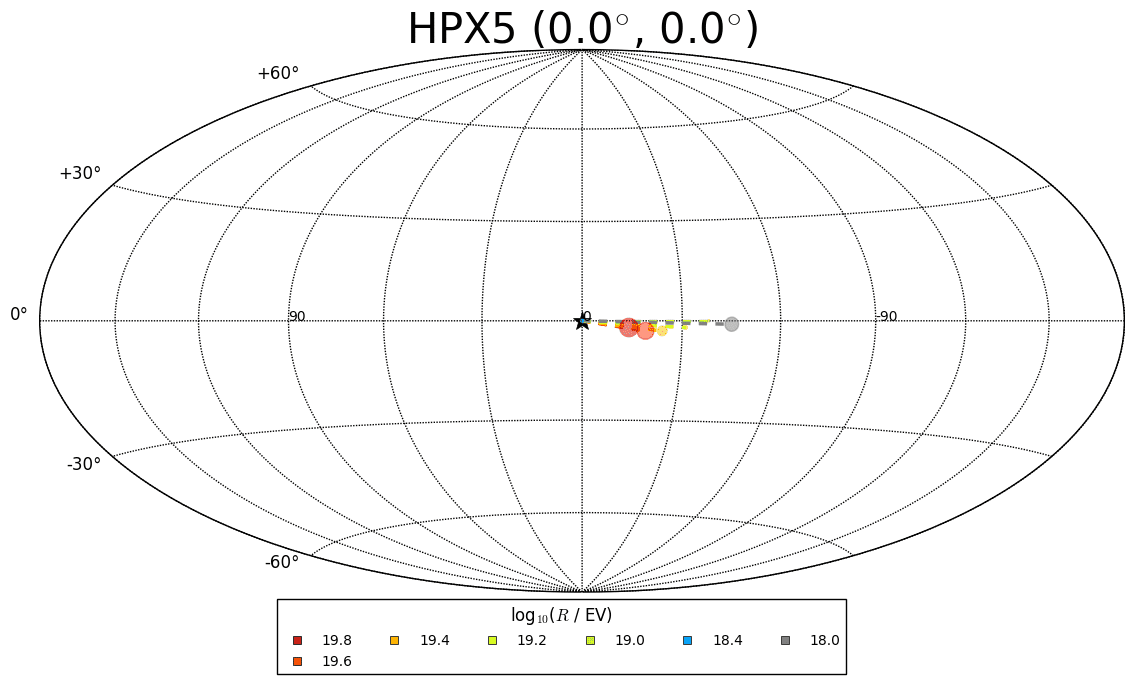}
\end{minipage}
\begin{minipage}[b]{0.32 \textwidth}
\includegraphics[width=1. \textwidth]{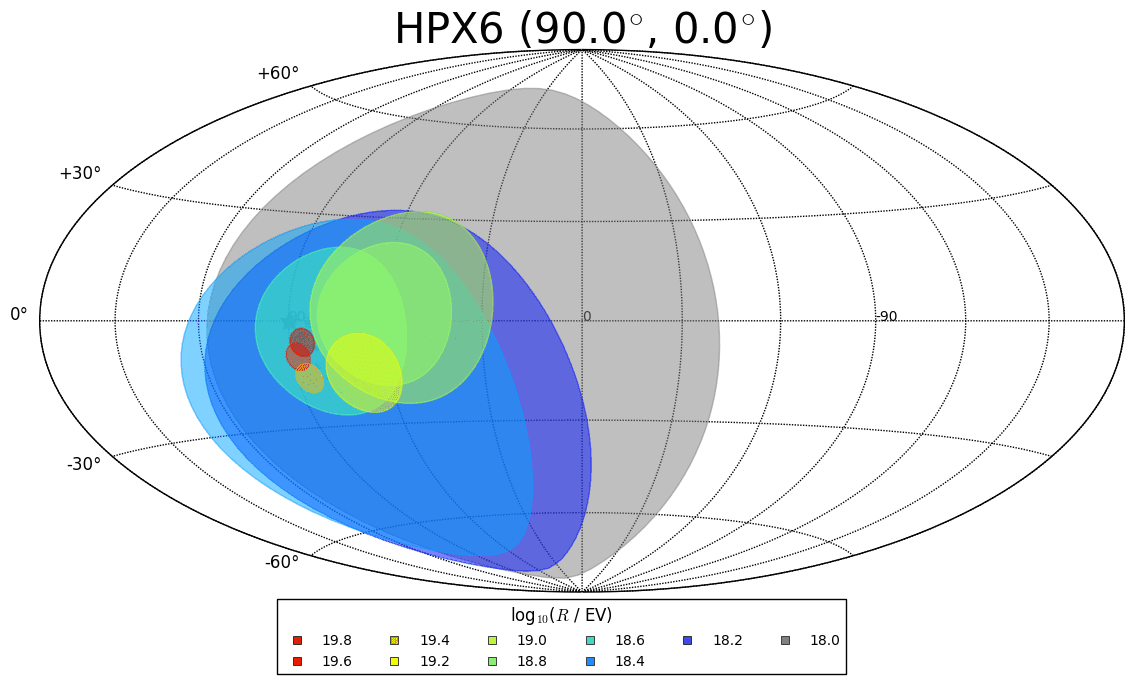}
\end{minipage}
\begin{minipage}[b]{0.32 \textwidth}
\includegraphics[width=1. \textwidth]{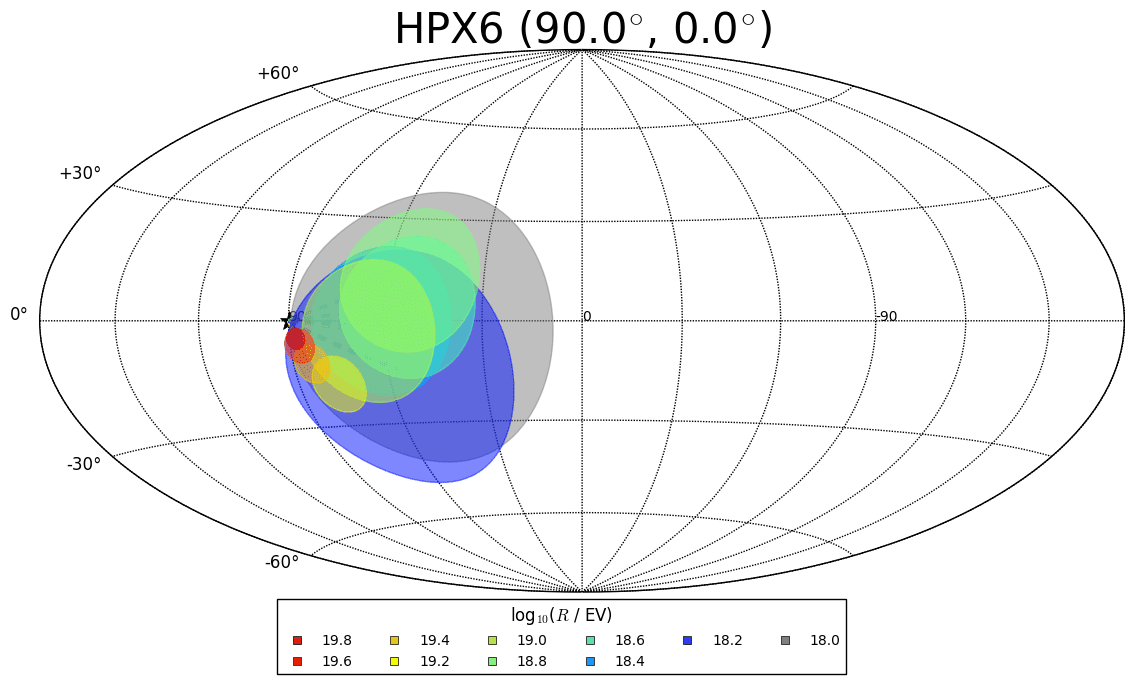}
\end{minipage}
\begin{minipage}[b]{0.32 \textwidth}
\includegraphics[width=1. \textwidth]{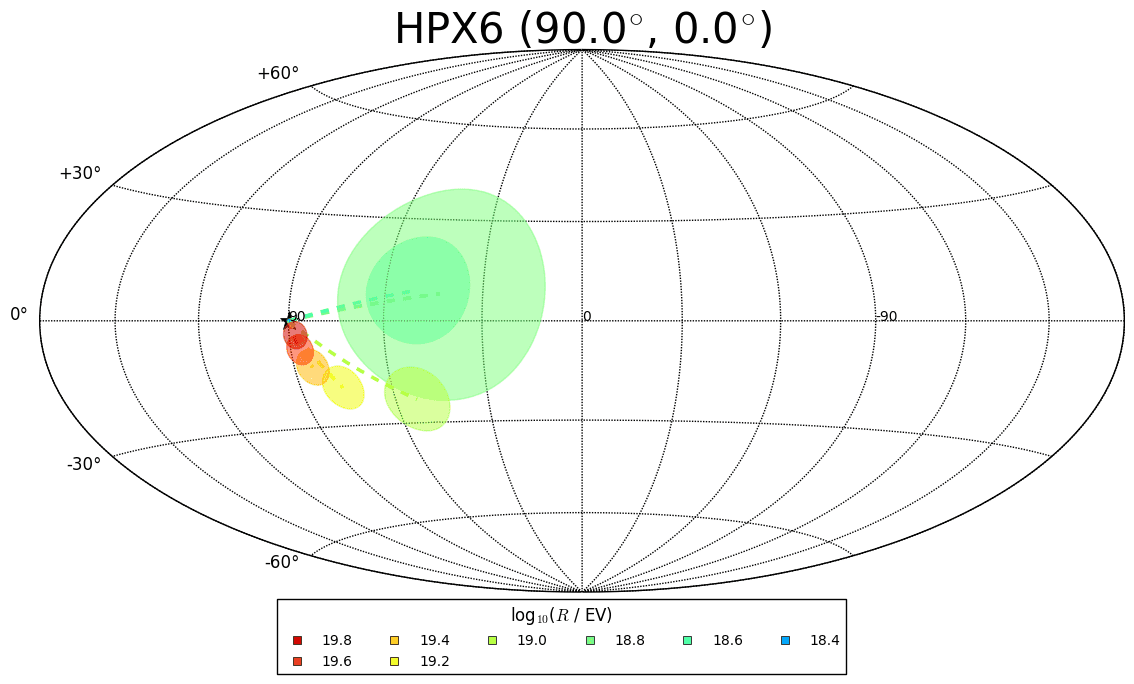}
\end{minipage}
\begin{minipage}[b]{0.32 \textwidth}
\includegraphics[width=1. \textwidth]{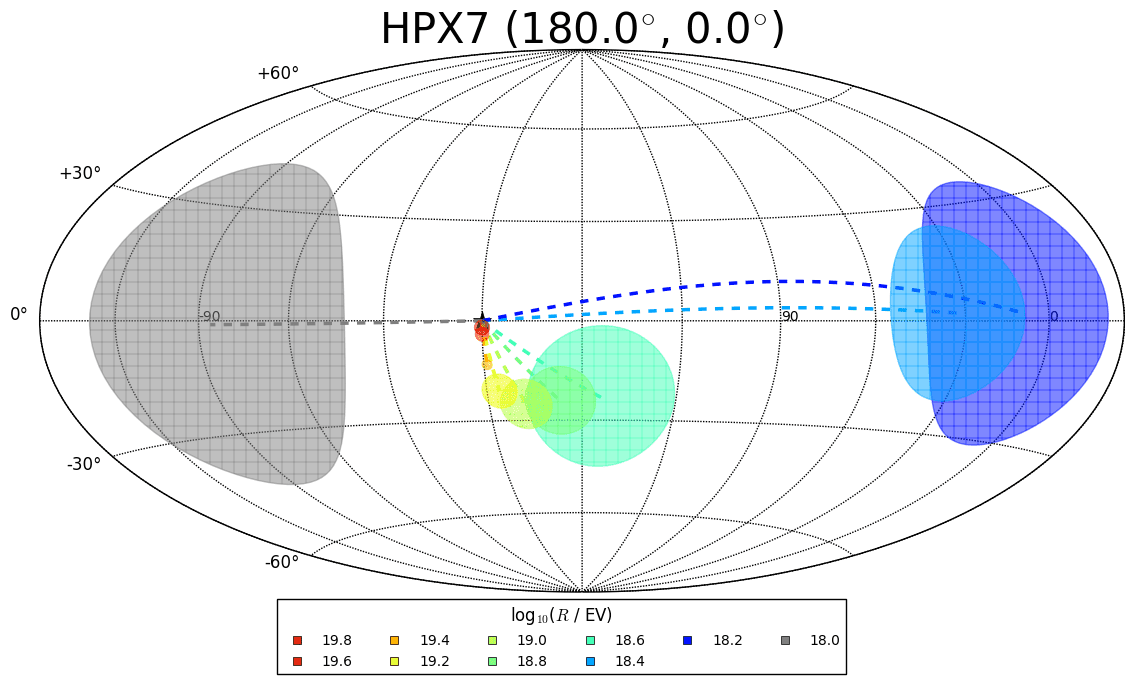}
\end{minipage}
\begin{minipage}[b]{0.32 \textwidth}
\includegraphics[width=1. \textwidth]{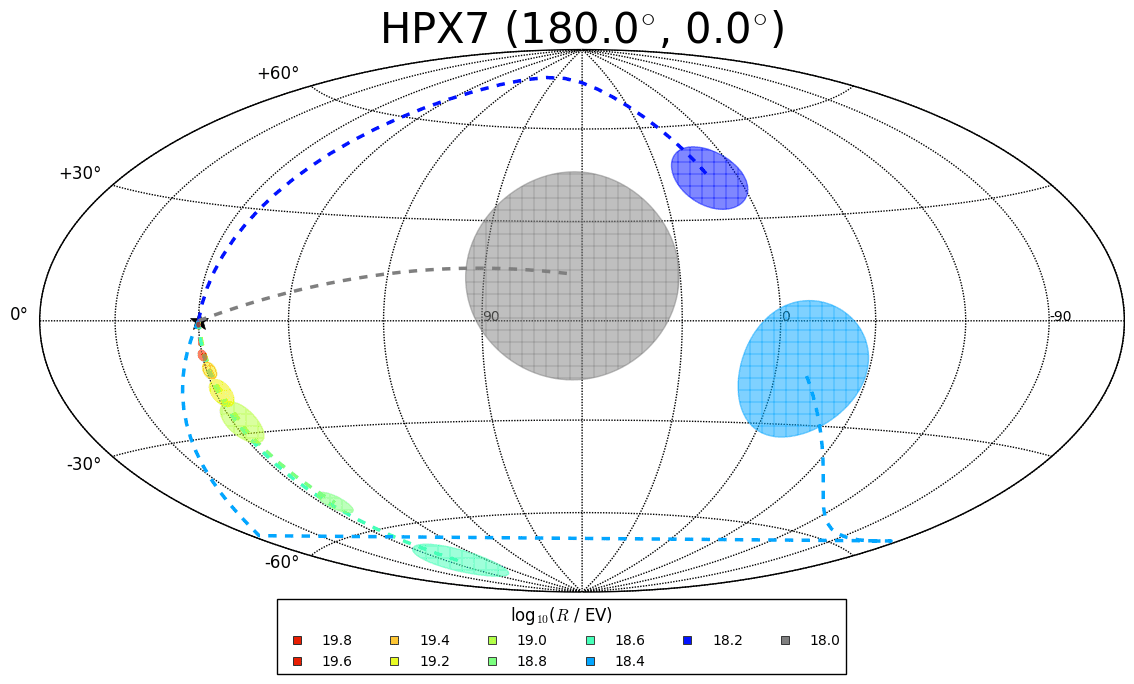}
\end{minipage}
\begin{minipage}[b]{0.32 \textwidth}
\includegraphics[width=1. \textwidth]{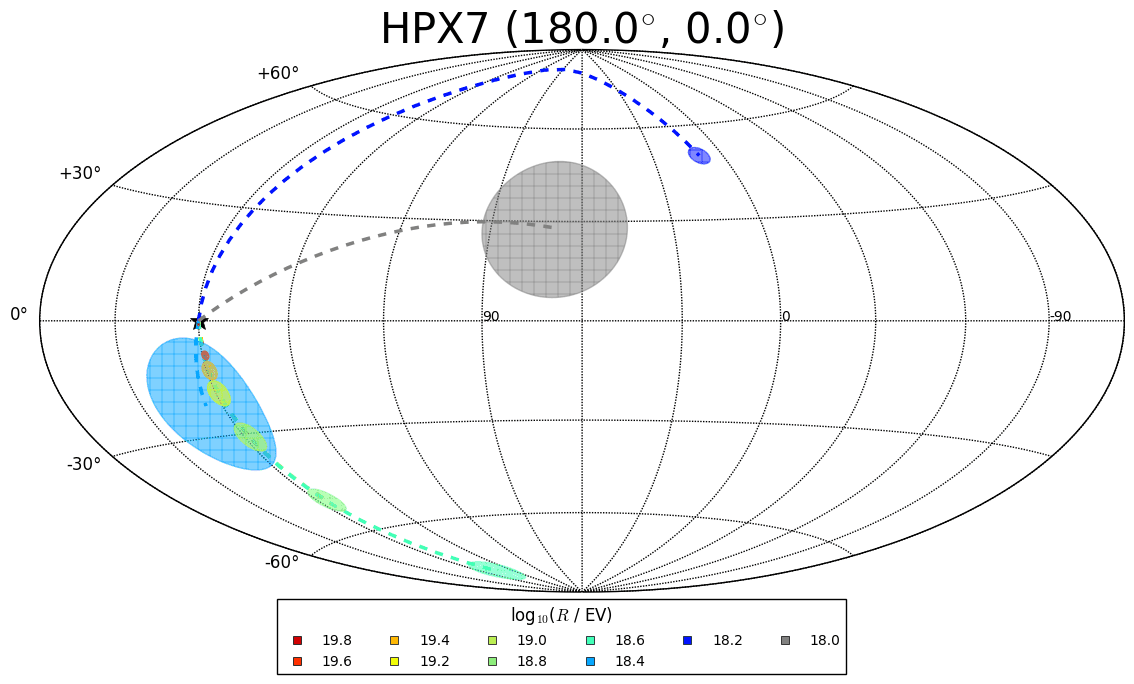}
\end{minipage}
\caption{{
\scriptsize Centroids and schematic RMS spread in arrival directions as a function of rigidity, for log$_{10}(R/V)$ values 18.0, 18.2, 18.4, 18.6, 18.8, 19.0, 19.2, 19.4, 19.6 and 19.8, for the first 7 HPX grid source directions.  Left, center, and right columns depict $L_{coh}=$ 100 and 30 pc, and coherent-only cases, respectively.   
The rainbow+grey color-rigidity mapping is shown in the legend.  Due to large median values, the disk radii in the HPX3 and HPZ7 plots are rescaled smaller by 90\%, and 50\% respectively.
Note that the coordinates are rotated for the HPX2 100pc and all 3 HPX7 plots for easier-to-grasp skyplots.  The dashed lines just connect the source to the centroids to aid proper identification -- they have nothing to do with the actual trajectory of any UHECR.  ``Missing" rigidity values are instances of blind directions for the given rigidity. }} 
\label{plt:cent1to7_allL}
\end{figure}

\begin{figure}[htb]
\hspace{-0.3in}
\centering
\begin{minipage}[b]{0.32 \textwidth}
\includegraphics[width=1. \textwidth]{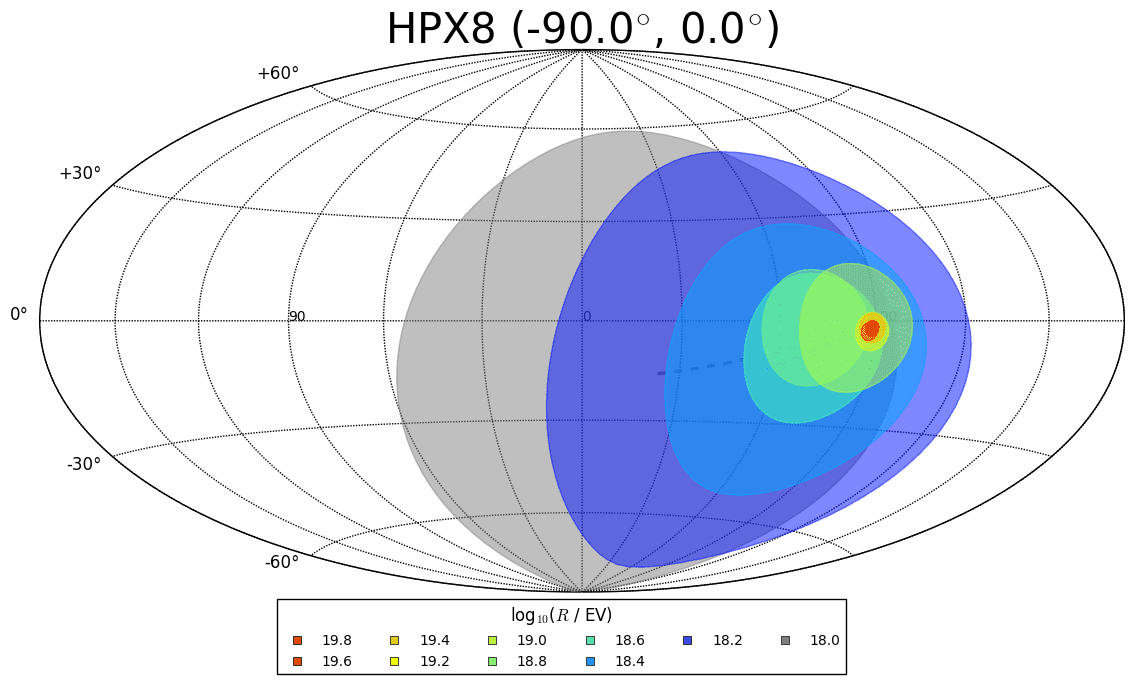}
\end{minipage}
\begin{minipage}[b]{0.32 \textwidth}
\includegraphics[width=1. \textwidth]{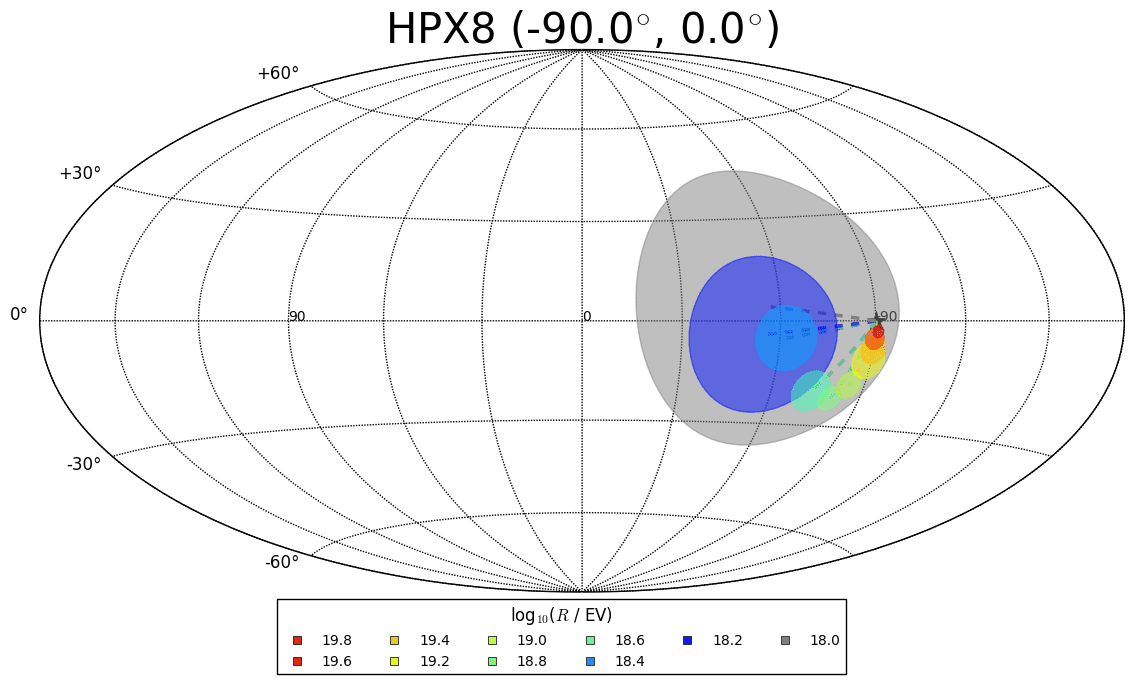}
\end{minipage}
\begin{minipage}[b]{0.32 \textwidth}
\includegraphics[width=1. \textwidth]{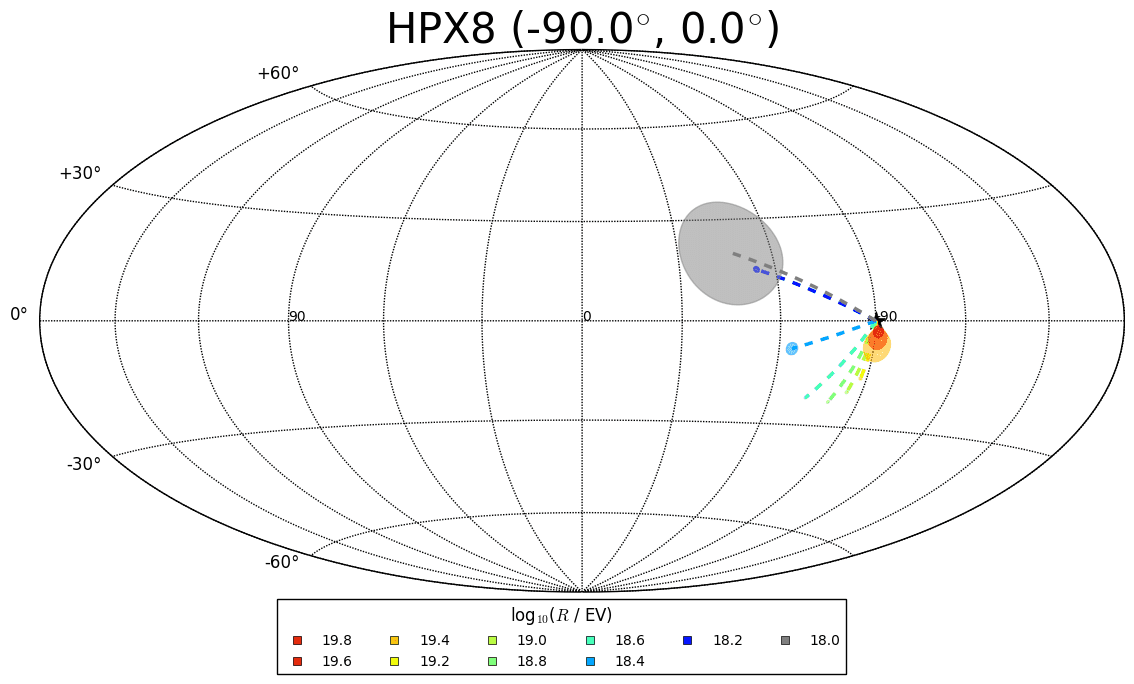}
\end{minipage}
\begin{minipage}[b]{0.32 \textwidth}
\includegraphics[width=1. \textwidth]{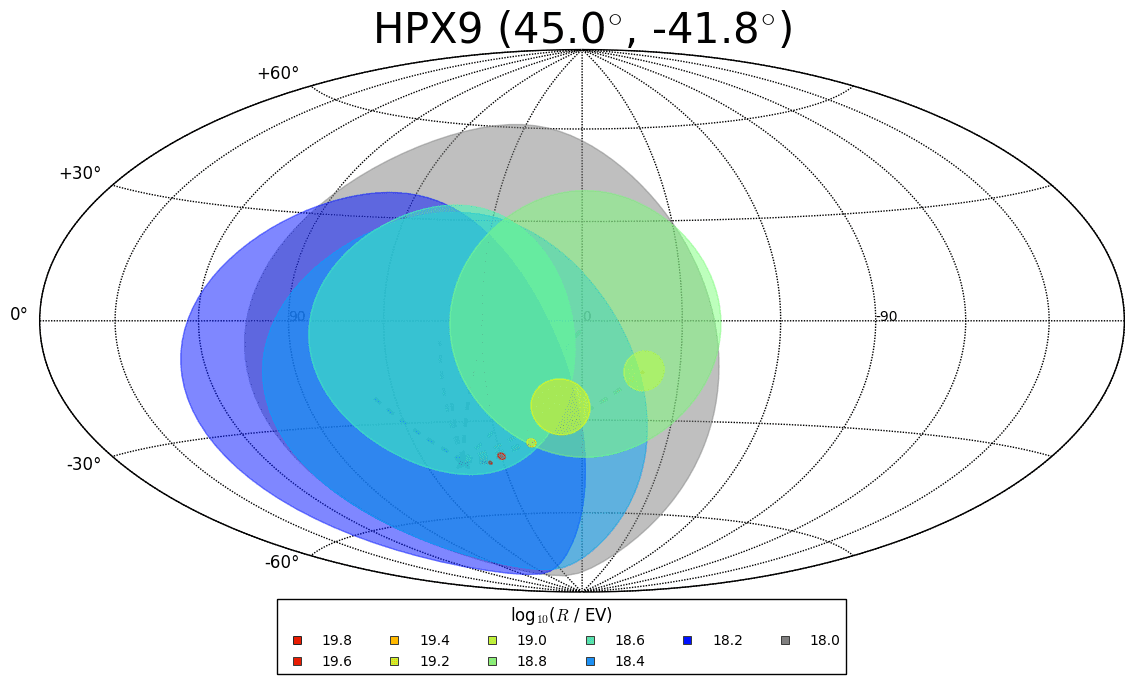}
\end{minipage}
\begin{minipage}[b]{0.32 \textwidth}
\includegraphics[width=1. \textwidth]{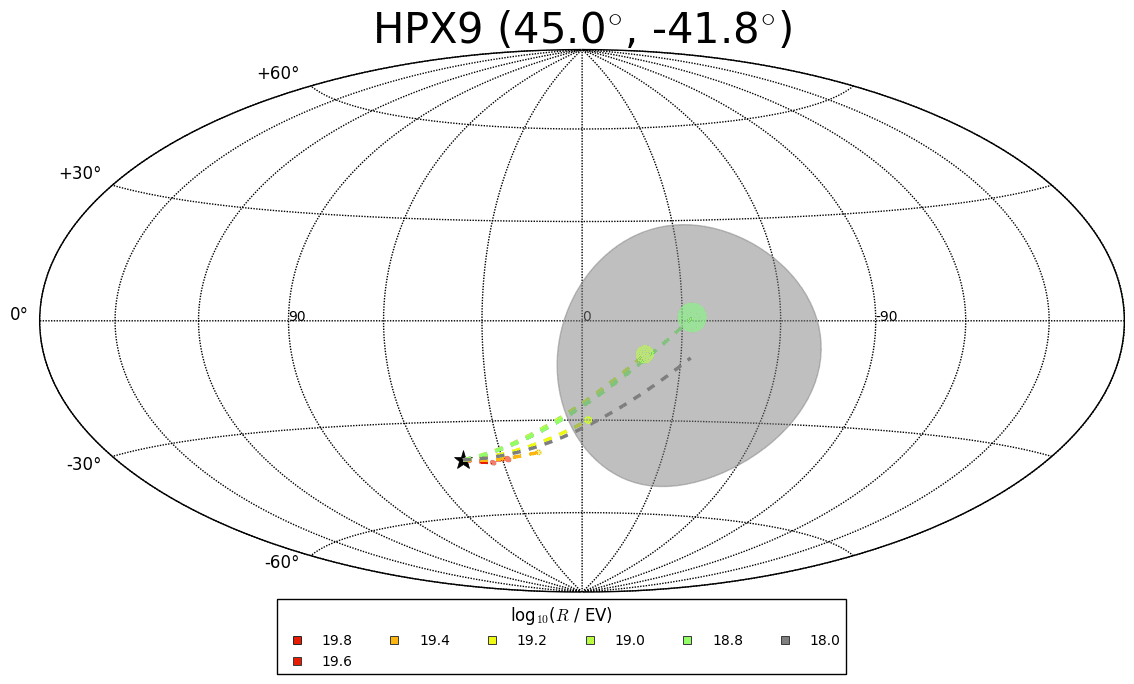}
\end{minipage}
\begin{minipage}[b]{0.32 \textwidth}
\includegraphics[width=1. \textwidth]{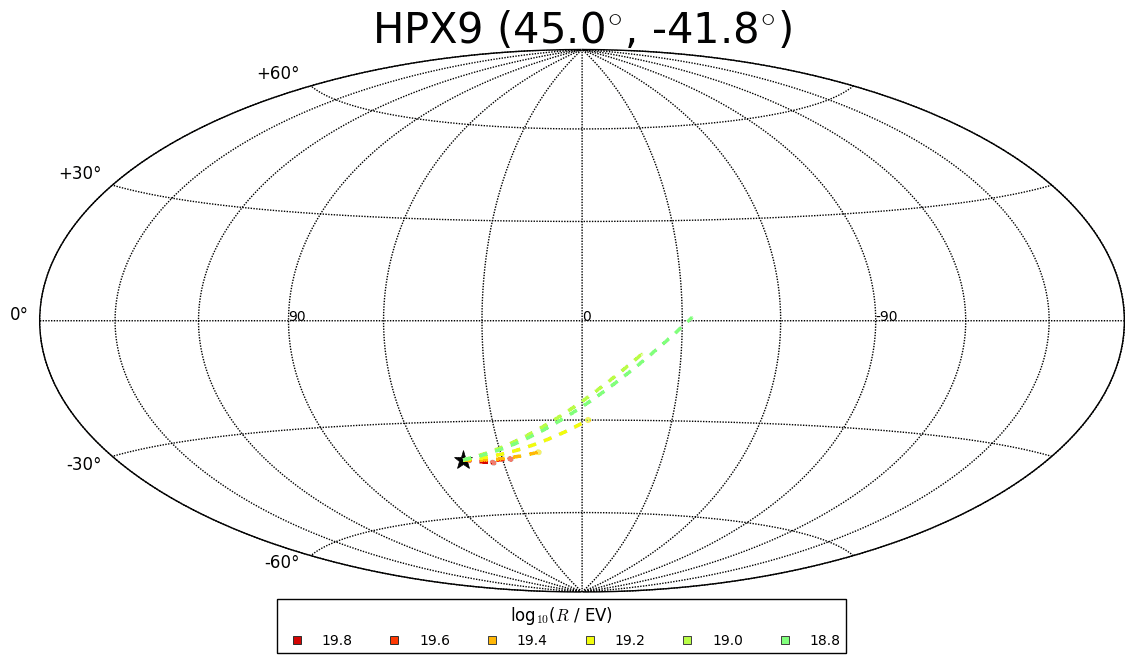}
\end{minipage}
\begin{minipage}[b]{0.32 \textwidth}
\includegraphics[width=1. \textwidth]{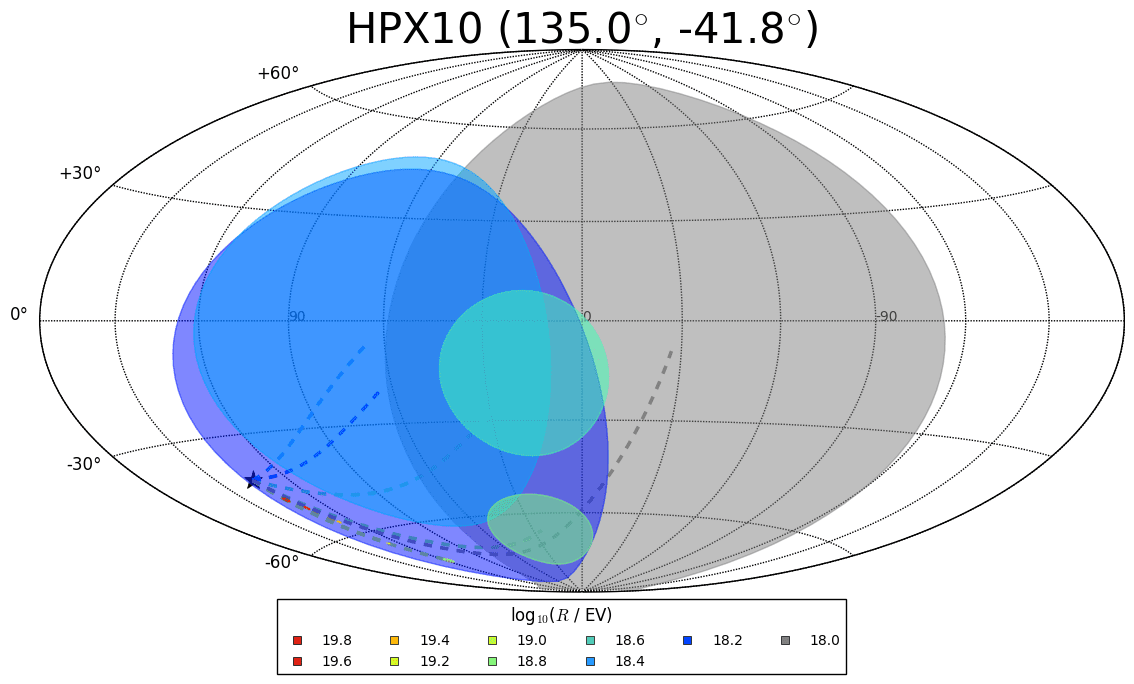}
\end{minipage}
\begin{minipage}[b]{0.32 \textwidth}
\includegraphics[width=1. \textwidth]{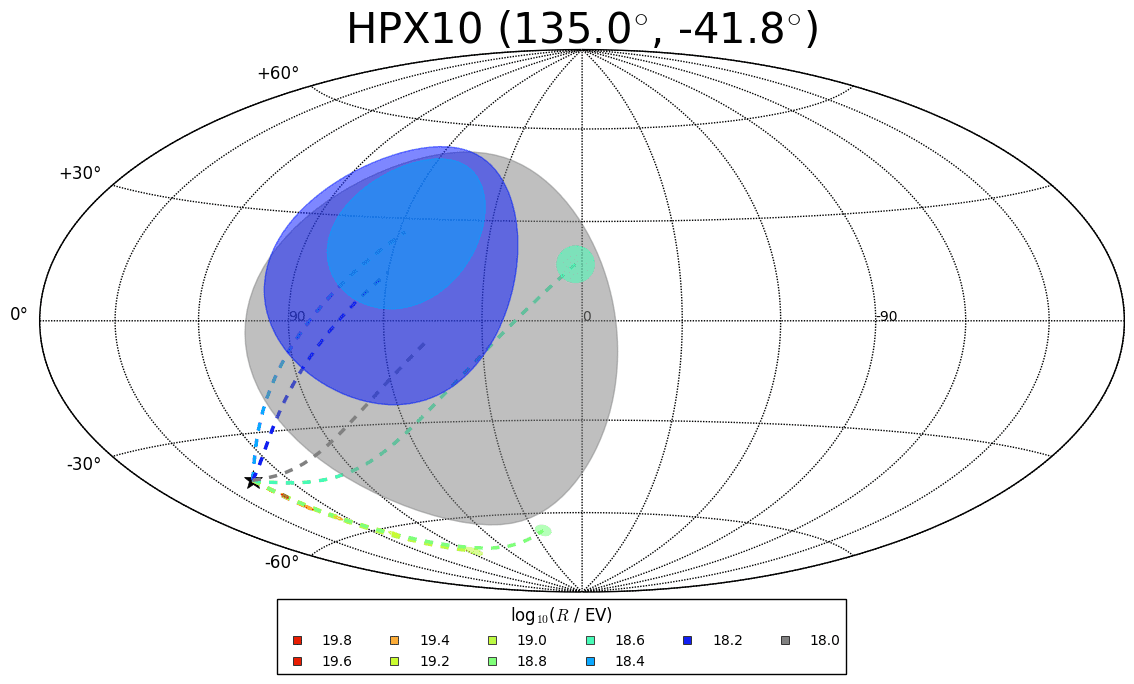}
\end{minipage}
\begin{minipage}[b]{0.32 \textwidth}
\includegraphics[width=1. \textwidth]{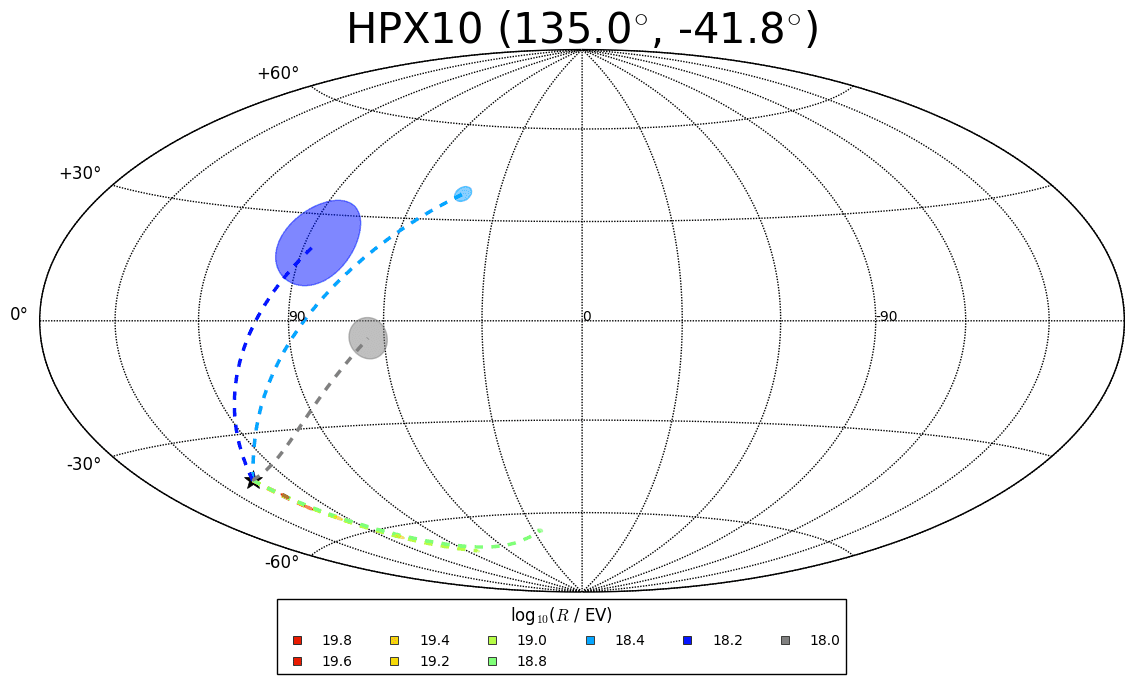}
\end{minipage}
\begin{minipage}[b]{0.32 \textwidth}
\includegraphics[width=1. \textwidth]{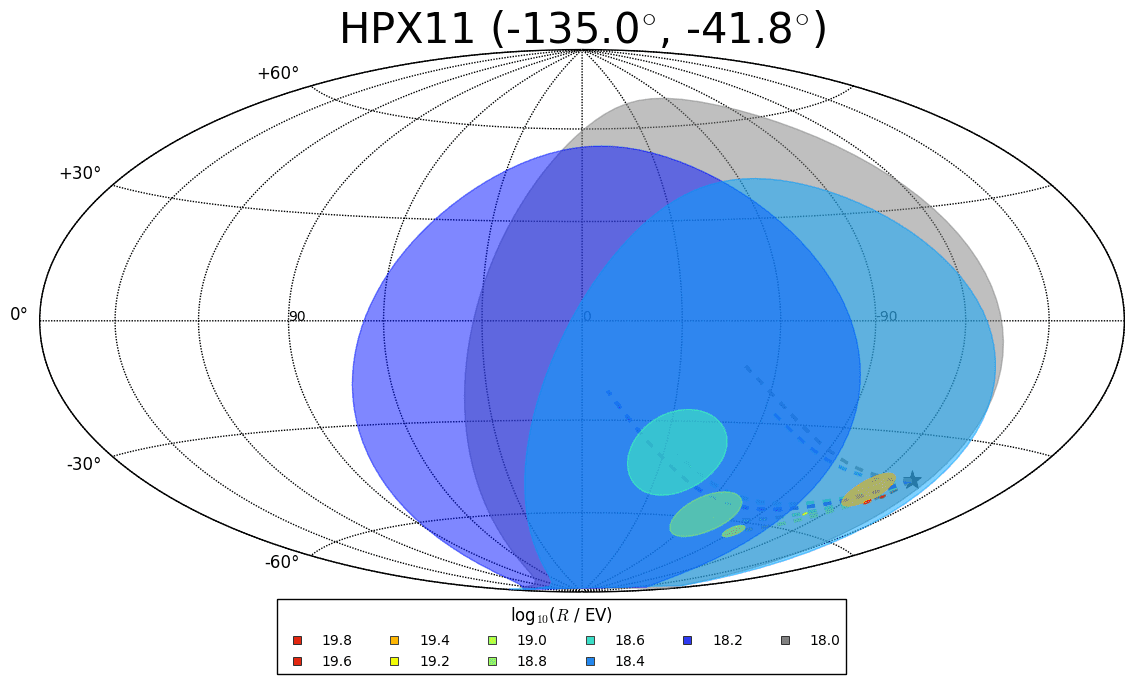}
\end{minipage}
\begin{minipage}[b]{0.32 \textwidth}
\includegraphics[width=1. \textwidth]{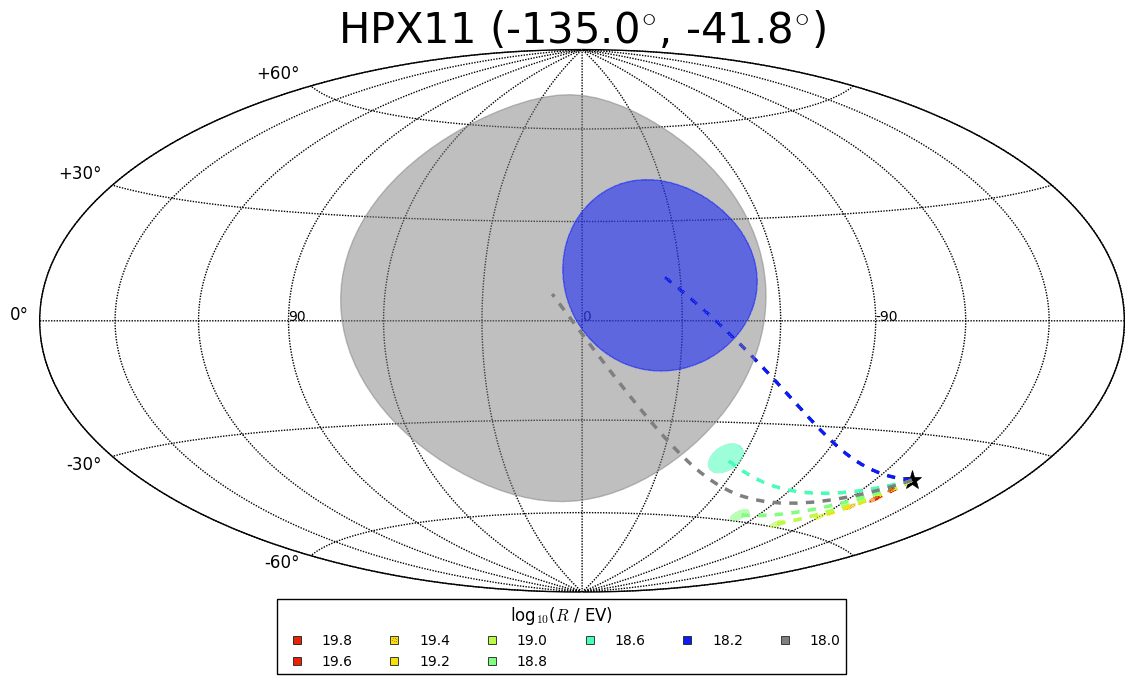}
\end{minipage}
\begin{minipage}[b]{0.32 \textwidth}
\includegraphics[width=1. \textwidth]{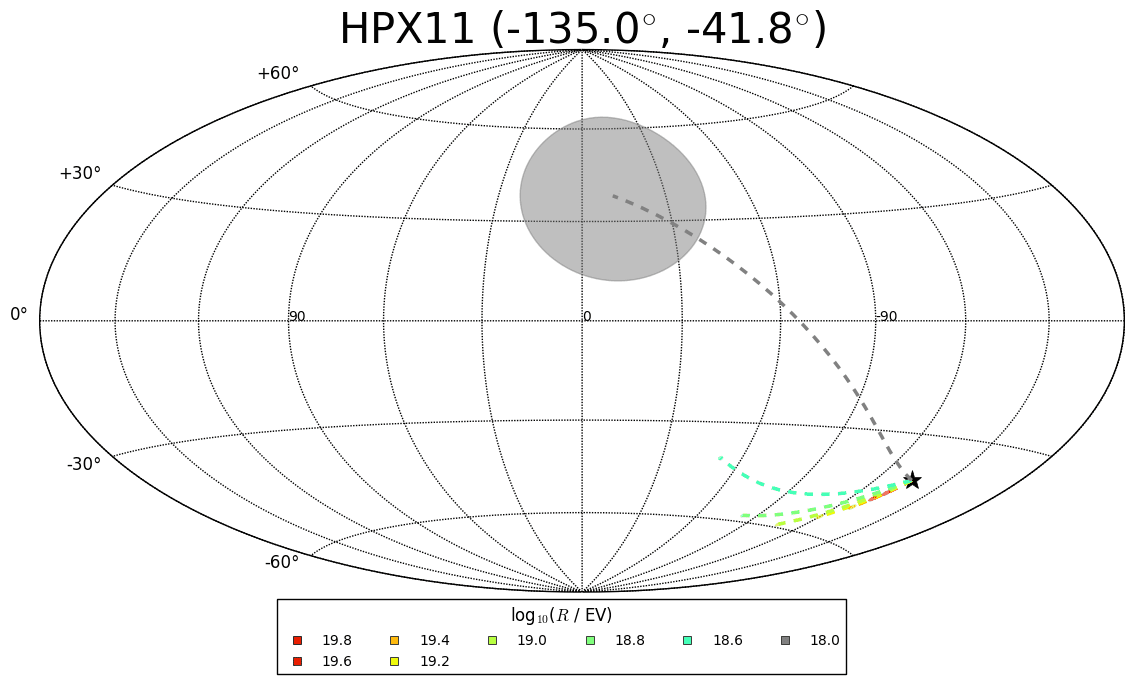}
\end{minipage}
\begin{minipage}[b]{0.32 \textwidth}
\includegraphics[width=1. \textwidth]{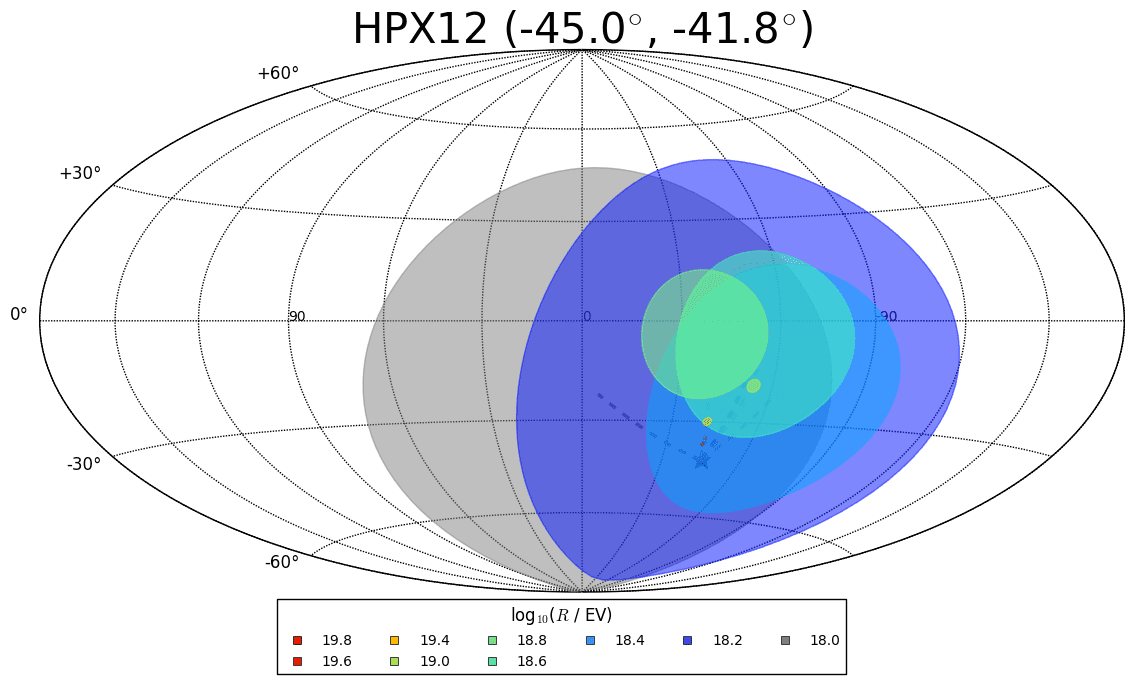}
\end{minipage}
\begin{minipage}[b]{0.32 \textwidth}
\includegraphics[width=1. \textwidth]{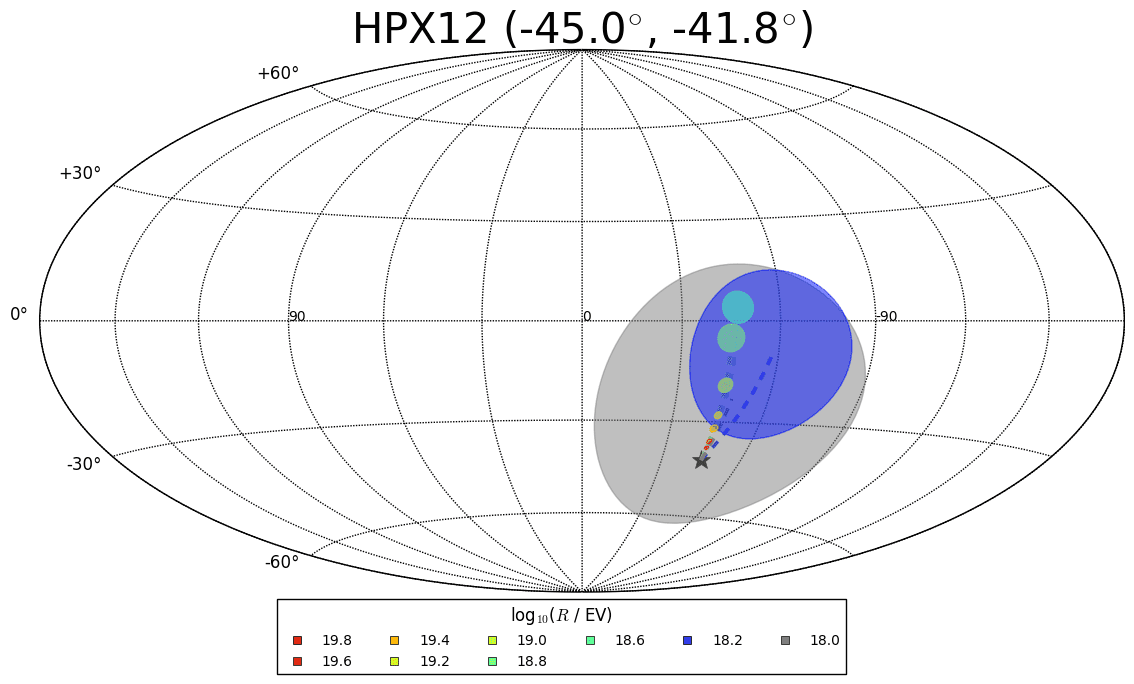}
\end{minipage}
\begin{minipage}[b]{0.32 \textwidth}
\includegraphics[width=1. \textwidth]{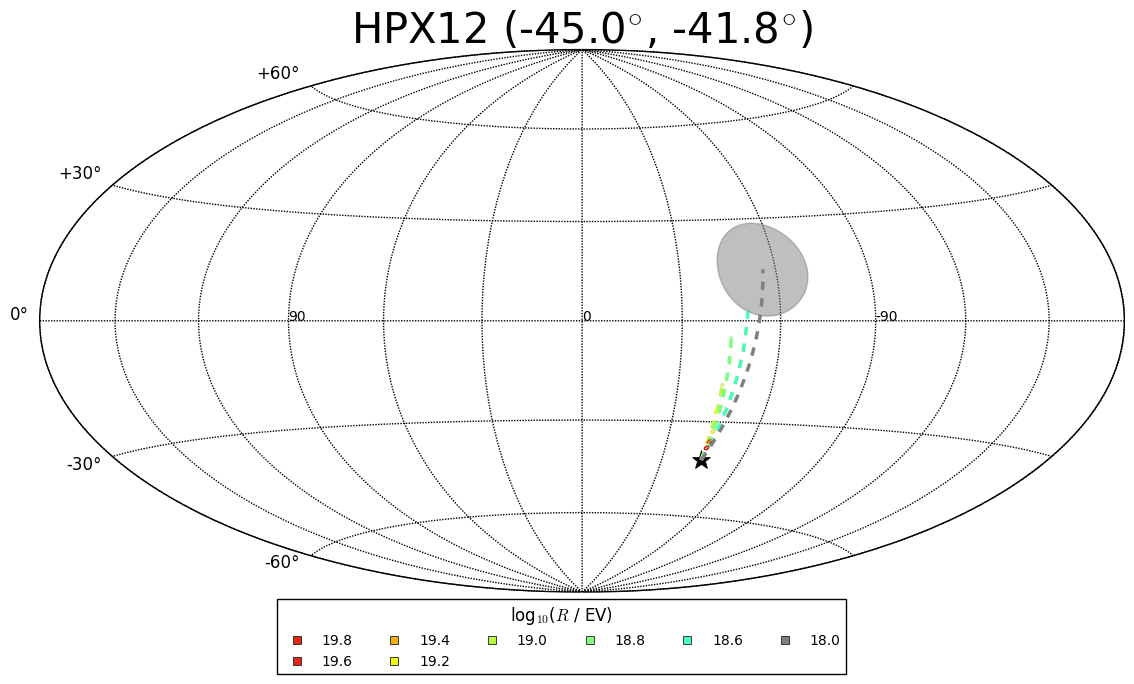}
\end{minipage}
\begin{minipage}[b]{0.32 \textwidth}
\includegraphics[width=1. \textwidth]{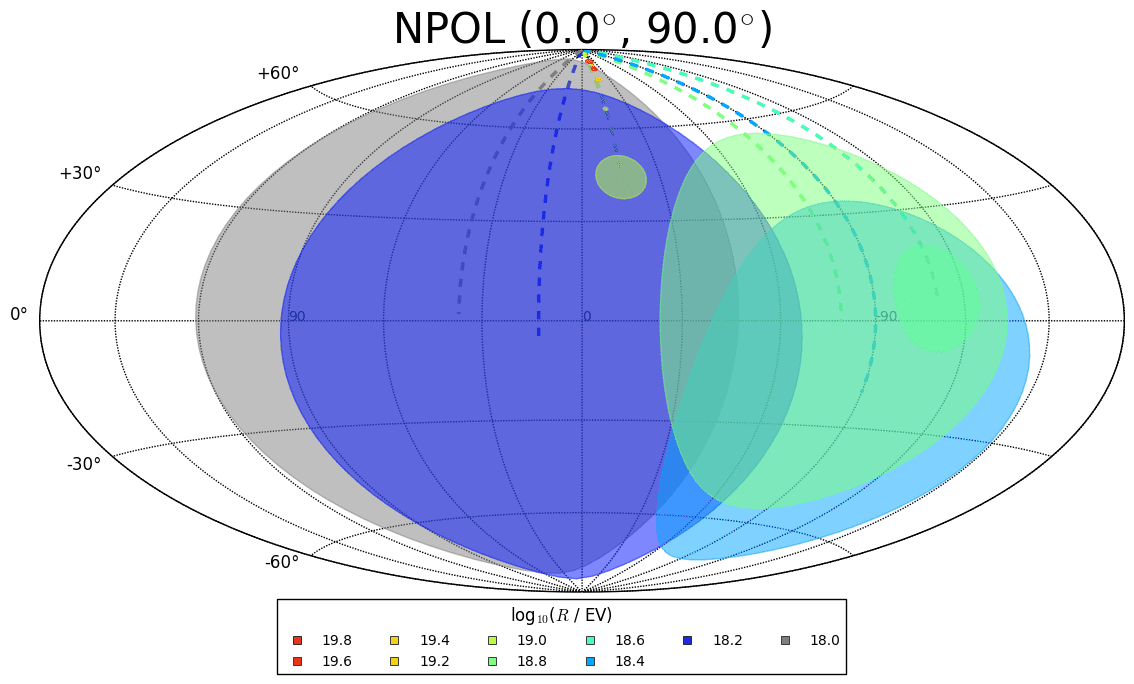}
\end{minipage}
\begin{minipage}[b]{0.32 \textwidth}
\includegraphics[width=1. \textwidth]{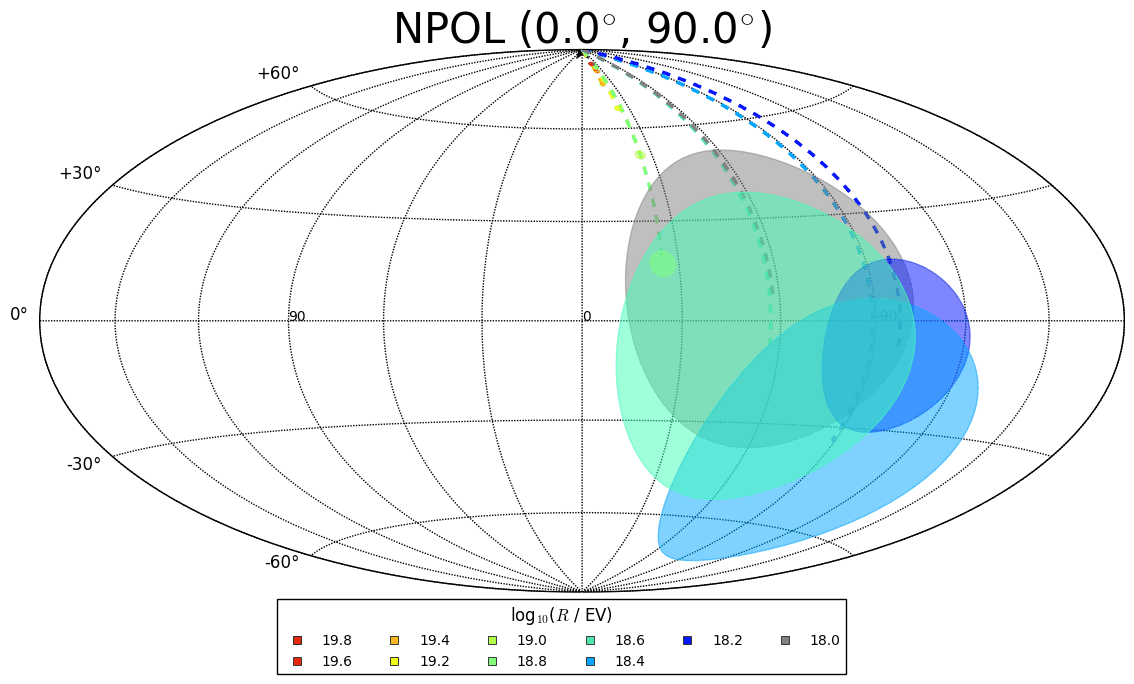}
\end{minipage}
\begin{minipage}[b]{0.32 \textwidth}
\includegraphics[width=1. \textwidth]{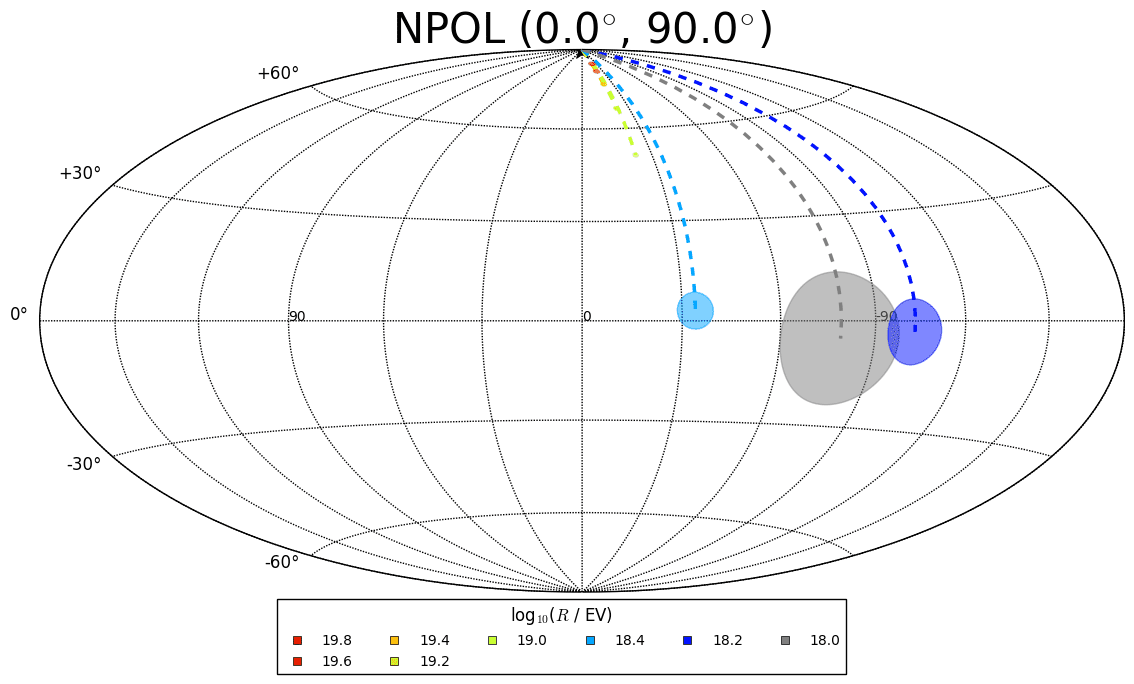}
\end{minipage}
\begin{minipage}[b]{0.32 \textwidth}
\includegraphics[width=1. \textwidth]{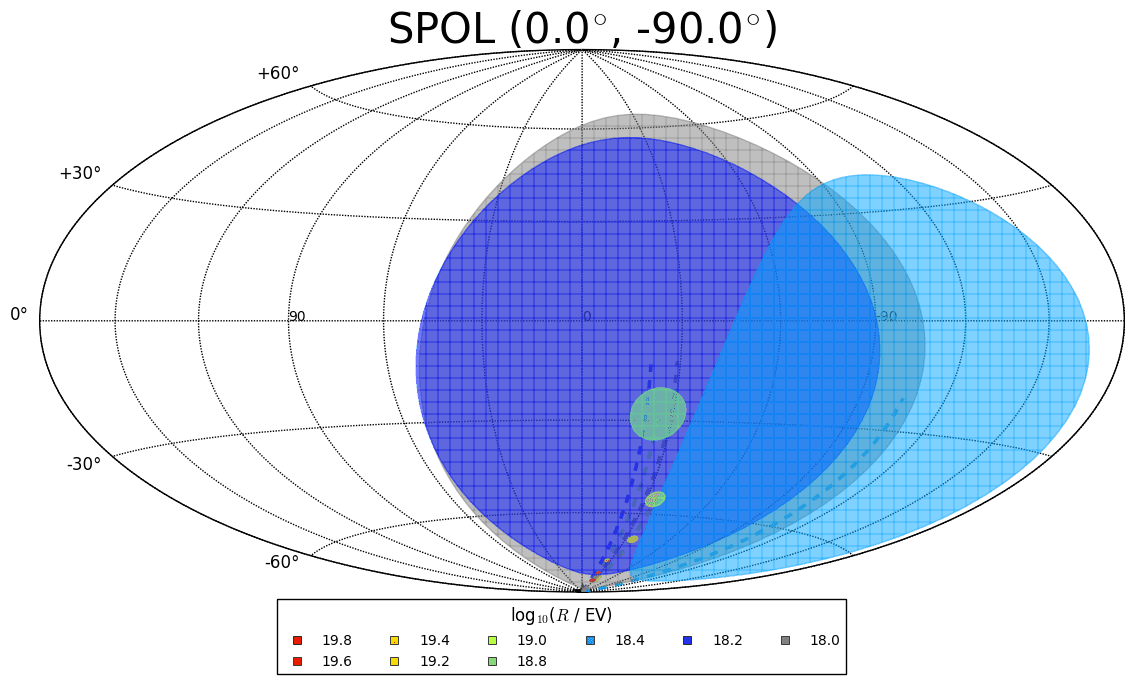}
\end{minipage}
\begin{minipage}[b]{0.32 \textwidth}
\includegraphics[width=1. \textwidth]{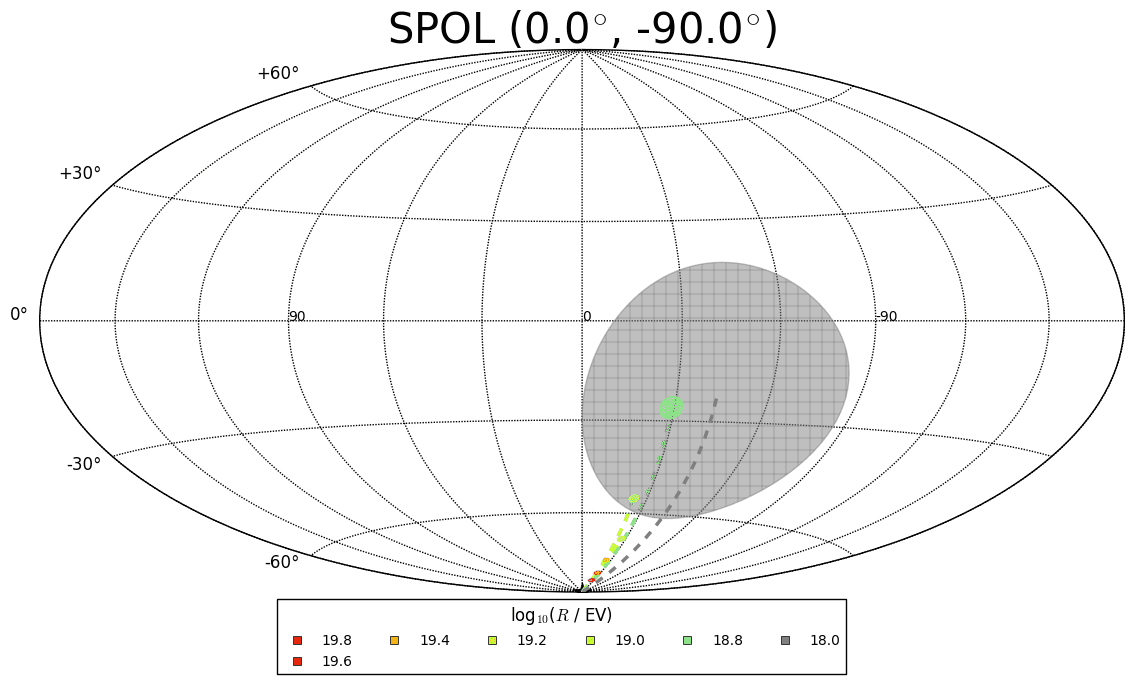}
\end{minipage}
\begin{minipage}[b]{0.32 \textwidth}
\includegraphics[width=1. \textwidth]{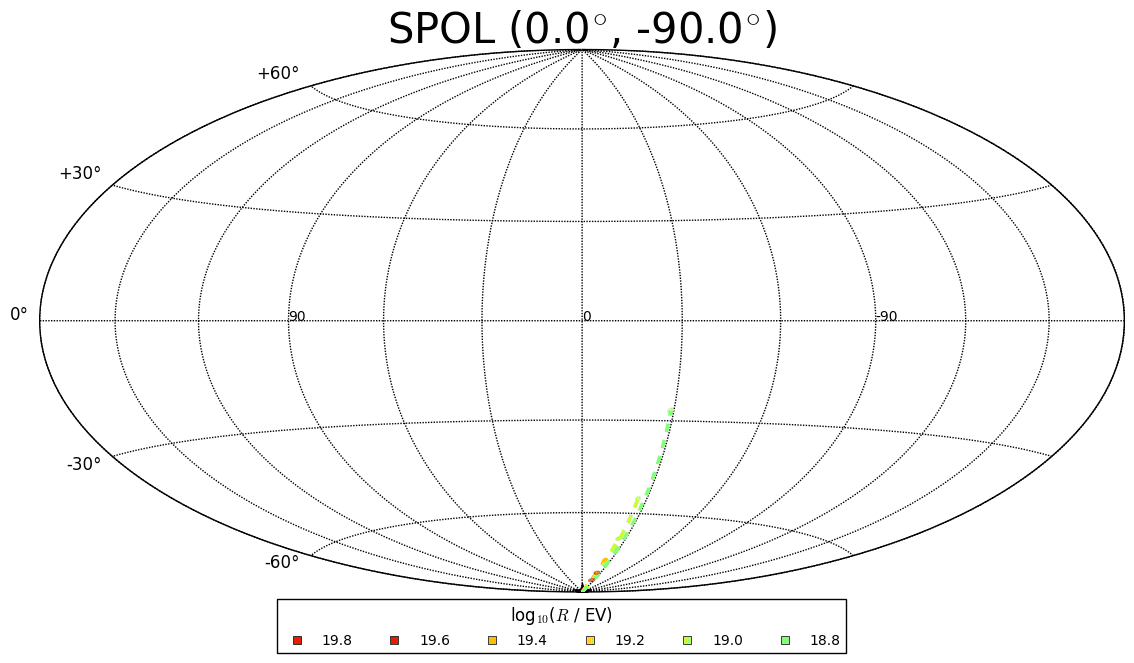}
\end{minipage}
\caption{Centroids and schematic RMS spread in arrival directions, as a function of rigidity as in Fig. \ref{plt:cent1to7_allL}, for the last 5 HPX grid source directions.  Due to large median values, the disk radii in the SPOL plots are rescaled smaller by 90\%. } 
\label{plt:cent8to12_NSP_allL}
\end{figure}

The very diversity of the deflection skyplots make them difficult to assimilate.   In this section we will simplify to extract some regularities.  
First, with a set of plots, we investigate how the centroid of the arrival direction distribution, and its RMS spread, evolve with rigidity.   Even in this simplified, schematic view, there is significant variation from one source direction to another.  Figs. \ref{plt:cent1to7_allL} and \ref{plt:cent8to12_NSP_allL} display this idealized picture of the mean deflections and spreading, as a function of source position and rigidity, for the 14 HPX grid source-directions.   We include only rigidities logR = 18, 18.2, 18.4, 18.6, 18.8, 19, 19.2, 19.4, 19.6, 19.8, as a compromise between legibility and completeness.  The spreading is represented as a disk of uniform density, whose radius is the median value of the angular separation distribution from the centroid position.  In some cases, not all rigidities are visible.  This is generally because of de-magnification.  For legibility we do not change the density to reflect magnification or demagnification; the reader can consult the tables of Appendix \ref{appdx:HEALPixTables} for this information.  
In the cases of HPX 7, the map is centered on $\ell=150^{\circ}$ for $L_{coh}=100$ pc and  $\ell=60^{\circ}$ for $L_{coh}=30$ pc and coherent-only.
For HPX2 $L_{coh}=100$ pc, the map is centered on 90$^{\circ}$.
Due to large median values, the disk radii in the HPX3 and SPOL plots are rescaled smaller by 90\% and by 50\% for HPX7.

The point of this series of skyplots is to show that the differences between arrival direction distributions, for different source directions and coherence lengths, is not just in their fine-structure.  In some cases, even the means and RMS spreads show significant differences.  Furthermore the mean deflection itself is affected by the random field, especially in some rigidity regions;  probably this is because for such rigidities the behavior is on the cusp of being quasi-chaotic and the random field pushes some trajectories over the edge.

We reiterate that the simple disks of these plots are NOT accurate representations of the actual images.  

The complexity and variation between sources notwithstanding, many phenomenological applications call for a qualitative description of the deflections and spreading.  Often deflections are ignored altogether, or taken to be independent of rigidity, or only spreading is considered while deflection is ignored.    Therefore, in the subsequent plots we use simple functional descriptions to serve as a general guide to their behavior.

\newpage

\clearpage 
\clearpage 
\newpage


\begin{figure}[H]
\hspace{-0.3in}
\centering
\begin{minipage}[b]{0.3 \textwidth}
\includegraphics[width=1. \textwidth]{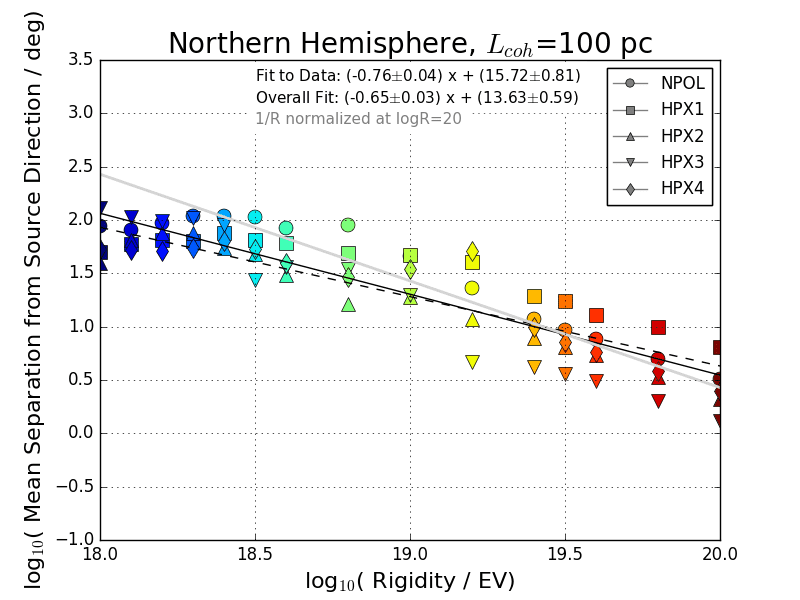}
\end{minipage}
\begin{minipage}[b]{0.3 \textwidth}
\includegraphics[width=1. \textwidth]{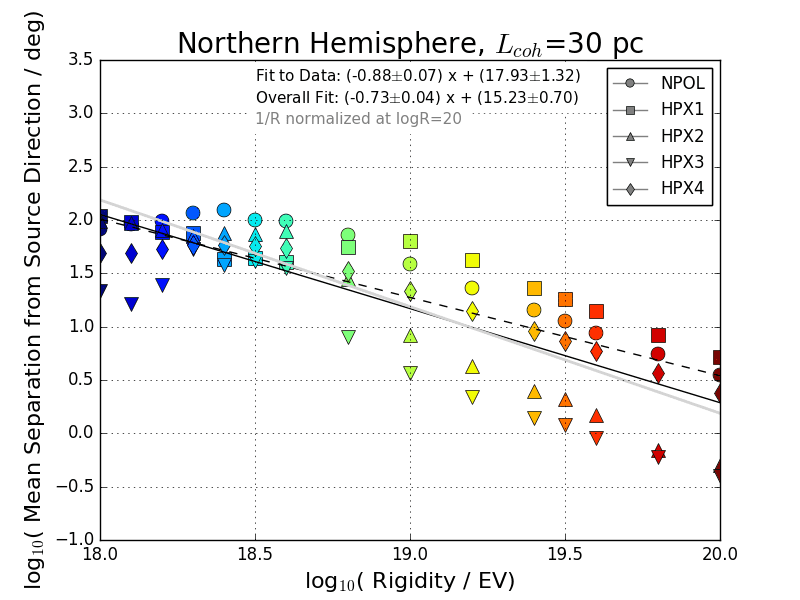}
\end{minipage}
\begin{minipage}[b]{0.3 \textwidth}
\includegraphics[width=1. \textwidth]{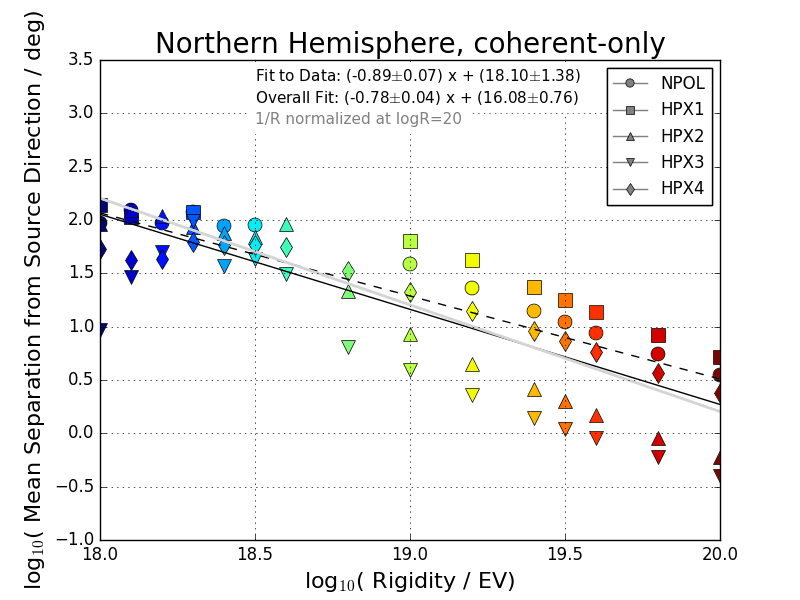}
\end{minipage}
\begin{minipage}[b]{0.3 \textwidth}
\includegraphics[width=1. \textwidth]{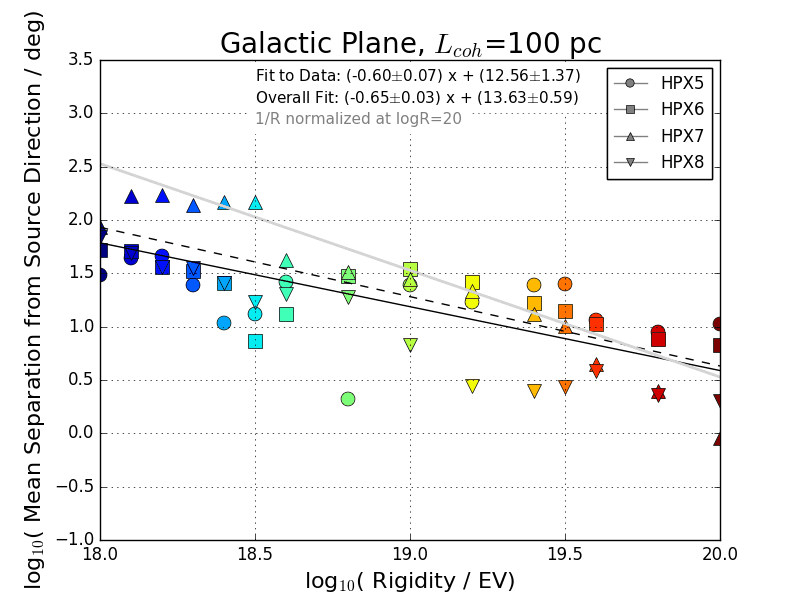}
\end{minipage}
\begin{minipage}[b]{0.3 \textwidth}
\includegraphics[width=1. \textwidth]{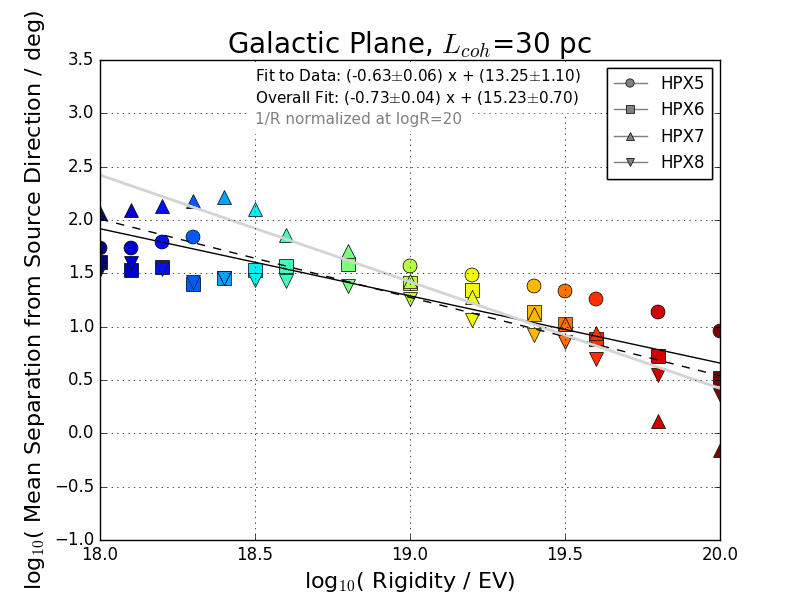}
\end{minipage}
\begin{minipage}[b]{0.3 \textwidth}
\includegraphics[width=1. \textwidth]{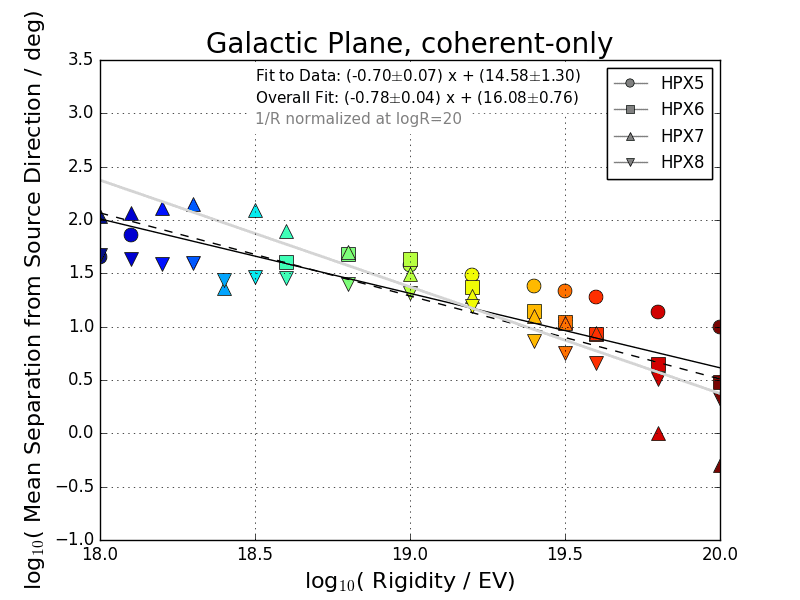}
\end{minipage}
\begin{minipage}[b]{0.3 \textwidth}
\includegraphics[width=1. \textwidth]{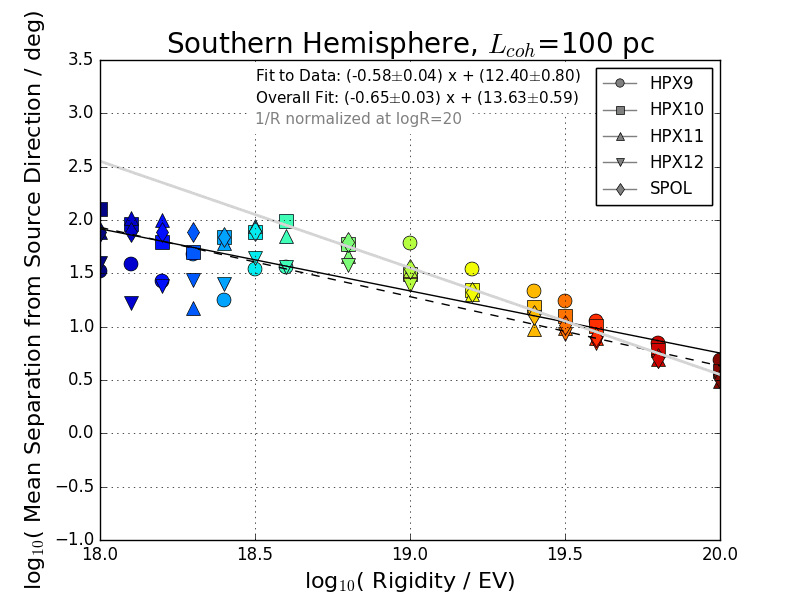}
\end{minipage}
\begin{minipage}[b]{0.3 \textwidth}
\includegraphics[width=1. \textwidth]{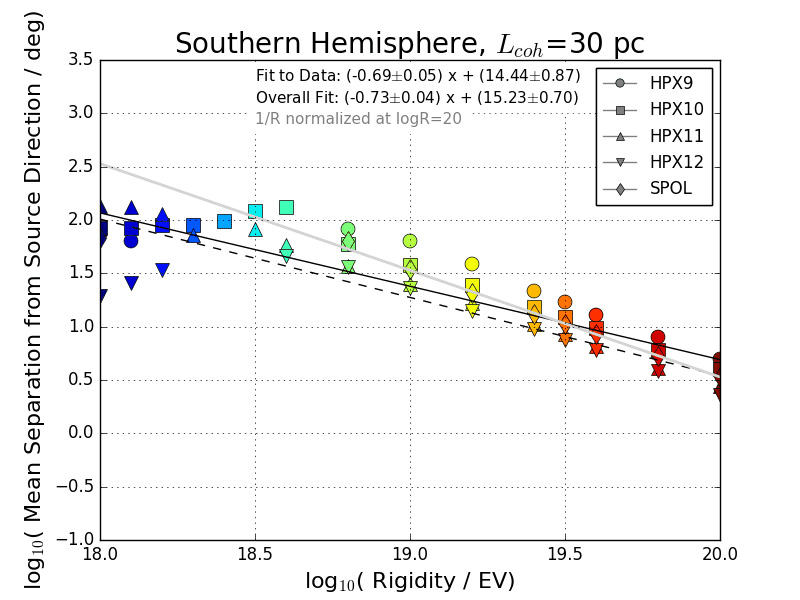}
\end{minipage}
\begin{minipage}[b]{0.3 \textwidth}
\includegraphics[width=1. \textwidth]{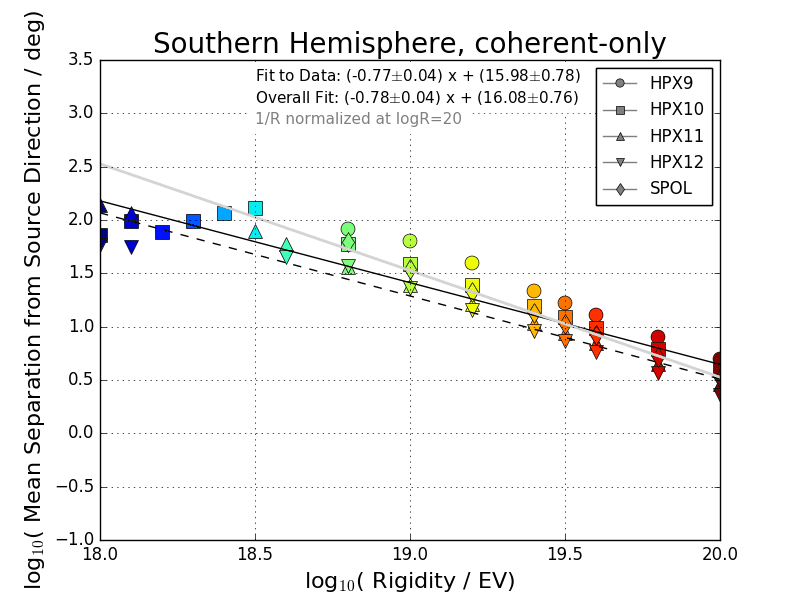}
\end{minipage}
\caption{Mean deflection of arrival directions versus rigidity, in different sky regions, for $L_{coh}=100$ pc (left),  $L_{coh}=30$ pc (center) and purely coherent field (right).  The solid black line is a linear (in log-log plot) fit with fit parameters indicated, for the given set of sources.  The dashed black line is a global fit to all 12 HPX plus pole sources, and the gray line shows $R^{-1}$ behavior for reference.  } 
\label{plt:defvR}
\vspace{-0.1in}
\end{figure}

Let us begin by considering the mean deflection $\vec{\Delta}$ as a function of rigidity.  Traversing a uniform field of size $D$, $\Delta \approx \frac{D}{R_L} \sim (Z/E) = R^{-1}$ in the small deflection limit, where $R_L$ is the Larmor radius.  For large rigidities the deflections may be so small that the trajectories from a given source are close enough that they all explore a similar enough volume of the Galaxy to see the same mean field strength.  In that case, at large rigidity we could imagine that a given source's deflections are roughly co-linear and increase as $\sim R^{-1}$.  

Fig. \ref{plt:defvR} shows the (log of the) mean deflection versus rigidity, grouped into three regions of the sky, with individual HPX source directions indicated by different symbols.  The solid black line is a power-law fit to the sources depicted in the plot, with parameters in a log$\Delta$ versus log$R$ fit given in the plot.  The dashed black line is a global fit to all 12 HPX plus pole sources, and the gray line shows $R^{-1}$ behavior for reference.  

\begin{figure}[H]
\hspace{-0.3in}
\centering
\begin{minipage}[b]{0.3 \textwidth}
\includegraphics[width=1. \textwidth]{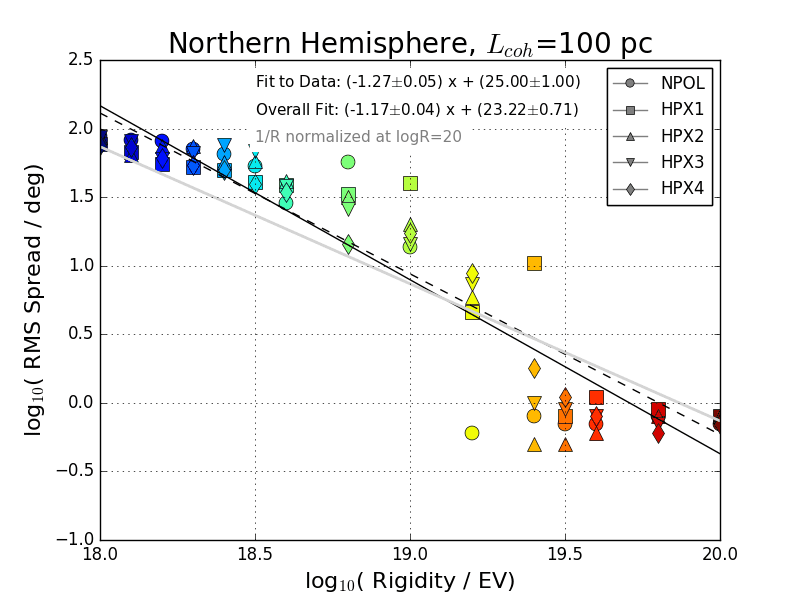}
\end{minipage}
\begin{minipage}[b]{0.3 \textwidth}
\includegraphics[width=1. \textwidth]{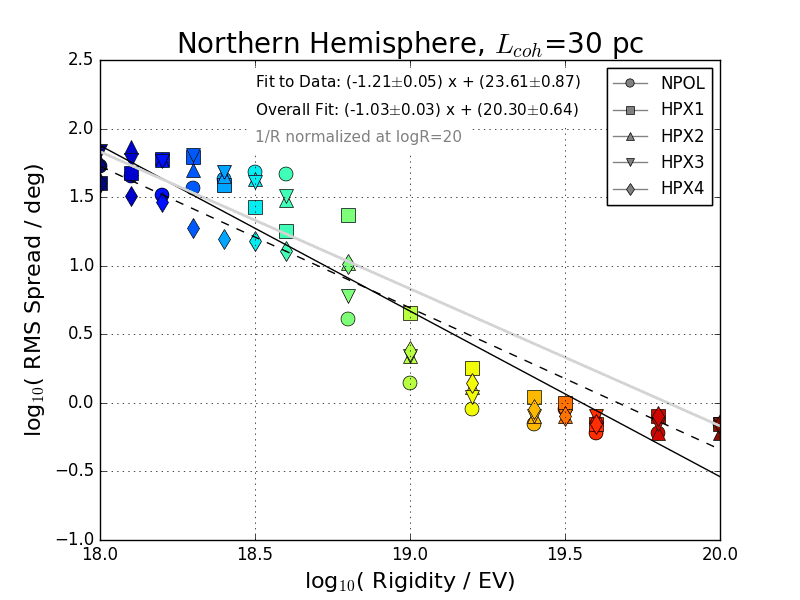}
\end{minipage}
\begin{minipage}[b]{0.3 \textwidth}
\includegraphics[width=1. \textwidth]{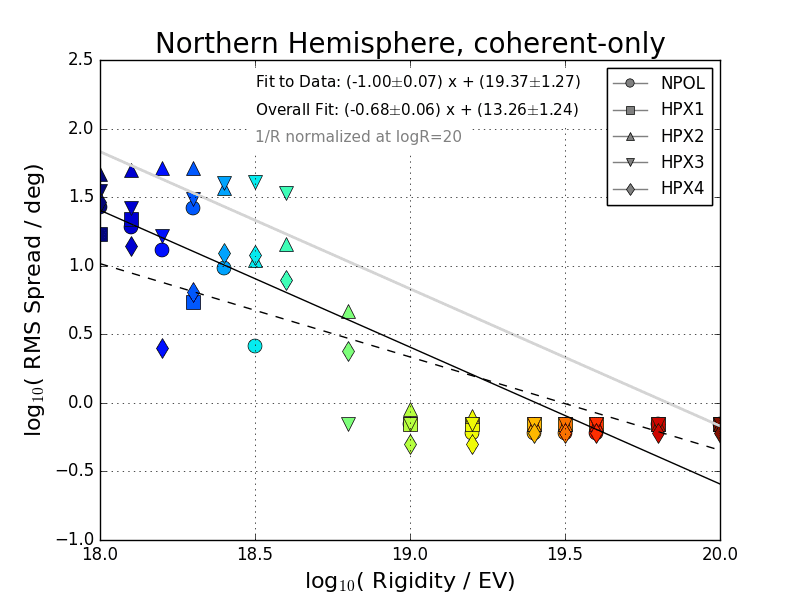}
\end{minipage}
\begin{minipage}[b]{0.3 \textwidth}
\includegraphics[width=1. \textwidth]{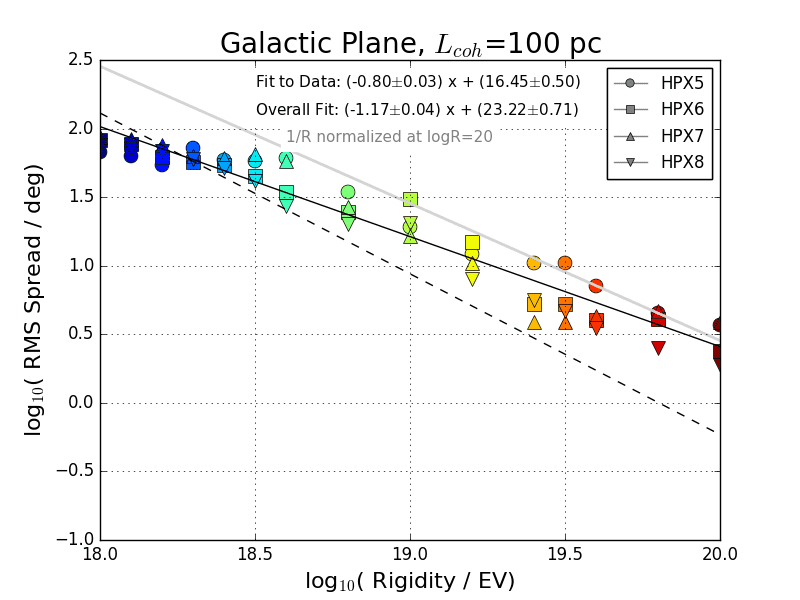}
\end{minipage}
\begin{minipage}[b]{0.3 \textwidth}
\includegraphics[width=1. \textwidth]{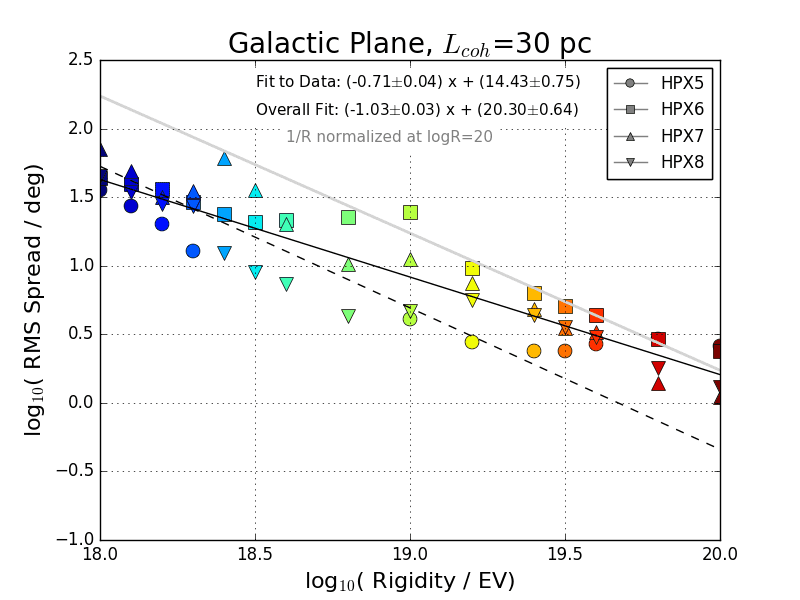}
\end{minipage}
\begin{minipage}[b]{0.3 \textwidth}
\includegraphics[width=1. \textwidth]{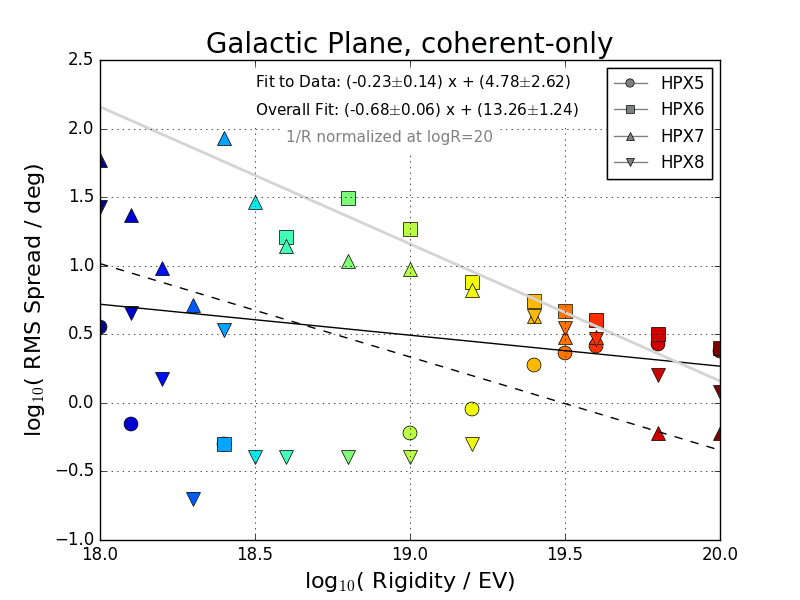}
\end{minipage}
\begin{minipage}[b]{0.3 \textwidth}
\includegraphics[width=1. \textwidth]{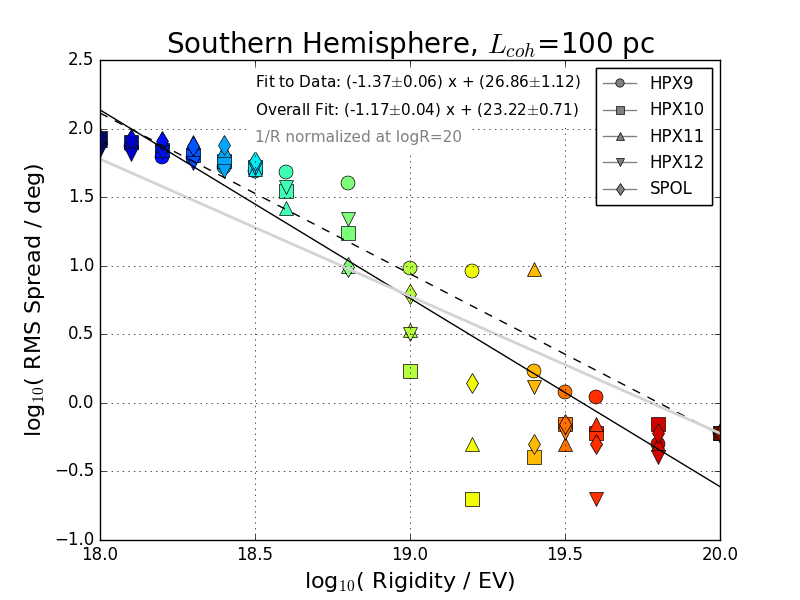}
\end{minipage}
\begin{minipage}[b]{0.3 \textwidth}
\includegraphics[width=1. \textwidth]{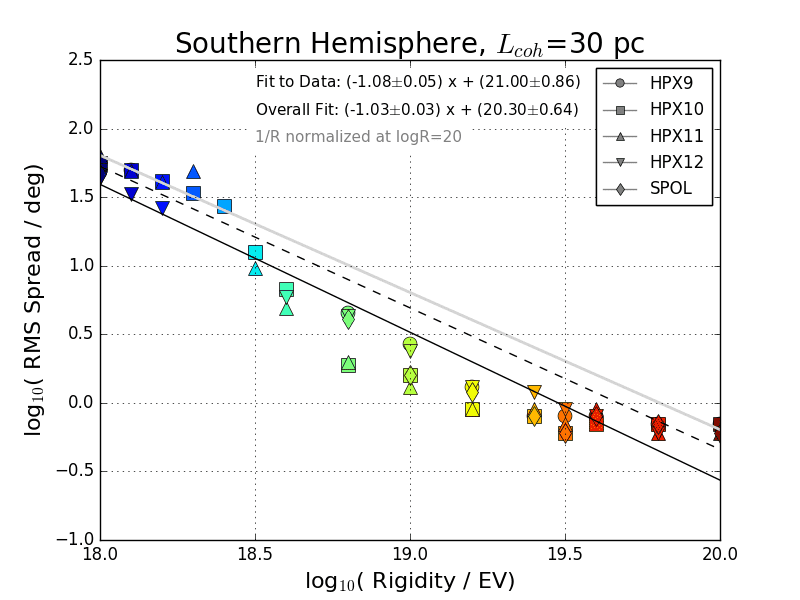}
\end{minipage}
\begin{minipage}[b]{0.3 \textwidth}
\includegraphics[width=1. \textwidth]{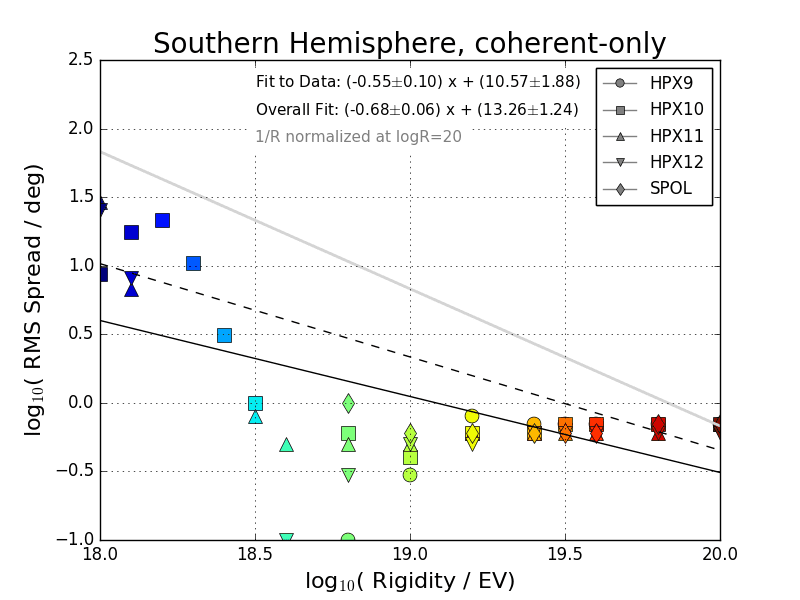}
\end{minipage}
\caption{RMS arrival direction spread versus rigidity, in different sky regions, for $L_{coh}=100$ pc (left),  $L_{coh}=30$ pc (center) and purely coherent field (right).  The solid black line is a linear (in log-log plot) fit for the given set of sources, dashed black line is a global fit to all 12 HPX plus pole sources, and the gray line shows $R^{-1}$ behavior for reference.} 
\label{plt:sigvR}
\vspace{-0.1in}
\end{figure}

We define the arrival direction \emph{spread} to be the median value of the angular distance between the centroid (i.e., mean arrival direction) and the individual arrival directions.  
 A similar simplistic treatment as above, but now of the arrival direction \emph{spread} in the limit of many small deflections, would suggest $\sigma_{\Delta}  \sim \sqrt{L_{coh}} ~R^{-1}$ as in eq. \ref{EGdefs}.  Fig. \ref{plt:sigvR} shows the arrival direction spread versus rigidity, for the three regions of the sky.  The spread is seen to be more regular in its behavior in the presence of a random field, than with a purely coherent field. In the latter case, the rigidity dependence of the spread does not follow a simple power law.  For the N and S hemisphere sources it decreases rapidly as the rigidity increases from 1 EV, there is a break at about $10^{18.5}$V, then it is roughly flat at higher rigidities.   A  power law fit is better for $L_{coh} = 30$ and 100 pc, but the slope does not follow the simple behavior of the above equation:  it is shallower for sources near the Galactic Plane, and steeper for those at high latitudes.    

\begin{figure}[htb]
\hspace{-0.3in}
\centering
\begin{minipage}[b]{0.48 \textwidth}
\includegraphics[width=1. \textwidth]{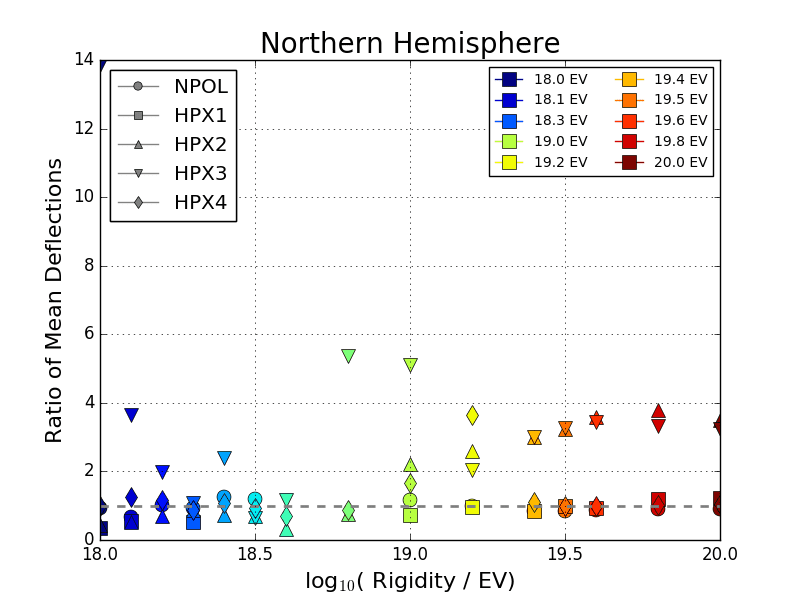}
\end{minipage}
\begin{minipage}[b]{0.48 \textwidth}
\includegraphics[width=1. \textwidth]{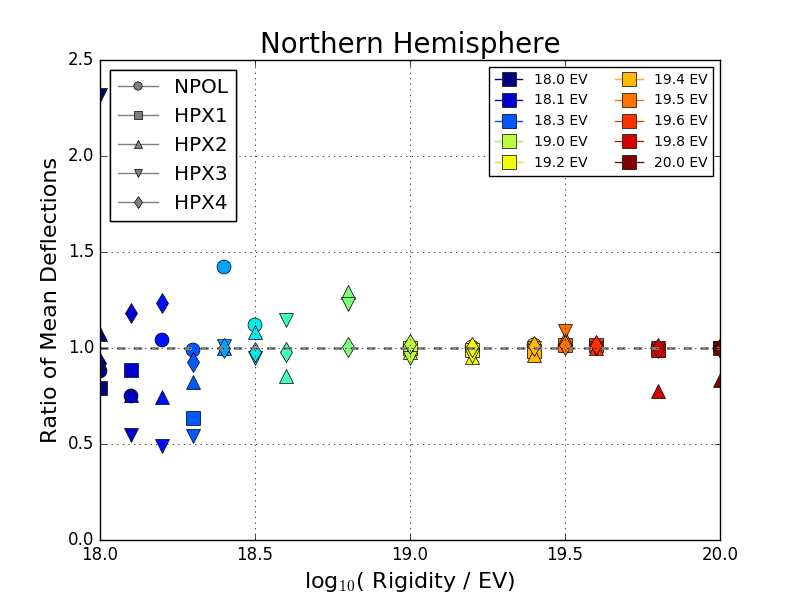}
\end{minipage}
\begin{minipage}[b]{0.48 \textwidth}
\includegraphics[width=1. \textwidth]{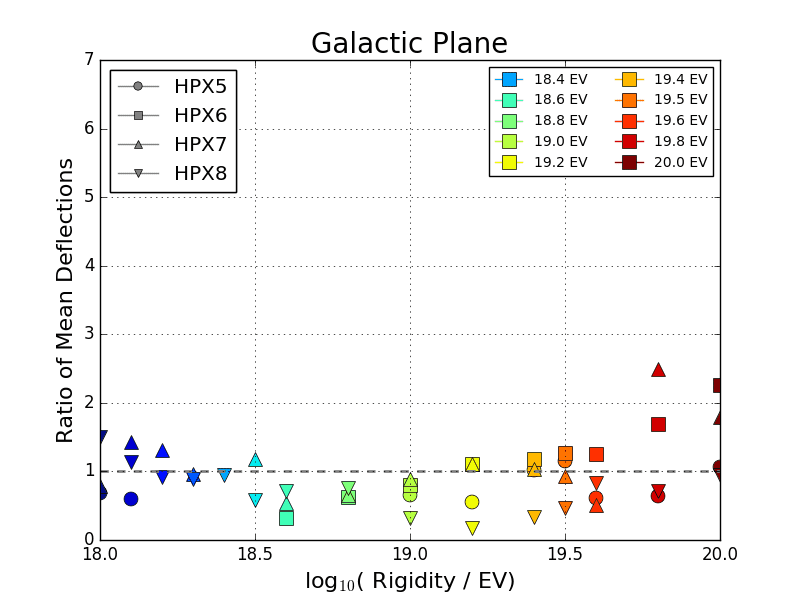}
\end{minipage}
\begin{minipage}[b]{0.48 \textwidth}
\includegraphics[width=1. \textwidth]{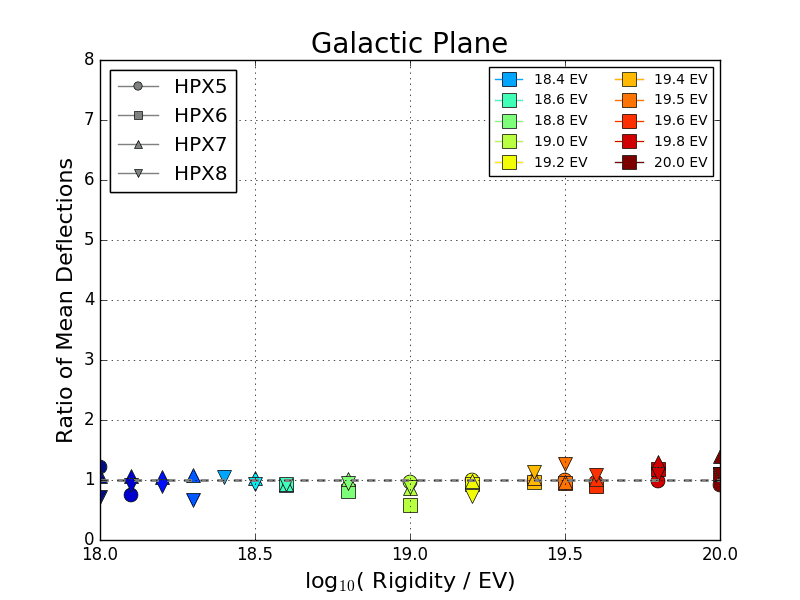}
\end{minipage}
\begin{minipage}[b]{0.48 \textwidth}
\includegraphics[width=1. \textwidth]{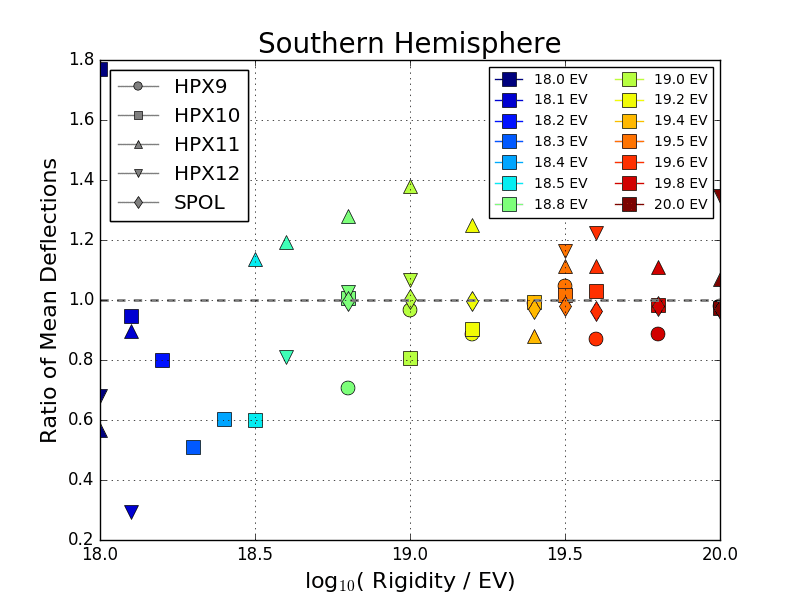}
\end{minipage}
\begin{minipage}[b]{0.48 \textwidth}
\includegraphics[width=1. \textwidth]{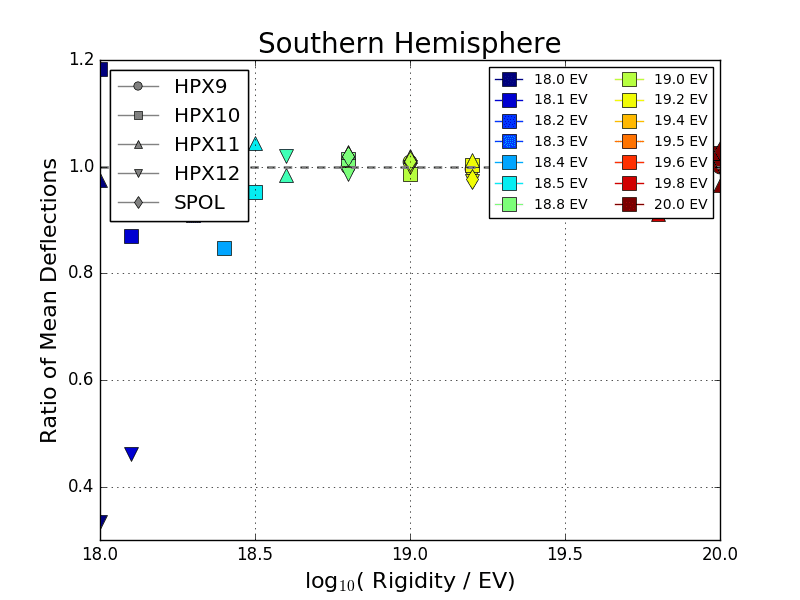}
\end{minipage}
\caption{Ratio of mean deflections for $L_{coh}=100$ pc versus $L_{coh}=30$ pc (left column), and for $L_{coh}=30$ pc versus coherent field only (right column), in different sky regions.  The horizontal lines at 1 shows the expectations in small-deflection approximation, which break down in some cases.} 
\label{plt:Meanratios}
\vspace{-0.1in}
\end{figure}

\begin{figure}[htb]
\hspace{-0.3in}
\centering
\begin{minipage}[b]{0.48 \textwidth}
\includegraphics[width=1. \textwidth]{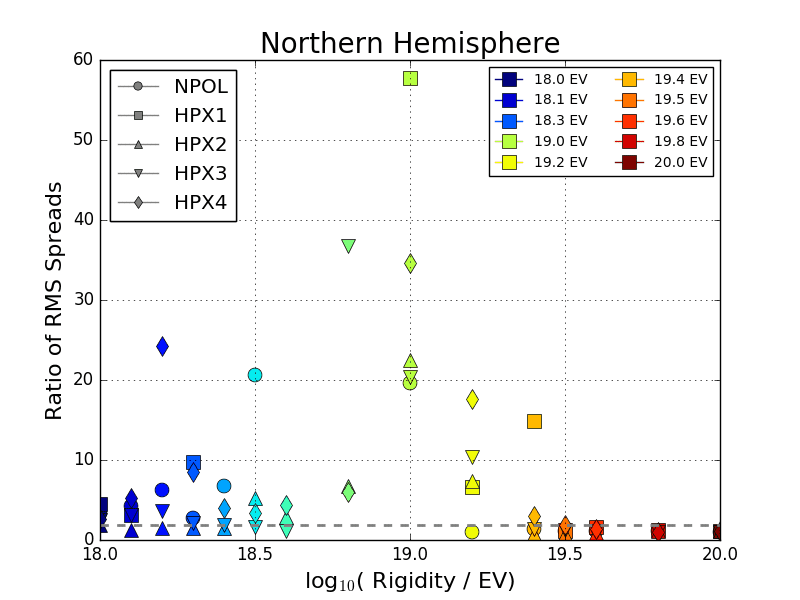}
\end{minipage}
\begin{minipage}[b]{0.48 \textwidth}
\includegraphics[width=1. \textwidth]{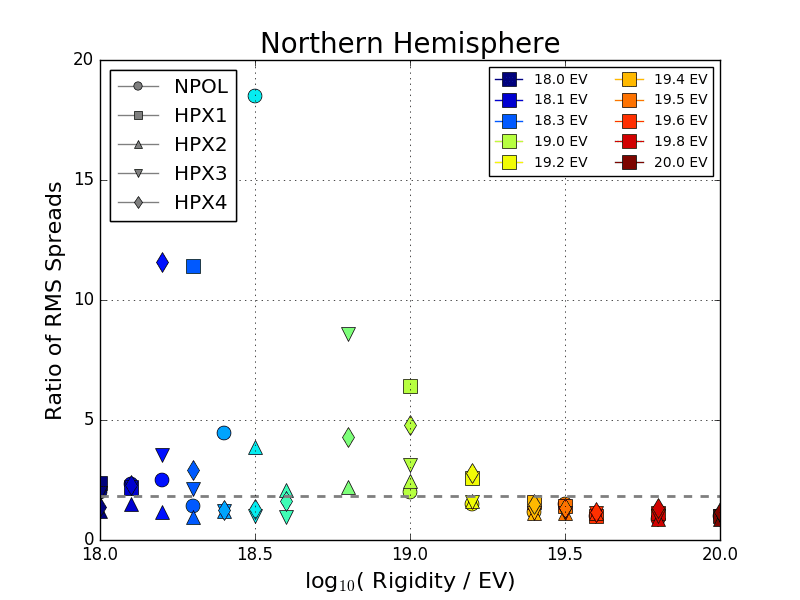}
\end{minipage}
\begin{minipage}[b]{0.48 \textwidth}
\includegraphics[width=1. \textwidth]{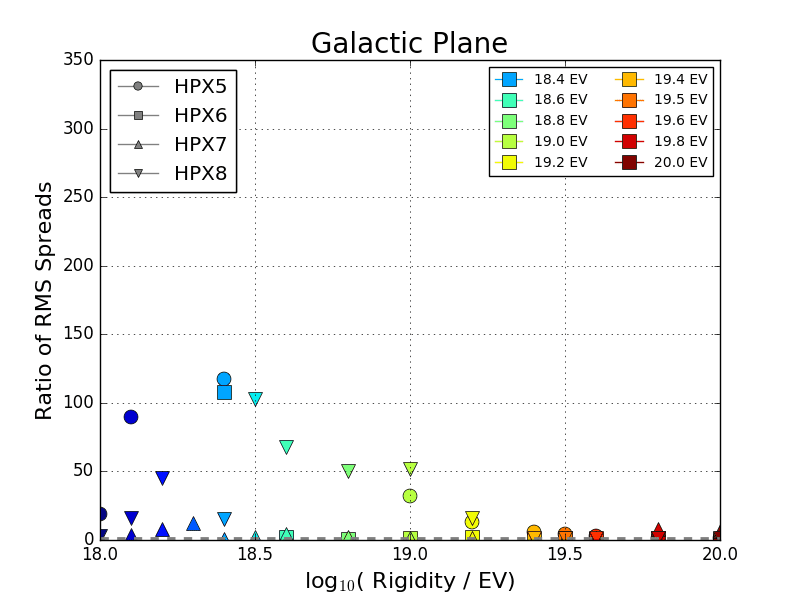}
\end{minipage}
\begin{minipage}[b]{0.48 \textwidth}
\includegraphics[width=1. \textwidth]{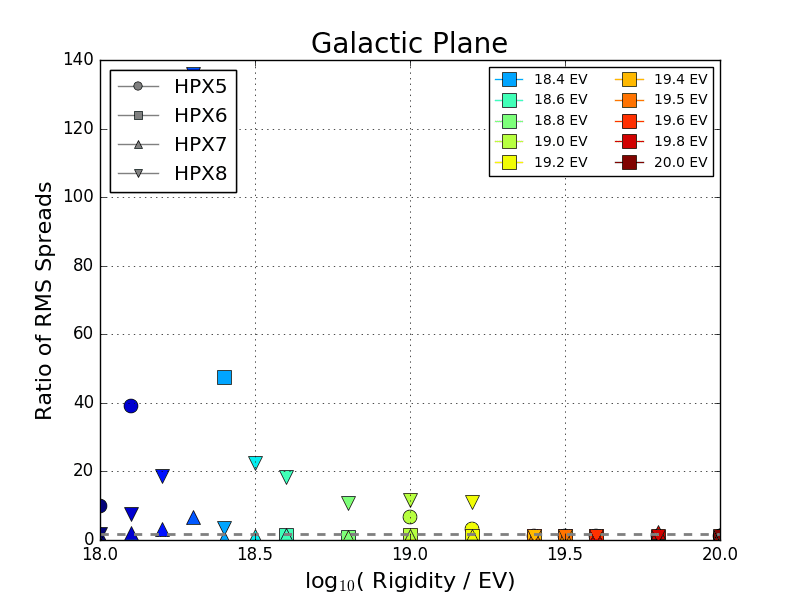}
\end{minipage}
\begin{minipage}[b]{0.48 \textwidth}
\includegraphics[width=1. \textwidth]{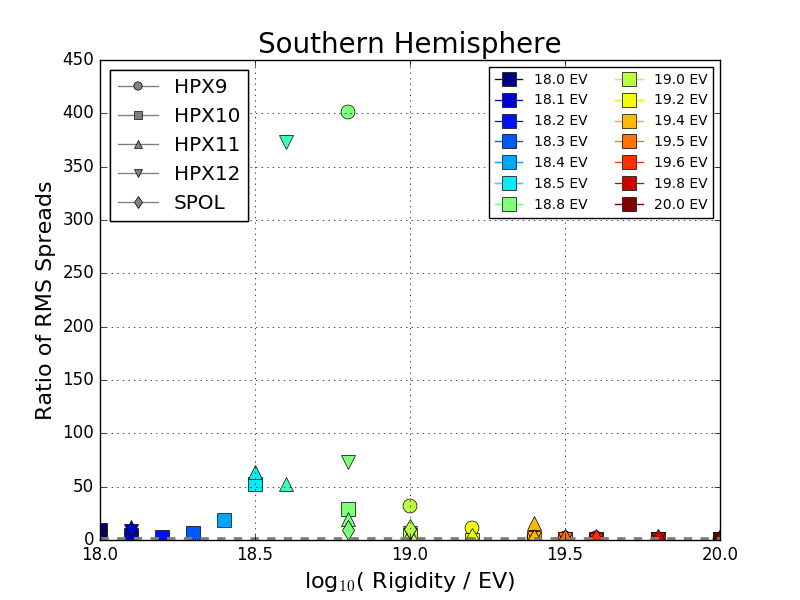}
\end{minipage}
\begin{minipage}[b]{0.48 \textwidth}
\includegraphics[width=1. \textwidth]{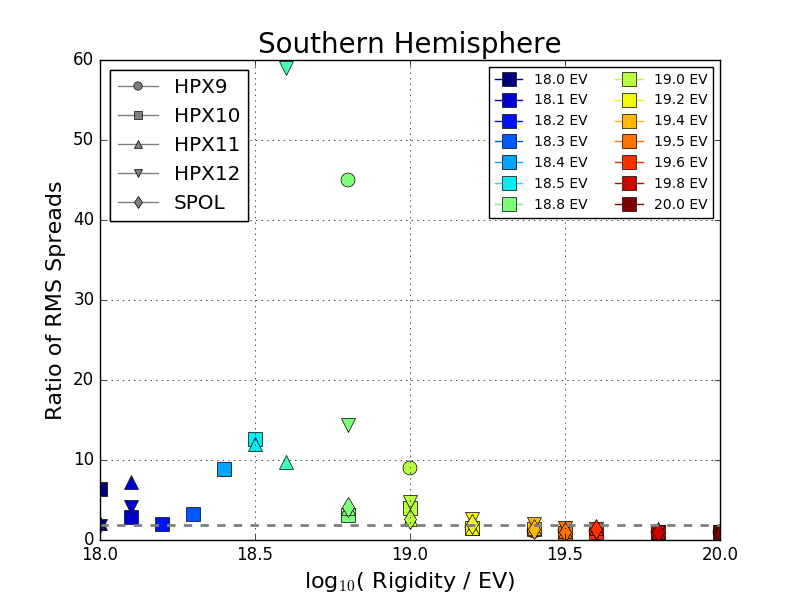}
\end{minipage}
\caption{Ratio of arrival direction spreads for $L_{coh}=100$ pc versus $L_{coh}=30$ pc (left column), and for $L_{coh}=30$ pc versus coherent field only (right column), in different sky regions.  The horizontal lines at 1.82 show the expectations in small-deflection approximation, which clearly break down severely in some cases.} 
\label{plt:Spreadratios}
\vspace{-0.1in}
\end{figure}

Figs. \ref{plt:Meanratios} and \ref{plt:Spreadratios} show the sensitivity to coherence length, of the mean deflection and the arrival direction spread.  The naive expectation in the commonly-used ``ballistic" or small-deflection approximation, is that mean deflections are not impacted by the coherence length.  This can be wrong, particularly at intermediate rigidities, for which UHECRs from sources at mid and polar latitudes pass through the Galactic plane region due to their deflections.   It should therefore not be surprising that the ratio of arrival direction spreads shown in Fig. \ref{plt:Spreadratios}, also does not always follow the small-deflection approximation expectation $\sim \sqrt{L_{coh}} $, which would give a ratio of 1.82.

\begin{figure}[htb]
\hspace{-0.3in}
\centering
\begin{minipage}[b]{0.3 \textwidth}
\includegraphics[width=1. \textwidth]{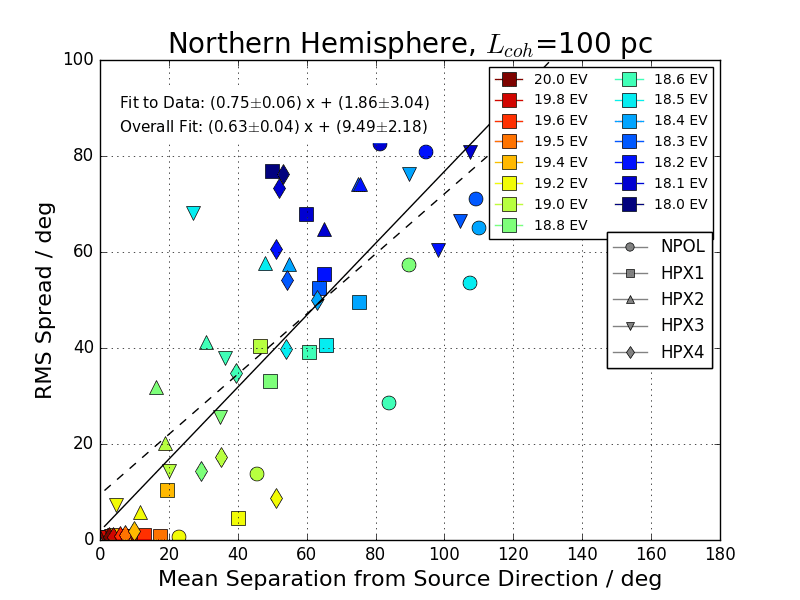}
\end{minipage}
\begin{minipage}[b]{0.3 \textwidth}
\includegraphics[width=1. \textwidth]{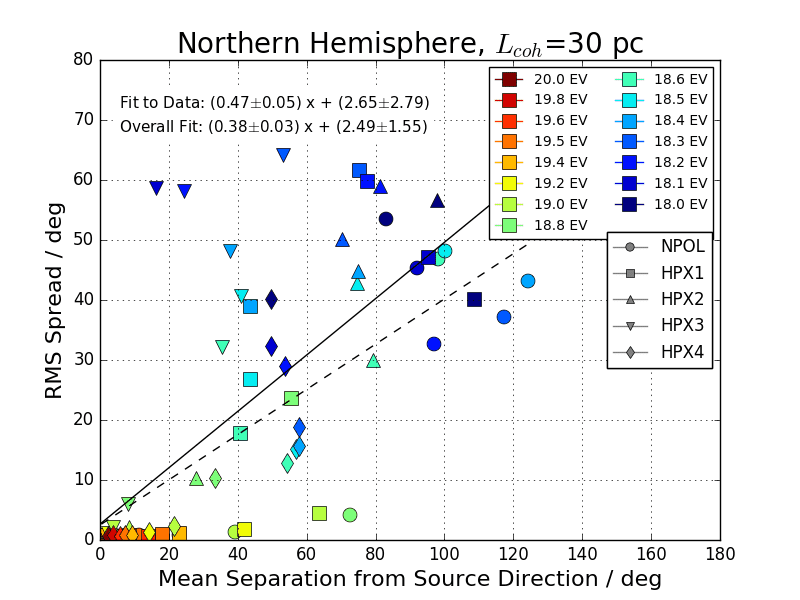}
\end{minipage}
\begin{minipage}[b]{0.3 \textwidth}
\includegraphics[width=1. \textwidth]{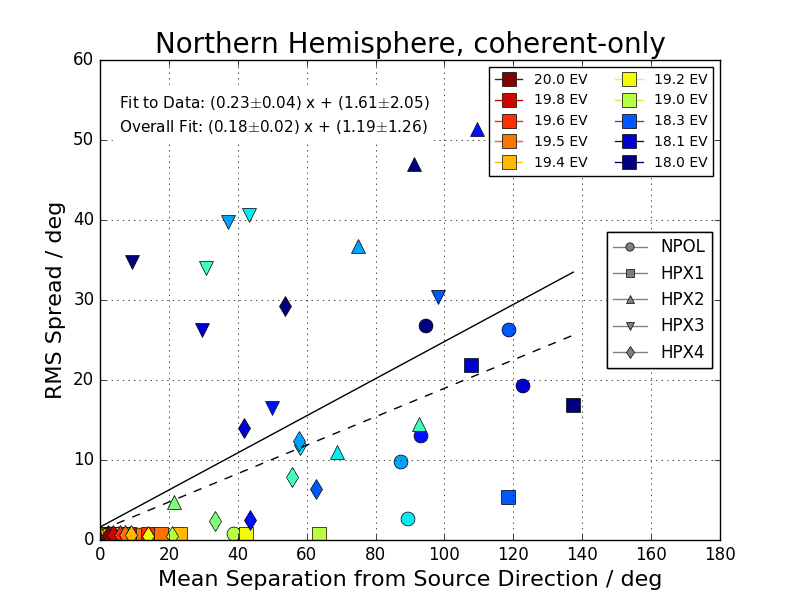}
\end{minipage}
\begin{minipage}[b]{0.3 \textwidth}
\includegraphics[width=1. \textwidth]{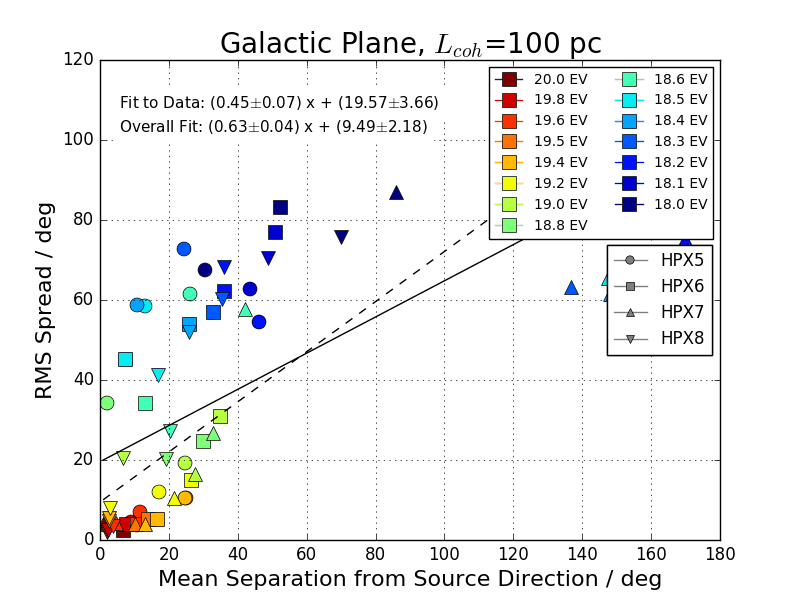}
\end{minipage}
\begin{minipage}[b]{0.3 \textwidth}
\includegraphics[width=1. \textwidth]{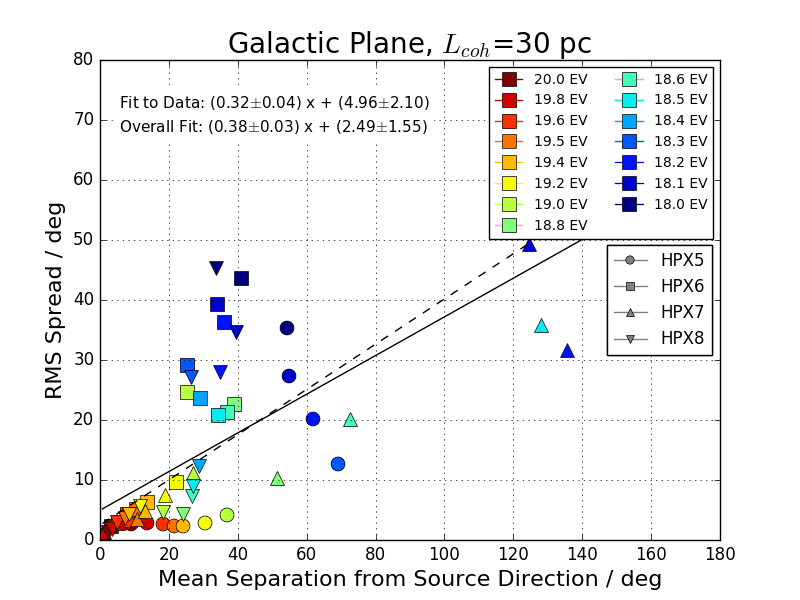}
\end{minipage}
\begin{minipage}[b]{0.3 \textwidth}
\includegraphics[width=1. \textwidth]{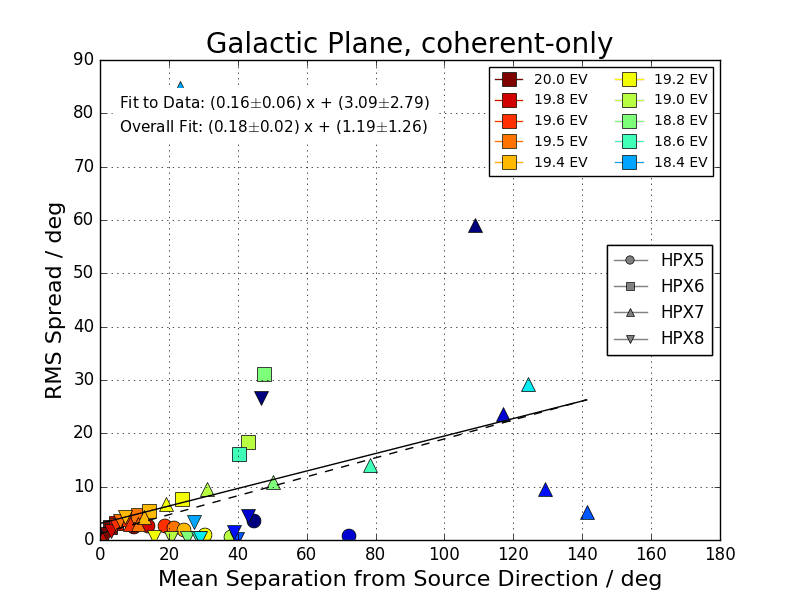}
\end{minipage}
\begin{minipage}[b]{0.3 \textwidth}
\includegraphics[width=1. \textwidth]{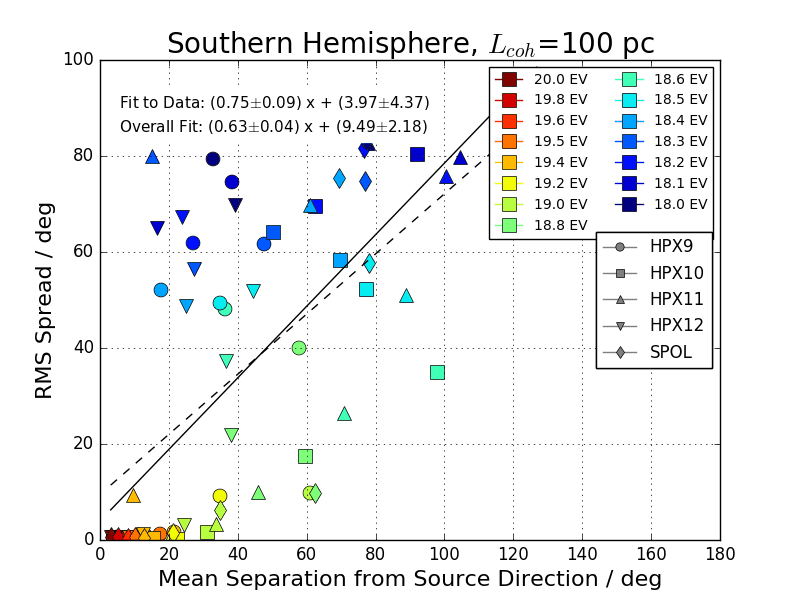}
\end{minipage}
\begin{minipage}[b]{0.3 \textwidth}
\includegraphics[width=1. \textwidth]{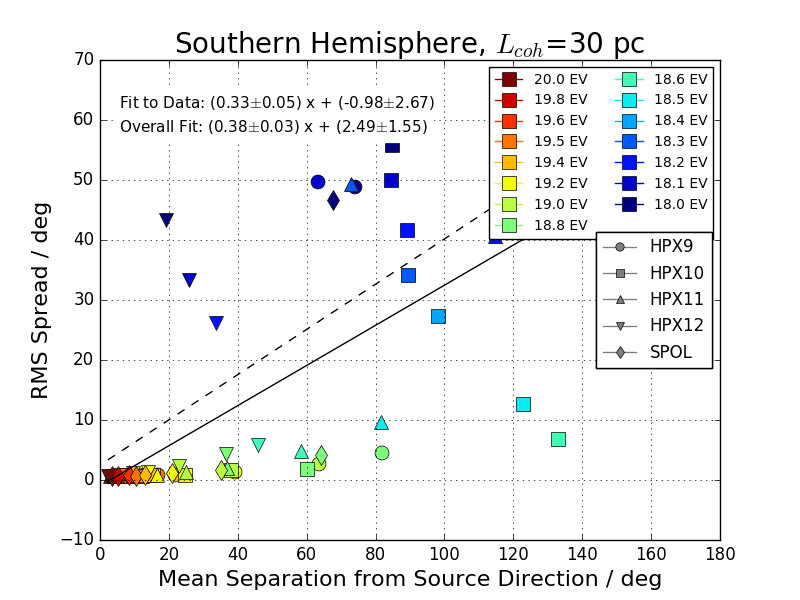}
\end{minipage}
\begin{minipage}[b]{0.3 \textwidth}
\includegraphics[width=1. \textwidth]{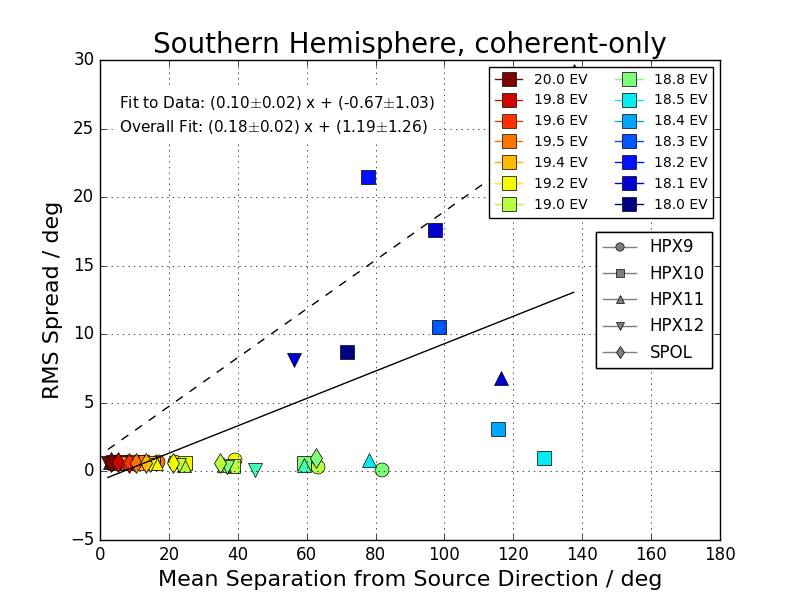}
\end{minipage}
\caption{Arrival direction spread versus mean deflection, in different sky regions, for $L_{coh}=100$ pc (left),  $L_{coh}=30$ pc (center) and purely coherent field (right).  } 
\label{plt:sigvdef}
\vspace{-0.1in}
\end{figure}
\newpage

Rather than looking at evolution in rigidity of the mean deflection and RMS spread, Fig. \ref{plt:sigvdef} shows the RMS spread \emph{versus} the mean deflection, grouped as above into different regions of the sky, with individual HPX source directions indicated by different symbols.   As can be seen, no clear relationship emerges.   

The results presented in this section should be considered only a first attempt to see what general features can be extracted, and it is far from comprehensive.  The reader should keep in mind that when we combined different rigidities and source direction, each separation and spreading value was given equal weight, whereas some instances have many more/fewer events due to magnification/demagnification.  We did not attempt to examine in any detail to what extent removing outliers (e.g., the most exceptional 5\% of the cases) would make qualitative differences in the results; using median rather than mean angular spread mitigates sensitivity to outliers.

\section{Summary and Conclusions}
We have explored how UHECR images develop as the rigidity is reduced from 100 EV to 1 EV, taking into account deflections in the Galactic magnetic field.
High statistics simulations have been performed for the Jansson-Farrar 2012 (JF12) model of coherent and random field, using several fixed realizations of the random field with different coherence lengths.  This work yields the image of any extragalactic source, for UHECRs with rigidities log$_{10} (E_{\rm EeV}/Z)$ = 18.0, 18.1, 18.2, 18.3, 18.4, 18.5, 18.6, 18.8, 19.0, 19.2, 19.4, 19.5, 19.6, 19.8, and 20.0. 

We provide in this paper the images for two exemplary sets of directions, for a representative subset of the simulated rigidities: \\
\noindent 1) M87, M82, and five radio galaxies (Cen A, UGC1841, NGC 1128, NGC 4782 and CGCG 114-025) identified as potential UHECR sources by virtue of satisfying the Hillas criterion for CR acceleration \cite{vV+RadioGals12}.  Skyplots of arrival directions from these 7 sources are shown in Figs. \ref{plt:cena}-\ref{plt:m87}.\\
\noindent 2) A selection of general sky directions probing lines-of-sight through varying Galactic magnetic environments, shown in Appendix \ref{appdx:HEALPixSkyplots}. 

Tables in the Appendix give quantitative properties of the distributions, for each of the images.  A schematic representation of the evolution of centroid and spread of the images with energy, is provided in Figs. \ref{plt:cent1to7_allL} and \ref{plt:cent8to12_NSP_allL}.  The rigidity dependence of the mean deflection and typical spreading for 14 representative source directions grouped by region of sky is shown in Figs. \ref{plt:defvR} and \ref{plt:sigvR};  a power-law fit is provided in the figures.   

Some results of our study are:
\begin{itemize}
\item Individual deflection magnitudes for rigidities at and below 10 EV can be large and the arrival direction distributions span a large fraction of the sky.  
\item The total flux arriving from a ``standard candle" source, varies strongly with its direction and the rigidity.
\item The r.m.s. spread in arrival directions varies strongly, depending on the source location and rigidity.
\item The distributions themselves exhibit considerable structure and variety, including concentrated and smooth hotspots and irregular shapes.
\item Characterizing smearing effects by, for example, a two dimensional Gaussian or spherical Kent distribution (as commonly used in anisotropy or point studies) is not a good representation of the arrival direction distributions seen here.
\item The coherence length of the random field has a significant impact on the arrival direction distributions.  Better constraining the coherence length, which may well vary from one region to another within the galaxy, is of urgent importance for UHECR deflection and source studies.
\item For a given coherence length, the particular realization of the random field has a visible but relatively small impact on the UHECR images of the source; see Figs.  \ref{plt:hpx19o5_krf10_krf11}-\ref{plt:hpx18o5_krf10_krf11} (Appendix \ref{appdx:realizations}).
\item The ``ballistic" or small-angle approximation to UHECR deflections is in general not valid for the field configurations and rigidities studied here.
\end{itemize}
  
Deflections of UHECRs are quite uncertain, especially at low rigidity, even when UHECR charges and hence rigidities are known, due to present uncertainties in the GMF.   The extent of the deflection uncertainty due to uncertainties in the coherent field, is illustrated in Fig. \ref{fig:UFICRC} and discussed in \cite{fCRAS14,ufICRC17,Erdmann+16}.   The uncertainty in the overall strength of the random field is even larger than the uncertainty in the coherent field, because the random field is constrained by the total synchrotron emission which is poorly determined.  The JF12 random field was inferred using the original WMAP synchrotron intensity maps \cite{jf12b}.  Now, alternate models of Galactic synchrotron emission are available from Planck and WMAP, and there are order-of-magnitude differences in the resultant synchrotron intensity ($I$) depending on the method used. (See \cite{ufICRC17} for a summary and references.)   

A factor-10 reduction in $I$ from the original WMAP value, would roughly speaking decrease the estimated RMS random field strength by a factor-3.  This would lessen the impact of the random field on the UHECR arrival directions, naively by a factor-3.  However as seen by the studies presented here, effects of coherent and random field on the deflections do not combine in a simple way, and much of the dispersion in arrival directions for source directions and rigidities probing the Galactic plane, can be due to separated images already present with the coherent field alone.  Thus detailed simulations will be needed to map out the arrival direction behavior in any alternate realistic, structured GMF model.  

Focussing on general features of UHECR deflections which are less sensitive to details of the GMF model, several conclusions can be  drawn:   
\begin{itemize}
\item
Images of the source are not simple except at highest rigidities.  
 \item
Predictions for UHECR arrival direction patterns can change significantly in the presence of a random field, especially if it has a large coherence length.   Sources which would be invisible (all their CRs being deflected away from Earth) can become visible due to the random field.
 \item
Deflections are large and helter-skelter for many source directions, even for rigidities of order 10 EV (as for a 60 EeV carbon nucleus); both net deflections and spreading are important. 
 \item
When there is no random field or the coherence length is small, multiple imaging is common.  For large coherence length the individual ``islands" fragment and are replaced by a more diffuse arrival pattern.    
 \item
At rigidities below about 10 EV, UHECRs from many potential source directions probably do not reach Earth.    
\end{itemize}

Especially at lower rigidities, the coherent field alone leads to multiple images, with the separation between these images in some cases dominating the spread of arrival directions about the mean arrival direction.  In such cases, the random field may mostly induce smearing of the individual images, with little change in the spreading about the global mean direction.   These and other effects depend strongly on source direction, and are best appreciated by studying the skyplots of Figs. \ref{plt:cena}-\ref{plt:m87} and in Appendix \ref{appdx:HEALPixSkyplots}.  

\section{Acknowledgements}

GRF acknowledges collaboration with Ronnie Jansson in the early stages of studying UHECR deflections in the GMF, and current collaboration with Michael Unger on developing a next-generation GMF model, one product of which is displayed in Fig. \ref{fig:UFICRC}.  We also recognize collaboration with Azadeh Keivani on development of the random field realizations, and colleagues in the Pierre Auger Collaboration for many helpful interactions and input.  GRF acknowledges support from the National Science Foundation and NASA under grants NSF-PHY-1212538, NSF-AST-1517319 and NNX10AC96G, and the James Simons Foundation.  Resources supporting this work were provided by the NASA High-End Computing (HEC) Program through the NASA Advanced Supercomputing (NAS) Division at Ames Research Center, consisting of time on the Pleiades supercomputing cluster awarded to GRF.   MS acknowledges support from the Department of Energy under grants DE-0009926 and DE-FG02-91-EF0617.  Some of the results in this paper have been derived using the HEALPix \cite{healpix} package.

\def\apj{Astrophys.\ J.}
\def\nat{Nature}
\def\apjl{Astrophys.\ J. Lett.}
\def\aap{Astron.\ Astrophys.}
\def\prd{Phys. Rev. D}
\def\physrep{Phys.\ Rep.}
\def\mnras{Month. Not. RAS }
\def\araa{Annual Rev. Astron. \& Astrophys.}
\def\aapr{Astron. \& Astrophys. Rev.}
\def\apss{Astrophys. \& Space Sci.}


\providecommand{\href}[2]{#2}\begingroup\raggedright\endgroup

\appendix

\section{Skyplots for Healpix Grid}
\label{appdx:HEALPixSkyplots}
\begin{figure}[t]
\hspace{-0.3in}
\centering
\begin{minipage}[b]{0.48 \textwidth}
\includegraphics[width=1. \textwidth]{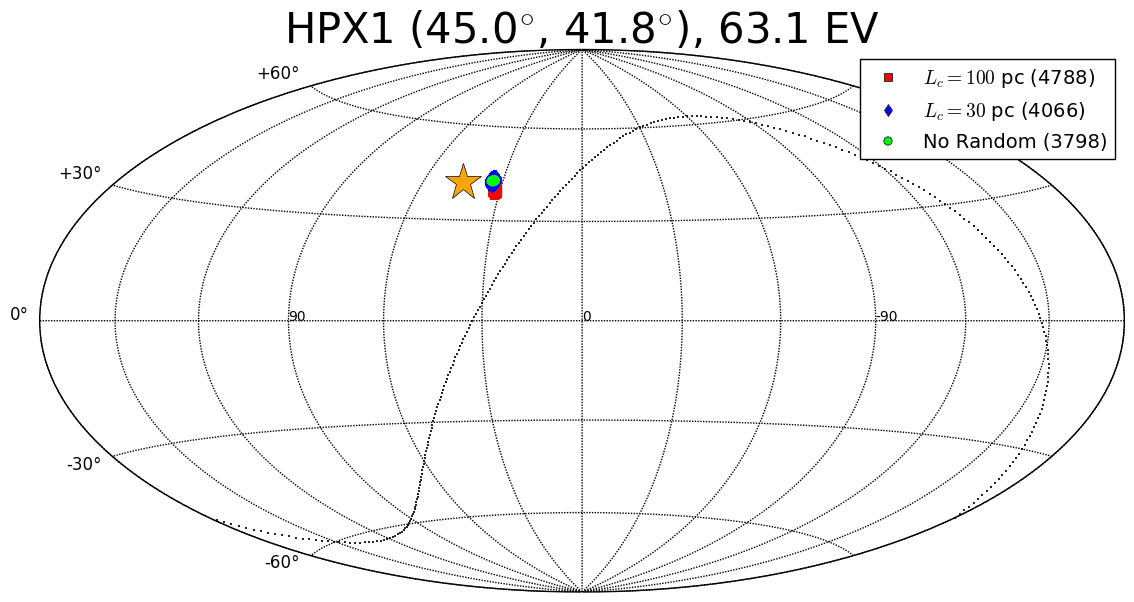}
\end{minipage}
\begin{minipage}[b]{0.48 \textwidth}
\includegraphics[width=1. \textwidth]{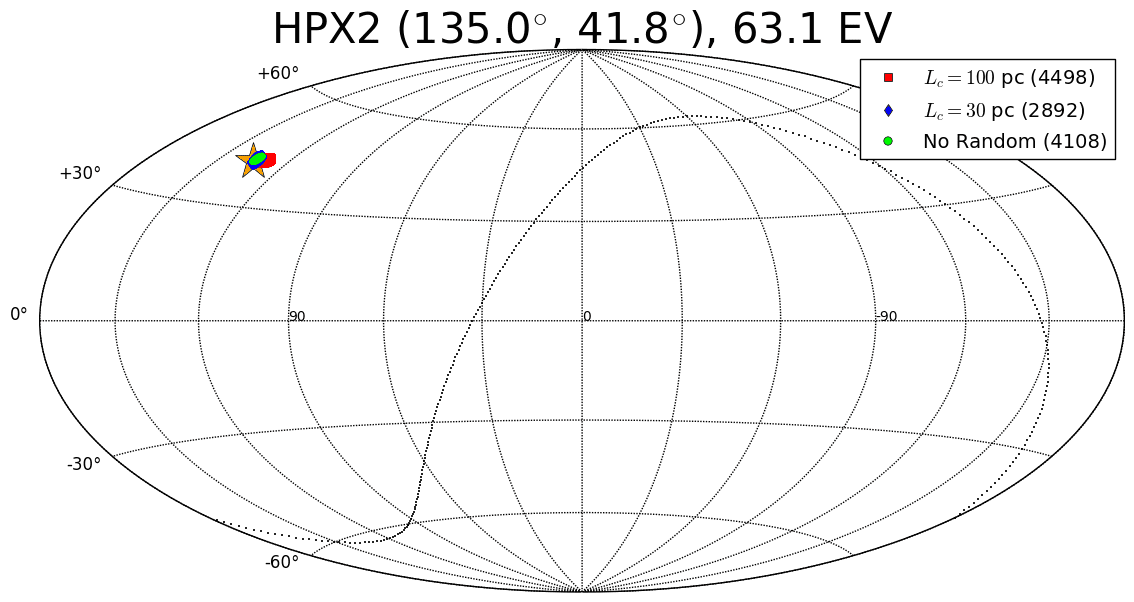}
\end{minipage}
\begin{minipage}[b]{0.48 \textwidth}
\includegraphics[width=1. \textwidth]{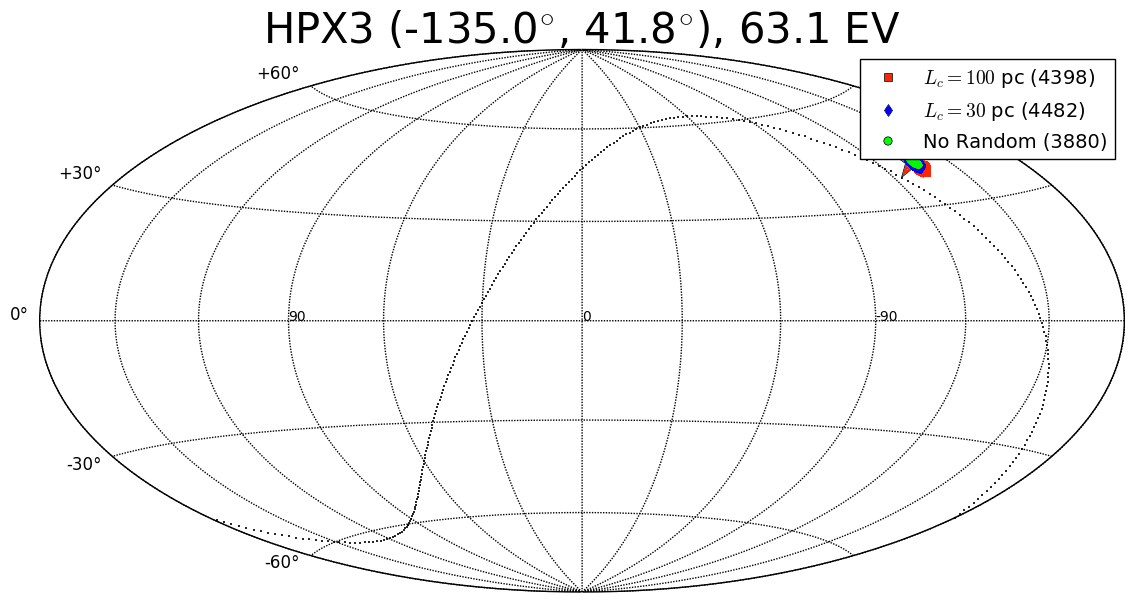}
\end{minipage}
\begin{minipage}[b]{0.48 \textwidth}
\includegraphics[width=1. \textwidth]{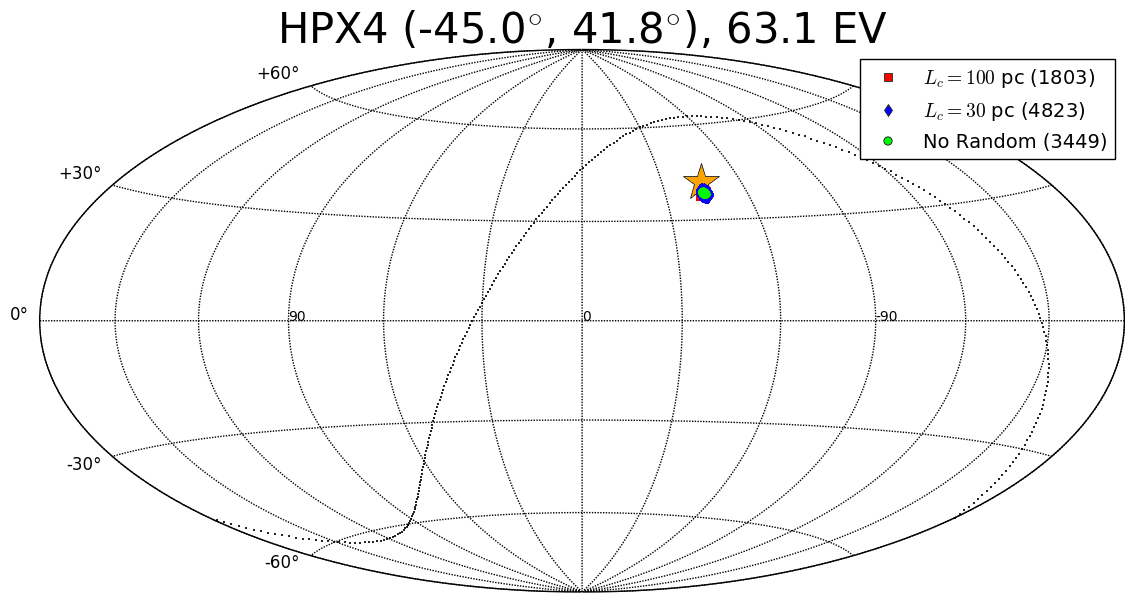}
\end{minipage}
\begin{minipage}[b]{0.48 \textwidth}
\includegraphics[width=1. \textwidth]{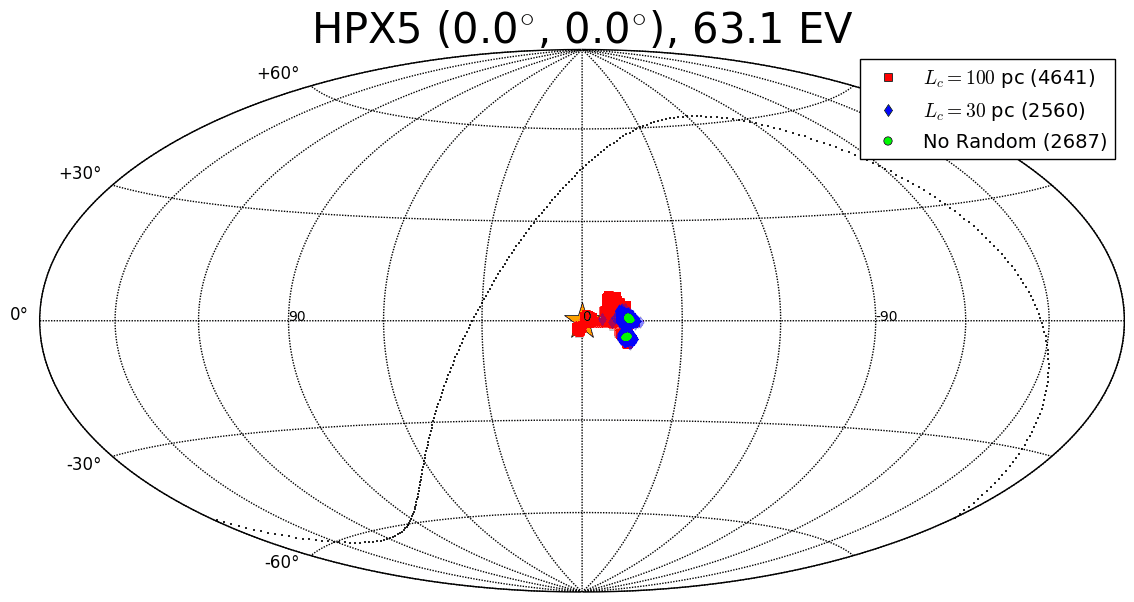}
\end{minipage}
\begin{minipage}[b]{0.48 \textwidth}
\includegraphics[width=1. \textwidth]{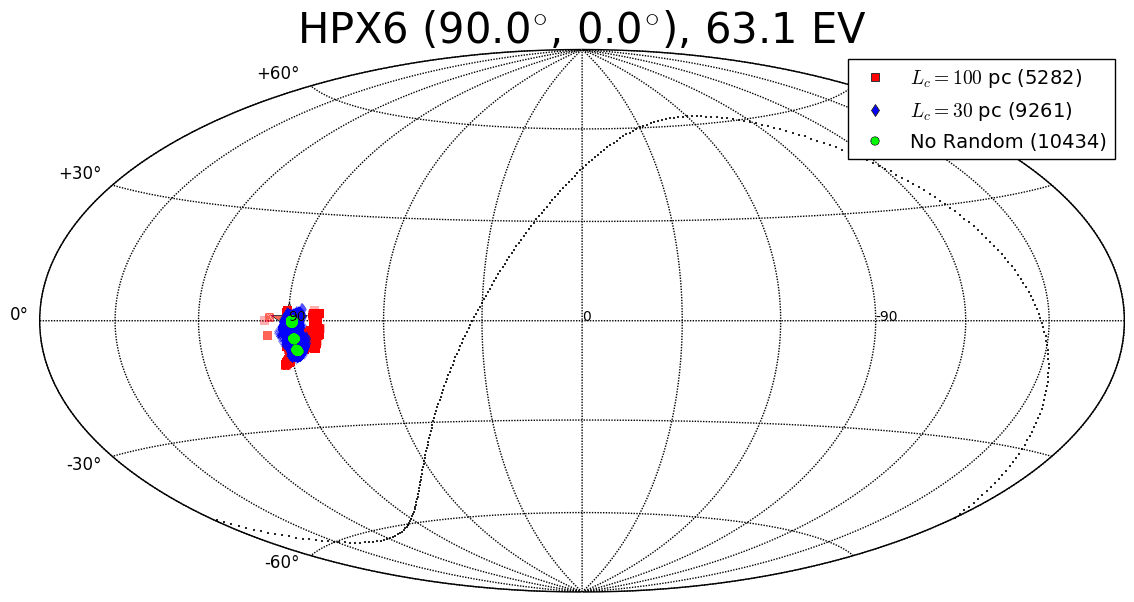}
\end{minipage}
\begin{minipage}[b]{0.48 \textwidth}
\includegraphics[width=1. \textwidth]{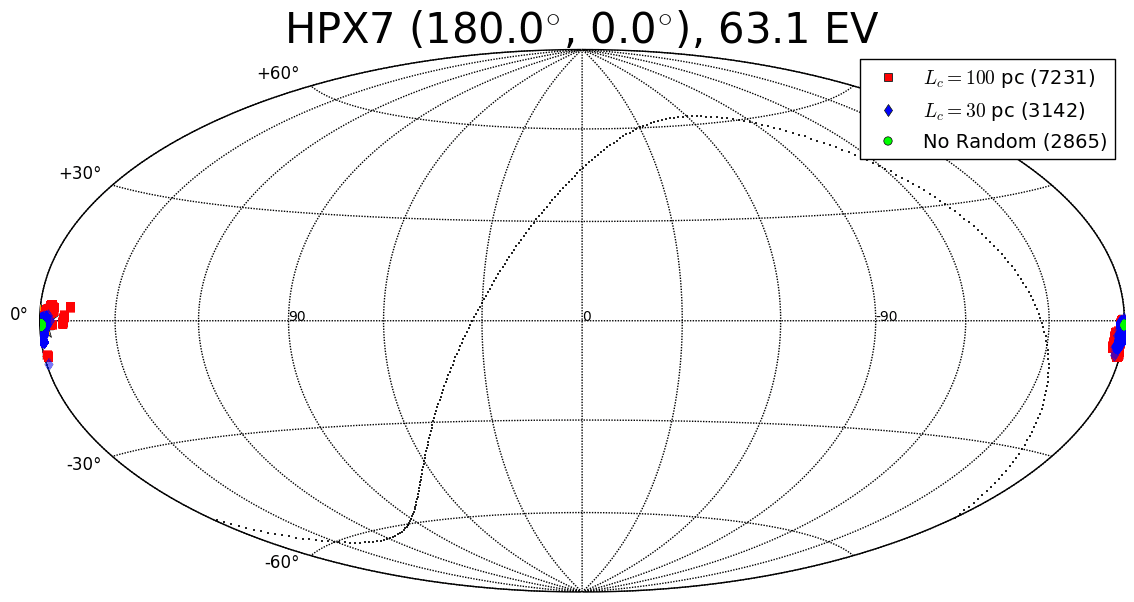}
\end{minipage}
\begin{minipage}[b]{0.48 \textwidth}
\includegraphics[width=1. \textwidth]{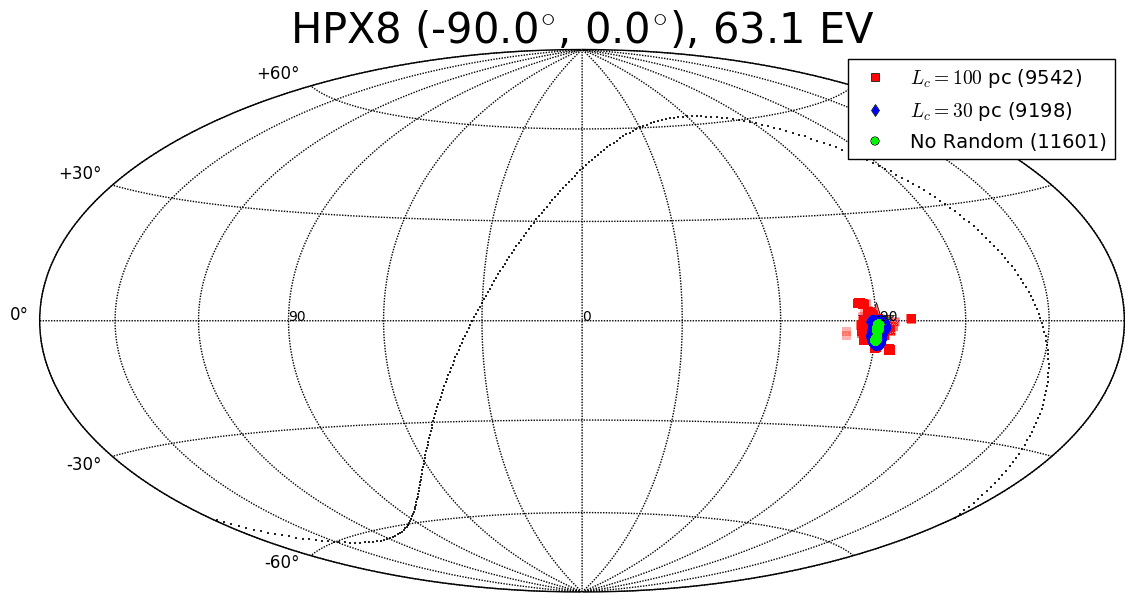}
\end{minipage}
\begin{minipage}[b]{0.48 \textwidth}
\includegraphics[width=1. \textwidth]{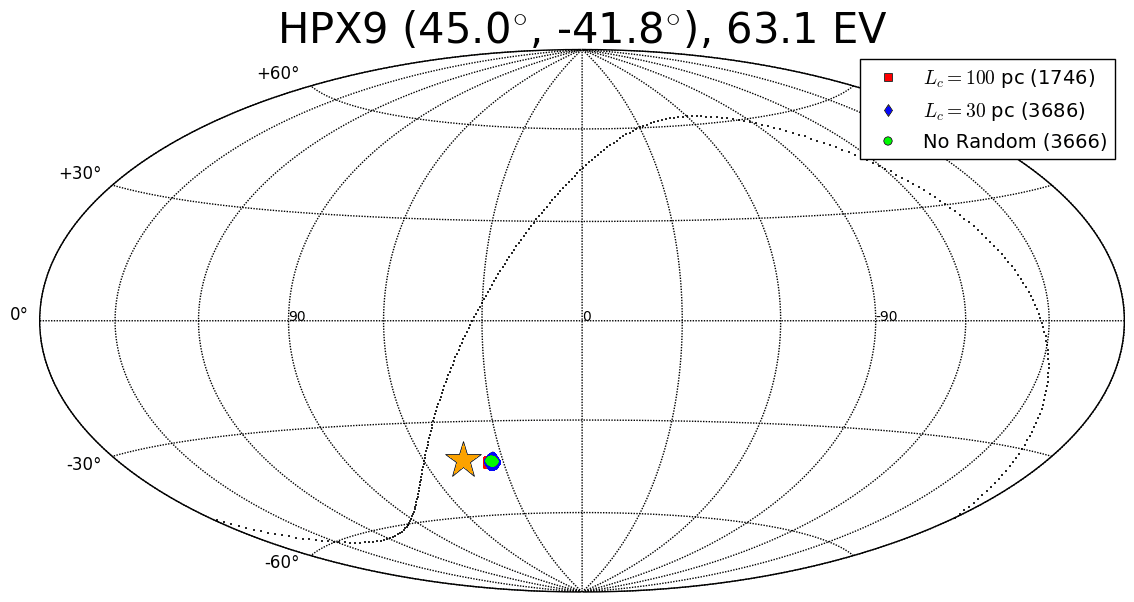}
\end{minipage}
\begin{minipage}[b]{0.48 \textwidth}
\includegraphics[width=1. \textwidth]{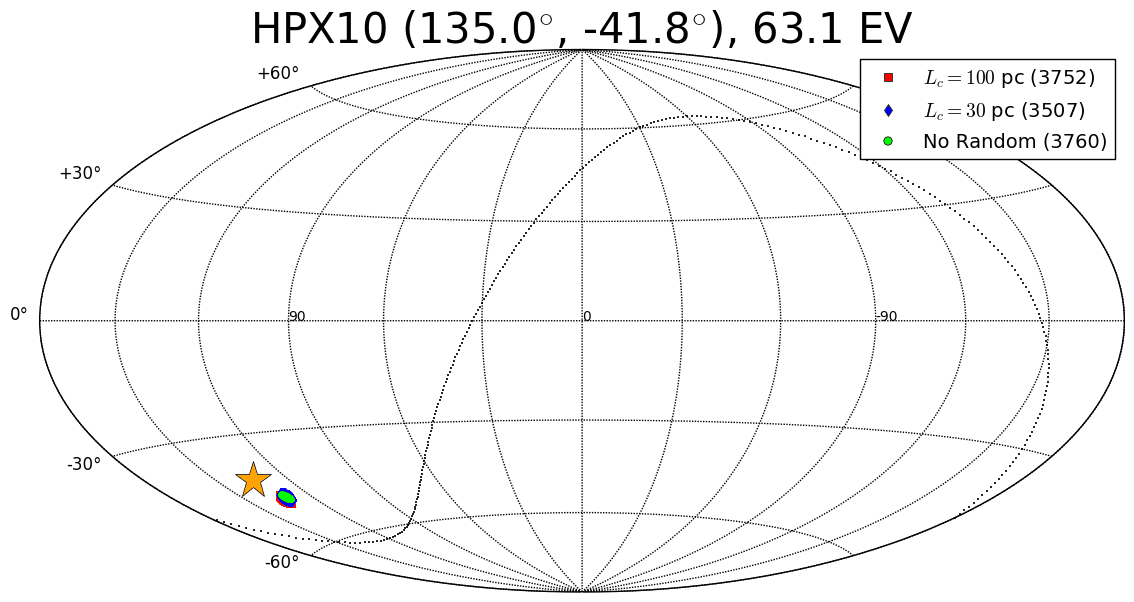}
\end{minipage}
\vspace{-0.3in}
\caption{Arrival direction distributions for log($R$ / V) = 19.8 for events from sources in the regularly-spaced grid (marked with orange stars) for the $L_{coh} = 100$ pc (KRF6), a $L_{coh} = 30$ pc (KRF10) realization, and the coherent-only field.
The source direction and name is listed in the plot title.
The sky map is in Galactic coordinates and the dotted line indicates decl. $\delta=0^{\circ}$.
The legend in each plot indicates the number of events arriving from the given source at the given rigidity for the given realization.} 
\label{plt:hpx19o8}
\vspace{-0.1in}
\end{figure}
\clearpage 
\begin{figure}[t]
\hspace{-0.3in}
\centering
\begin{minipage}[b]{0.48 \textwidth}
\includegraphics[width=1. \textwidth]{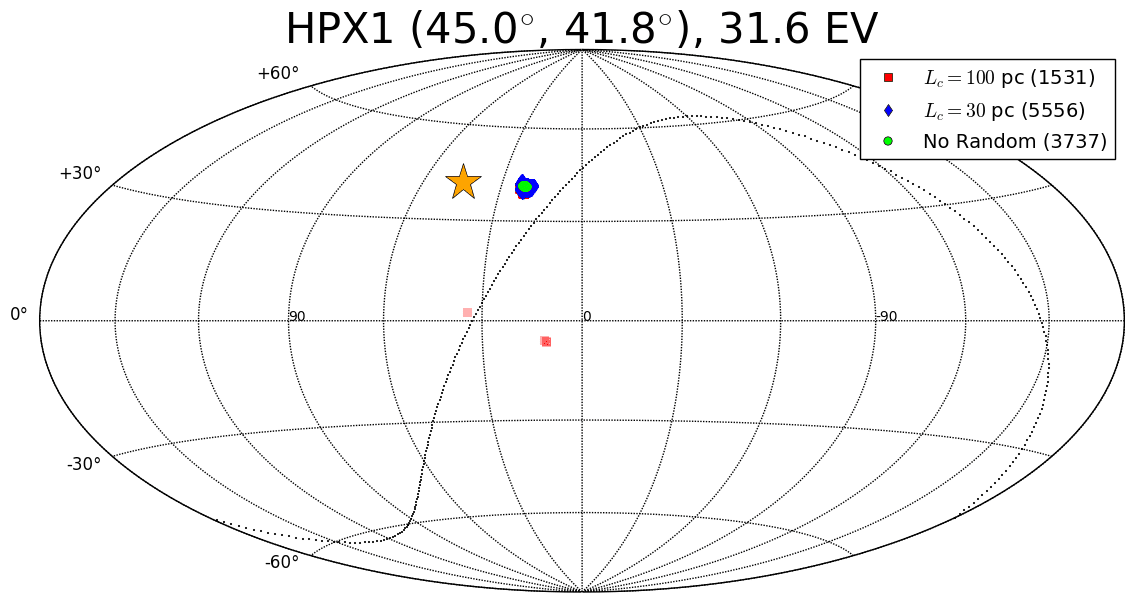}
\end{minipage}
\begin{minipage}[b]{0.48 \textwidth}
\includegraphics[width=1. \textwidth]{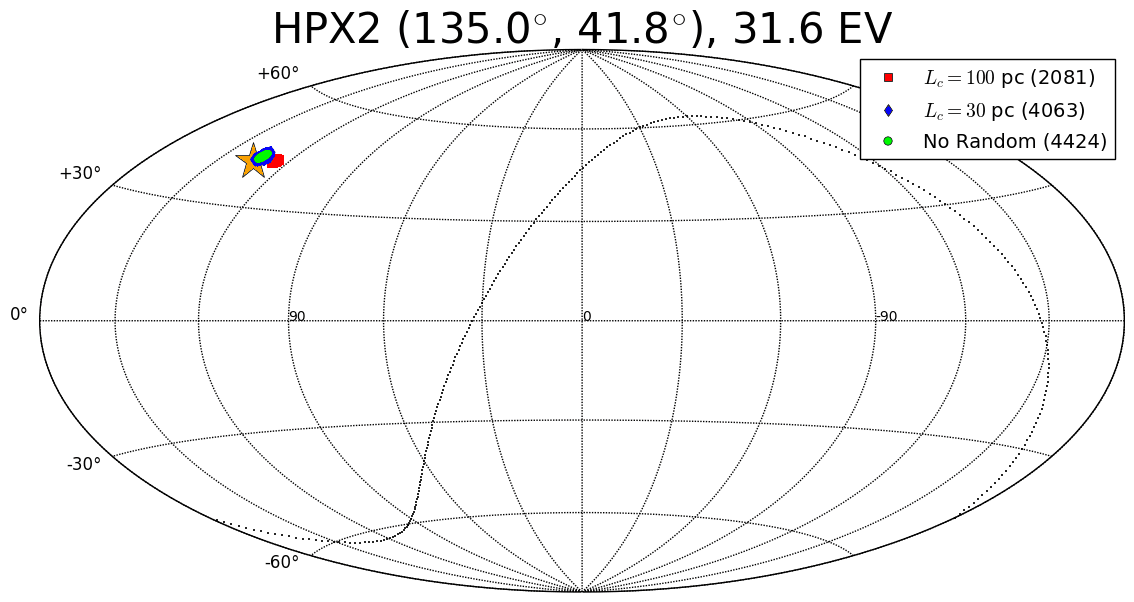}
\end{minipage}
\begin{minipage}[b]{0.48 \textwidth}
\includegraphics[width=1. \textwidth]{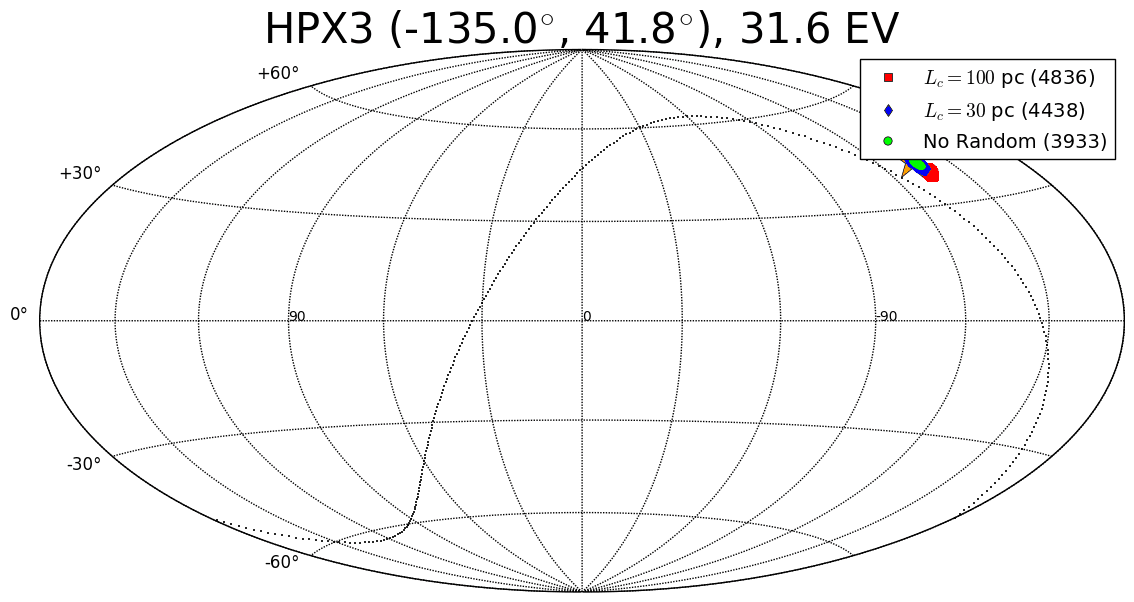}
\end{minipage}
\begin{minipage}[b]{0.48 \textwidth}
\includegraphics[width=1. \textwidth]{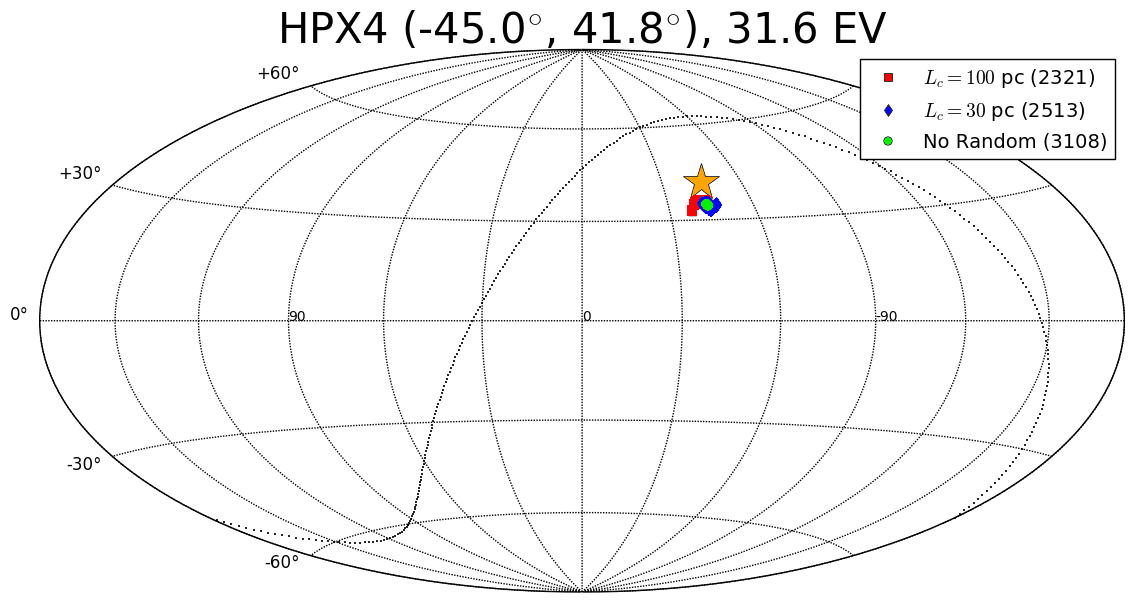}
\end{minipage}
\begin{minipage}[b]{0.48 \textwidth}
\includegraphics[width=1. \textwidth]{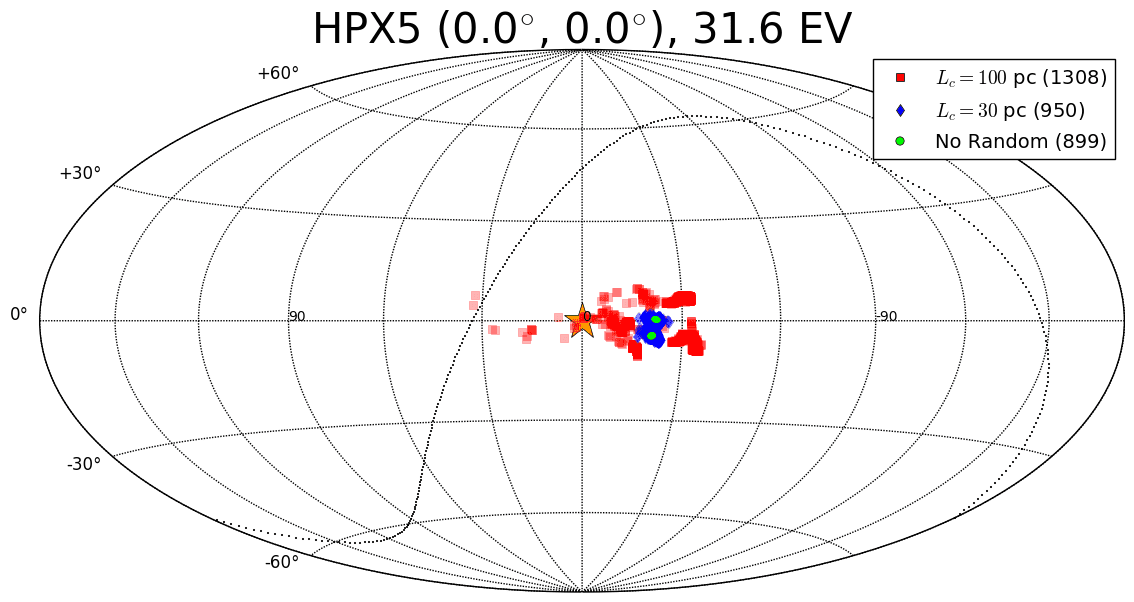}
\end{minipage}
\begin{minipage}[b]{0.48 \textwidth}
\includegraphics[width=1. \textwidth]{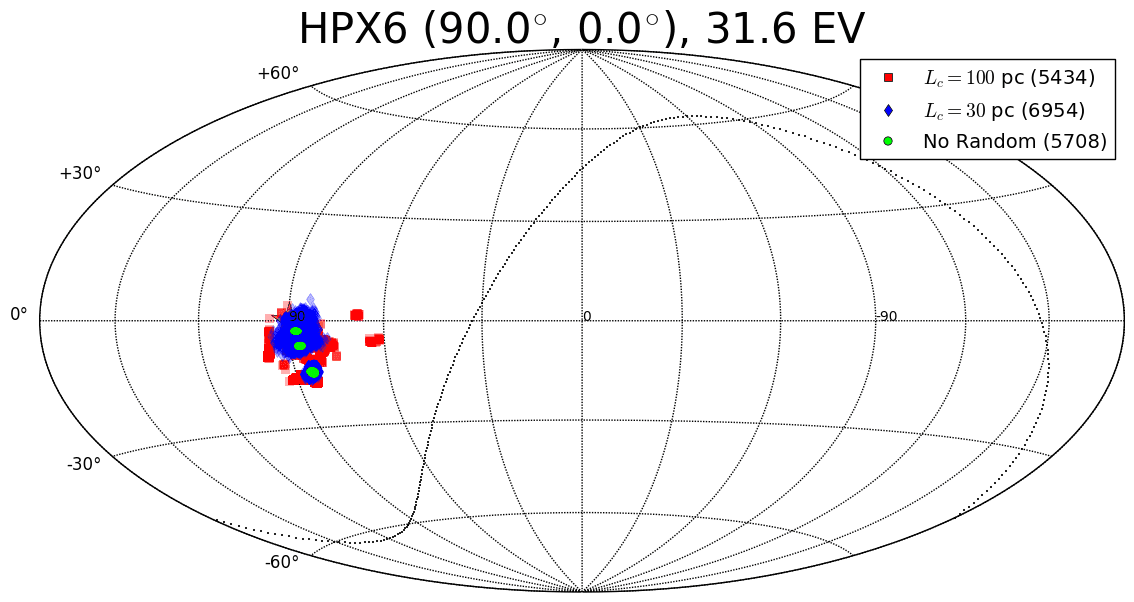}
\end{minipage}
\begin{minipage}[b]{0.48 \textwidth}
\includegraphics[width=1. \textwidth]{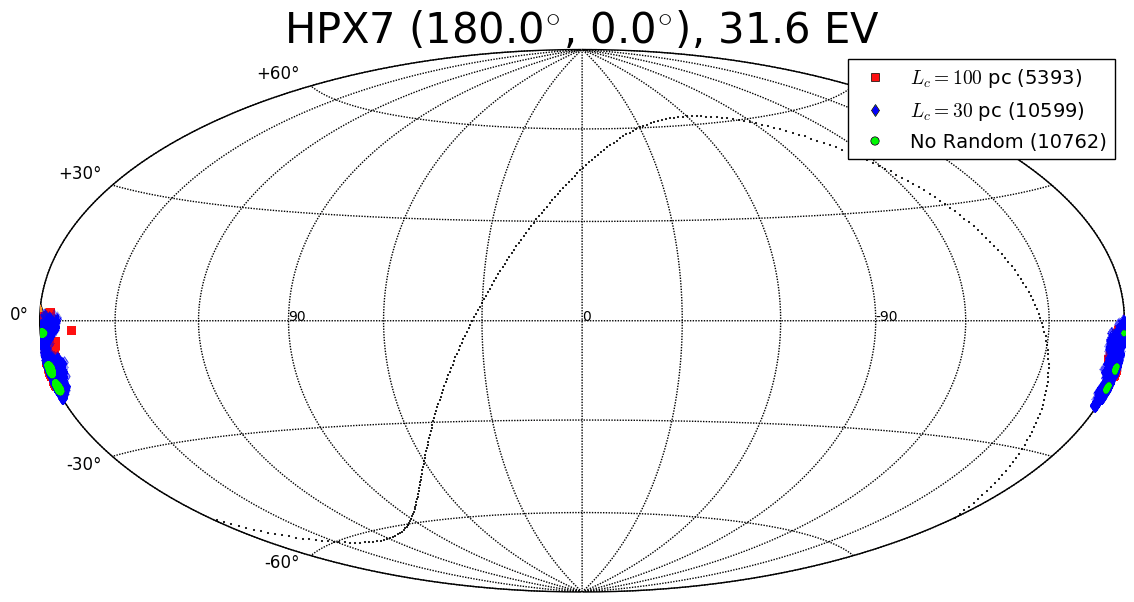}
\end{minipage}
\begin{minipage}[b]{0.48 \textwidth}
\includegraphics[width=1. \textwidth]{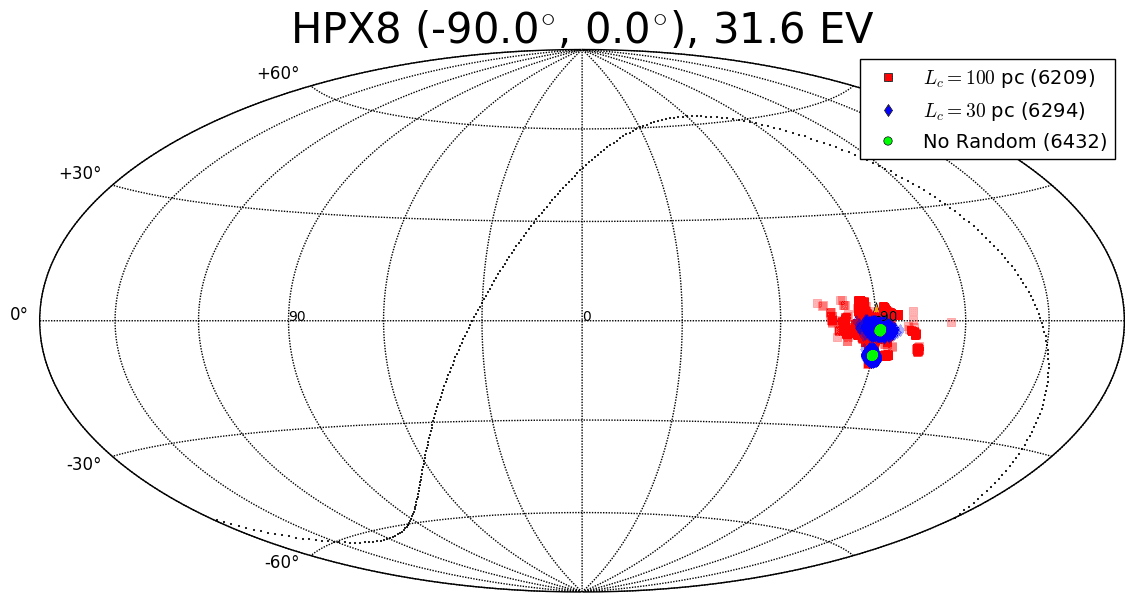}
\end{minipage}
\begin{minipage}[b]{0.48 \textwidth}
\includegraphics[width=1. \textwidth]{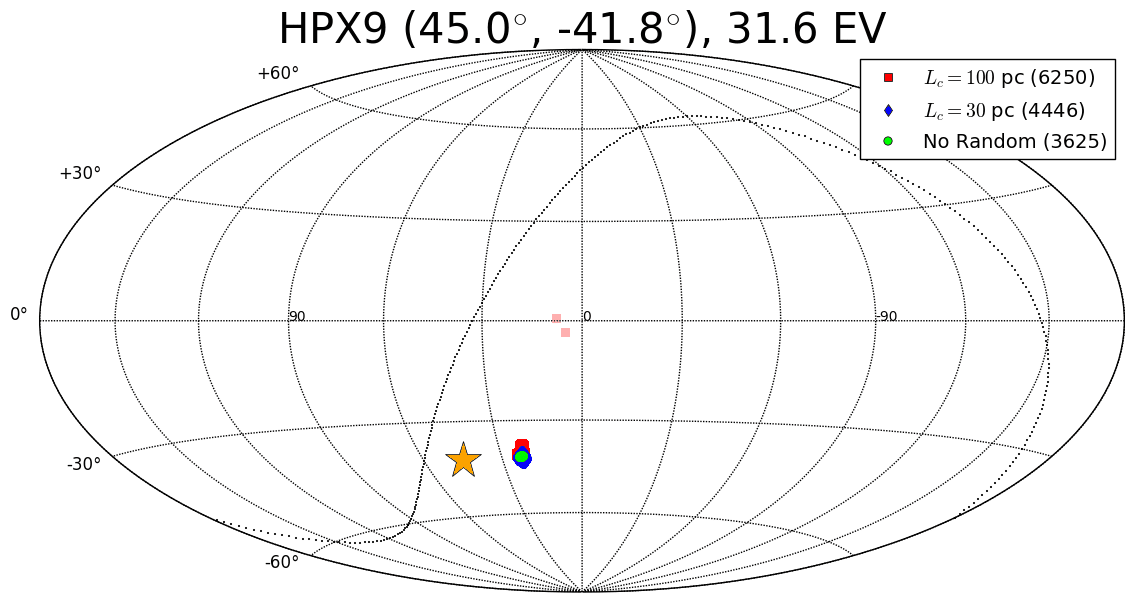}
\end{minipage}
\begin{minipage}[b]{0.48 \textwidth}
\includegraphics[width=1. \textwidth]{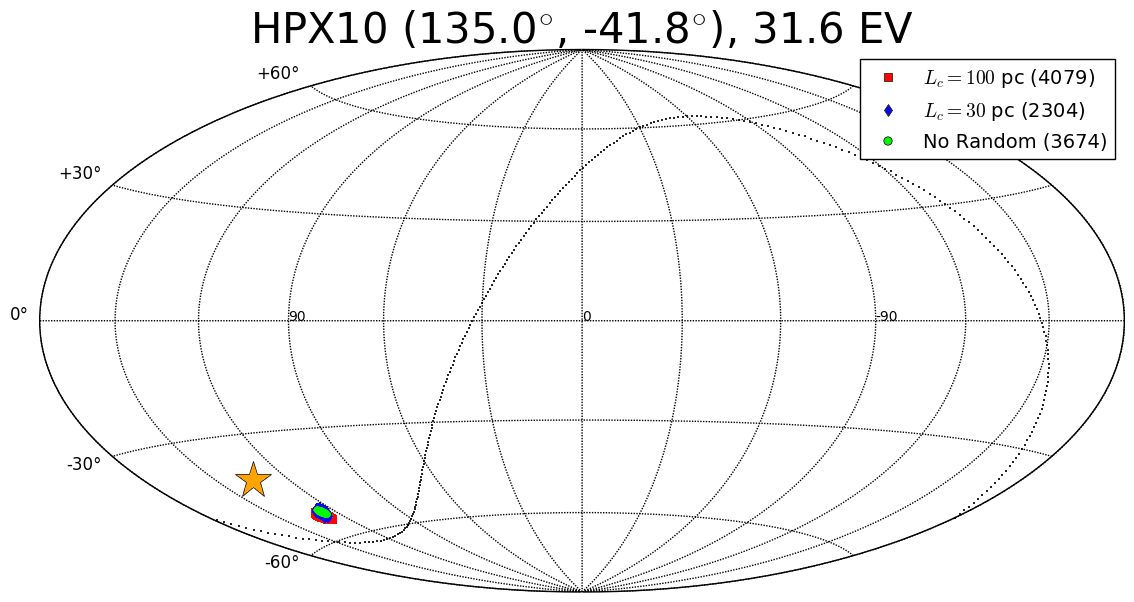}
\end{minipage}
\vspace{-0.3in}
\caption{As in Fig. \ref{plt:hpx19o8}, arrival direction distributions for log($R$ / V) = 19.5.} 
\label{plt:hpx19o5}
\vspace{-0.1in}
\end{figure}
\clearpage 
\begin{figure}[t]
\hspace{-0.3in}
\centering
\begin{minipage}[b]{0.48 \textwidth}
\includegraphics[width=1. \textwidth]{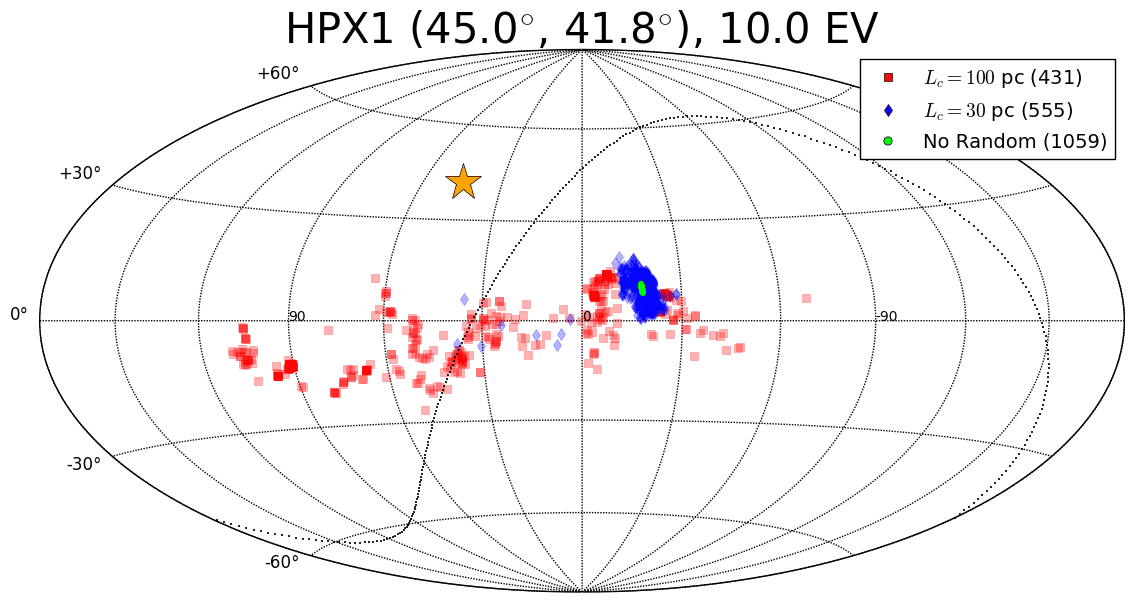}
\end{minipage}
\begin{minipage}[b]{0.48 \textwidth}
\includegraphics[width=1. \textwidth]{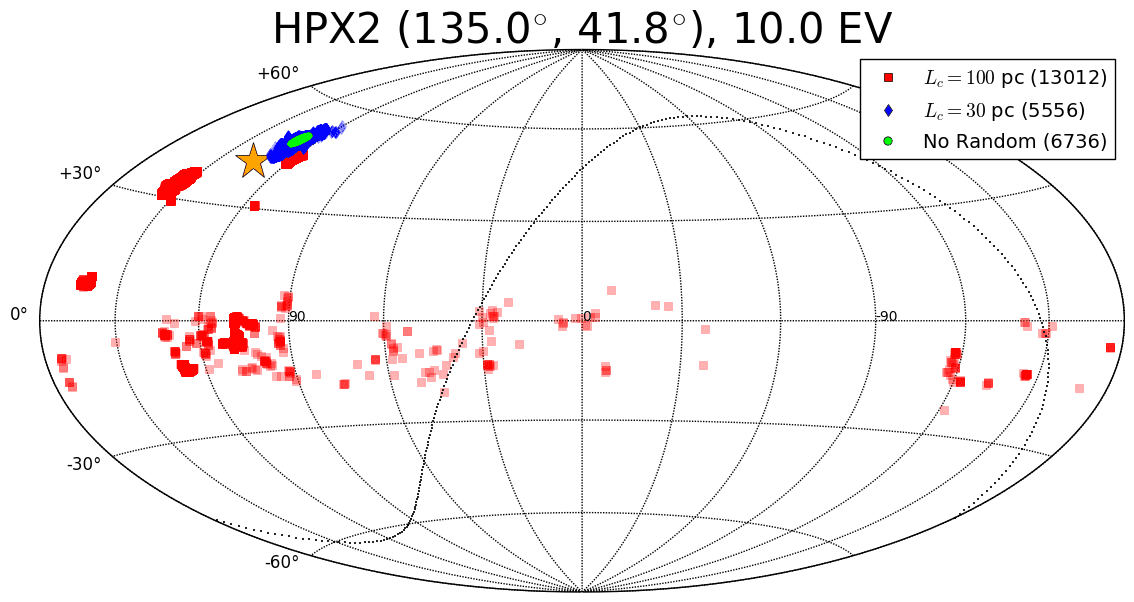}
\end{minipage}
\begin{minipage}[b]{0.48 \textwidth}
\includegraphics[width=1. \textwidth]{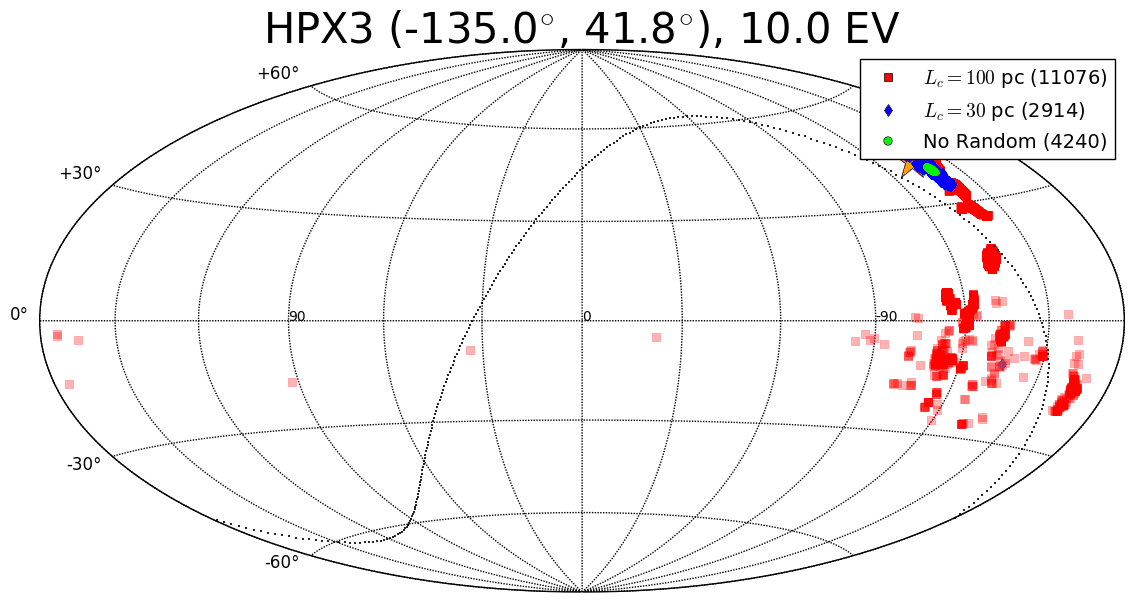}
\end{minipage}
\begin{minipage}[b]{0.48 \textwidth}
\includegraphics[width=1. \textwidth]{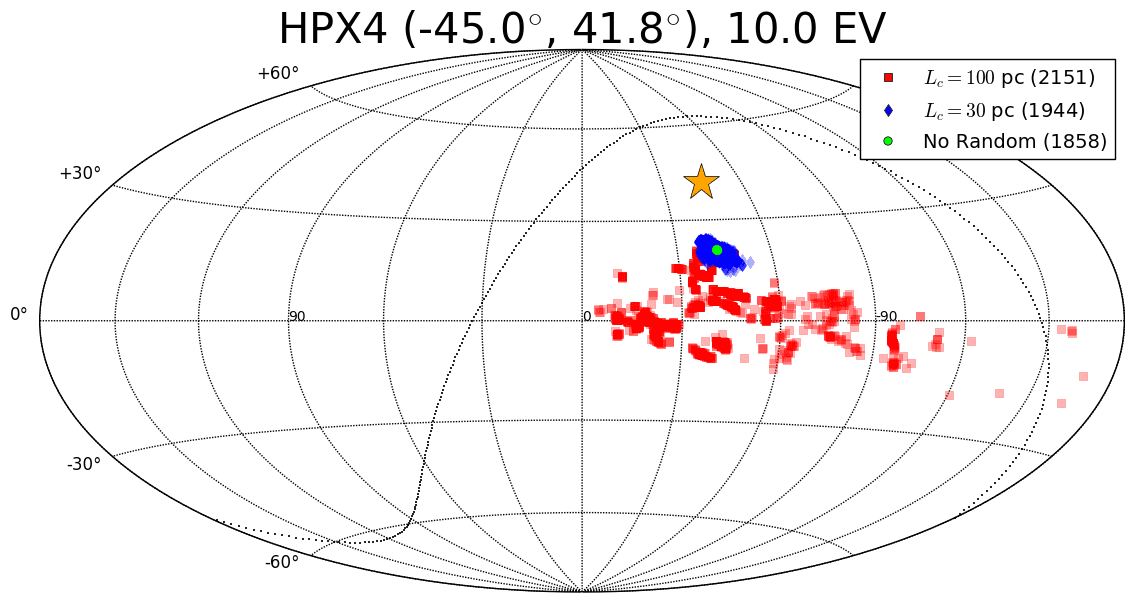}
\end{minipage}
\begin{minipage}[b]{0.48 \textwidth}
\includegraphics[width=1. \textwidth]{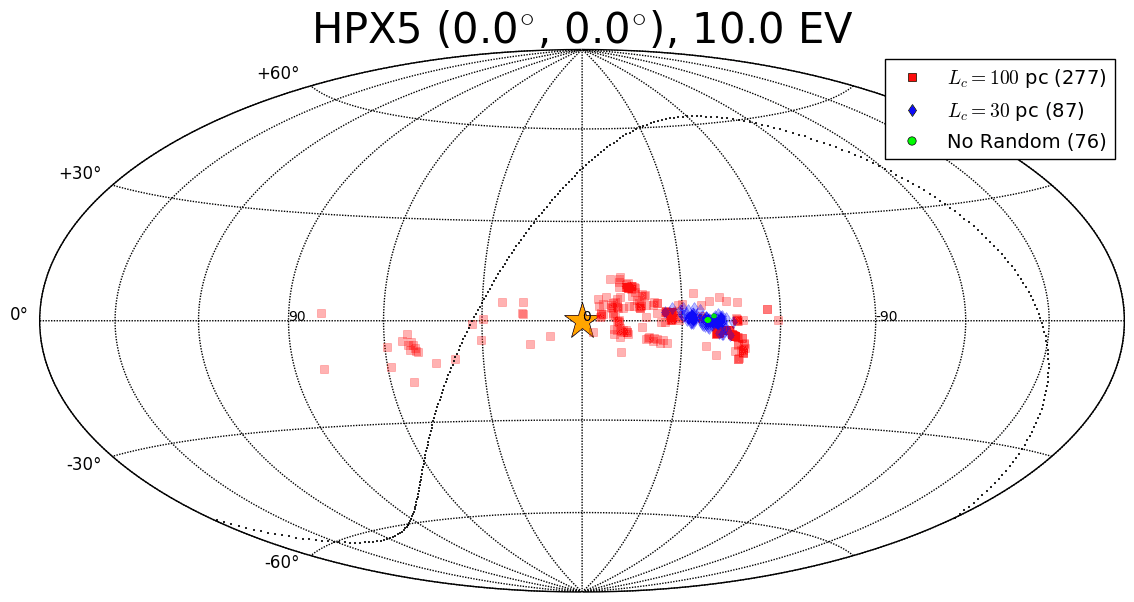}
\end{minipage}
\begin{minipage}[b]{0.48 \textwidth}
\includegraphics[width=1. \textwidth]{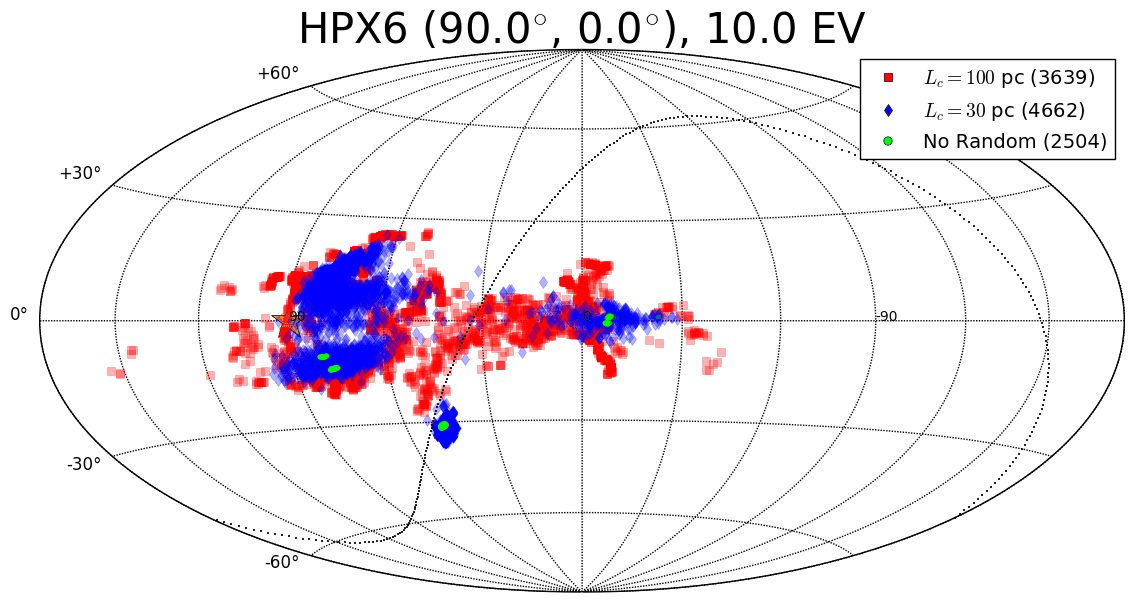}
\end{minipage}
\begin{minipage}[b]{0.48 \textwidth}
\includegraphics[width=1. \textwidth]{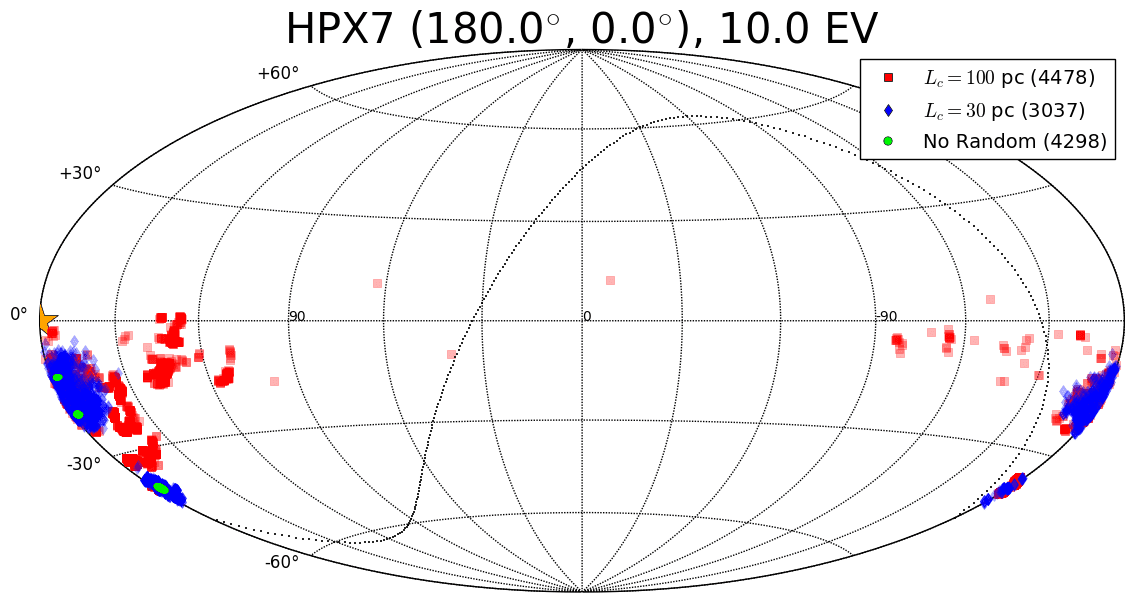}
\end{minipage}
\begin{minipage}[b]{0.48 \textwidth}
\includegraphics[width=1. \textwidth]{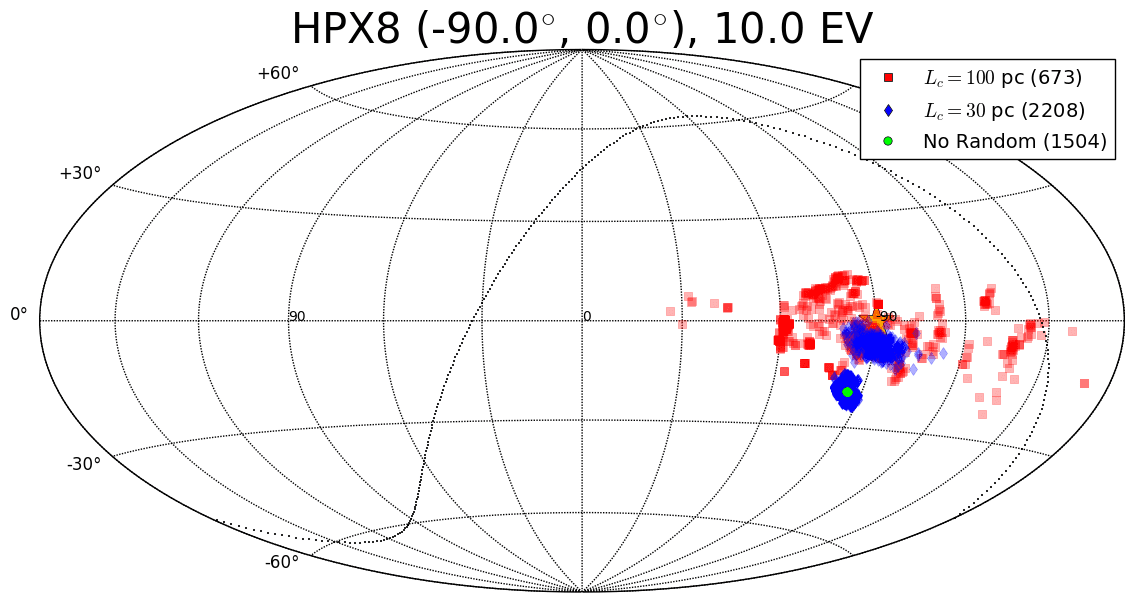}
\end{minipage}
\begin{minipage}[b]{0.48 \textwidth}
\includegraphics[width=1. \textwidth]{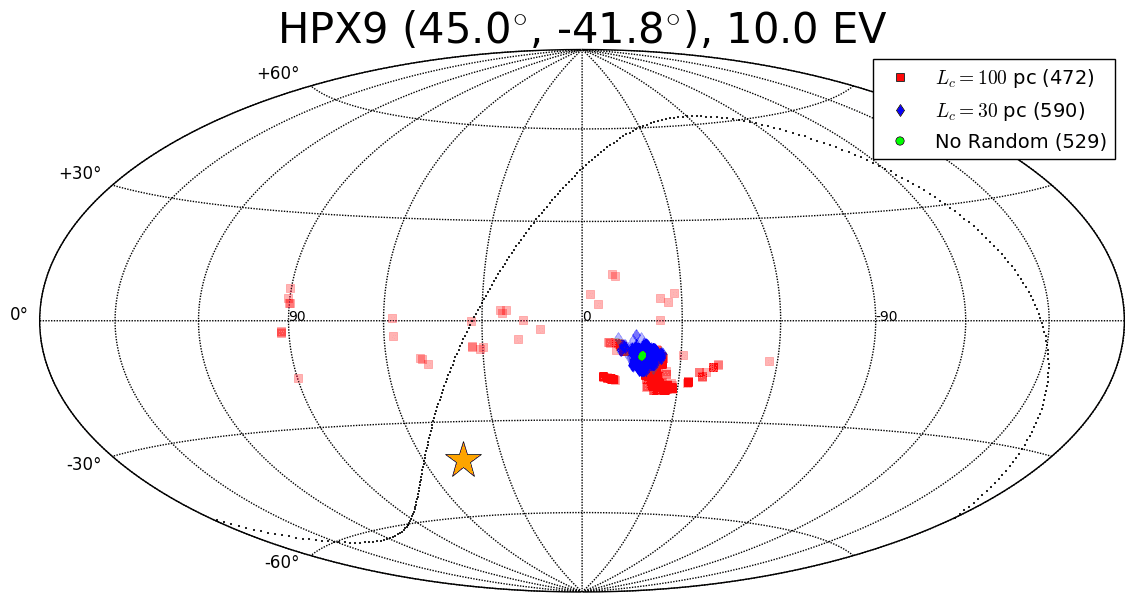}
\end{minipage}
\begin{minipage}[b]{0.48 \textwidth}
\includegraphics[width=1. \textwidth]{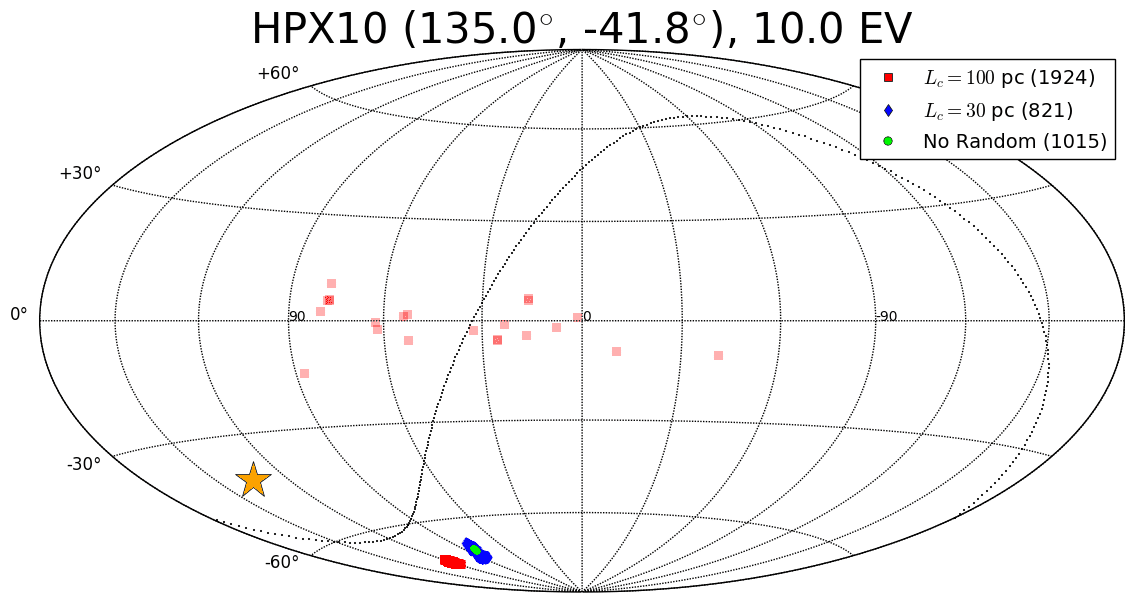}
\end{minipage}
\vspace{-0.3in}
\caption{As in Fig. \ref{plt:hpx19o8}, arrival direction distributions for log($R$ / V) = 19.0.} 
\label{plt:hpx19o0}
\vspace{-0.1in}
\end{figure}
\clearpage 
\begin{figure}[t]
\hspace{-0.3in}
\centering
\begin{minipage}[b]{0.48 \textwidth}
\includegraphics[width=1. \textwidth]{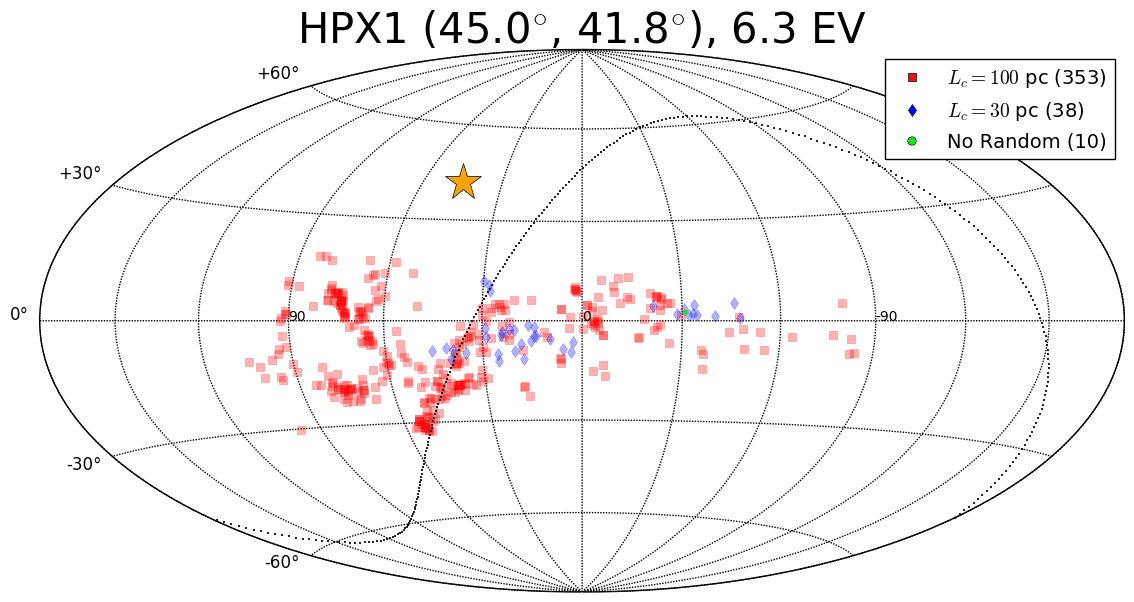}
\end{minipage}
\begin{minipage}[b]{0.48 \textwidth}
\includegraphics[width=1. \textwidth]{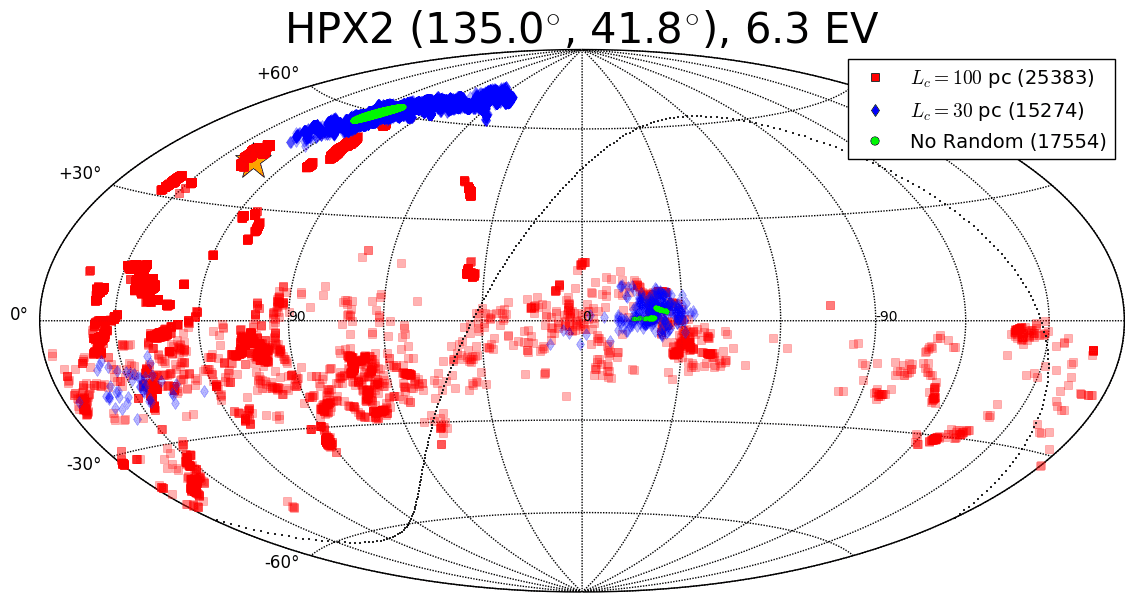}
\end{minipage}
\begin{minipage}[b]{0.48 \textwidth}
\includegraphics[width=1. \textwidth]{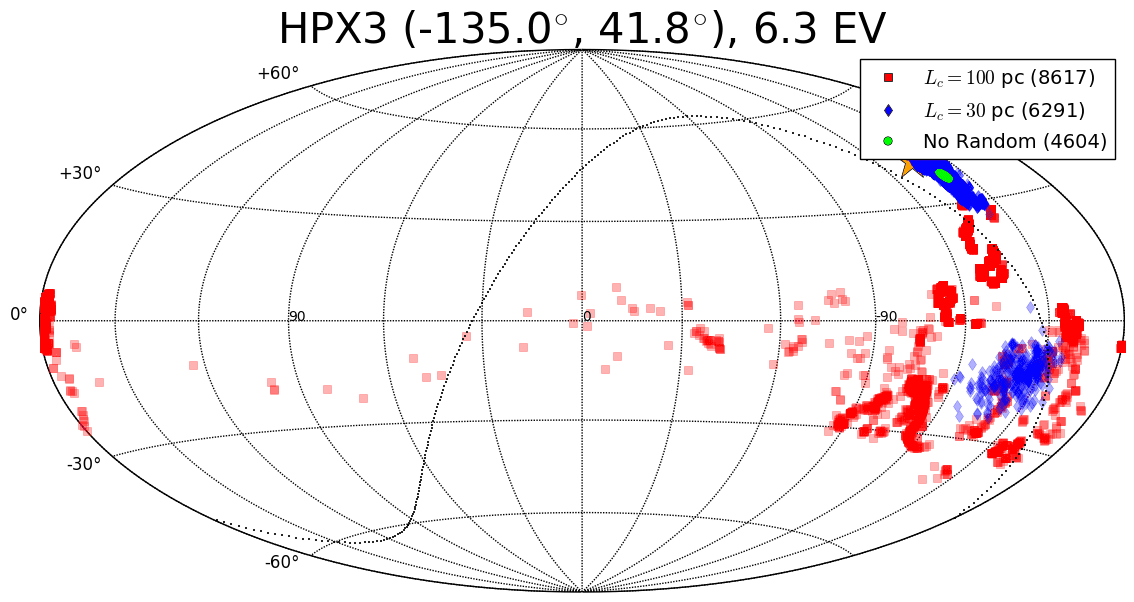}
\end{minipage}
\begin{minipage}[b]{0.48 \textwidth}
\includegraphics[width=1. \textwidth]{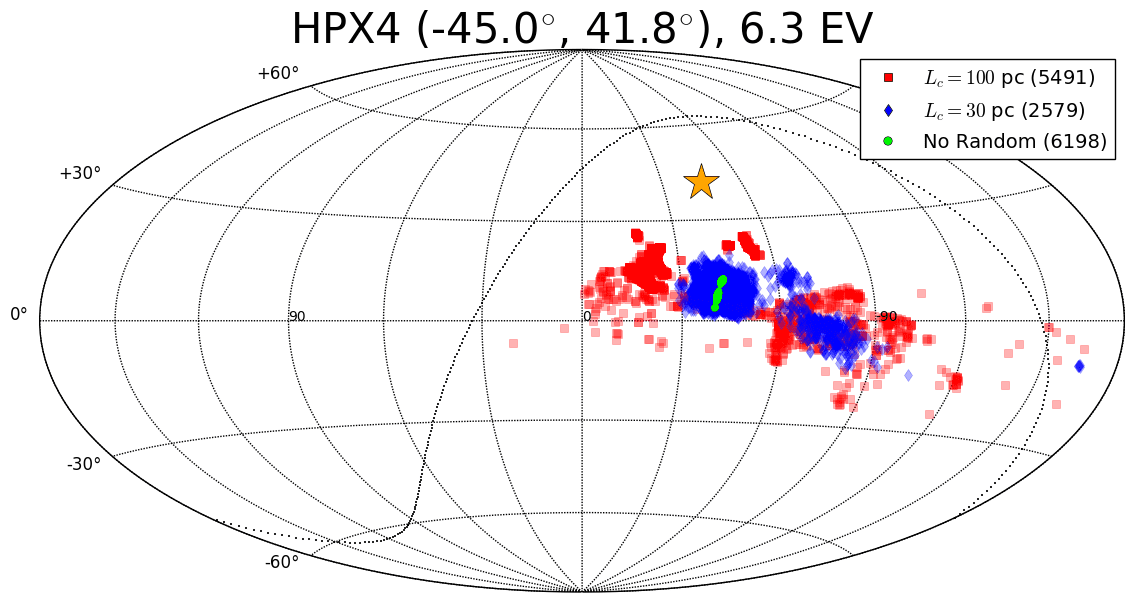}
\end{minipage}
\begin{minipage}[b]{0.48 \textwidth}
\includegraphics[width=1. \textwidth]{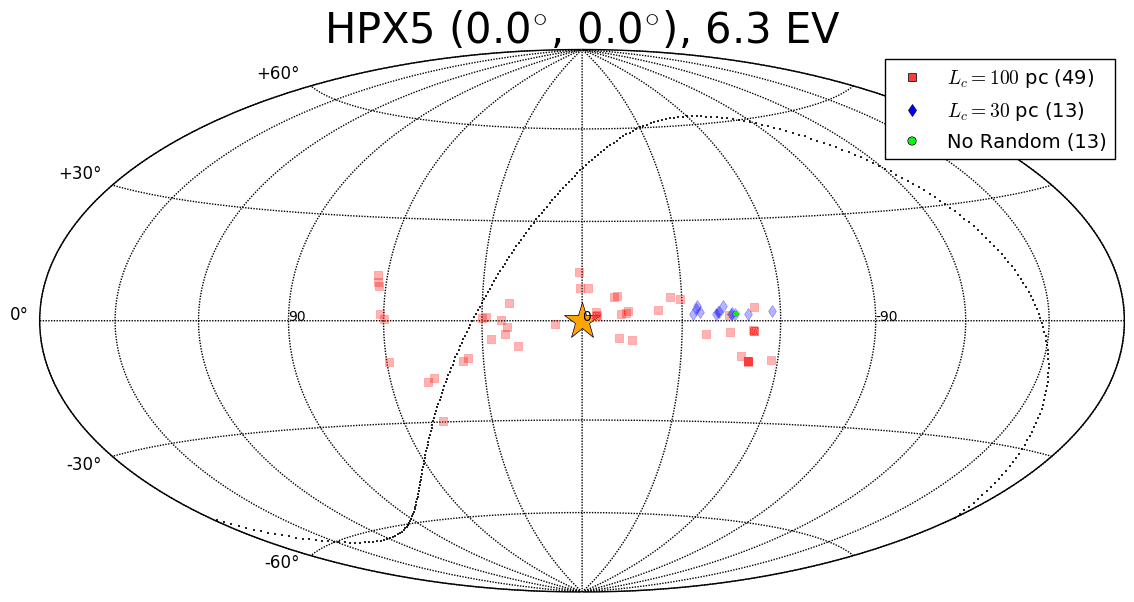}
\end{minipage}
\begin{minipage}[b]{0.48 \textwidth}
\includegraphics[width=1. \textwidth]{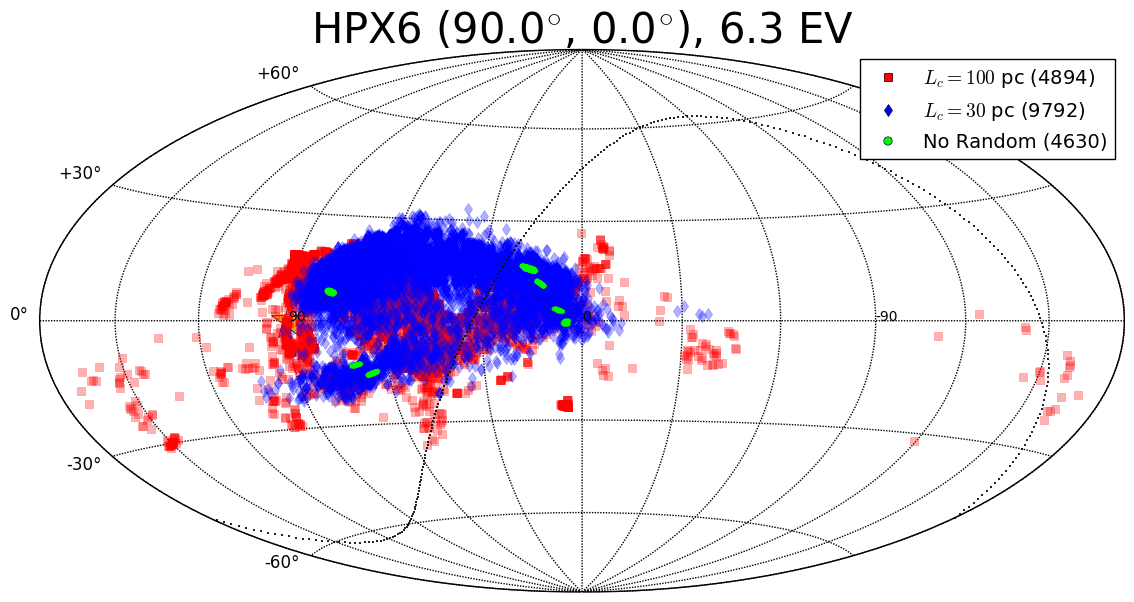}
\end{minipage}
\begin{minipage}[b]{0.48 \textwidth}
\includegraphics[width=1. \textwidth]{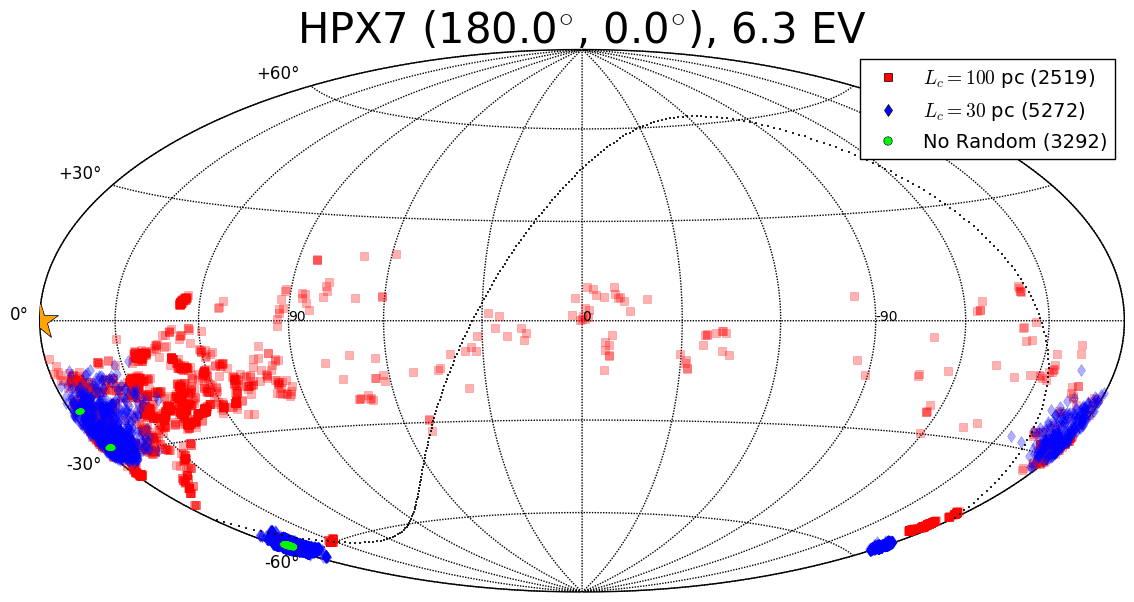}
\end{minipage}
\begin{minipage}[b]{0.48 \textwidth}
\includegraphics[width=1. \textwidth]{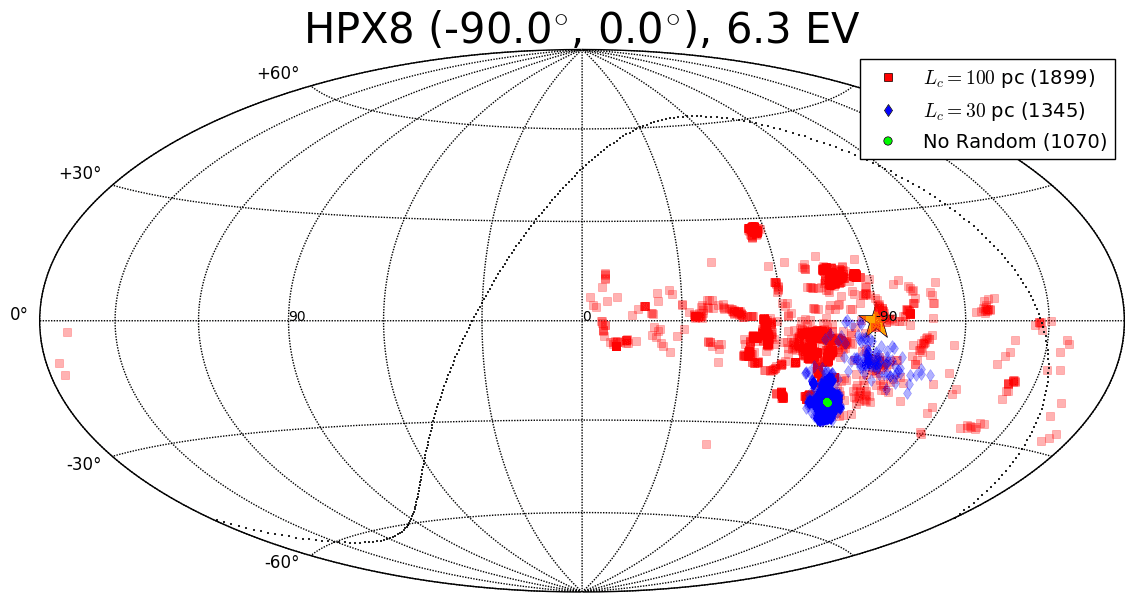}
\end{minipage}
\begin{minipage}[b]{0.48 \textwidth}
\includegraphics[width=1. \textwidth]{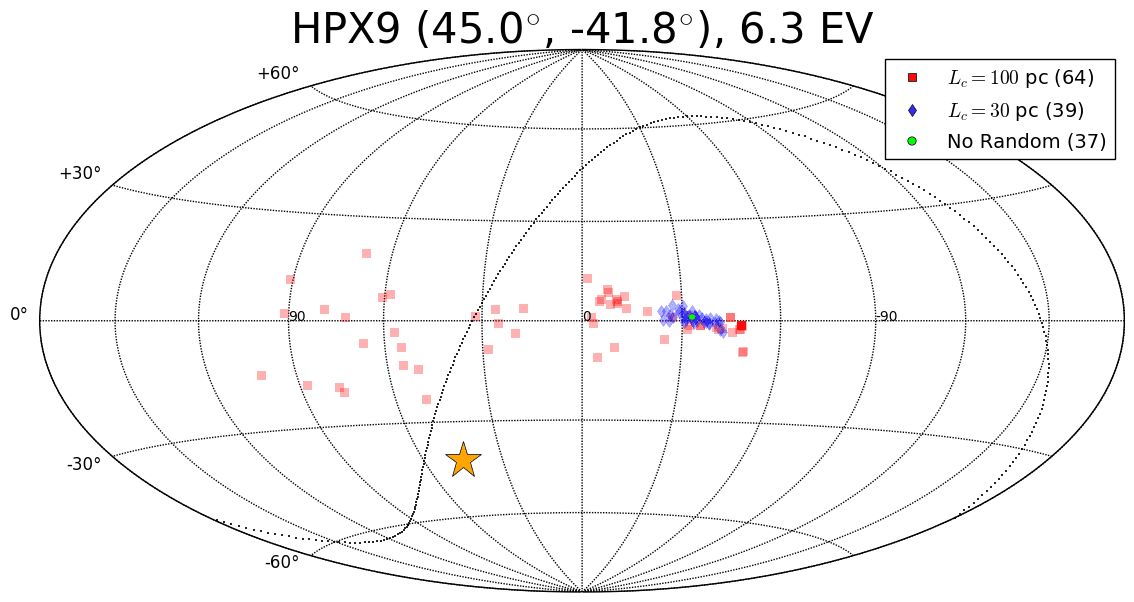}
\end{minipage}
\begin{minipage}[b]{0.48 \textwidth}
\includegraphics[width=1. \textwidth]{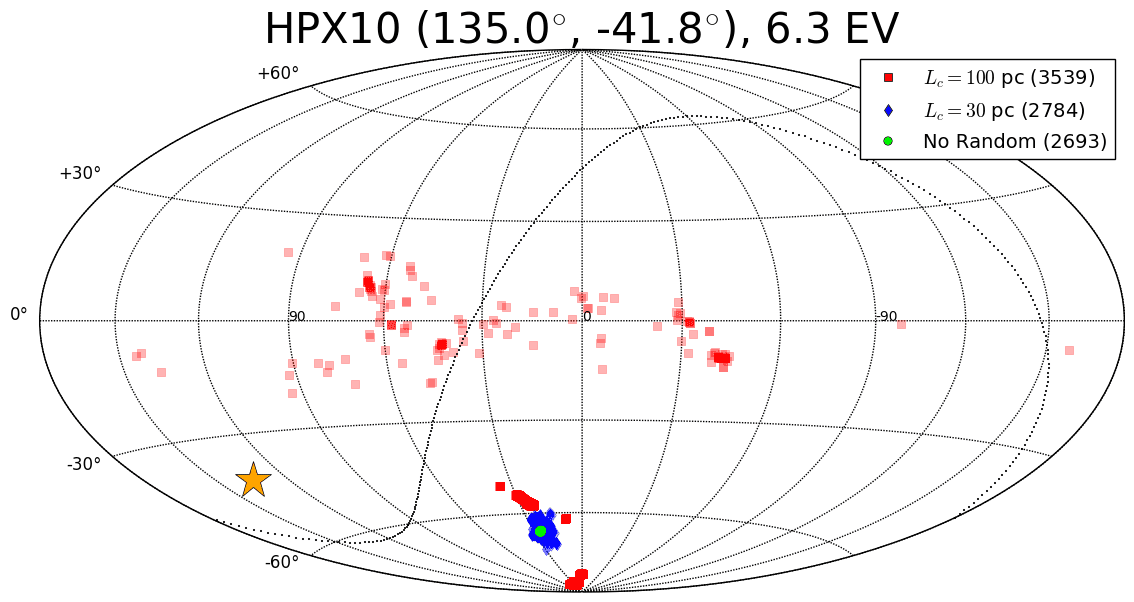}
\end{minipage}
\vspace{-0.3in}
\caption{As in Fig. \ref{plt:hpx19o8}, arrival direction distributions for log($R$ / V) = 18.8.} 
\label{plt:hpx18o8}
\vspace{-0.1in}
\end{figure}
\clearpage 
\begin{figure}[t]
\hspace{-0.3in}
\centering
\begin{minipage}[b]{0.48 \textwidth}
\includegraphics[width=1. \textwidth]{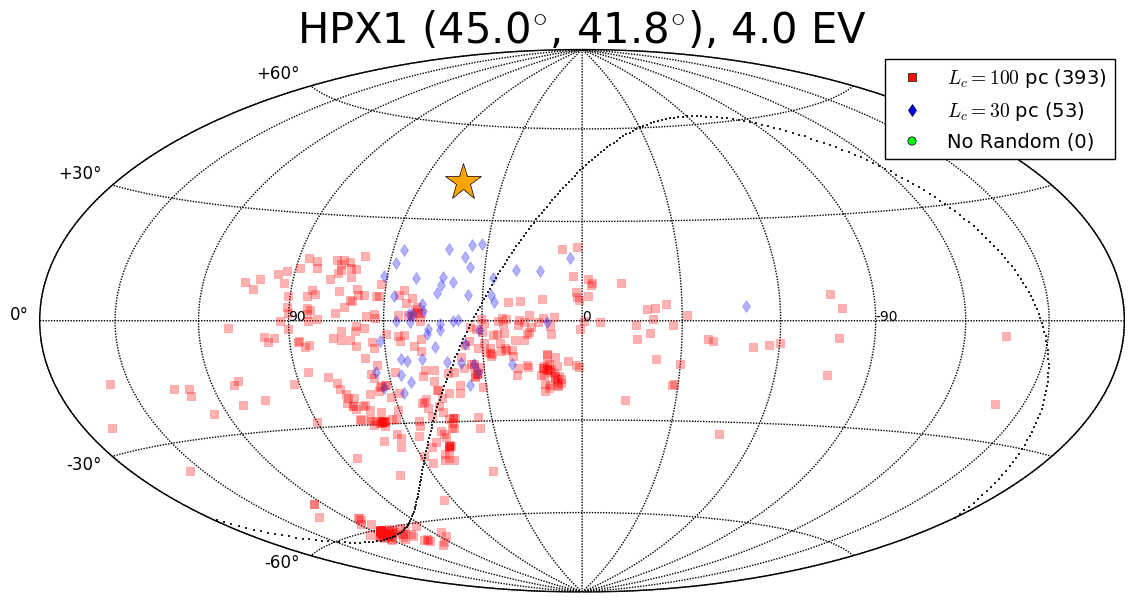}
\end{minipage}
\begin{minipage}[b]{0.48 \textwidth}
\includegraphics[width=1. \textwidth]{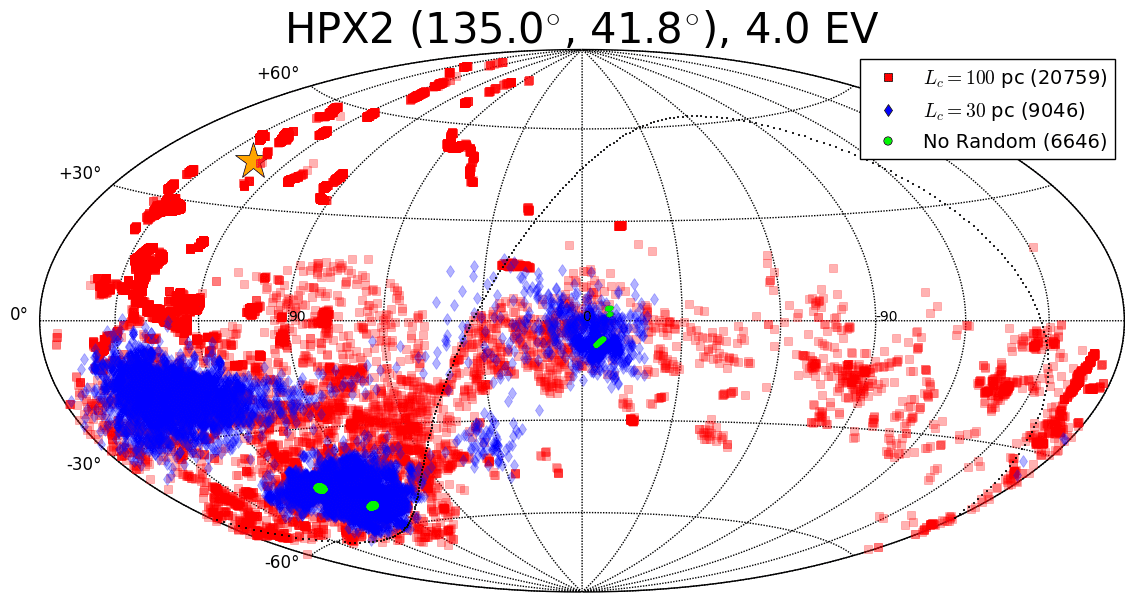}
\end{minipage}
\begin{minipage}[b]{0.48 \textwidth}
\includegraphics[width=1. \textwidth]{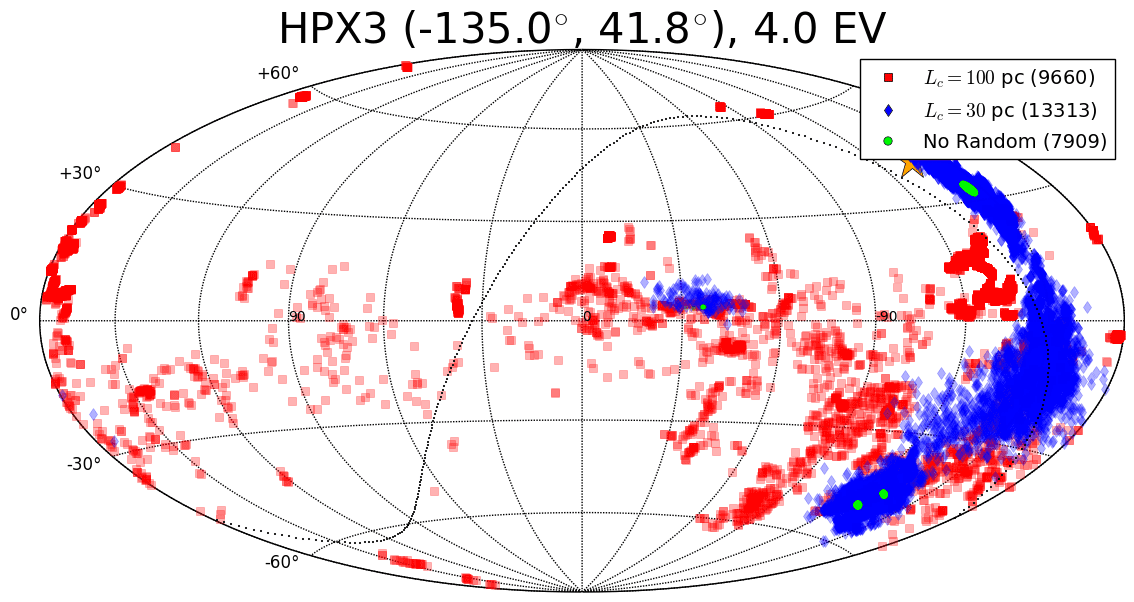}
\end{minipage}
\begin{minipage}[b]{0.48 \textwidth}
\includegraphics[width=1. \textwidth]{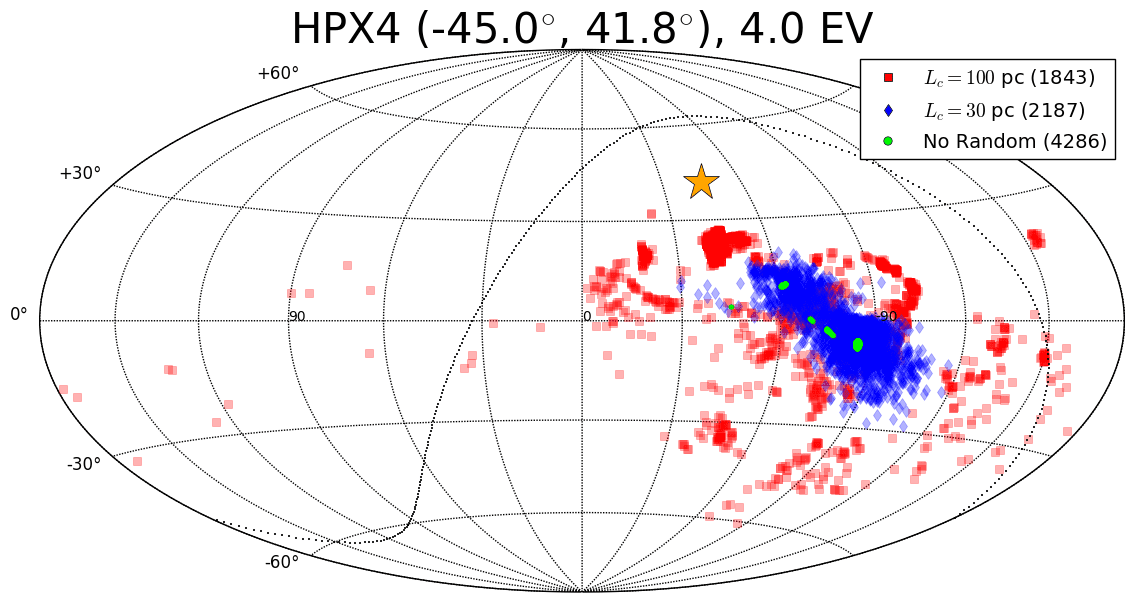}
\end{minipage}
\begin{minipage}[b]{0.48 \textwidth}
\includegraphics[width=1. \textwidth]{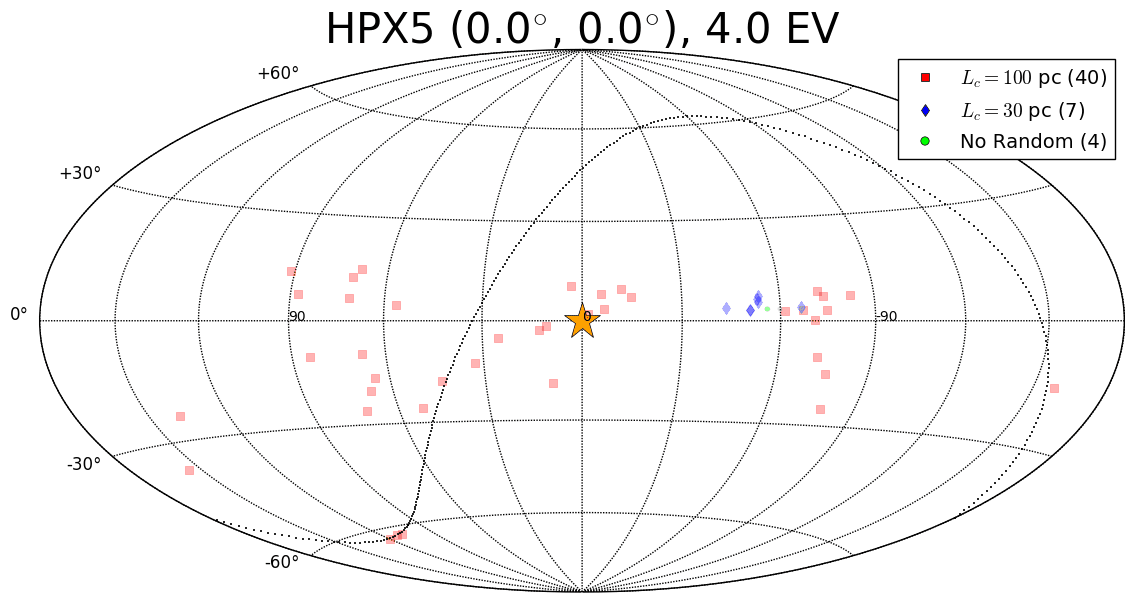}
\end{minipage}
\begin{minipage}[b]{0.48 \textwidth}
\includegraphics[width=1. \textwidth]{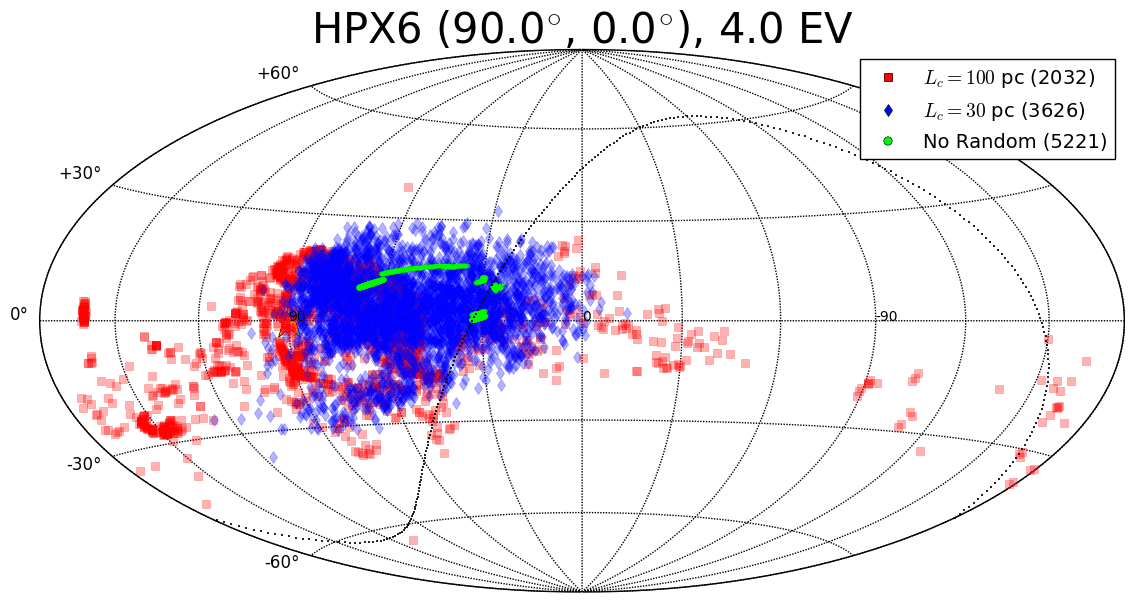}
\end{minipage}
\begin{minipage}[b]{0.48 \textwidth}
\includegraphics[width=1. \textwidth]{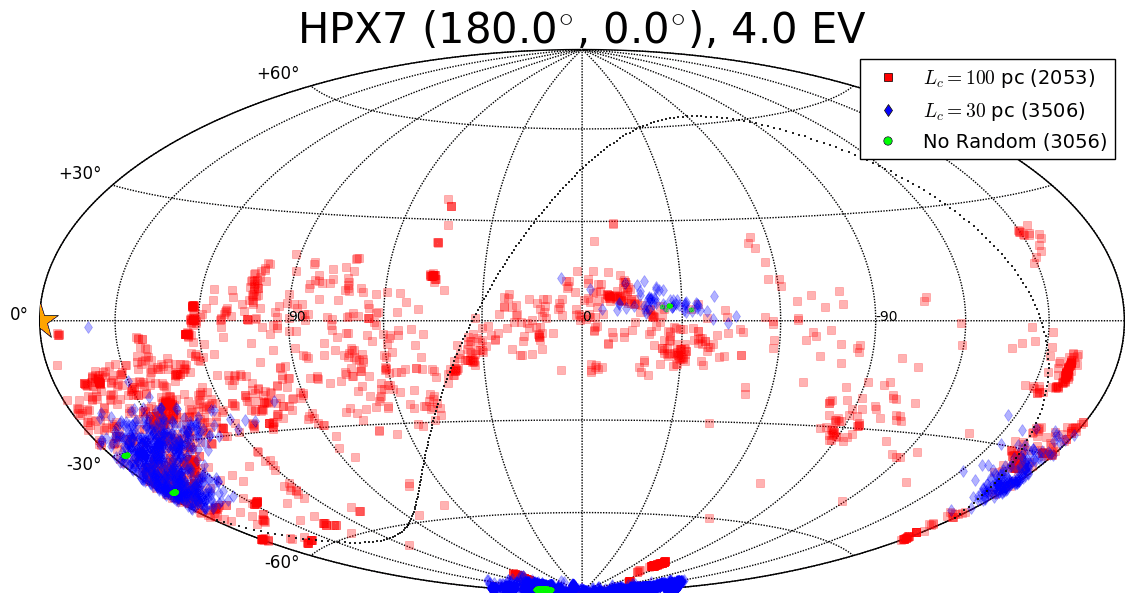}
\end{minipage}
\begin{minipage}[b]{0.48 \textwidth}
\includegraphics[width=1. \textwidth]{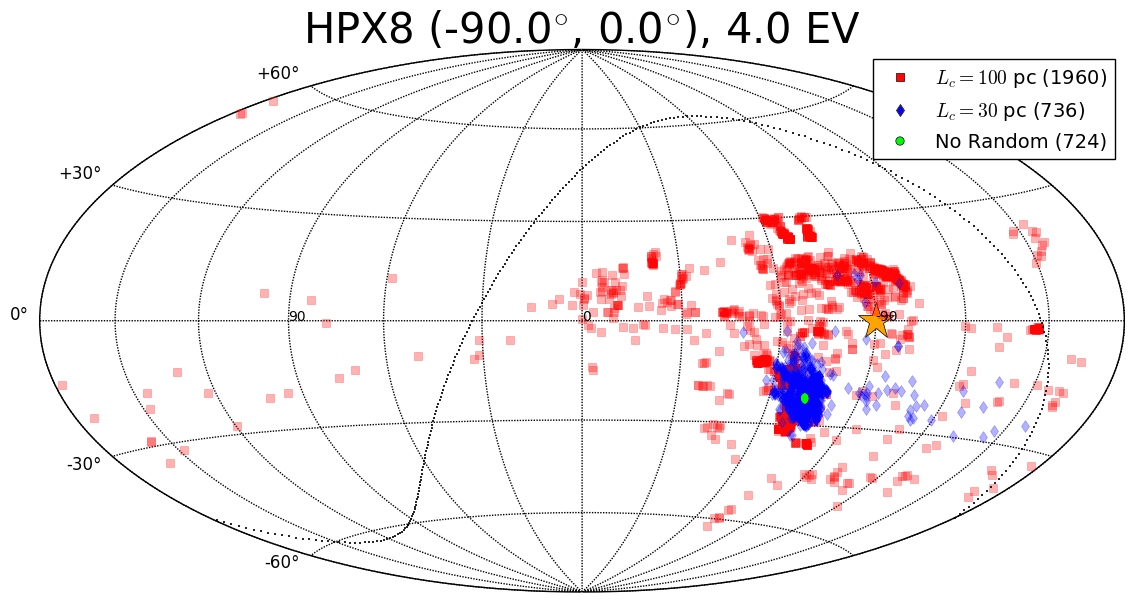}
\end{minipage}
\begin{minipage}[b]{0.48 \textwidth}
\includegraphics[width=1. \textwidth]{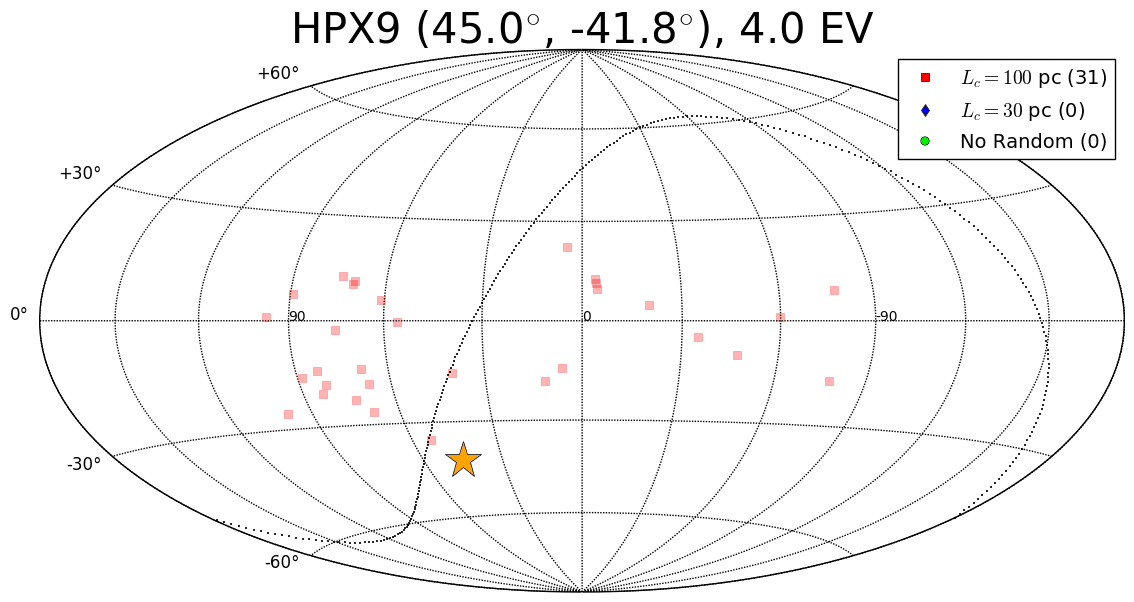}
\end{minipage}
\begin{minipage}[b]{0.48 \textwidth}
\includegraphics[width=1. \textwidth]{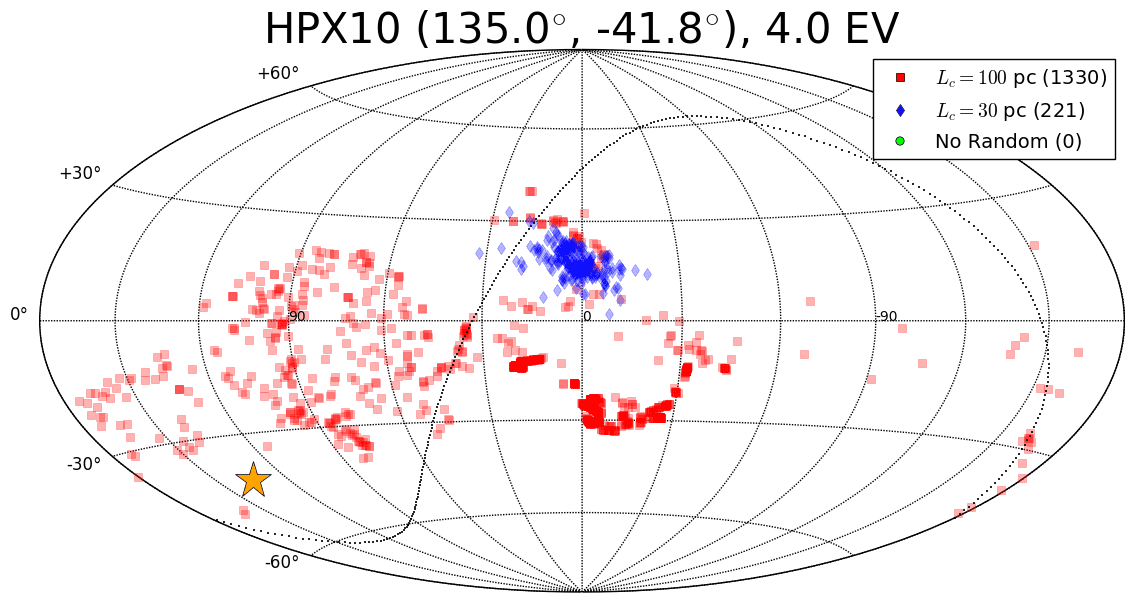}
\end{minipage}
\vspace{-0.3in}
\caption{As in Fig. \ref{plt:hpx19o8}, arrival direction distributions for log($R$ / V) = 18.6.} 
\label{plt:hpx18o6}
\vspace{-0.1in}
\end{figure}
\clearpage 
\begin{figure}[t]
\hspace{-0.3in}
\centering
\begin{minipage}[b]{0.48 \textwidth}
\includegraphics[width=1. \textwidth]{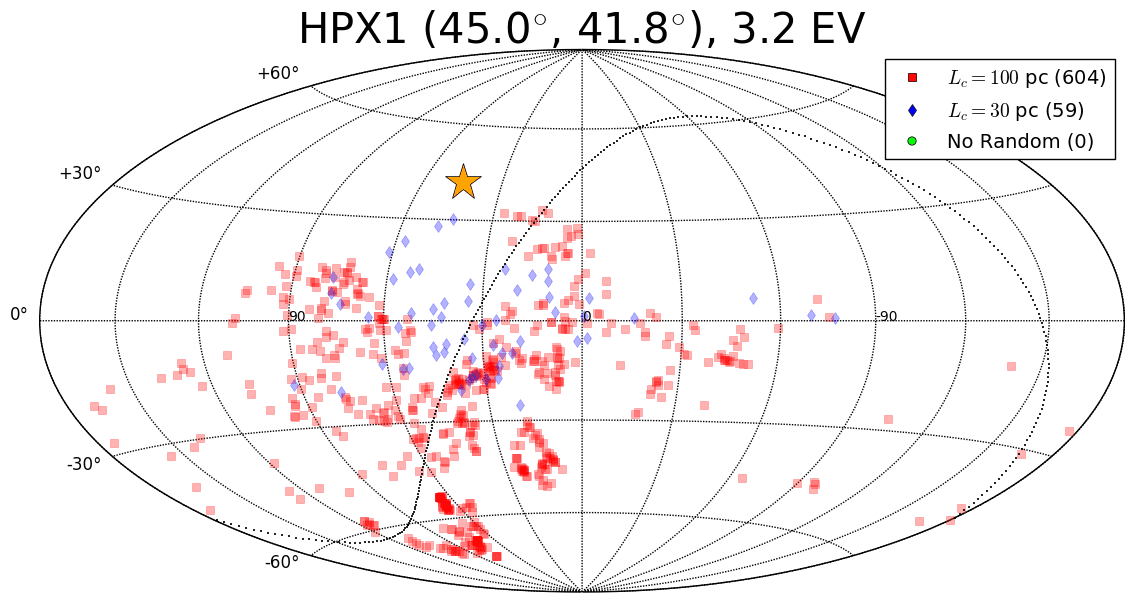}
\end{minipage}
\begin{minipage}[b]{0.48 \textwidth}
\includegraphics[width=1. \textwidth]{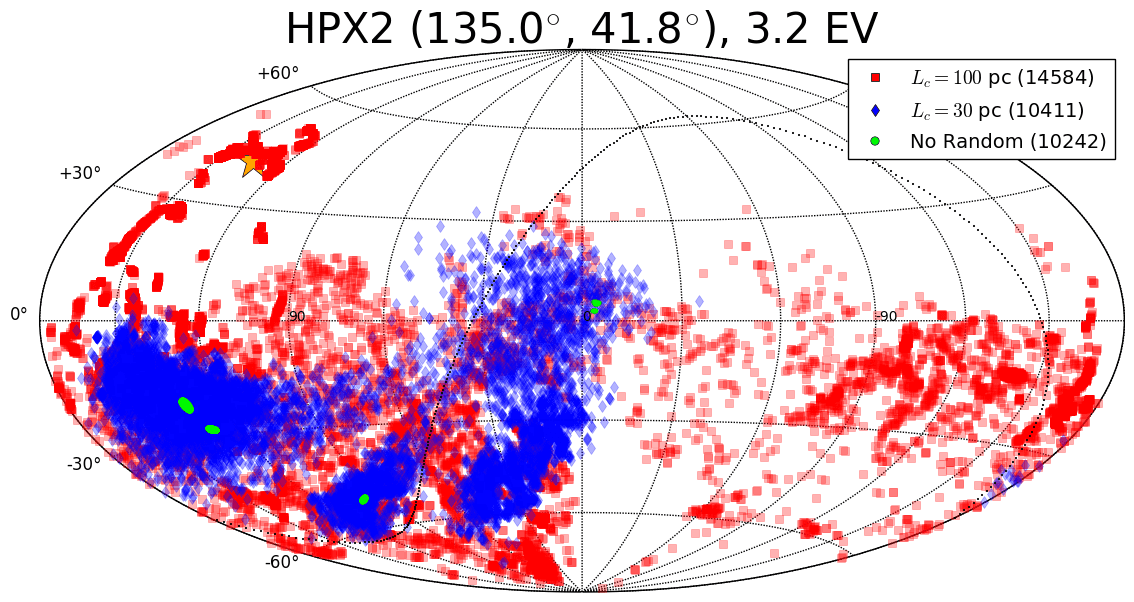}
\end{minipage}
\begin{minipage}[b]{0.48 \textwidth}
\includegraphics[width=1. \textwidth]{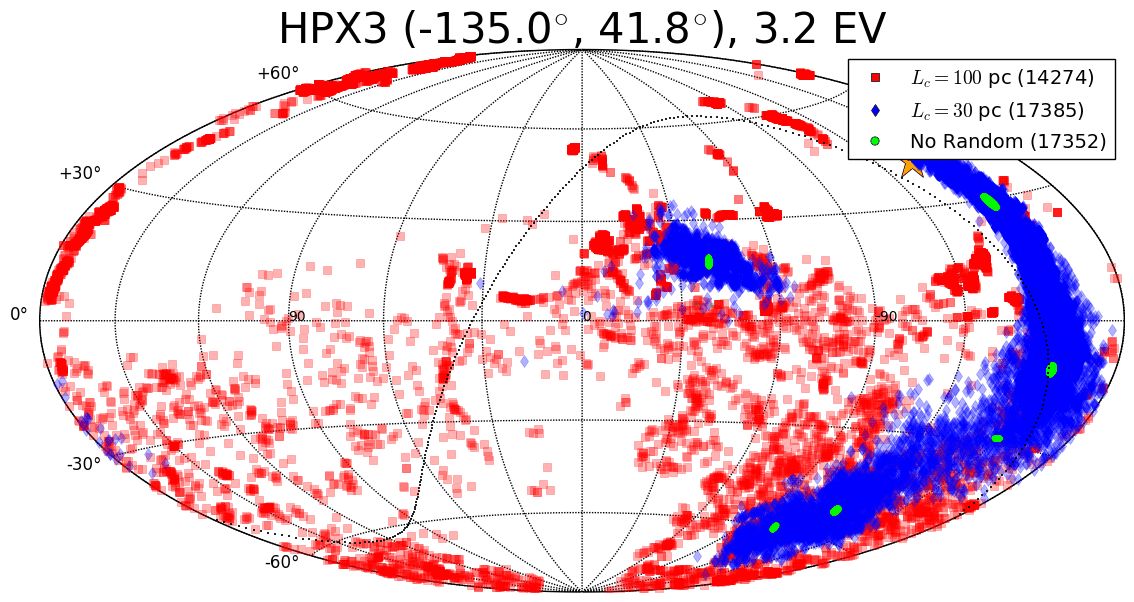}
\end{minipage}
\begin{minipage}[b]{0.48 \textwidth}
\includegraphics[width=1. \textwidth]{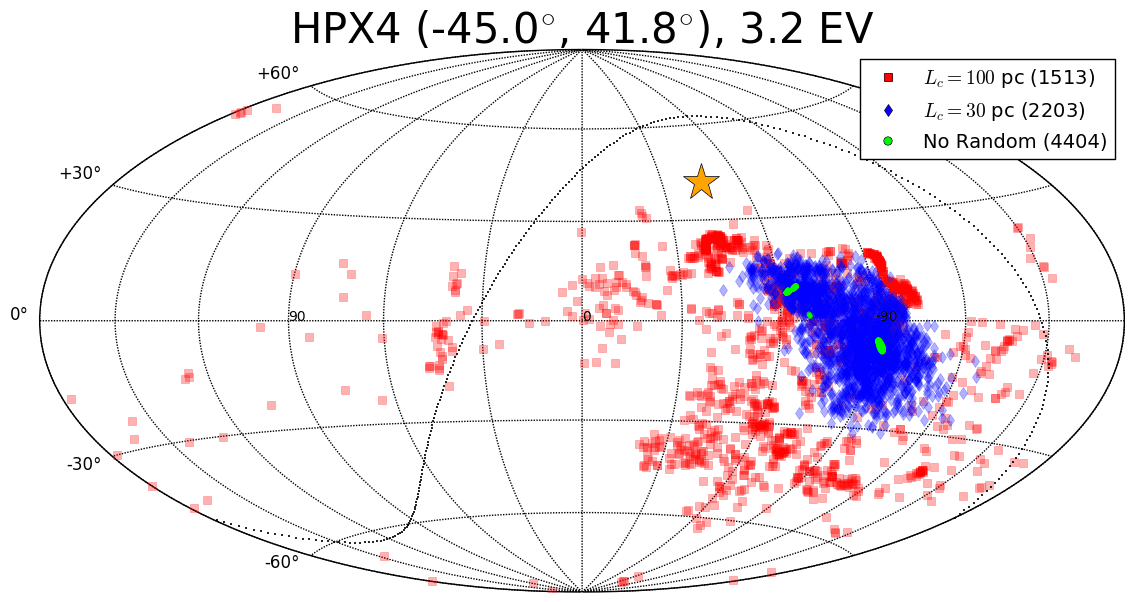}
\end{minipage}
\begin{minipage}[b]{0.48 \textwidth}
\includegraphics[width=1. \textwidth]{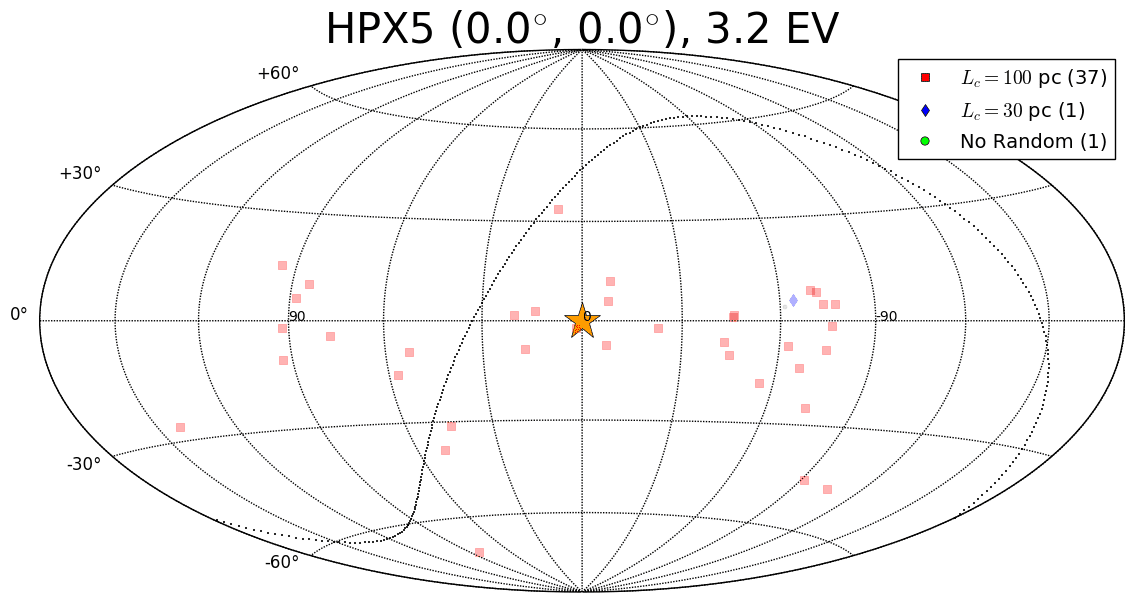}
\end{minipage}
\begin{minipage}[b]{0.48 \textwidth}
\includegraphics[width=1. \textwidth]{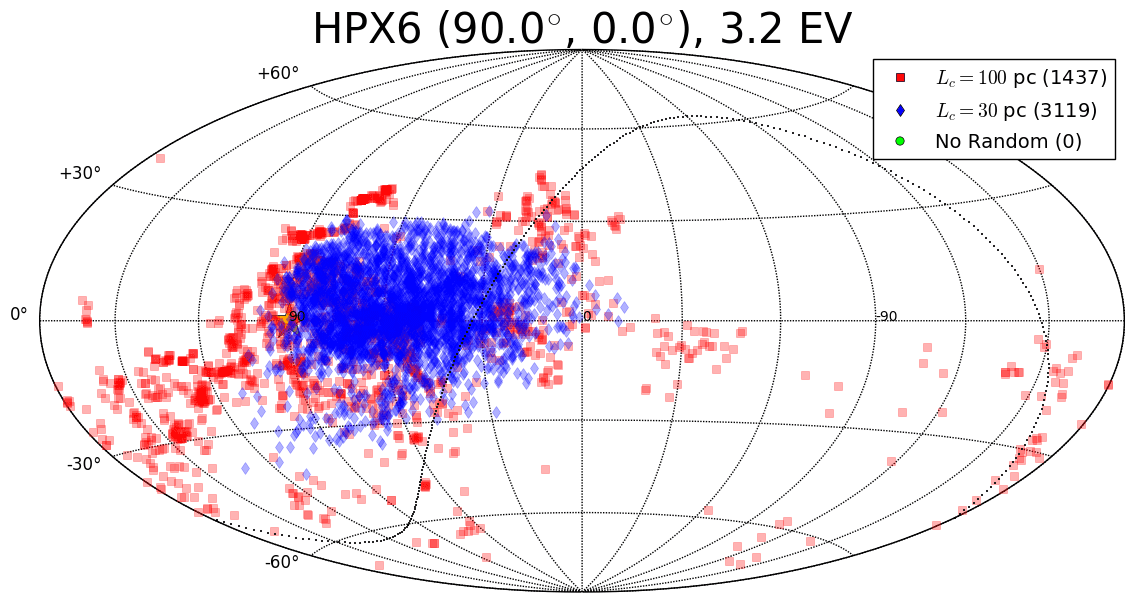}
\end{minipage}
\begin{minipage}[b]{0.48 \textwidth}
\includegraphics[width=1. \textwidth]{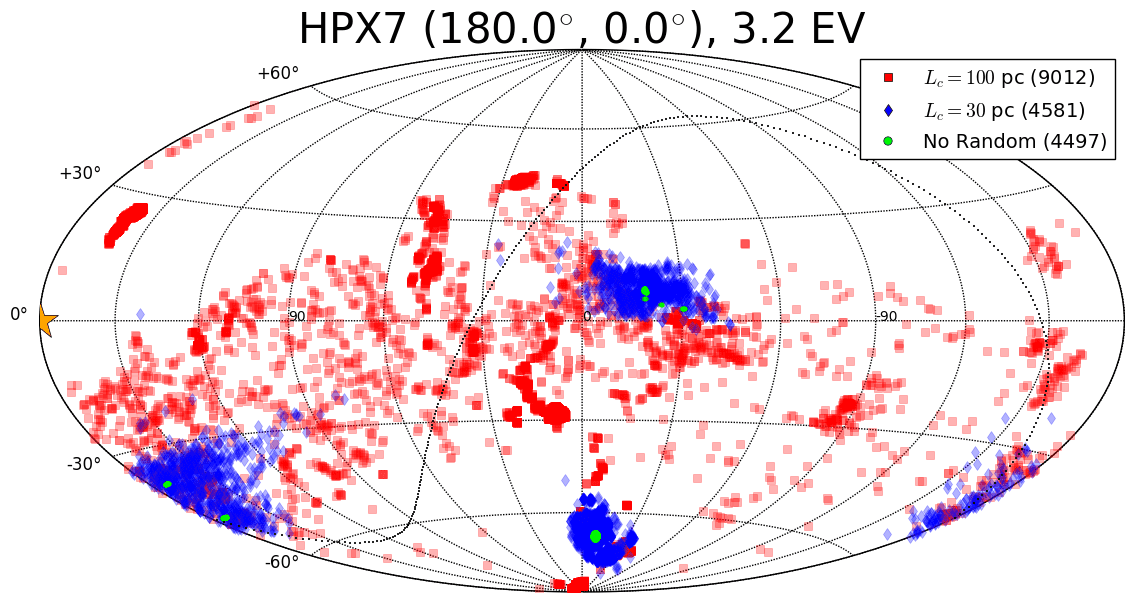}
\end{minipage}
\begin{minipage}[b]{0.48 \textwidth}
\includegraphics[width=1. \textwidth]{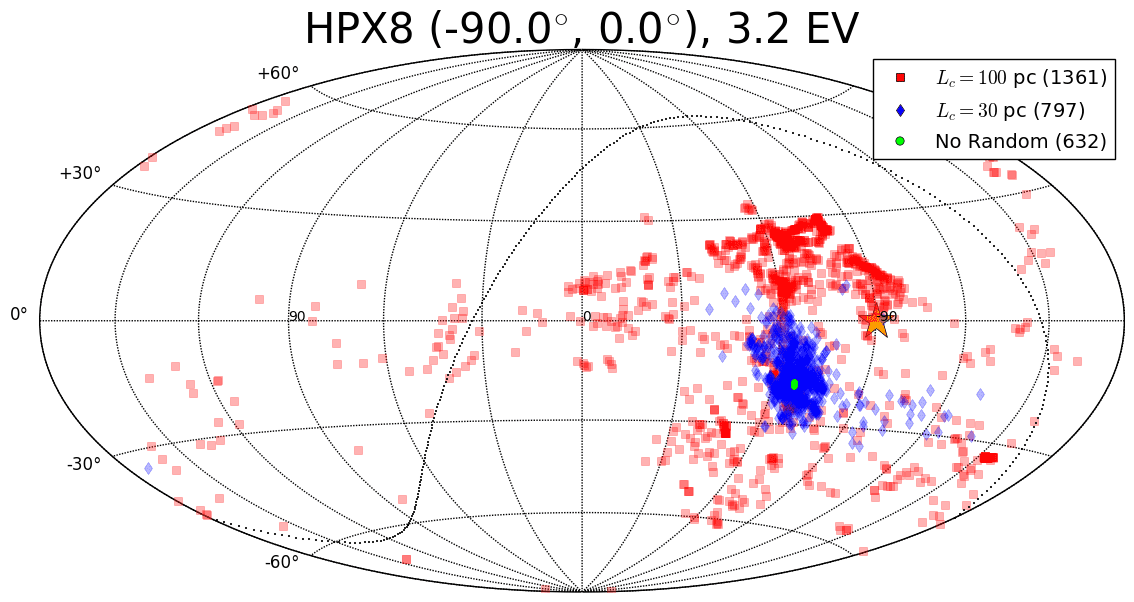}
\end{minipage}
\begin{minipage}[b]{0.48 \textwidth}
\includegraphics[width=1. \textwidth]{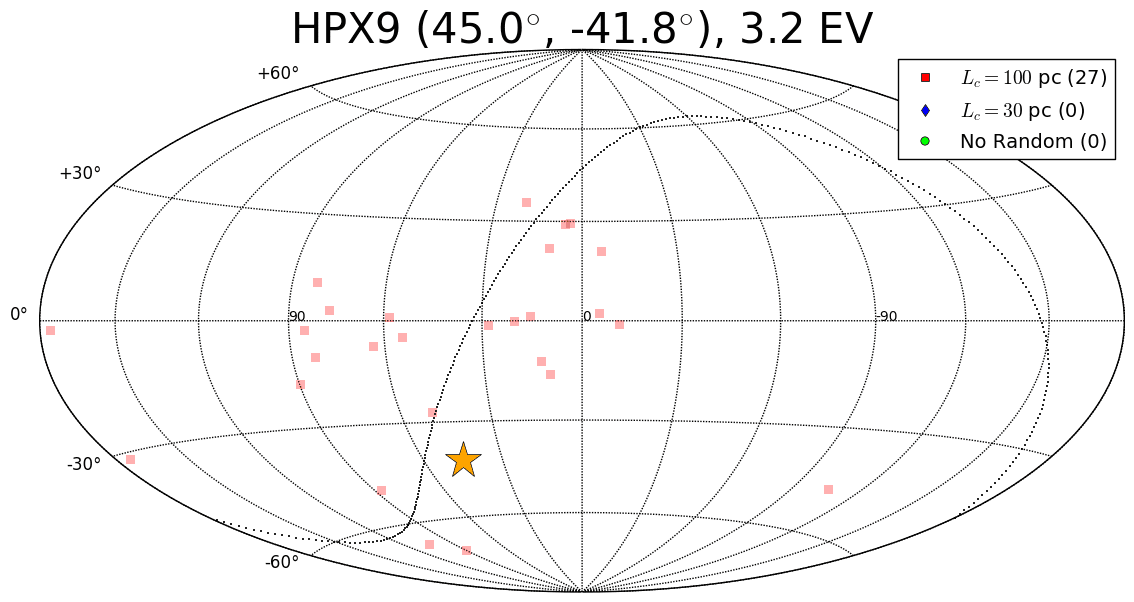}
\end{minipage}
\begin{minipage}[b]{0.48 \textwidth}
\includegraphics[width=1. \textwidth]{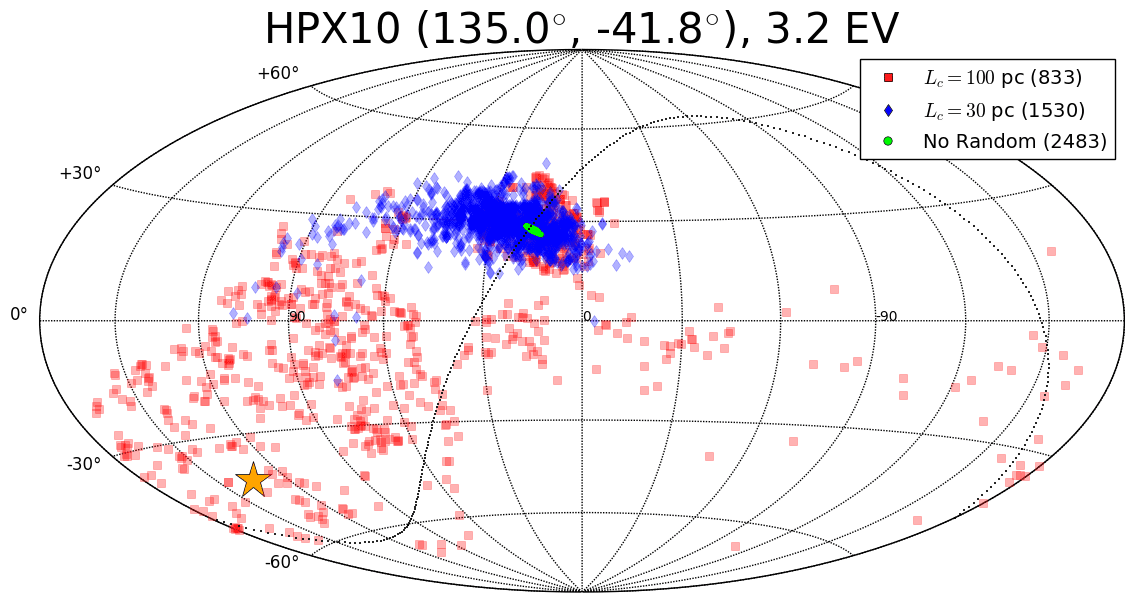}
\end{minipage}
\vspace{-0.3in}
\caption{As in Fig. \ref{plt:hpx19o8}, arrival direction distributions for log($R$ / V) = 18.5.} 
\label{plt:hpx18o5}
\vspace{-0.1in}
\end{figure}
\clearpage 
\begin{figure}[t]
\hspace{-0.3in}
\centering
\begin{minipage}[b]{0.48 \textwidth}
\includegraphics[width=1. \textwidth]{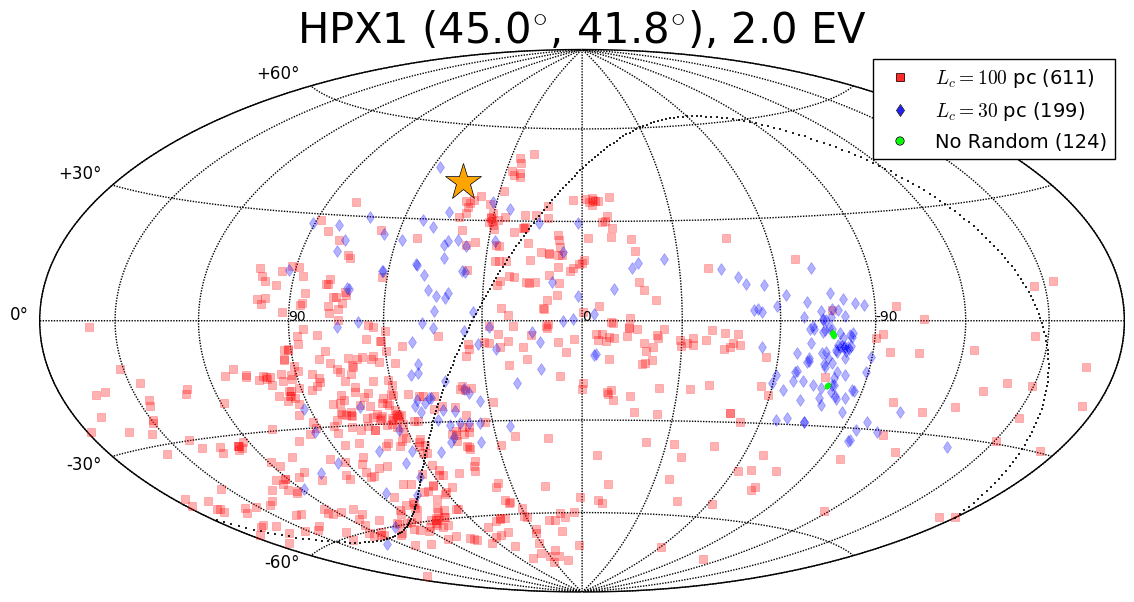}
\end{minipage}
\begin{minipage}[b]{0.48 \textwidth}
\includegraphics[width=1. \textwidth]{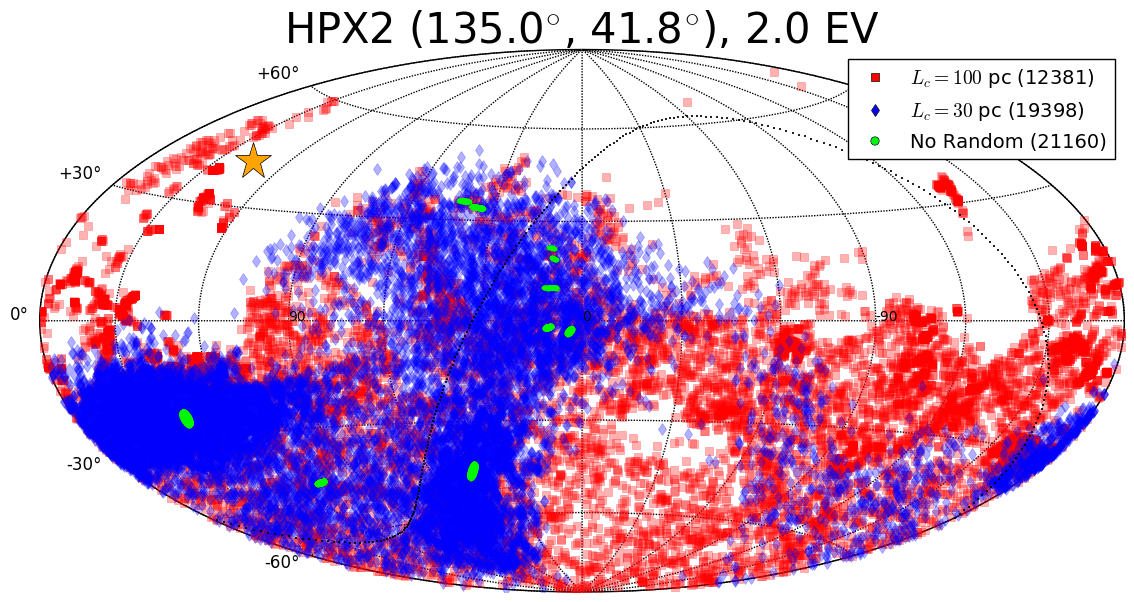}
\end{minipage}
\begin{minipage}[b]{0.48 \textwidth}
\includegraphics[width=1. \textwidth]{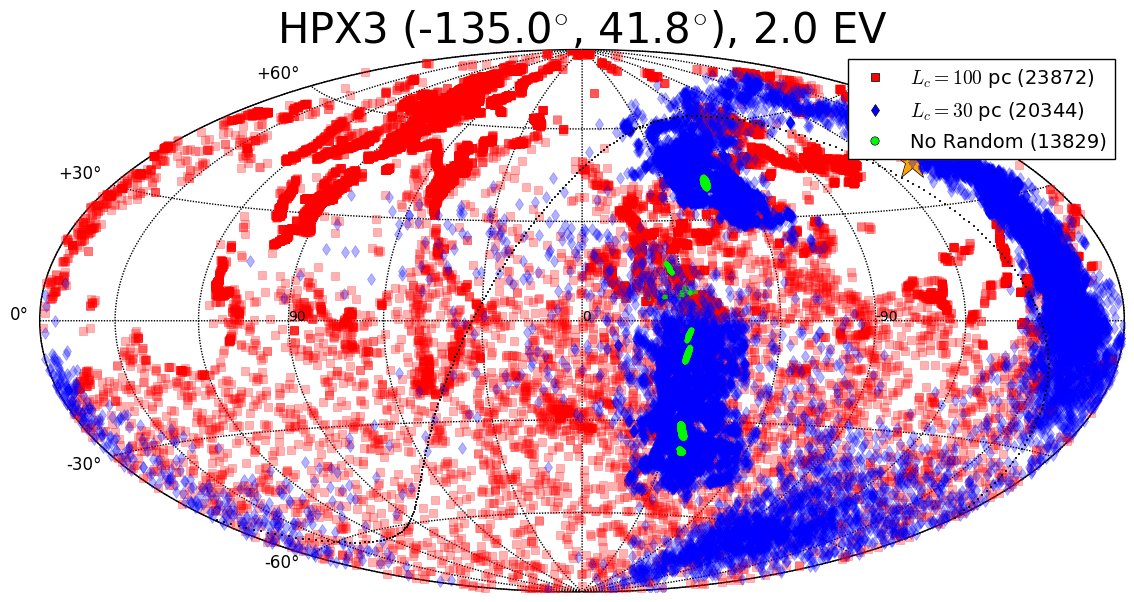}
\end{minipage}
\begin{minipage}[b]{0.48 \textwidth}
\includegraphics[width=1. \textwidth]{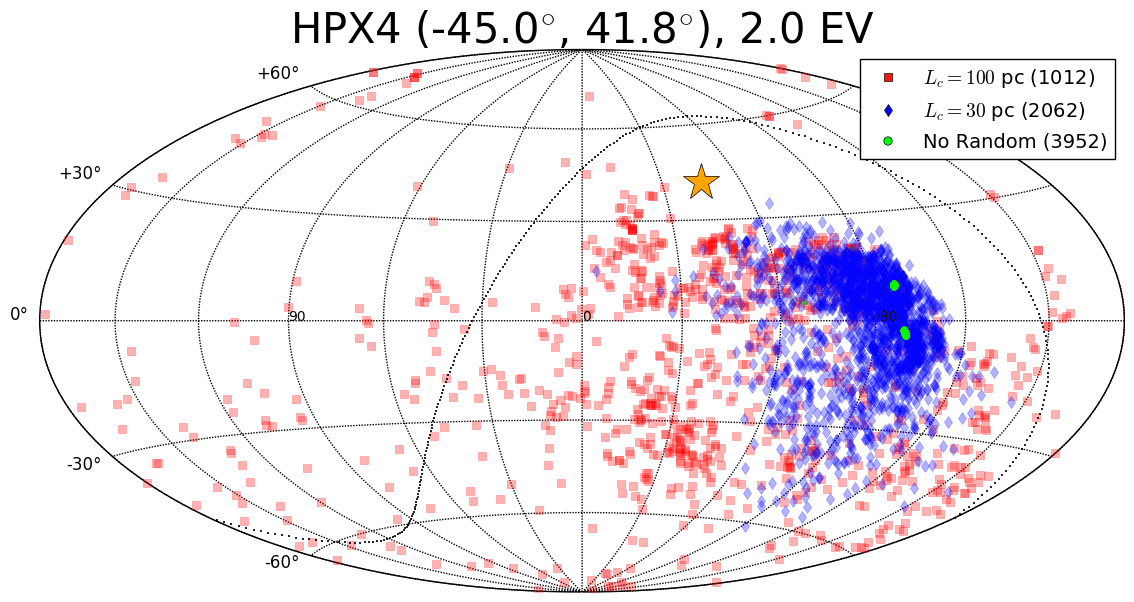}
\end{minipage}
\begin{minipage}[b]{0.48 \textwidth}
\includegraphics[width=1. \textwidth]{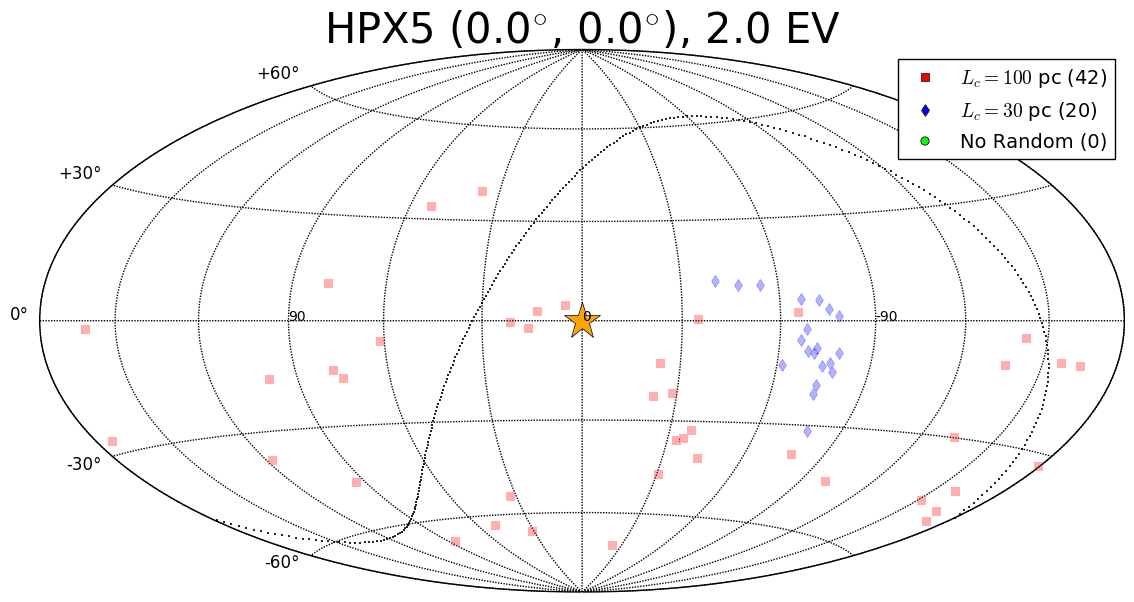}
\end{minipage}
\begin{minipage}[b]{0.48 \textwidth}
\includegraphics[width=1. \textwidth]{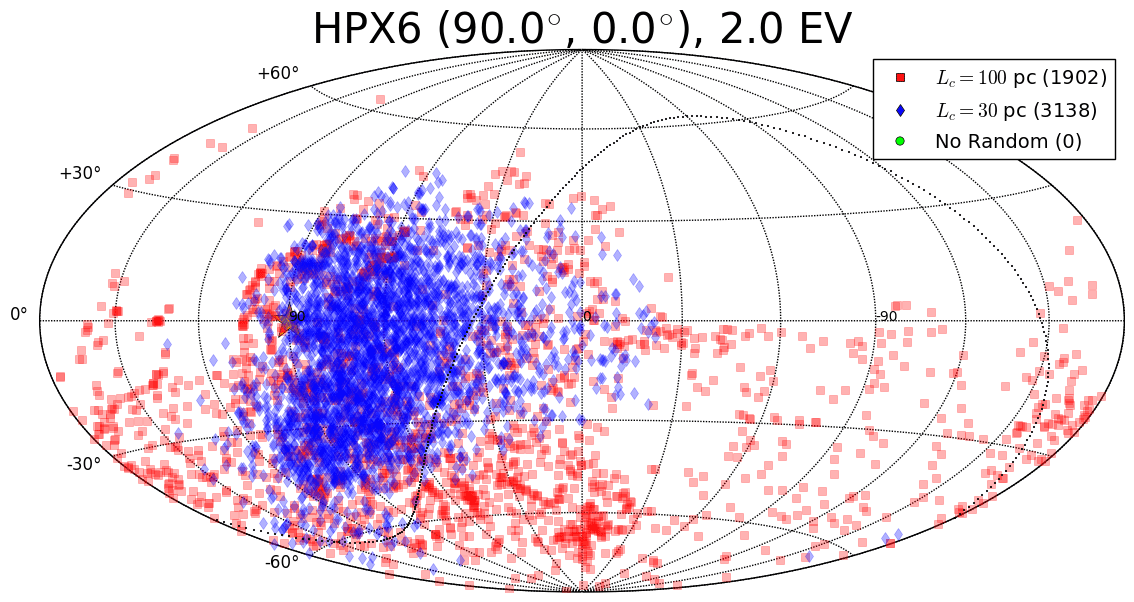}
\end{minipage}
\begin{minipage}[b]{0.48 \textwidth}
\includegraphics[width=1. \textwidth]{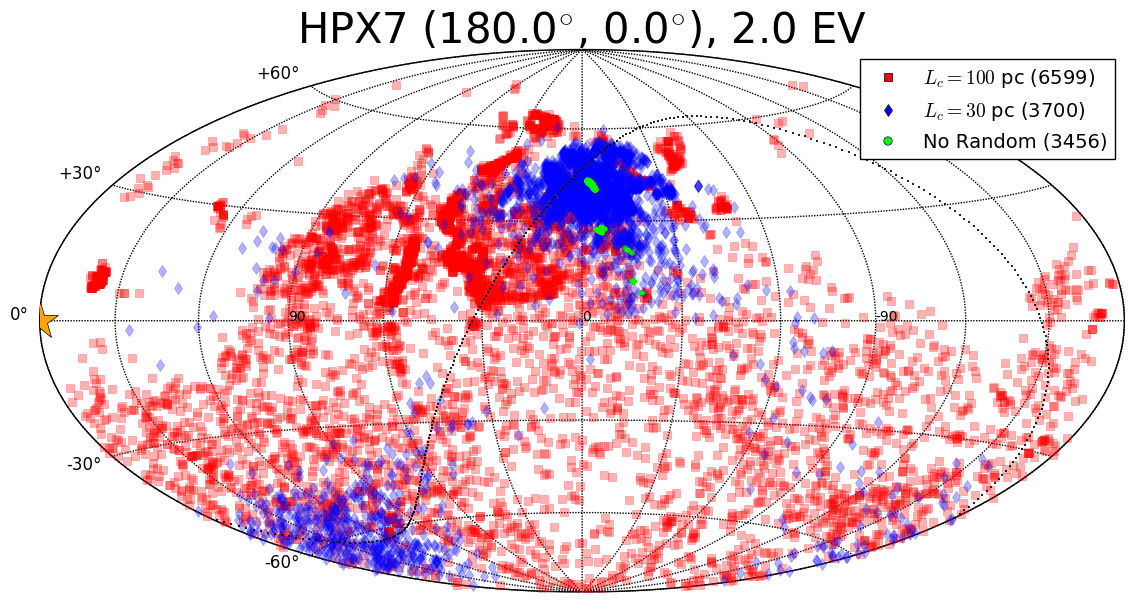}
\end{minipage}
\begin{minipage}[b]{0.48 \textwidth}
\includegraphics[width=1. \textwidth]{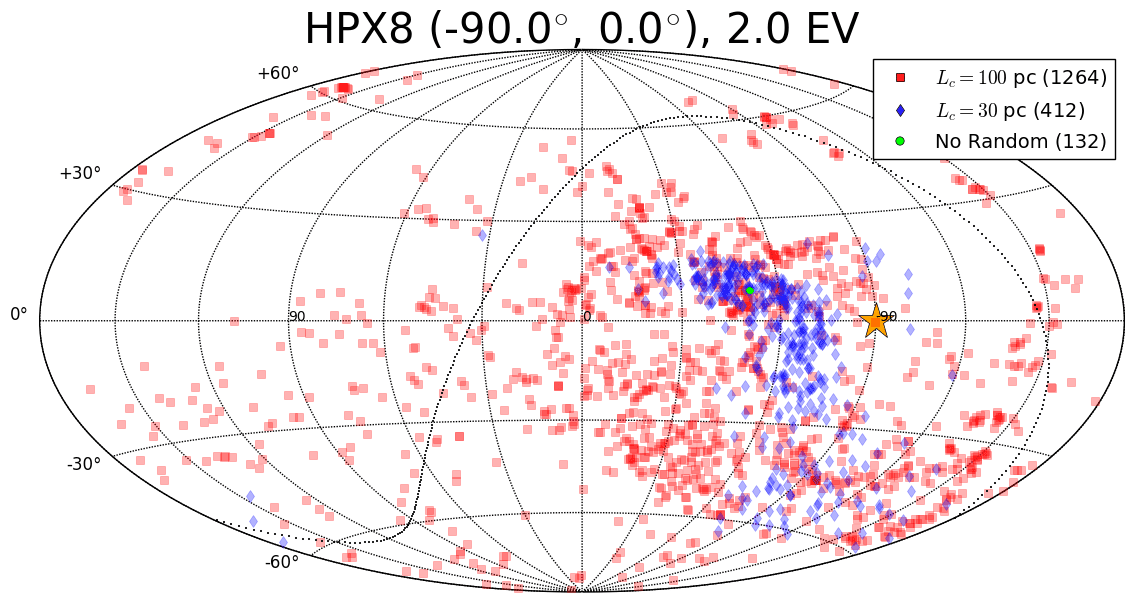}
\end{minipage}
\begin{minipage}[b]{0.48 \textwidth}
\includegraphics[width=1. \textwidth]{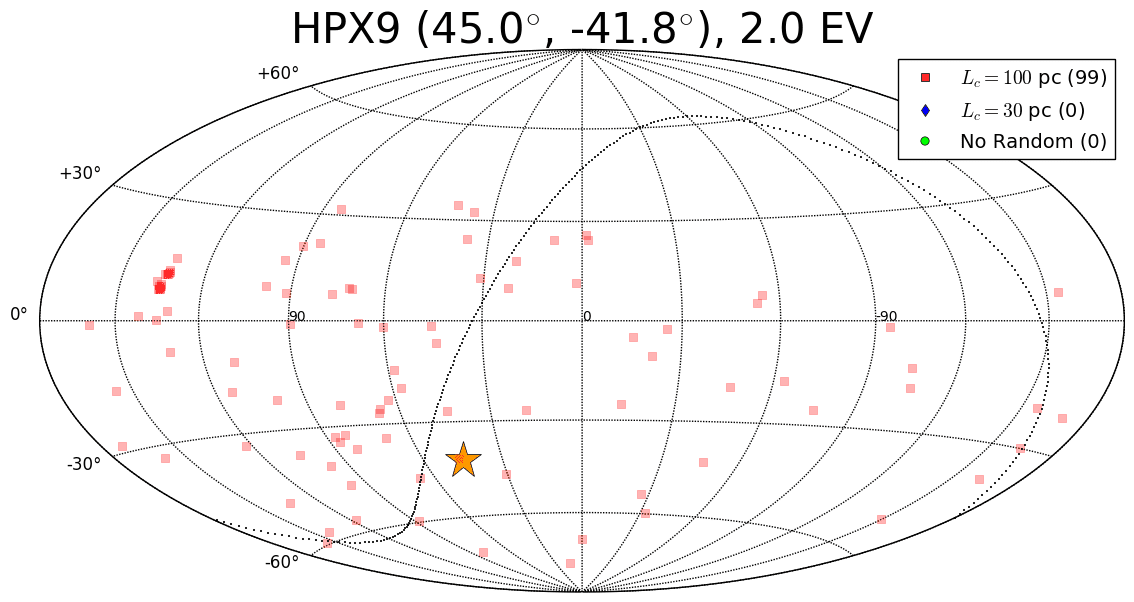}
\end{minipage}
\begin{minipage}[b]{0.48 \textwidth}
\includegraphics[width=1. \textwidth]{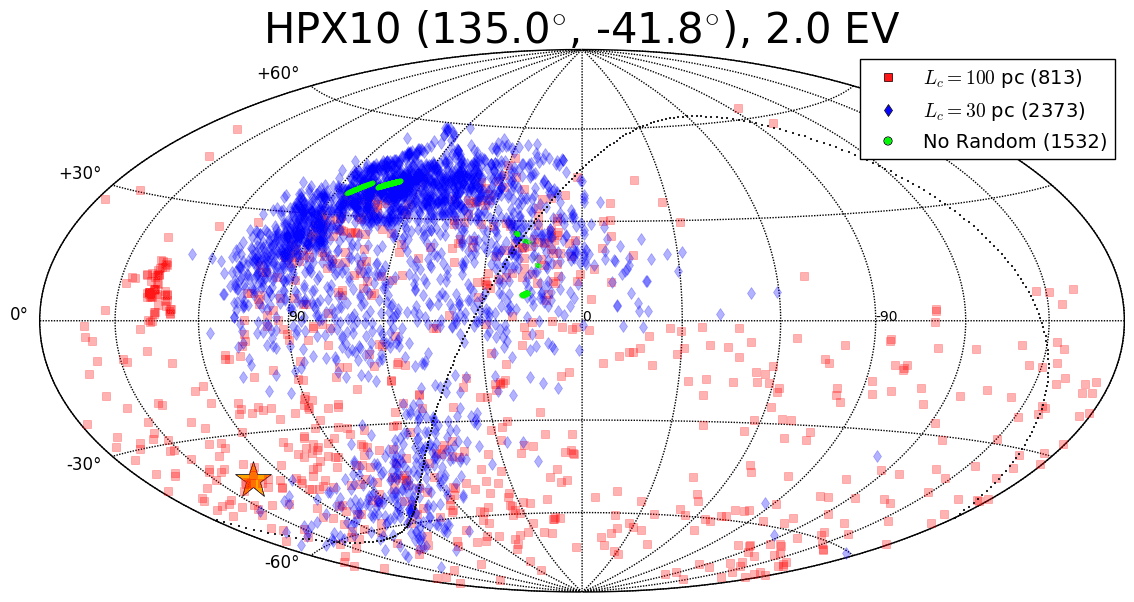}
\end{minipage}
\vspace{-0.3in}
\caption{As in Fig. \ref{plt:hpx19o8}, arrival direction distributions for log($R$ / V) = 18.3.} 
\label{plt:hpx18o3}
\vspace{-0.1in}
\end{figure}
\clearpage

\begin{figure}[t]
\hspace{-0.3in}
\centering
\begin{minipage}[b]{0.48 \textwidth}
\includegraphics[width=1. \textwidth]{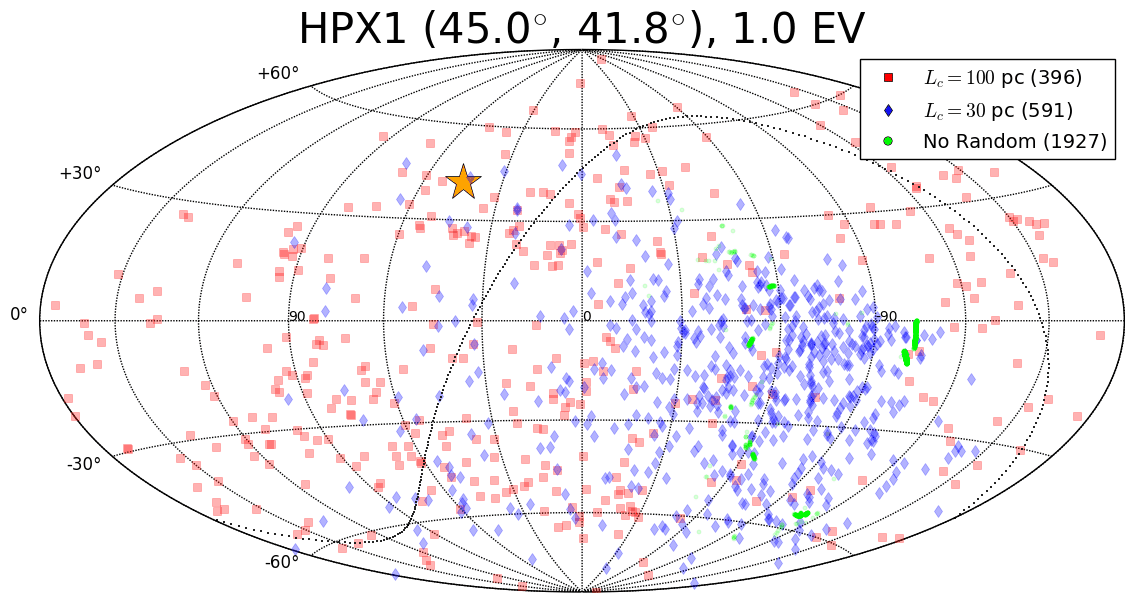}
\end{minipage}
\begin{minipage}[b]{0.48 \textwidth}
\includegraphics[width=1. \textwidth]{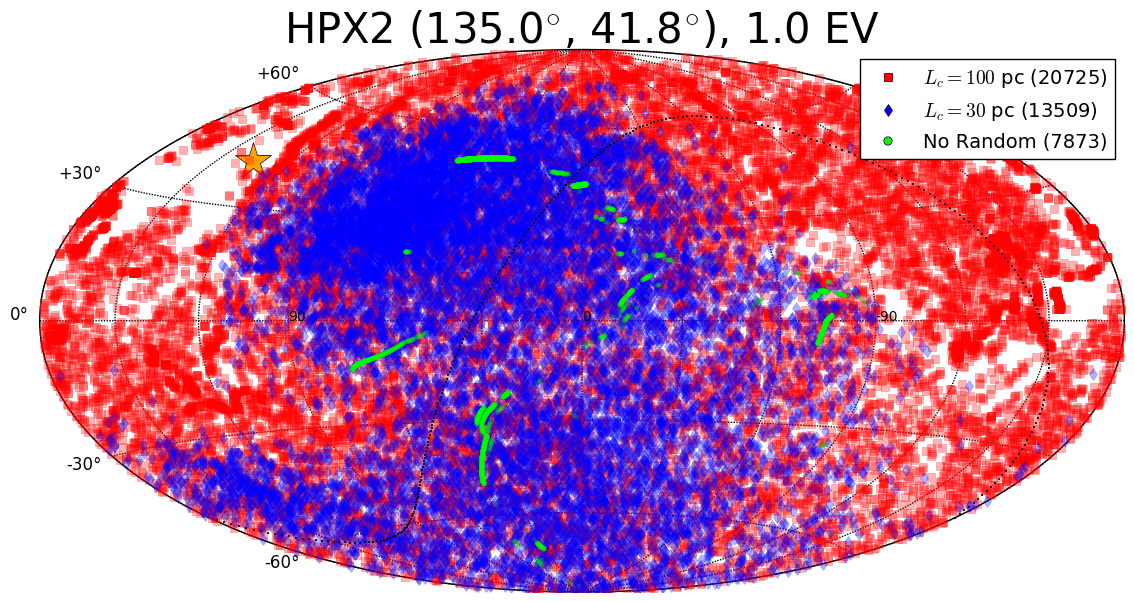}
\end{minipage}
\begin{minipage}[b]{0.48 \textwidth}
\includegraphics[width=1. \textwidth]{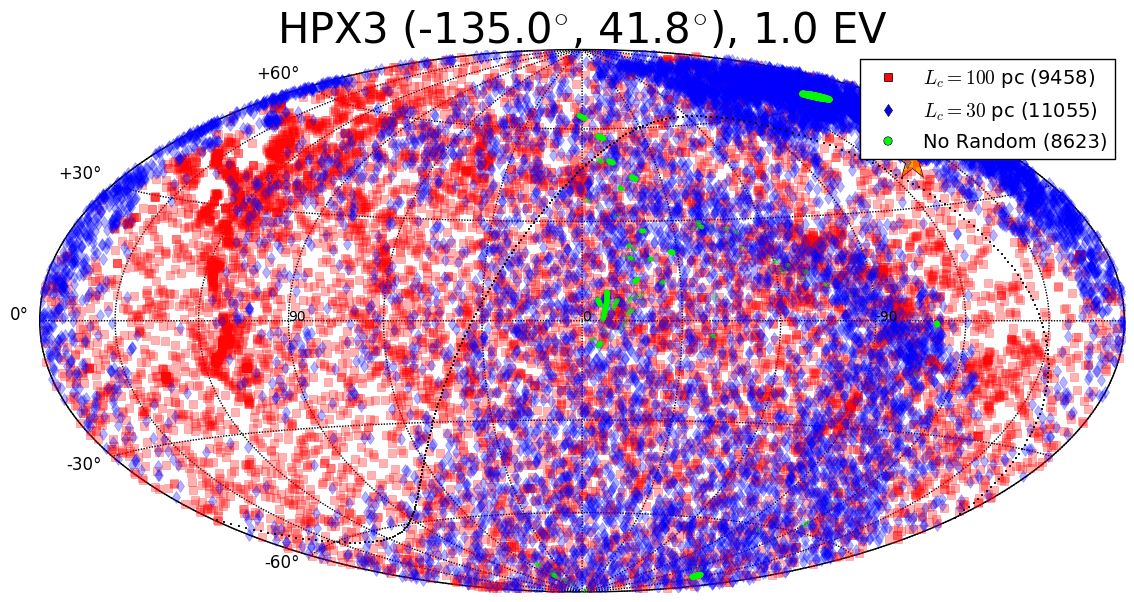}
\end{minipage}
\begin{minipage}[b]{0.48 \textwidth}
\includegraphics[width=1. \textwidth]{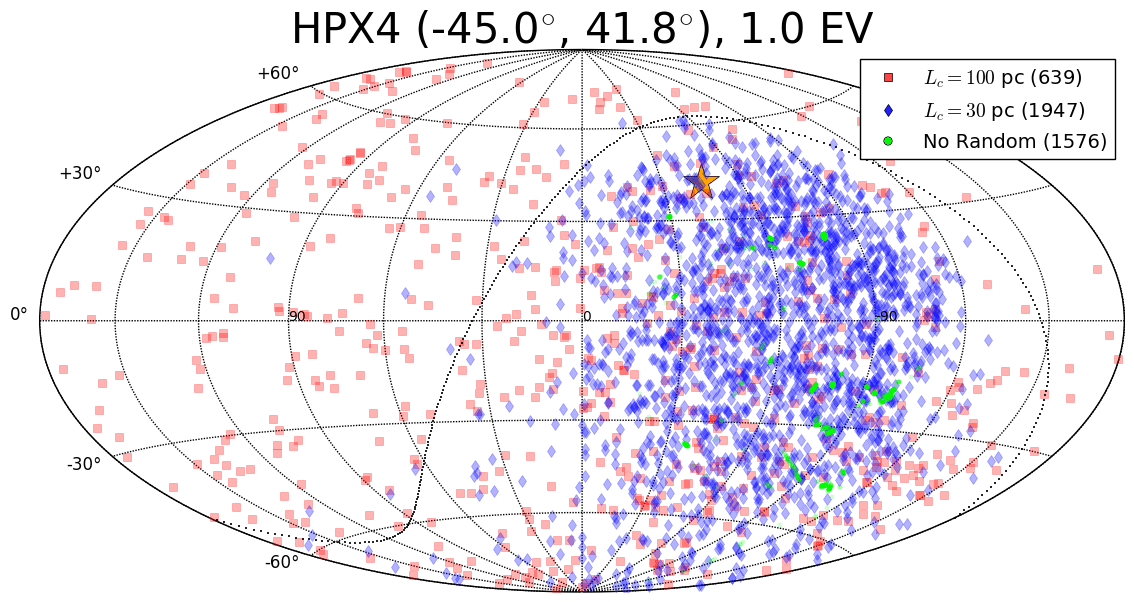}
\end{minipage}
\begin{minipage}[b]{0.48 \textwidth}
\includegraphics[width=1. \textwidth]{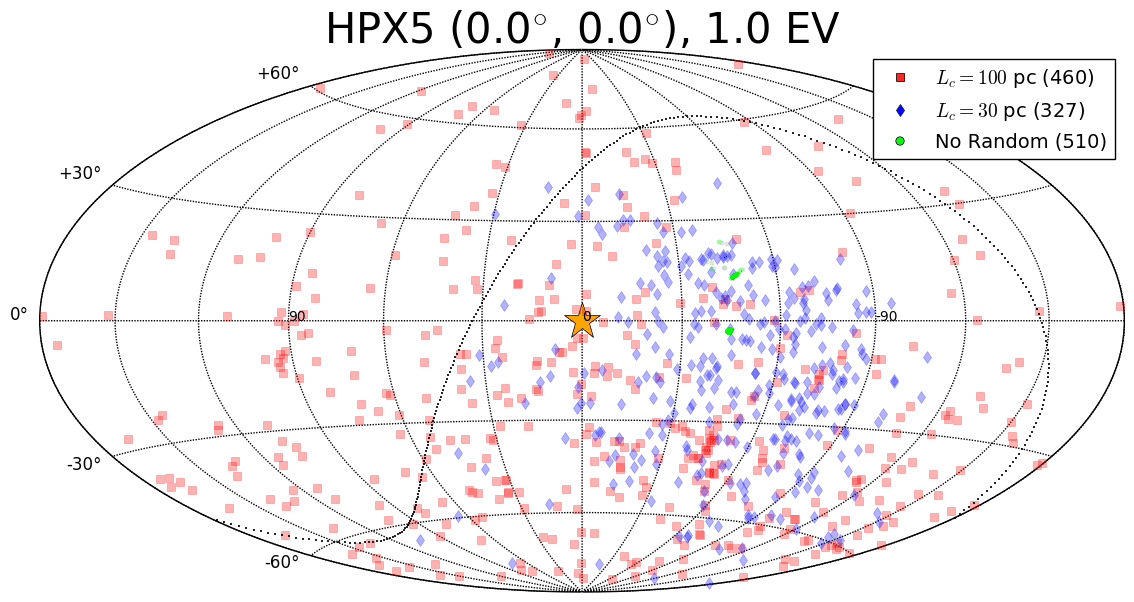}
\end{minipage}
\begin{minipage}[b]{0.48 \textwidth}
\includegraphics[width=1. \textwidth]{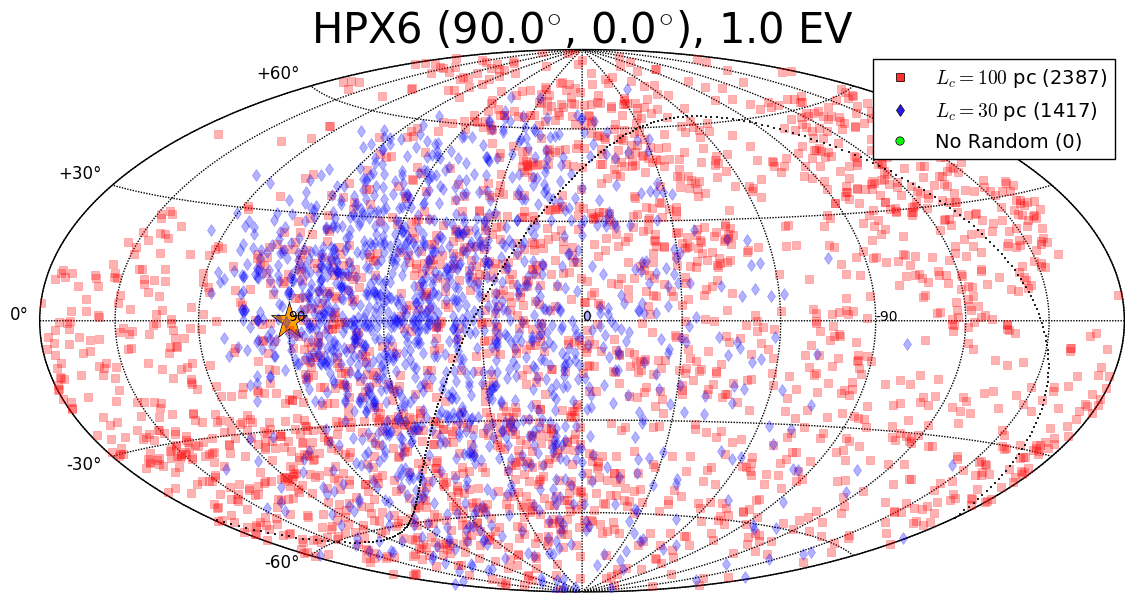}
\end{minipage}
\begin{minipage}[b]{0.48 \textwidth}
\includegraphics[width=1. \textwidth]{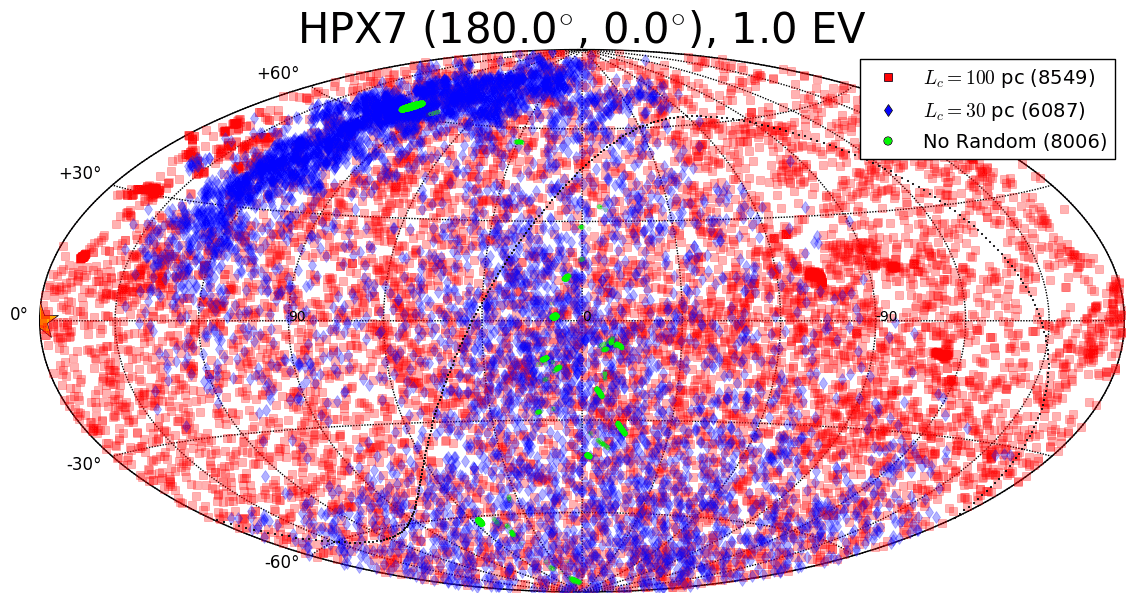}
\end{minipage}
\begin{minipage}[b]{0.48 \textwidth}
\includegraphics[width=1. \textwidth]{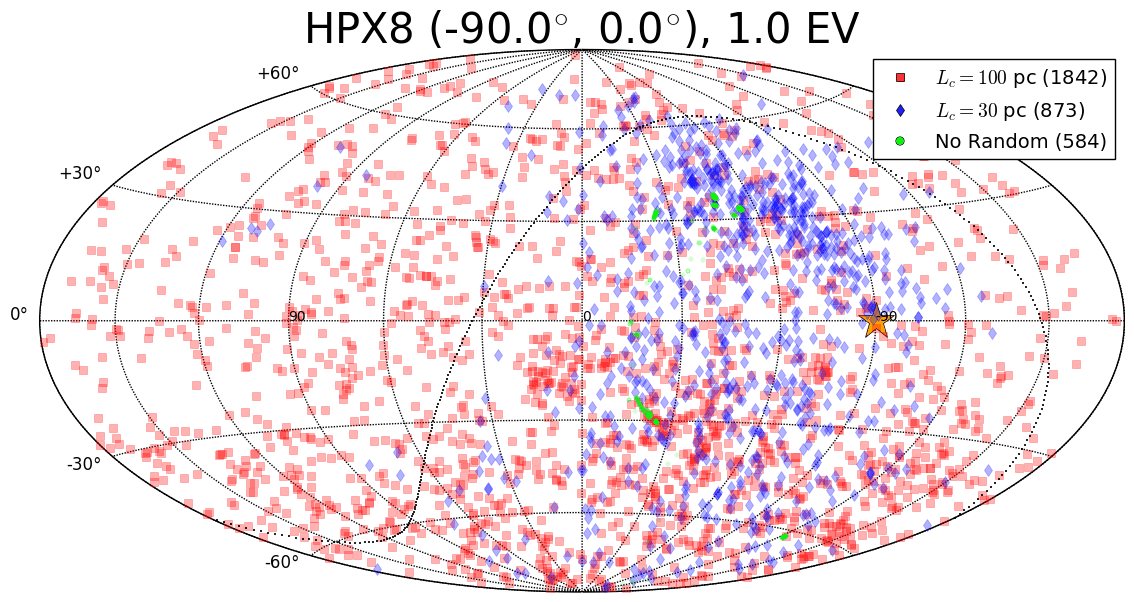}
\end{minipage}
\begin{minipage}[b]{0.48 \textwidth}
\includegraphics[width=1. \textwidth]{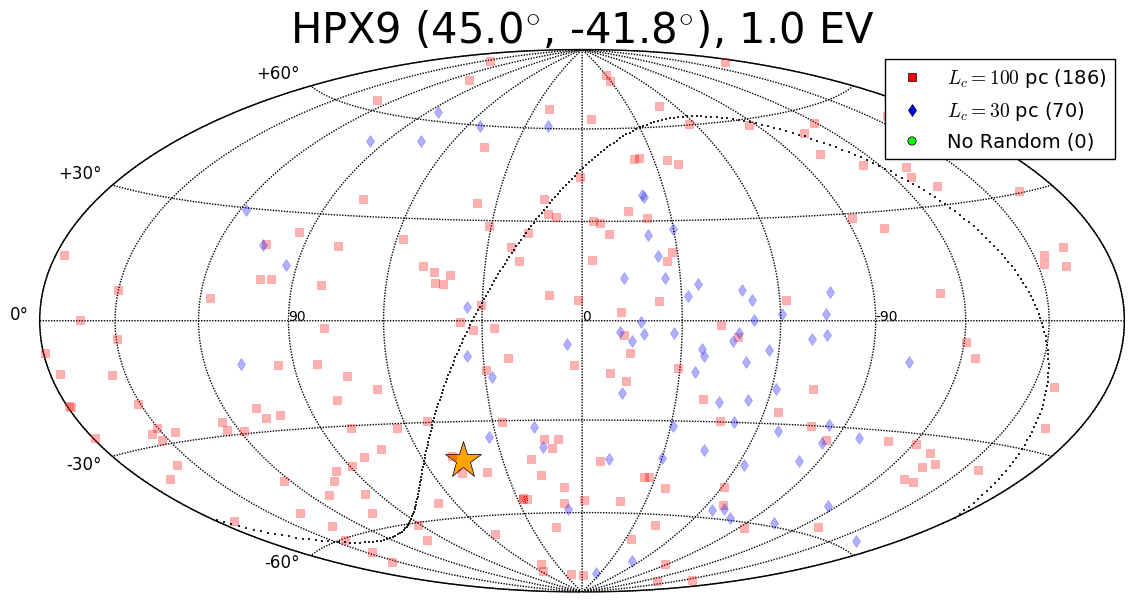}
\end{minipage}
\begin{minipage}[b]{0.48 \textwidth}
\includegraphics[width=1. \textwidth]{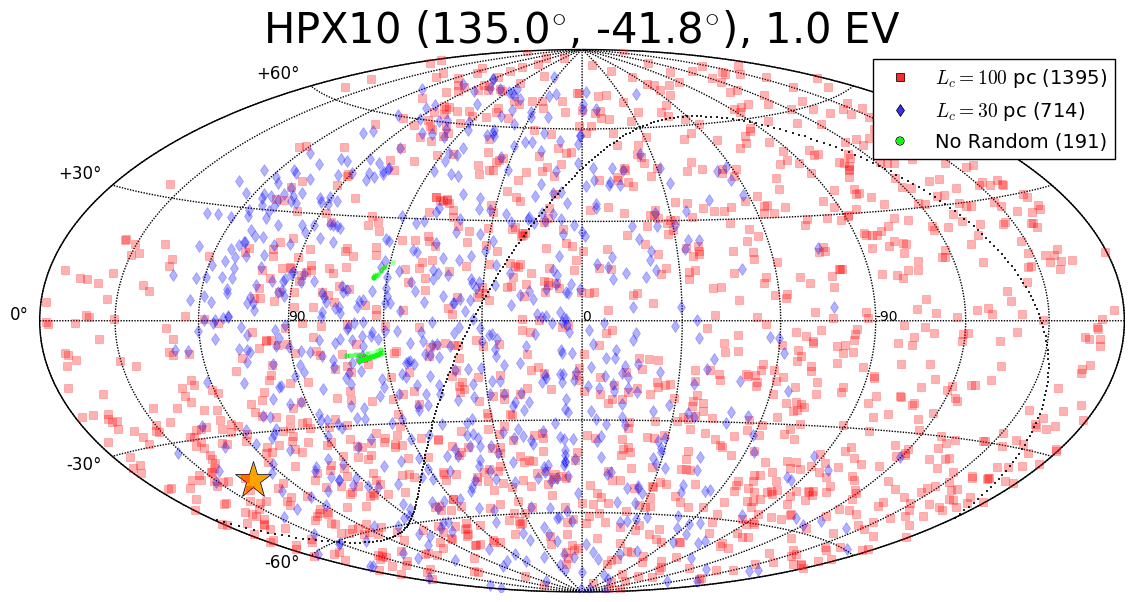}
\end{minipage}
\vspace{-0.3in}
\caption{As in Fig. \ref{plt:hpx19o8}, arrival direction distributions for log($R$ / V) = 18.0.} 
\label{plt:hpx18o0}
\vspace{-0.1in}
\end{figure}
\clearpage 
\begin{figure}[t]
\hspace{-0.3in}
\centering
\begin{minipage}[b]{0.48 \textwidth}
\includegraphics[width=1. \textwidth]{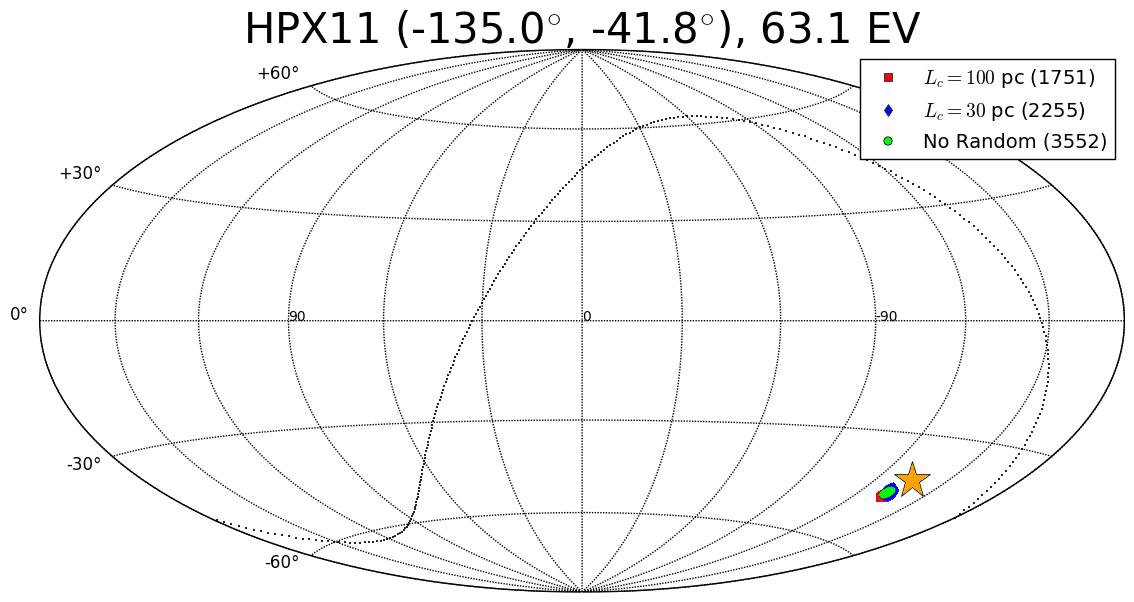}
\end{minipage}
\begin{minipage}[b]{0.48 \textwidth}
\includegraphics[width=1. \textwidth]{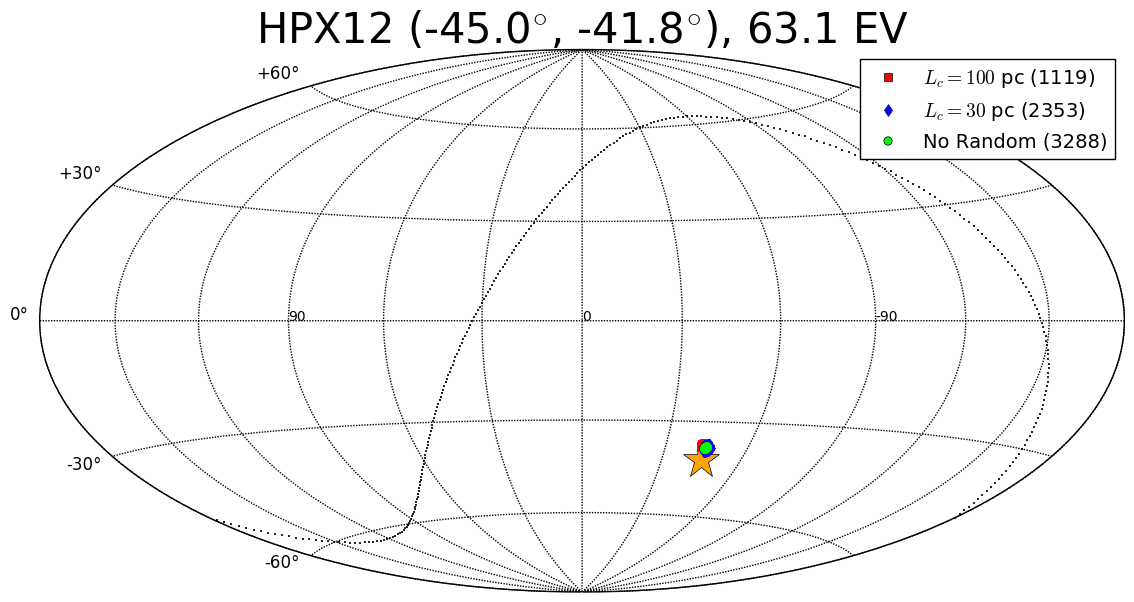}
\end{minipage}
\begin{minipage}[b]{0.48 \textwidth}
\includegraphics[width=1. \textwidth]{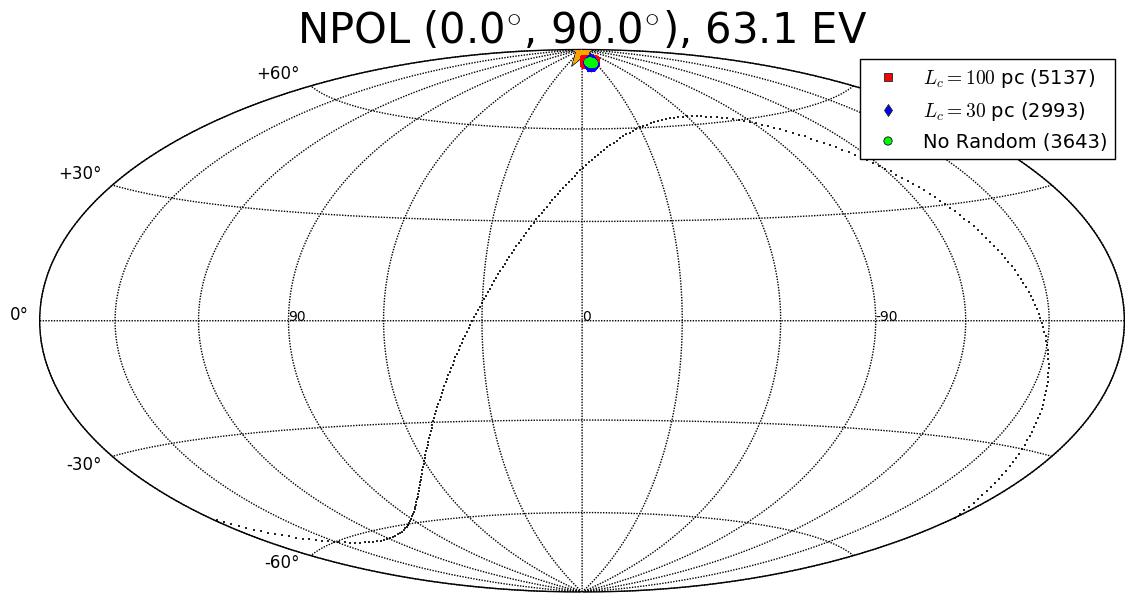}
\end{minipage}
\begin{minipage}[b]{0.48 \textwidth}
\includegraphics[width=1. \textwidth]{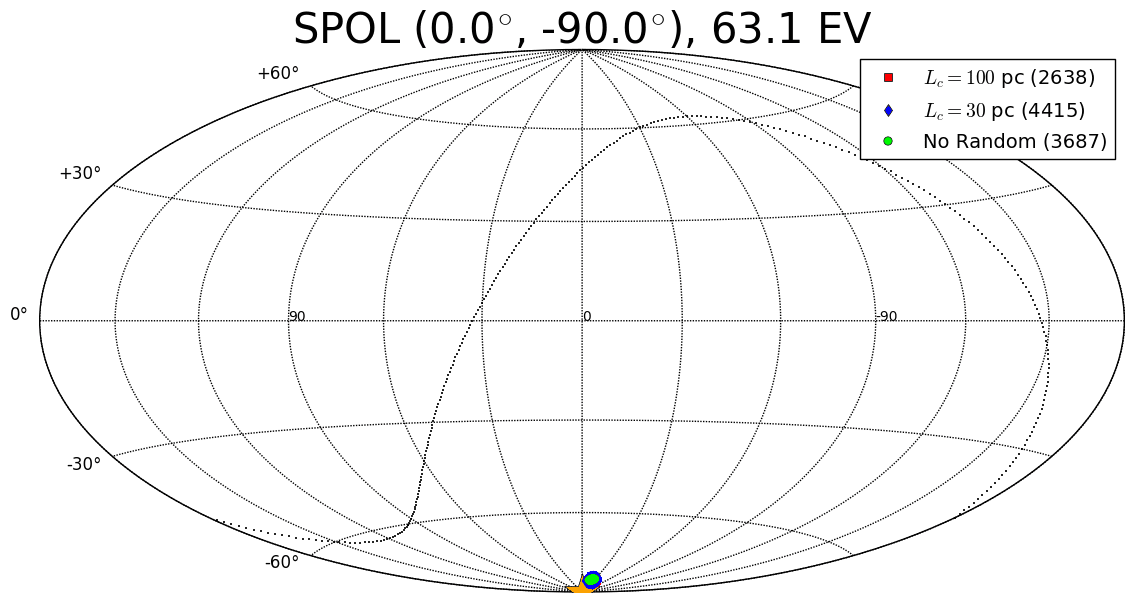}
\end{minipage}
\begin{minipage}[b]{0.48 \textwidth}
\includegraphics[width=1. \textwidth]{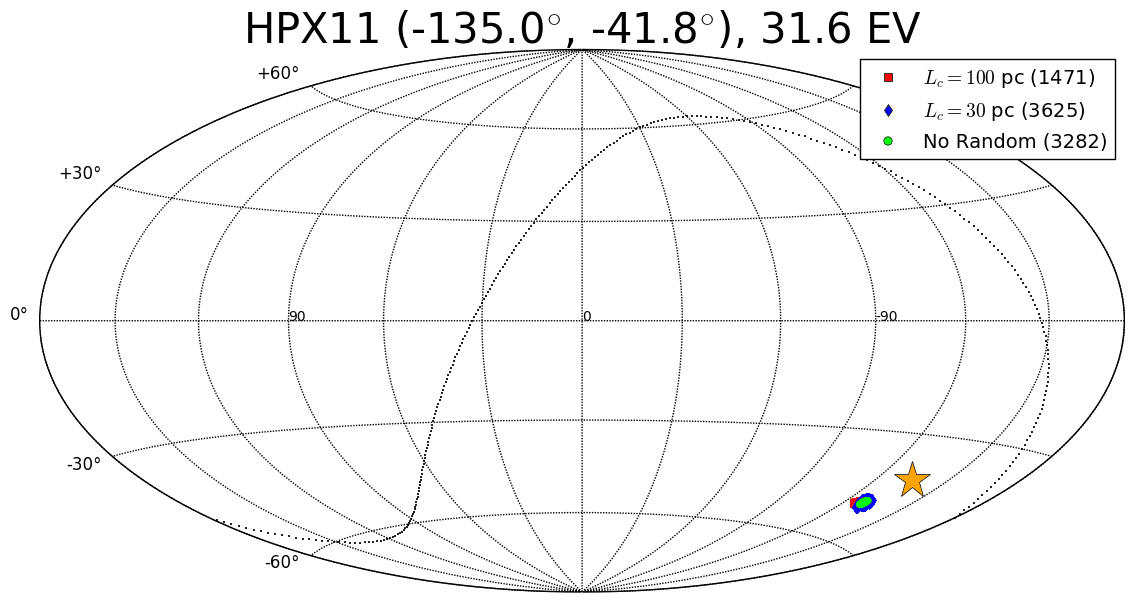}
\end{minipage}
\begin{minipage}[b]{0.48 \textwidth}
\includegraphics[width=1. \textwidth]{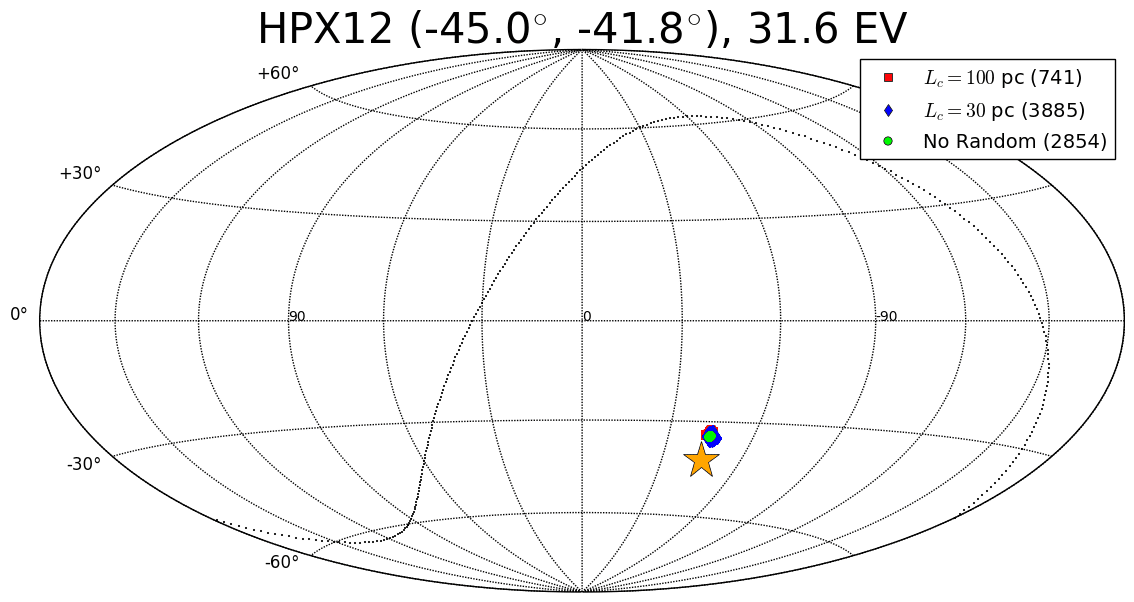}
\end{minipage}
\begin{minipage}[b]{0.48 \textwidth}
\includegraphics[width=1. \textwidth]{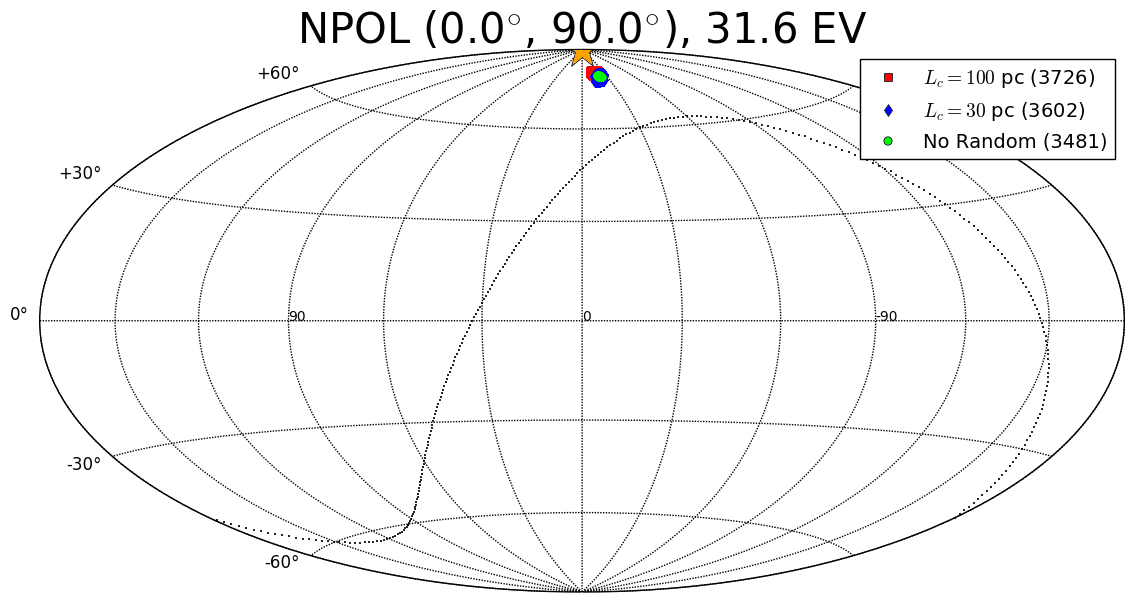}
\end{minipage}
\begin{minipage}[b]{0.48 \textwidth}
\includegraphics[width=1. \textwidth]{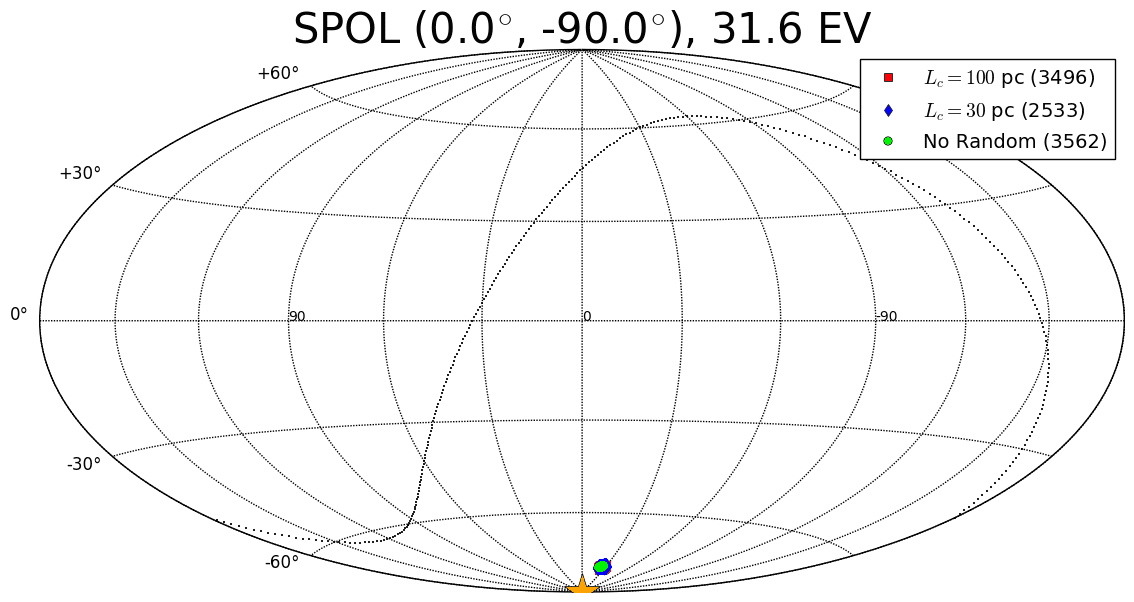}
\end{minipage}
\vspace{-0.3in}
\caption{Arrival direction distributions for log($R$ / V) = 19.8 (top 2 rows) and 19.5 (bottom two rows), for events  from sources in the regularly-spaced grid (marked with orange stars) for the $L_{coh} = 100$ pc (KRF6), a $L_{coh} = 30$ pc (KRF10) realization, and the coherent-only field.
The source direction and name is listed in the plot title.
The sky map is in Galactic coordinates and the dotted line indicates decl. $\delta=0^{\circ}$.
The legend in each plot indicates the number of events arriving from the given source at the given rigidity for the given realization.}
\label{plt:hpx_xtra_19o8_19o5}
\vspace{-0.1in}
\end{figure}
\clearpage 
\begin{figure}[t]
\hspace{-0.3in}
\centering
\begin{minipage}[b]{0.48 \textwidth}
\includegraphics[width=1. \textwidth]{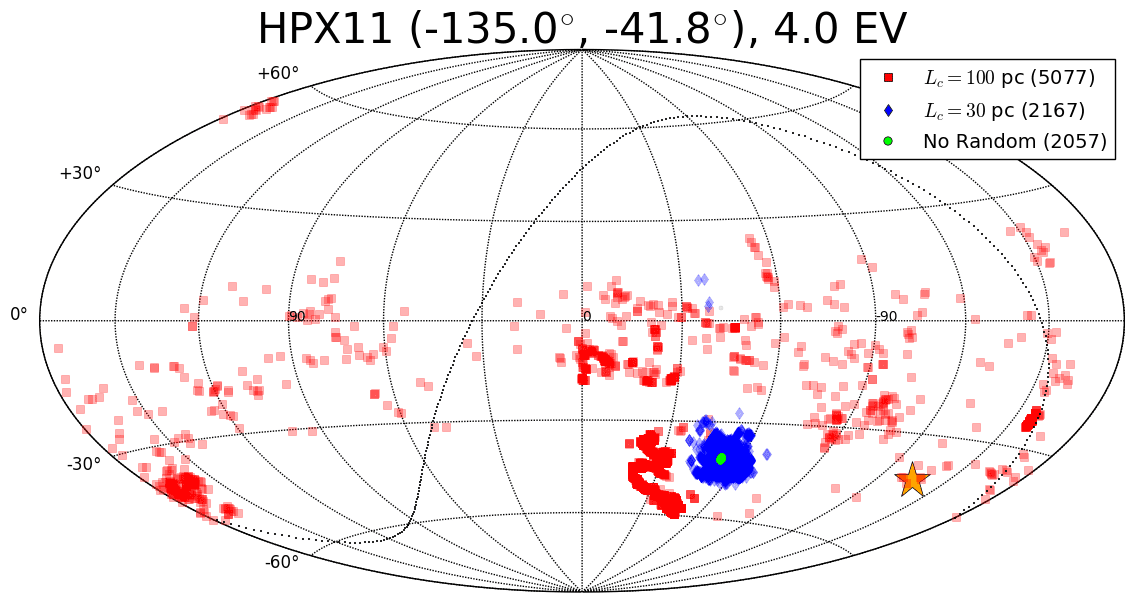}
\end{minipage}
\begin{minipage}[b]{0.48 \textwidth}
\includegraphics[width=1. \textwidth]{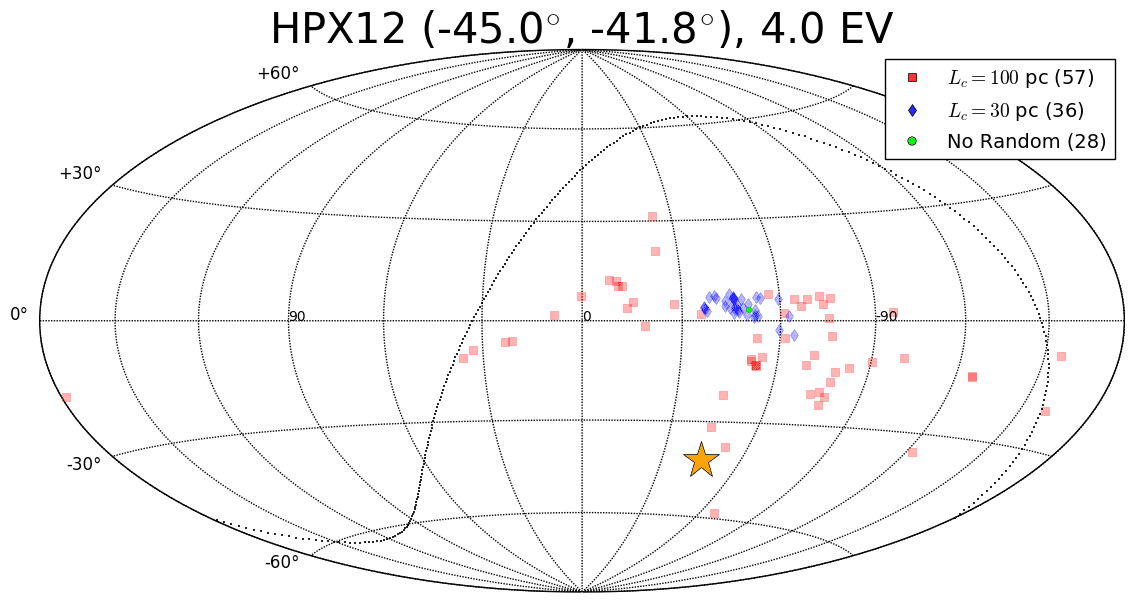}
\end{minipage}
\begin{minipage}[b]{0.48 \textwidth}
\includegraphics[width=1. \textwidth]{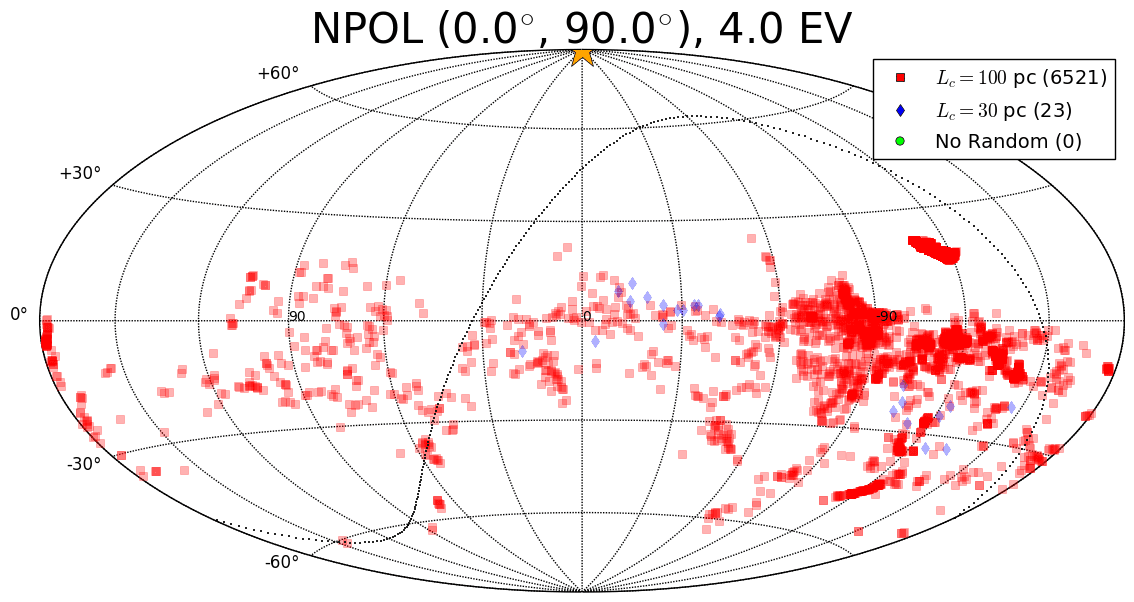}
\end{minipage}
\begin{minipage}[b]{0.48 \textwidth}
\includegraphics[width=1. \textwidth]{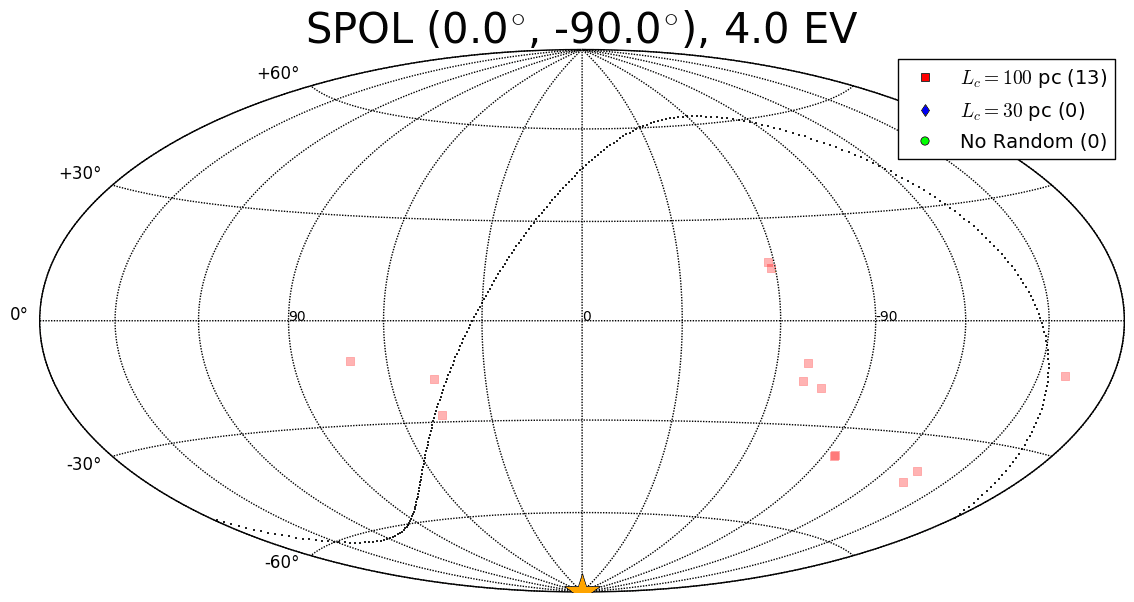}
\end{minipage}
\begin{minipage}[b]{0.48 \textwidth}
\includegraphics[width=1. \textwidth]{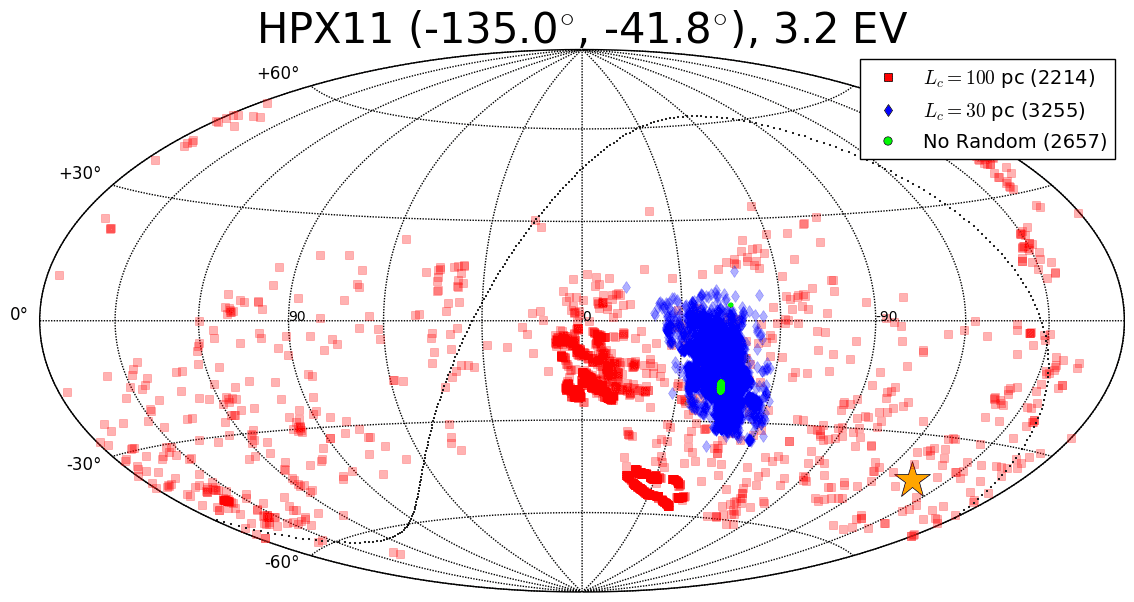}
\end{minipage}
\begin{minipage}[b]{0.48 \textwidth}
\includegraphics[width=1. \textwidth]{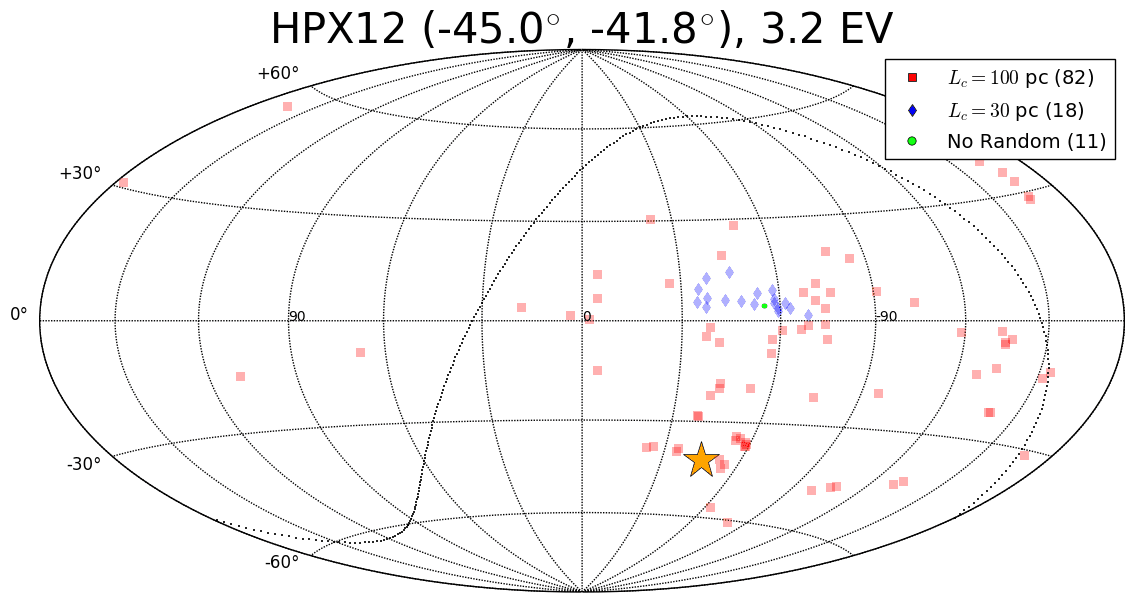}
\end{minipage}
\begin{minipage}[b]{0.48 \textwidth}
\includegraphics[width=1. \textwidth]{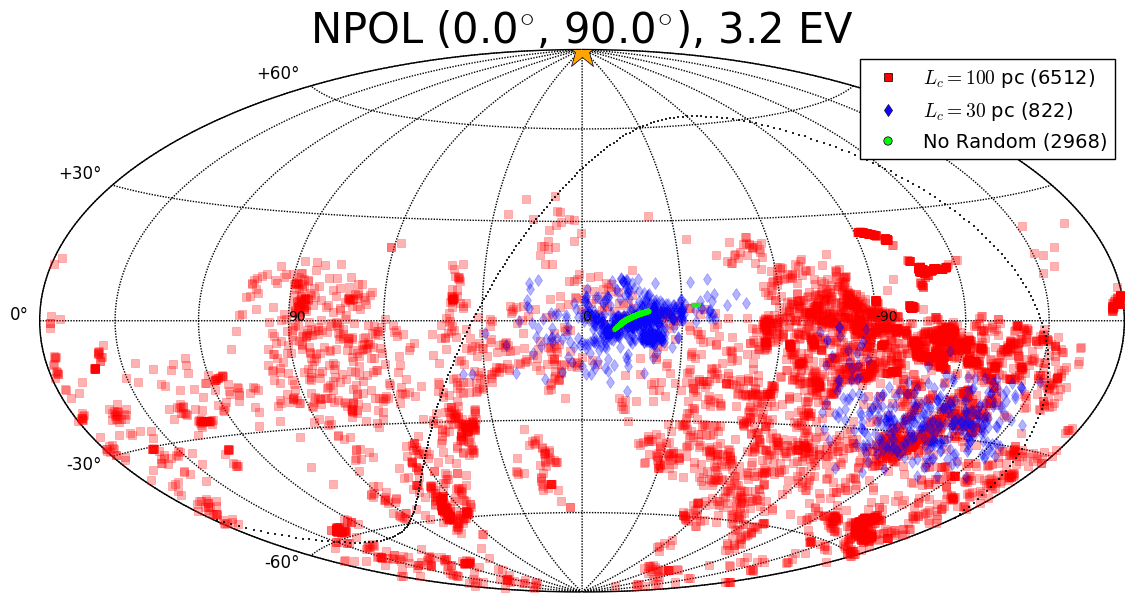}
\end{minipage}
\begin{minipage}[b]{0.48 \textwidth}
\includegraphics[width=1. \textwidth]{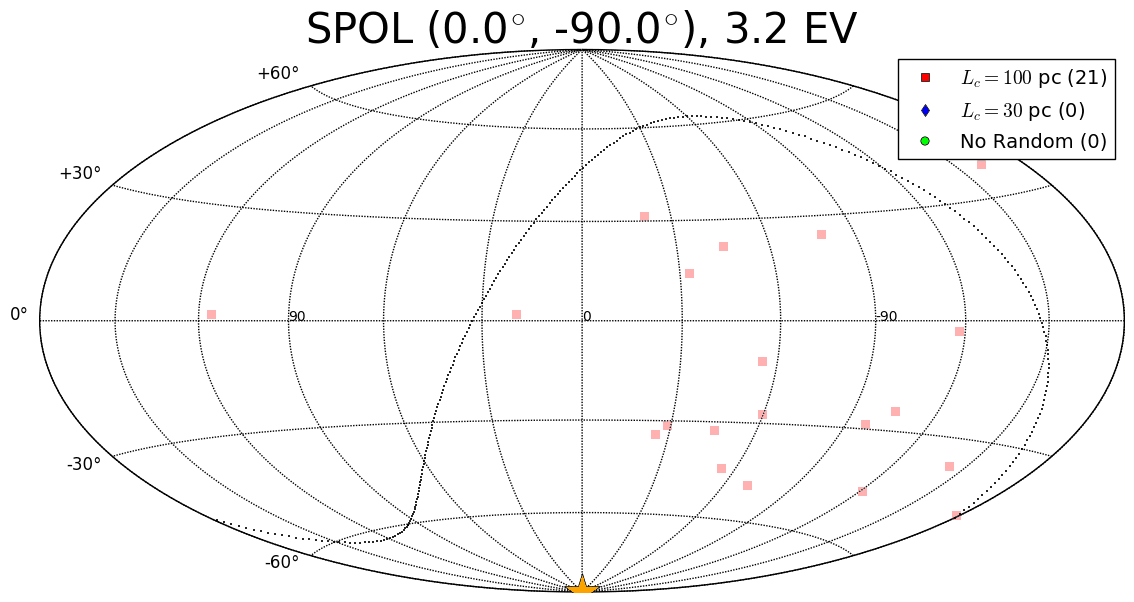}
\end{minipage}
\vspace{-0.3in}
\caption{As in Fig. \ref{plt:hpx_xtra_19o8_19o5}, arrival direction distributions for log($R$ / V) = 18.6 (top 2 rows) and 18.5 (bottom two rows).}
\label{plt:hpx_xtra_18o6_18o5}
\vspace{-0.1in}
\end{figure}
\clearpage 
\begin{figure}[t]
\hspace{-0.3in}
\centering
\begin{minipage}[b]{0.48 \textwidth}
\includegraphics[width=1. \textwidth]{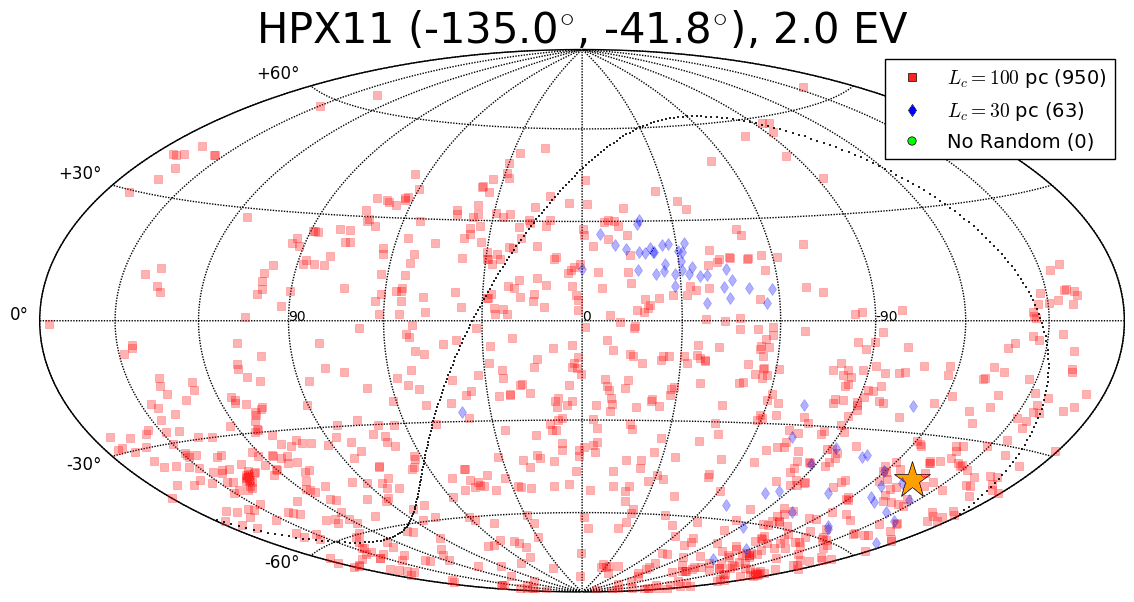}
\end{minipage}
\begin{minipage}[b]{0.48 \textwidth}
\includegraphics[width=1. \textwidth]{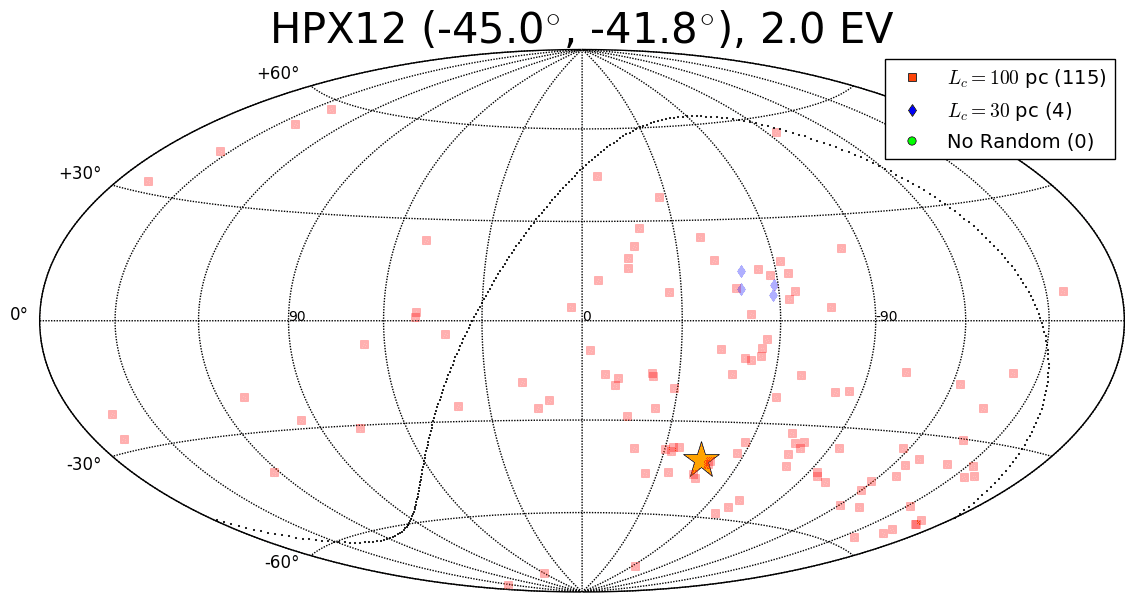}
\end{minipage}
\begin{minipage}[b]{0.48 \textwidth}
\includegraphics[width=1. \textwidth]{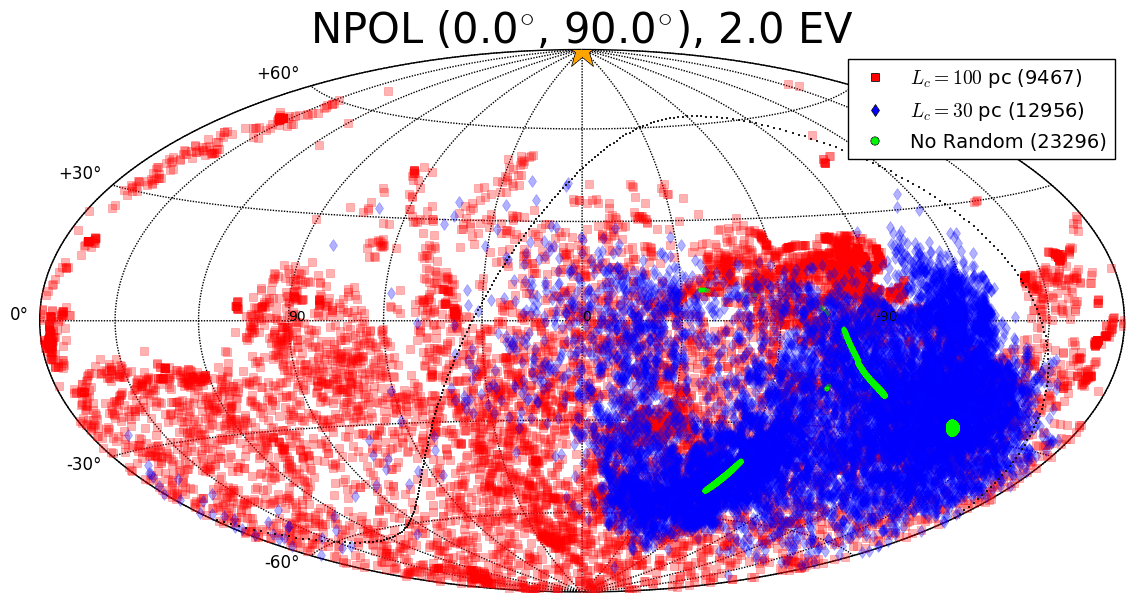}
\end{minipage}
\begin{minipage}[b]{0.48 \textwidth}
\includegraphics[width=1. \textwidth]{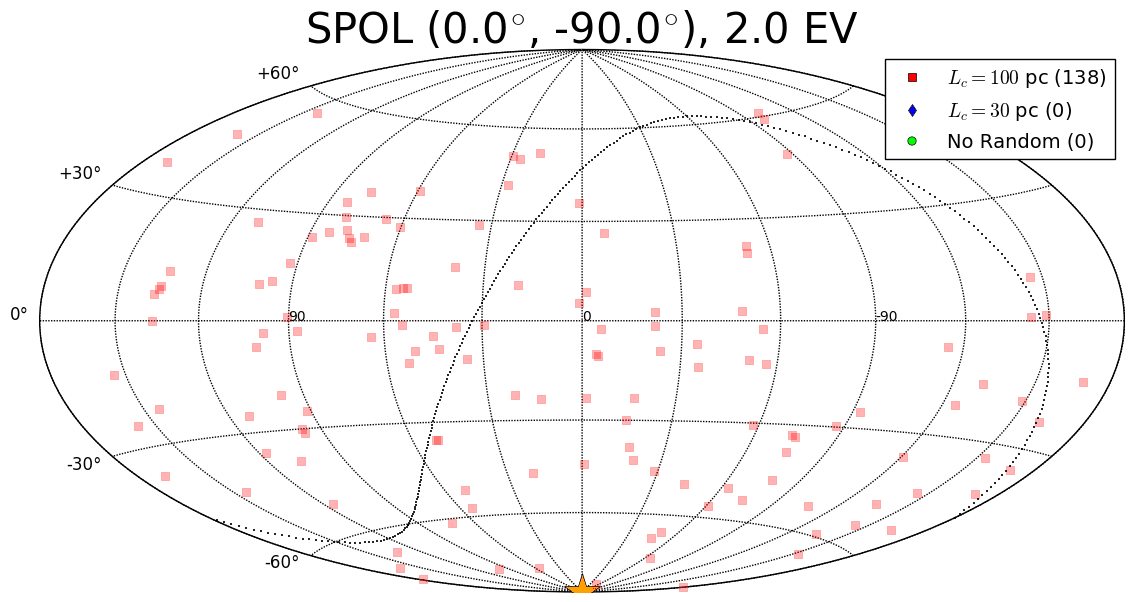}
\end{minipage}
\begin{minipage}[b]{0.48 \textwidth}
\includegraphics[width=1. \textwidth]{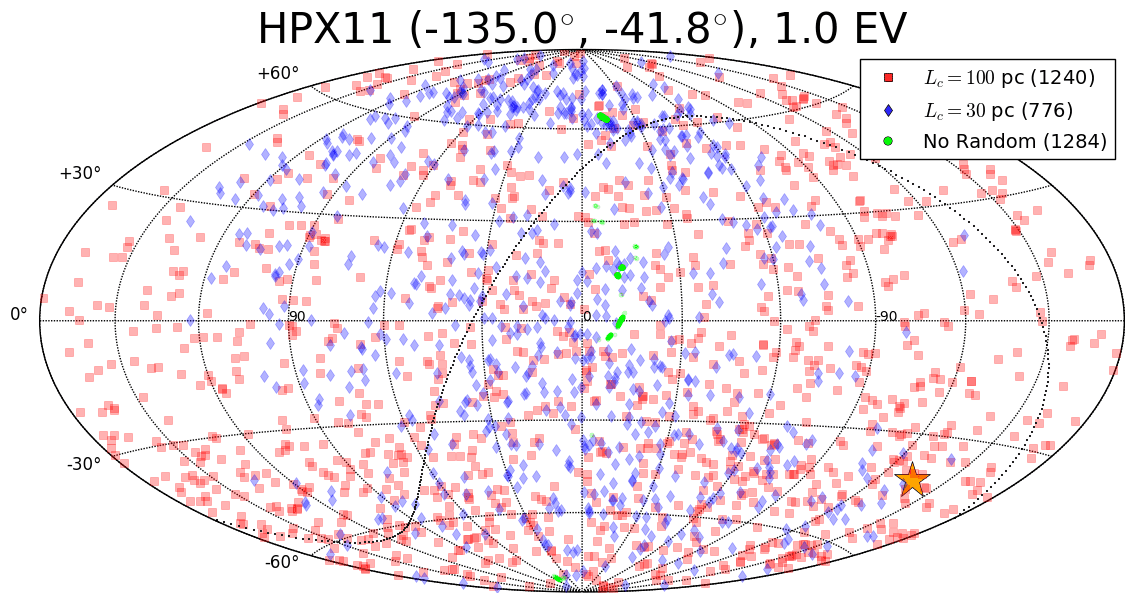}
\end{minipage}
\begin{minipage}[b]{0.48 \textwidth}
\includegraphics[width=1. \textwidth]{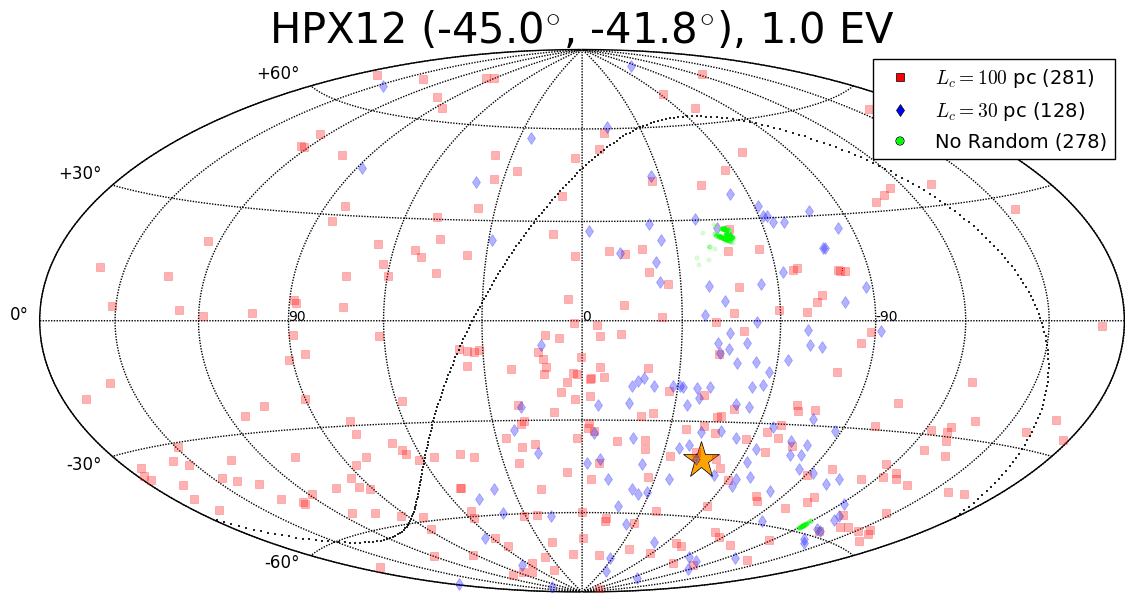}
\end{minipage}
\begin{minipage}[b]{0.48 \textwidth}
\includegraphics[width=1. \textwidth]{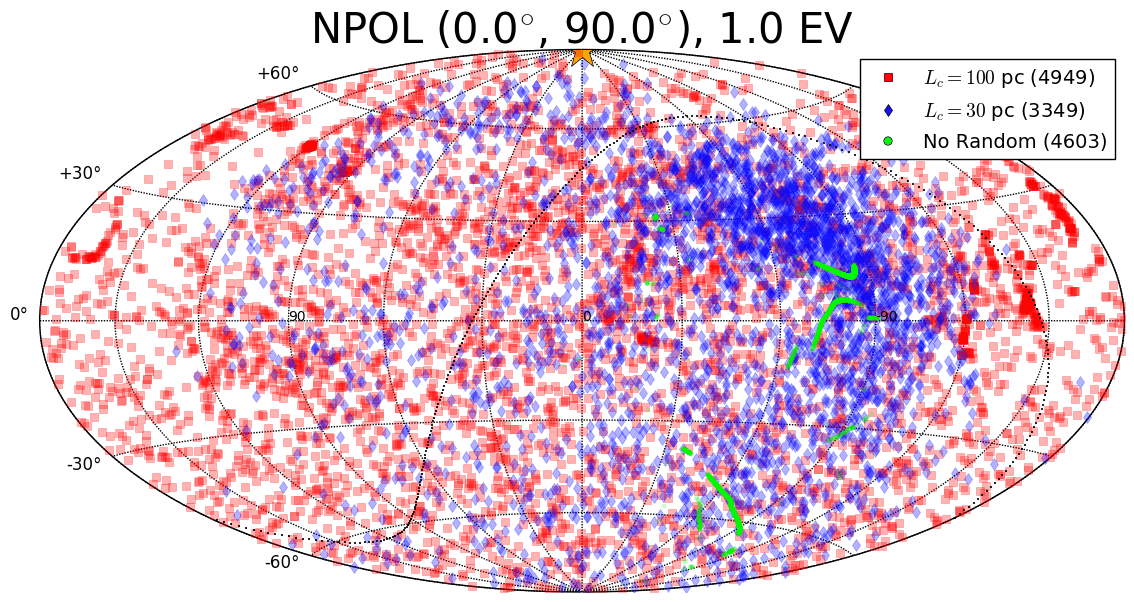}
\end{minipage}
\begin{minipage}[b]{0.48 \textwidth}
\includegraphics[width=1. \textwidth]{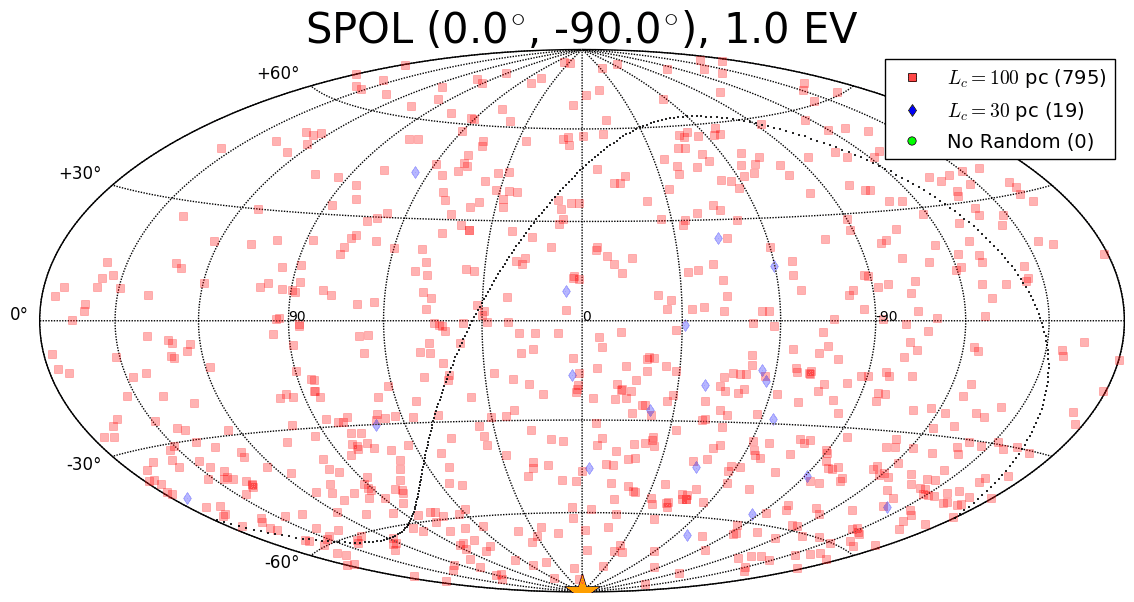}
\end{minipage}
\vspace{-0.3in}
\caption{As in Fig. \ref{plt:hpx_xtra_19o8_19o5}, arrival direction distributions for log($R$ / V) = 18.3 (top 2 rows) and 18.0 (bottom two rows).}
\label{plt:hpx_xtra_18o3_18o0}
\vspace{-0.1in}
\end{figure}
\clearpage

\section{Skyplots showing sensitivity to different realizations of the random field}
\label{appdx:realizations}
\begin{figure}[t]
\hspace{-0.3in}
\centering
\begin{minipage}[b]{0.48 \textwidth}
\includegraphics[width=1. \textwidth]{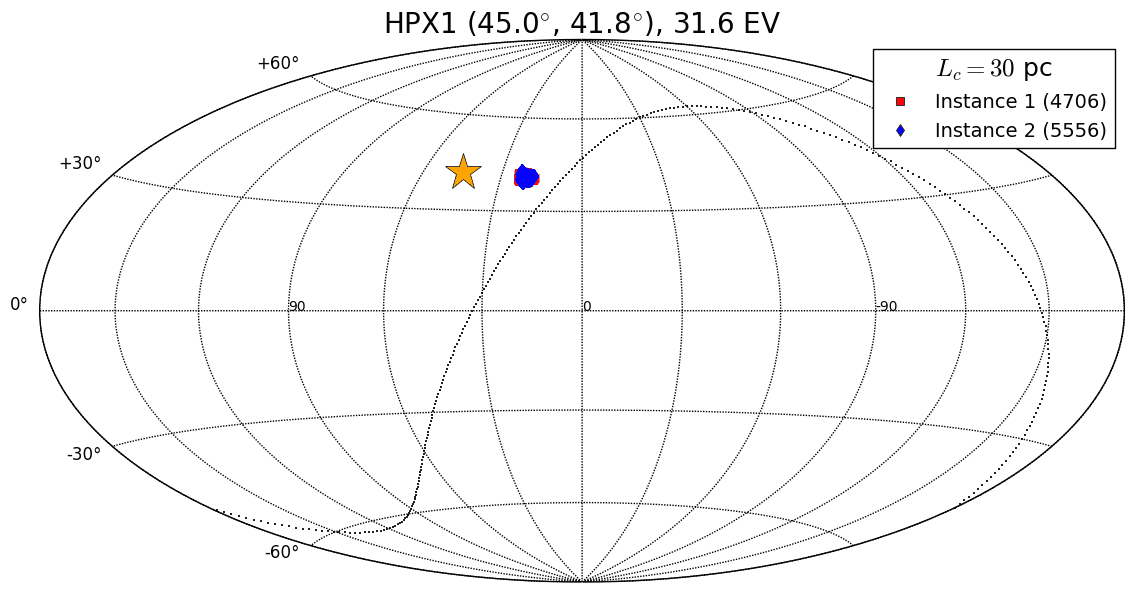}
\end{minipage}
\begin{minipage}[b]{0.48 \textwidth}
\includegraphics[width=1. \textwidth]{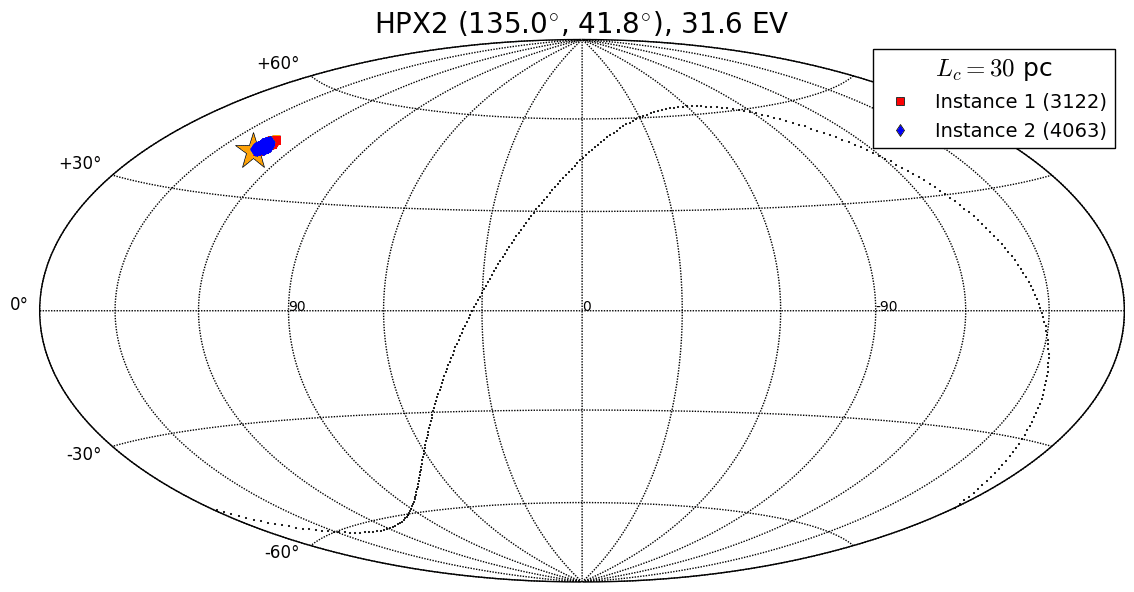}
\end{minipage}
\begin{minipage}[b]{0.48 \textwidth}
\includegraphics[width=1. \textwidth]{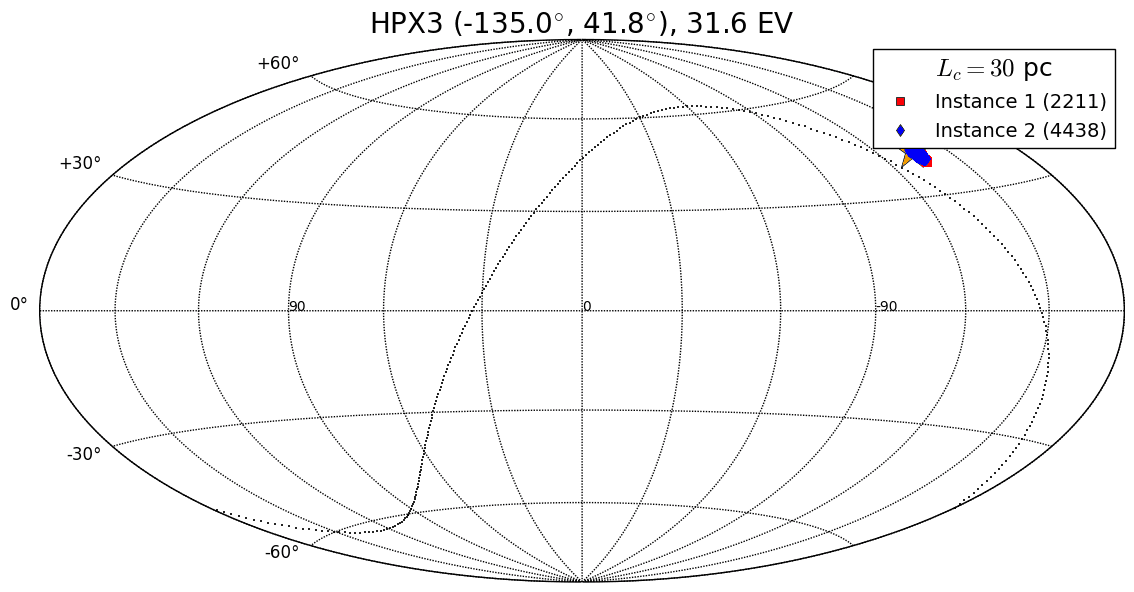}
\end{minipage}
\begin{minipage}[b]{0.48 \textwidth}
\includegraphics[width=1. \textwidth]{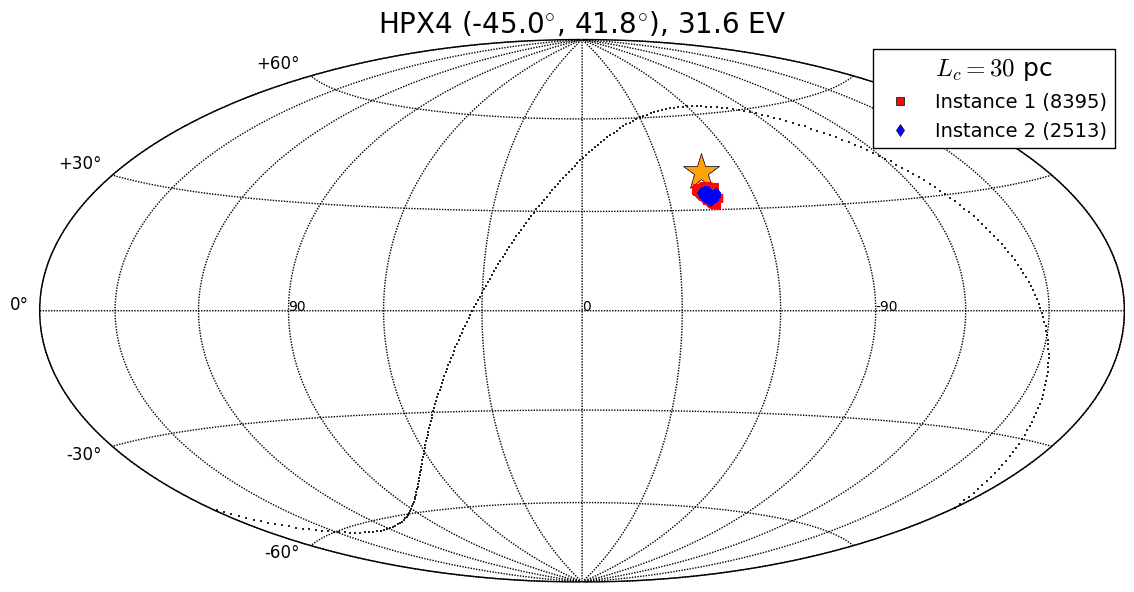}
\end{minipage}
\begin{minipage}[b]{0.48 \textwidth}
\includegraphics[width=1. \textwidth]{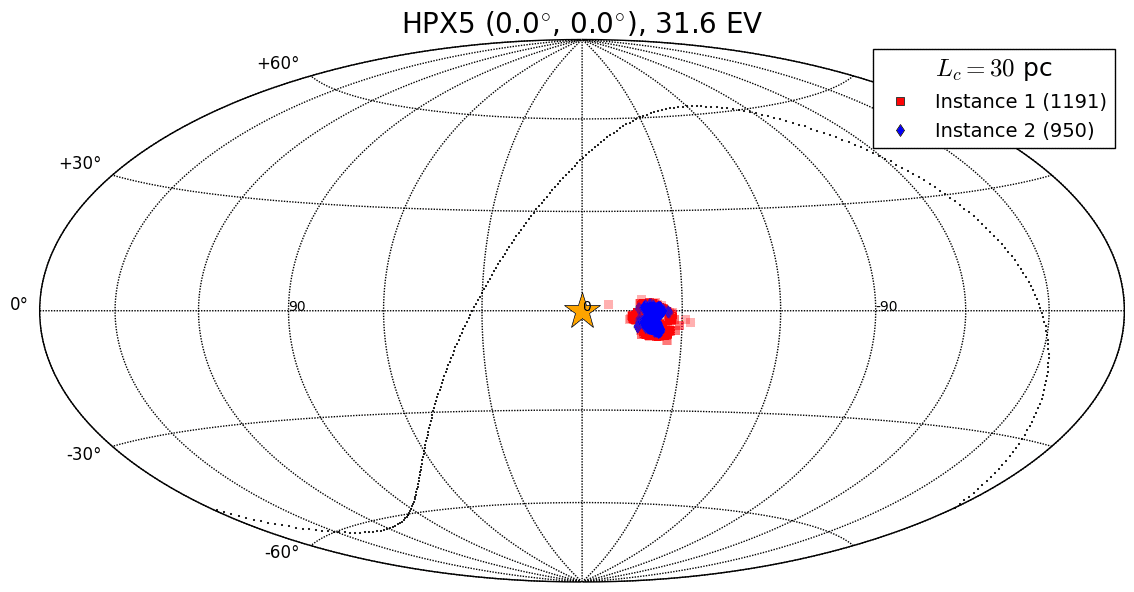}
\end{minipage}
\begin{minipage}[b]{0.48 \textwidth}
\includegraphics[width=1. \textwidth]{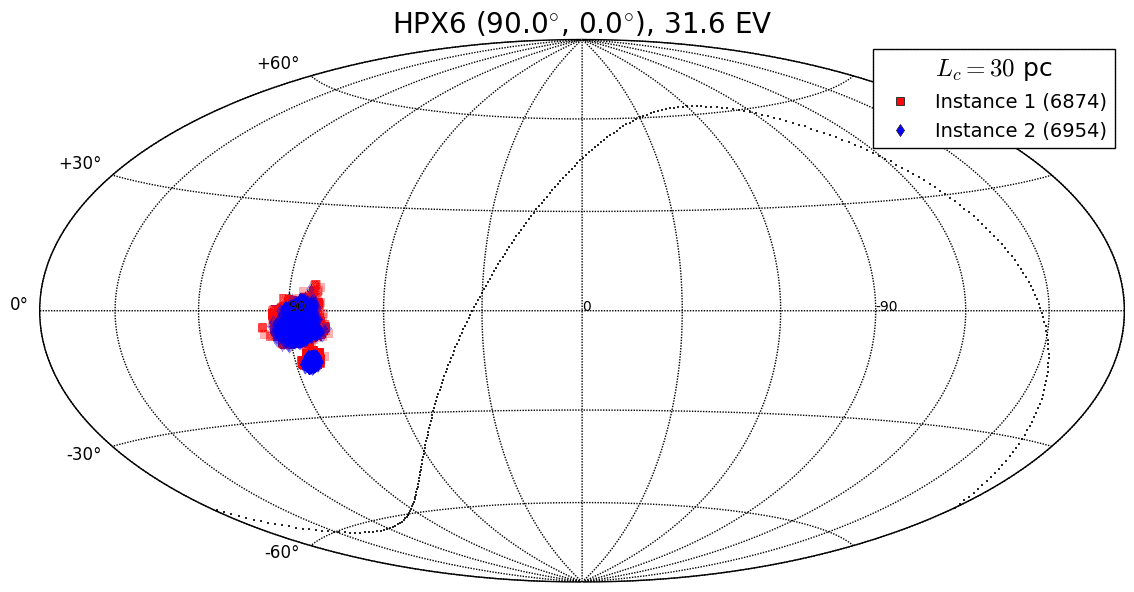}
\end{minipage}
\begin{minipage}[b]{0.48 \textwidth}
\includegraphics[width=1. \textwidth]{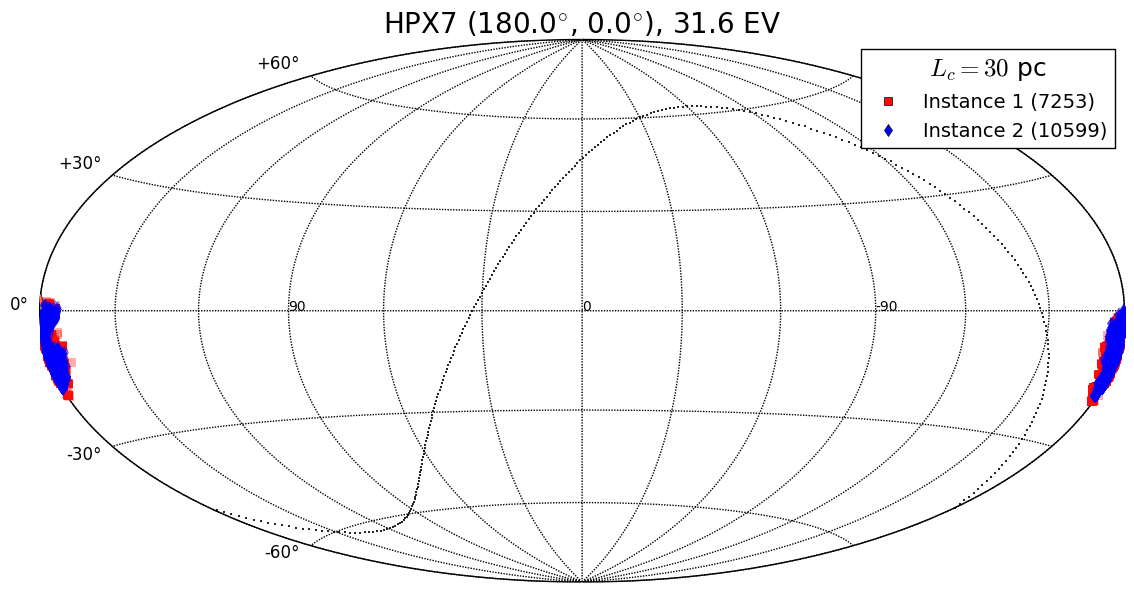}
\end{minipage}
\begin{minipage}[b]{0.48 \textwidth}
\includegraphics[width=1. \textwidth]{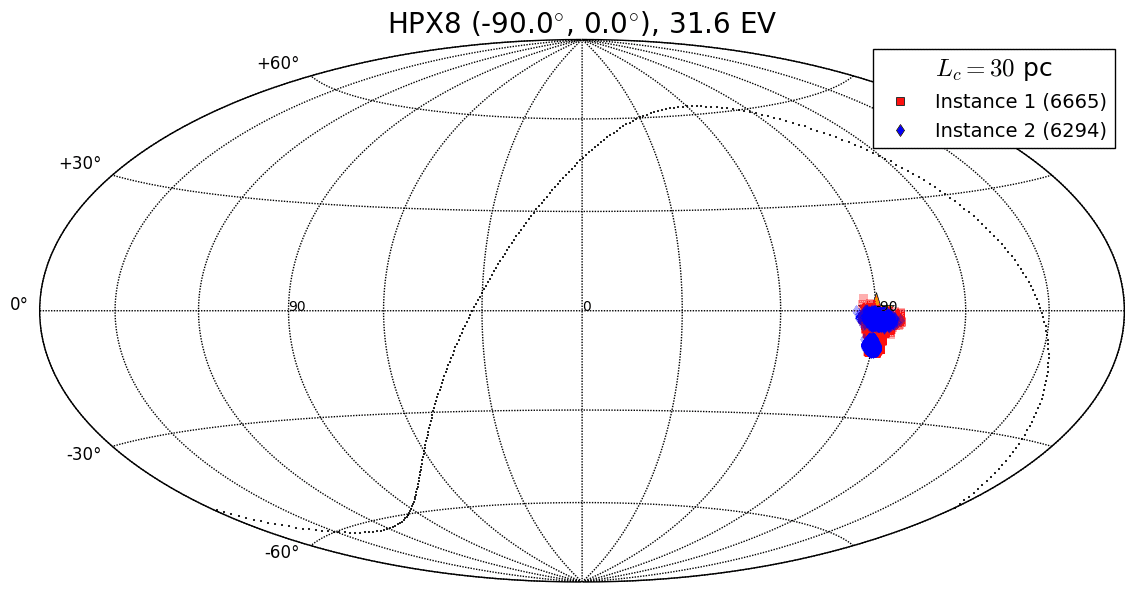}
\end{minipage}
\begin{minipage}[b]{0.48 \textwidth}
\includegraphics[width=1. \textwidth]{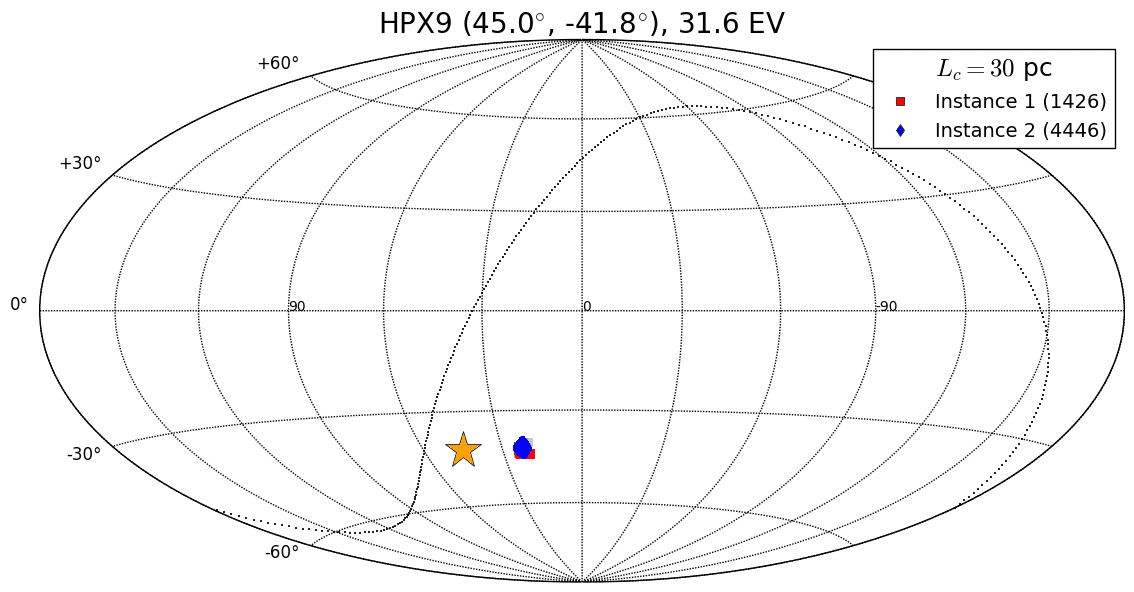}
\end{minipage}
\begin{minipage}[b]{0.48 \textwidth}
\includegraphics[width=1. \textwidth]{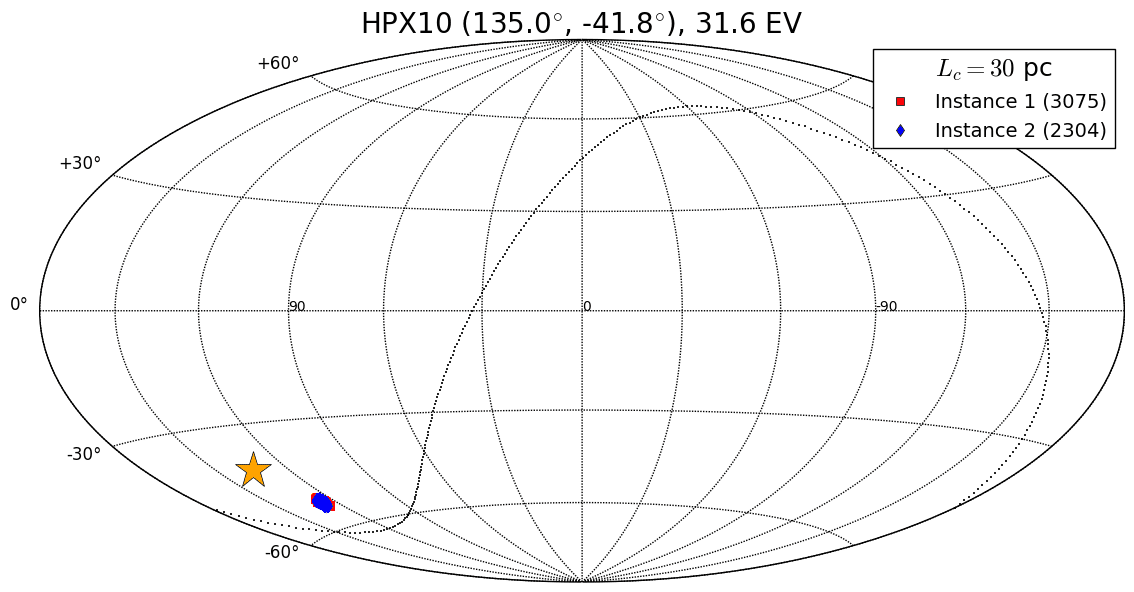}
\end{minipage}
\vspace{-0.3in}
\caption{Comparison of arrival direction distributions using different realizations of the $L_{coh} = 30$ pc random field (KRF 10 and KRF11), at log($R$ / V) = 18.5, from various sources in the regularly-spaced grid (marked with orange stars).
The source direction and name is listed in the plot title.
The sky map is in Galactic coordinates and the dotted line indicates decl. $\delta=0^{\circ}$.
The legend in each plot indicates the number of events arriving from the given source at the given rigidity for the given realization.} 
\label{plt:hpx19o5_krf10_krf11}
\vspace{-0.1in}
\end{figure}
\clearpage 
\begin{figure}[t]
\hspace{-0.3in}
\centering
\begin{minipage}[b]{0.48 \textwidth}
\includegraphics[width=1. \textwidth]{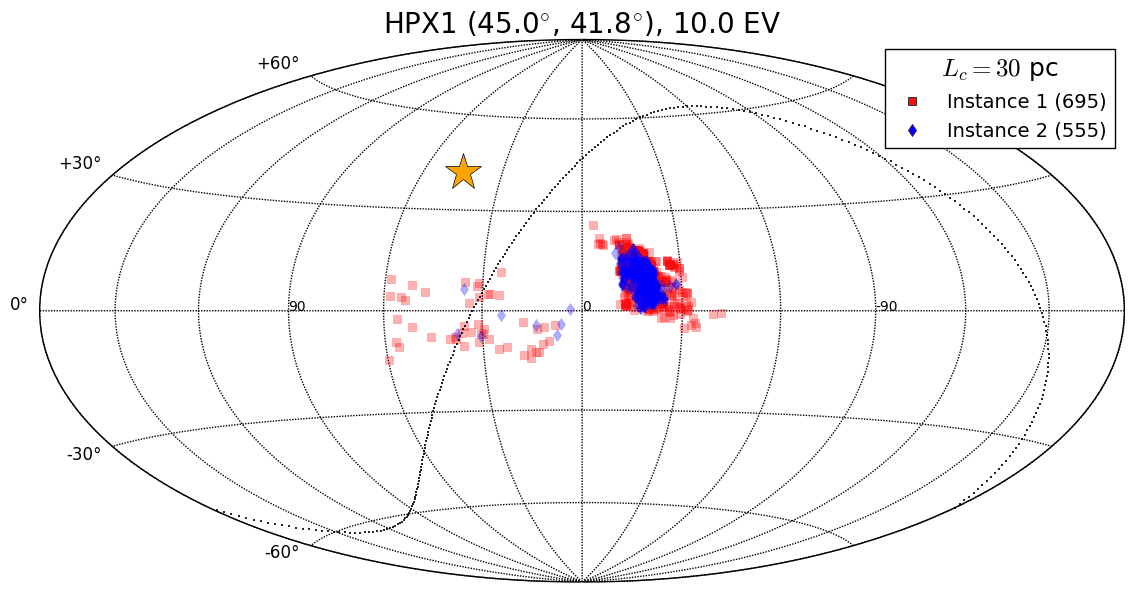}
\end{minipage}
\begin{minipage}[b]{0.48 \textwidth}
\includegraphics[width=1. \textwidth]{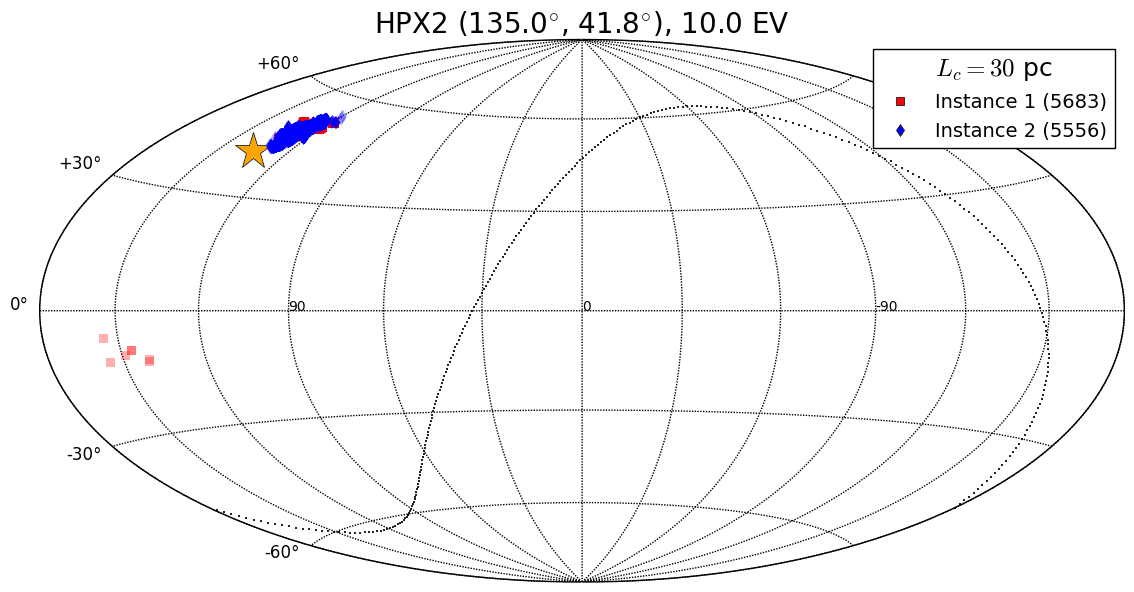}
\end{minipage}
\begin{minipage}[b]{0.48 \textwidth}
\includegraphics[width=1. \textwidth]{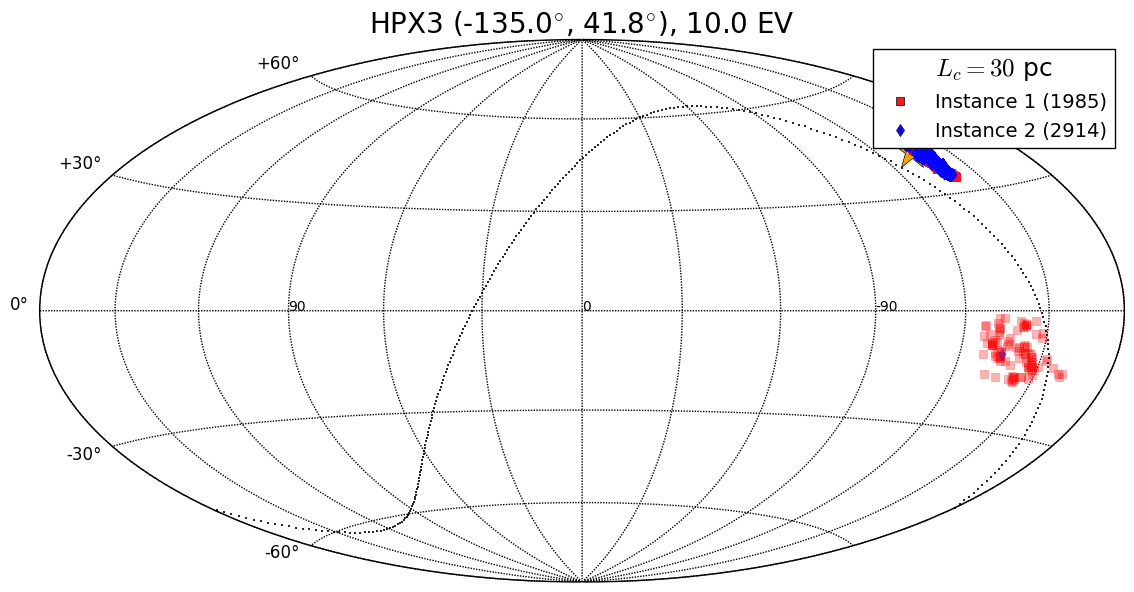}
\end{minipage}
\begin{minipage}[b]{0.48 \textwidth}
\includegraphics[width=1. \textwidth]{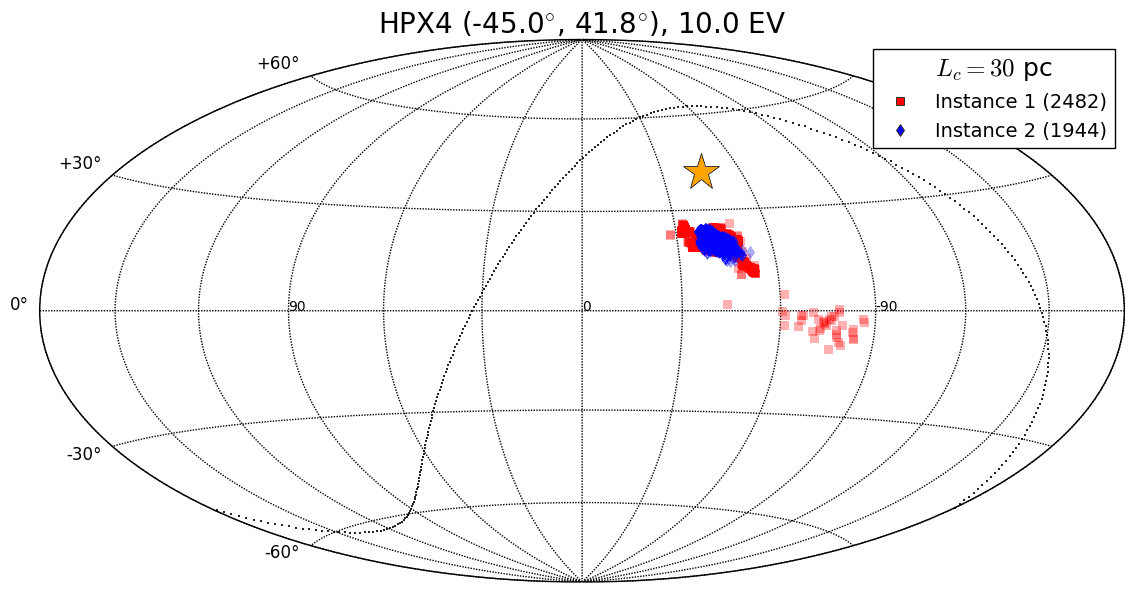}
\end{minipage}
\begin{minipage}[b]{0.48 \textwidth}
\includegraphics[width=1. \textwidth]{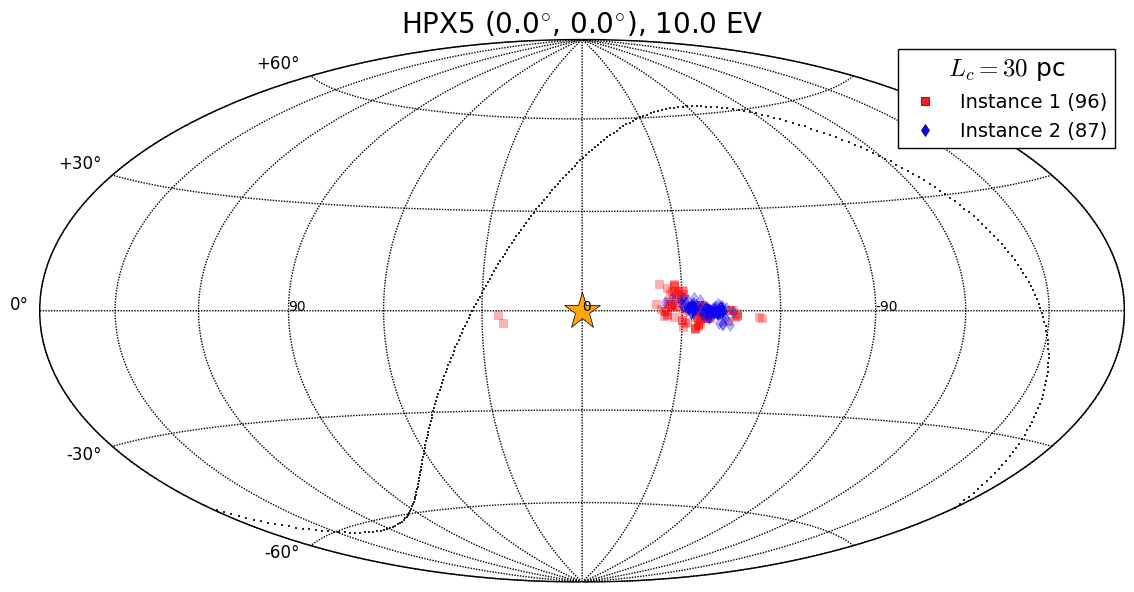}
\end{minipage}
\begin{minipage}[b]{0.48 \textwidth}
\includegraphics[width=1. \textwidth]{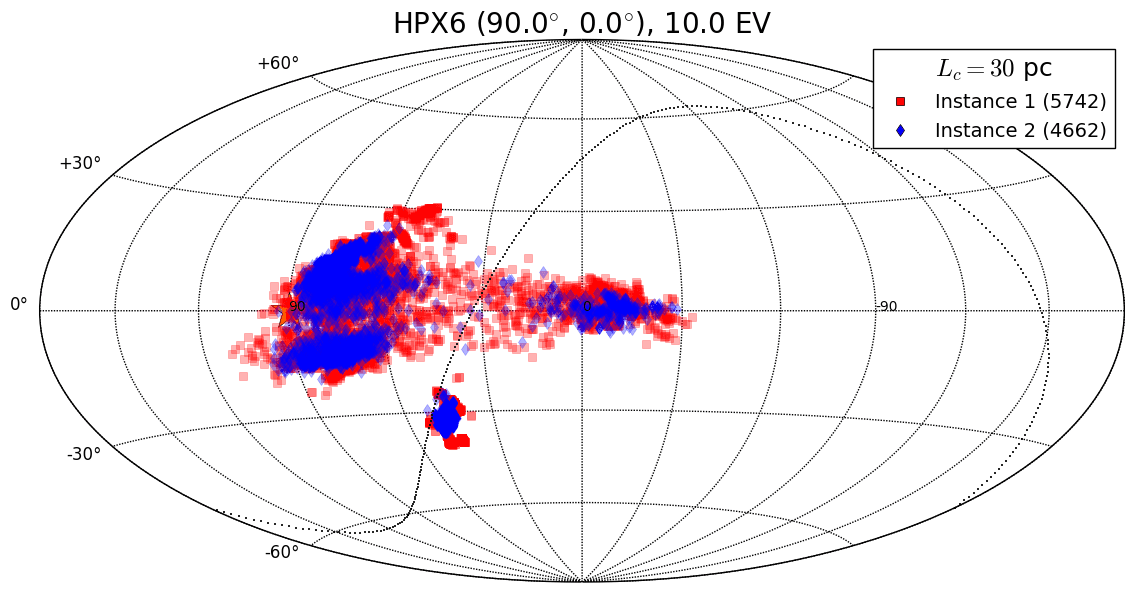}
\end{minipage}
\begin{minipage}[b]{0.48 \textwidth}
\includegraphics[width=1. \textwidth]{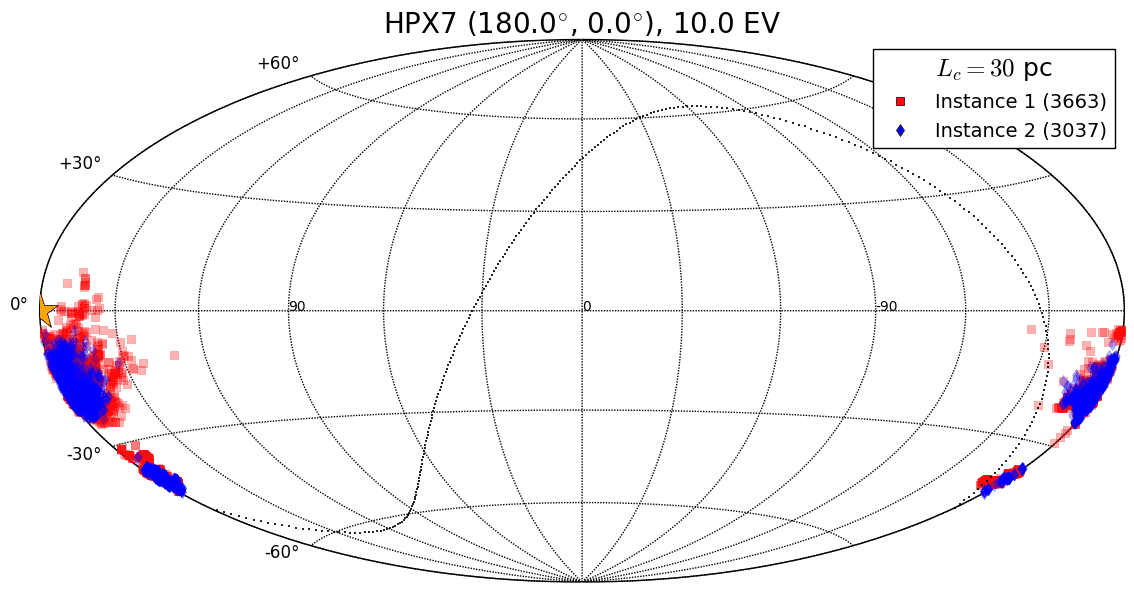}
\end{minipage}
\begin{minipage}[b]{0.48 \textwidth}
\includegraphics[width=1. \textwidth]{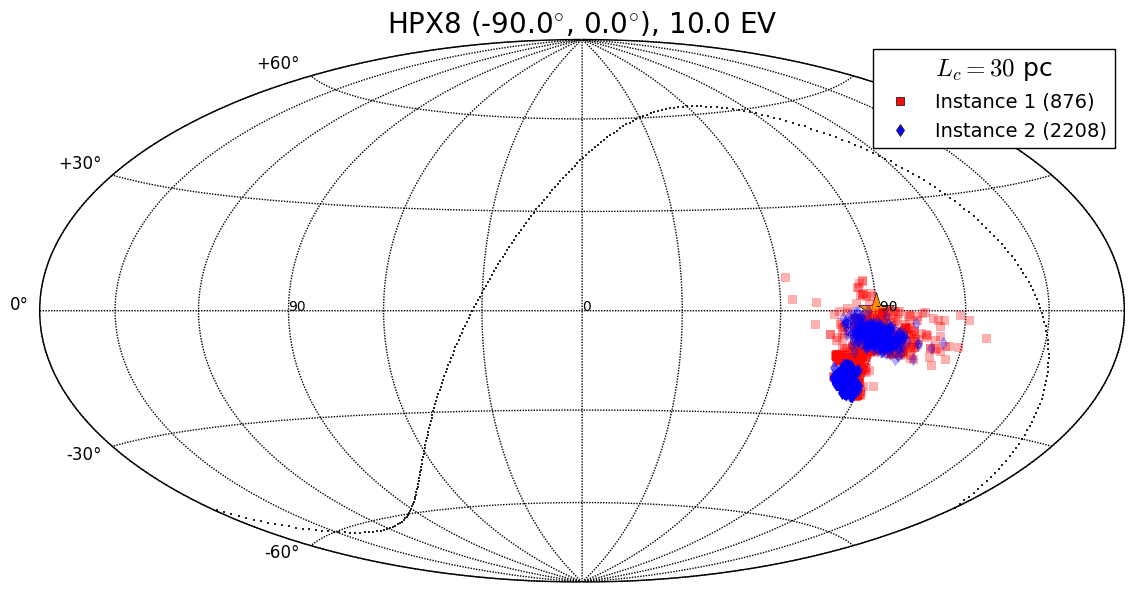}
\end{minipage}
\begin{minipage}[b]{0.48 \textwidth}
\includegraphics[width=1. \textwidth]{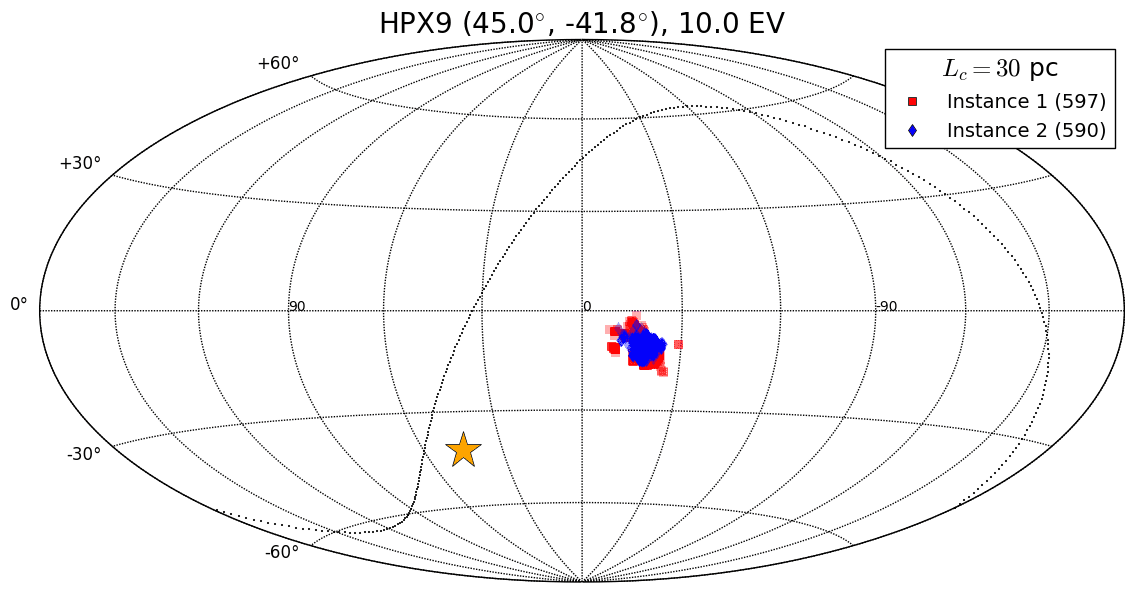}
\end{minipage}
\begin{minipage}[b]{0.48 \textwidth}
\includegraphics[width=1. \textwidth]{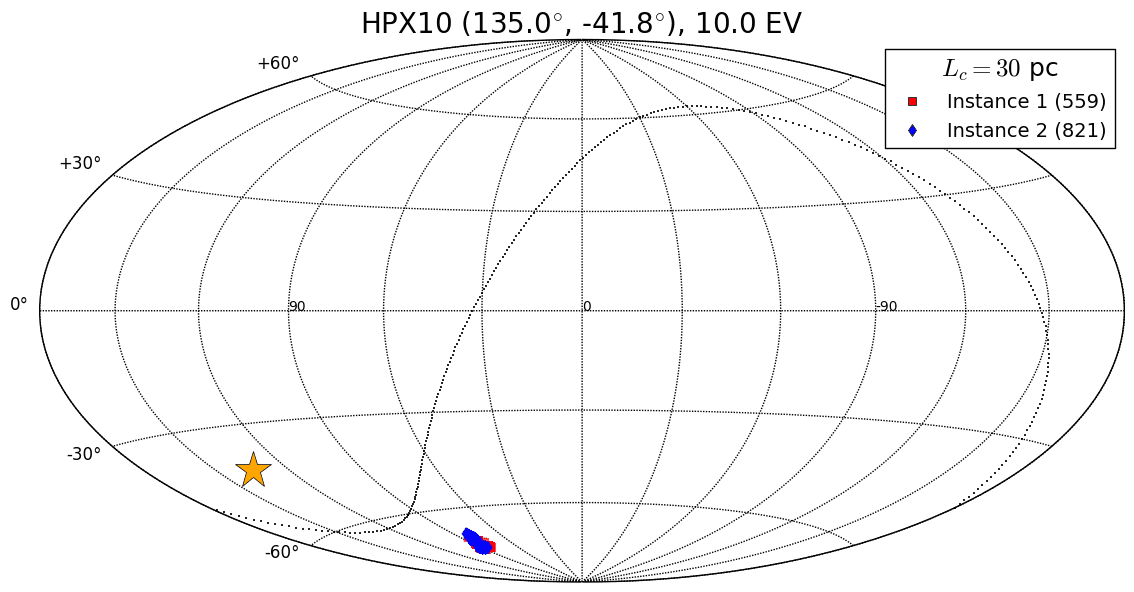}
\end{minipage}
\vspace{-0.3in}
\caption{Comparison of arrival direction distributions using different realizations of the $L_{coh} = 30$ pc random field, as in Fig. \ref{plt:hpx19o5_krf10_krf11}, but for log($R$ / V) = 19.0.} 
\label{plt:hpx19o0_krf10_krf11}
\vspace{-0.1in}
\end{figure}
\clearpage 
\begin{figure}[t!]
\hspace{-0.3in}
\centering
\begin{minipage}[b]{0.48 \textwidth}
\includegraphics[width=1. \textwidth]{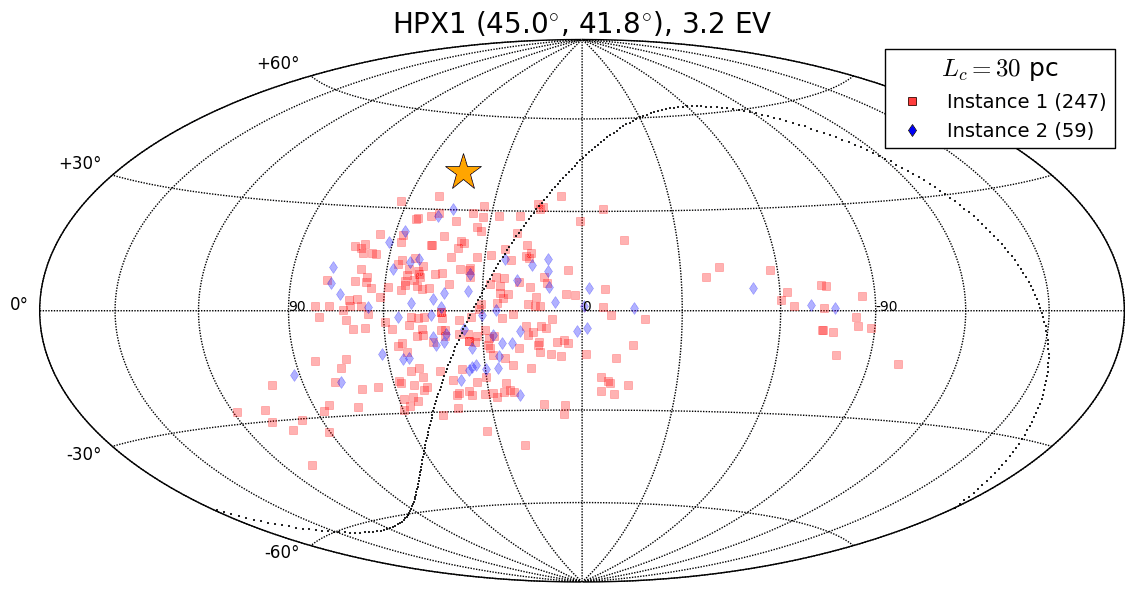}
\end{minipage}
\begin{minipage}[b]{0.48 \textwidth}
\includegraphics[width=1. \textwidth]{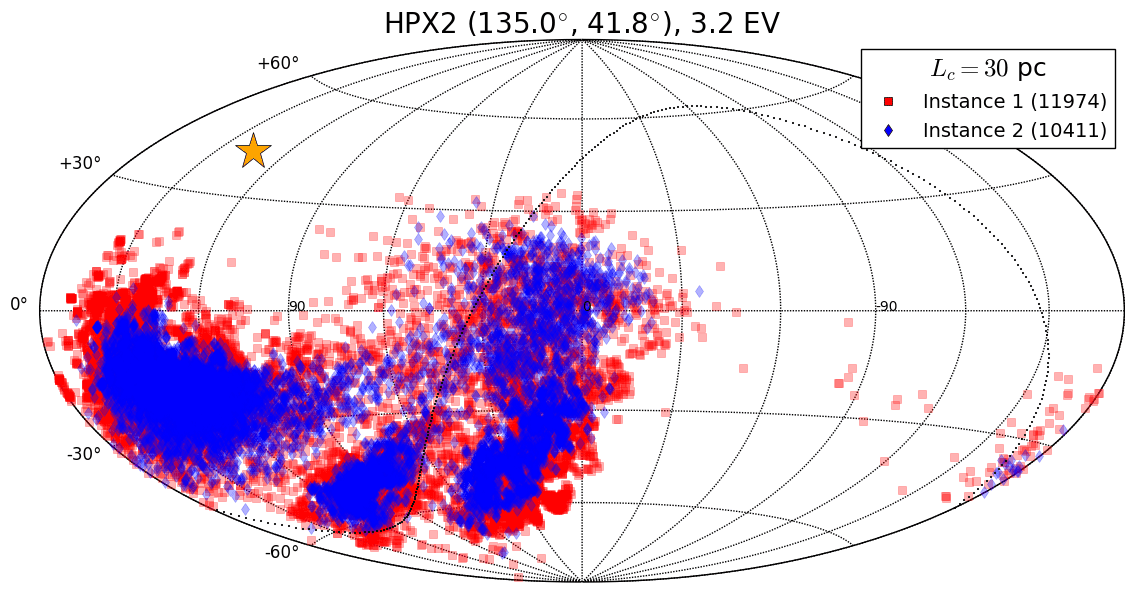}
\end{minipage}
\begin{minipage}[b]{0.48 \textwidth}
\includegraphics[width=1. \textwidth]{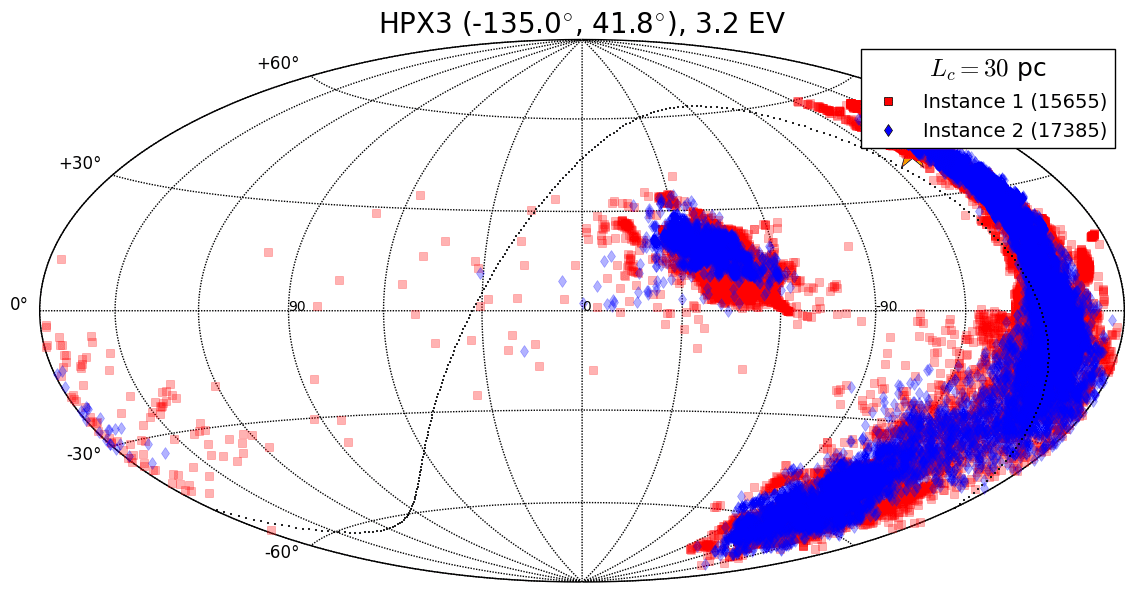}
\end{minipage}
\begin{minipage}[b]{0.48 \textwidth}
\includegraphics[width=1. \textwidth]{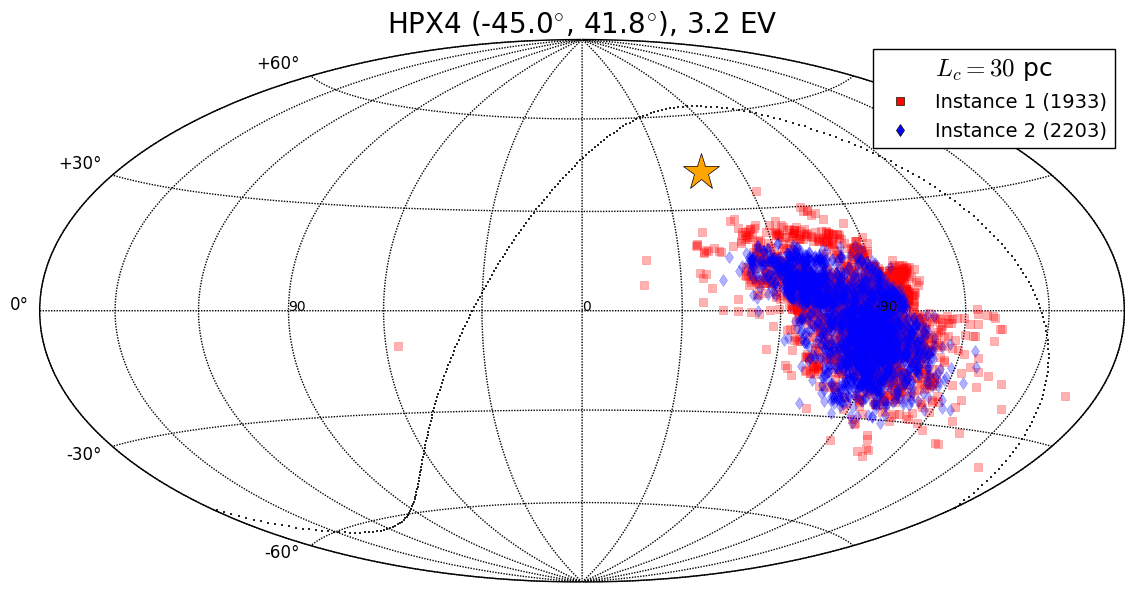}
\end{minipage}
\begin{minipage}[b]{0.48 \textwidth}
\includegraphics[width=1. \textwidth]{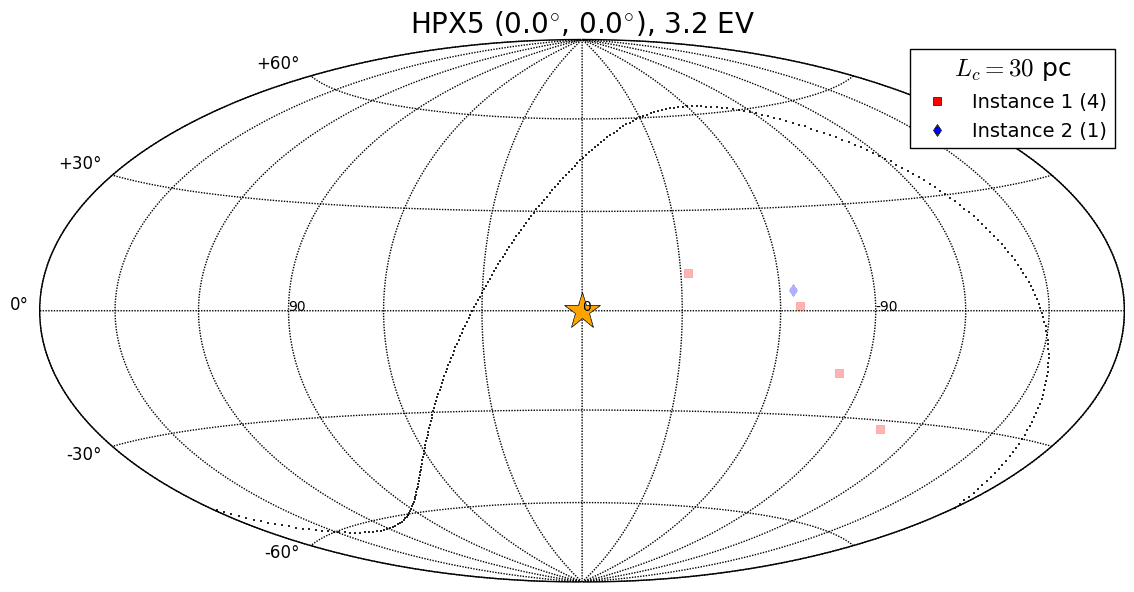}
\end{minipage}
\begin{minipage}[b]{0.48 \textwidth}
\includegraphics[width=1. \textwidth]{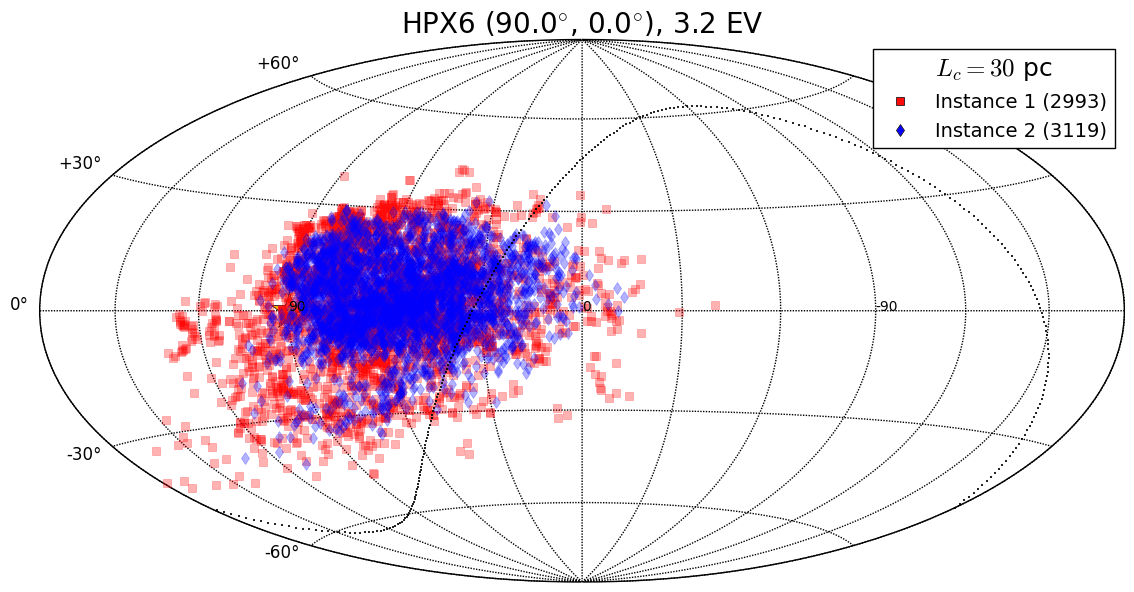}
\end{minipage}
\begin{minipage}[b]{0.48 \textwidth}
\includegraphics[width=1. \textwidth]{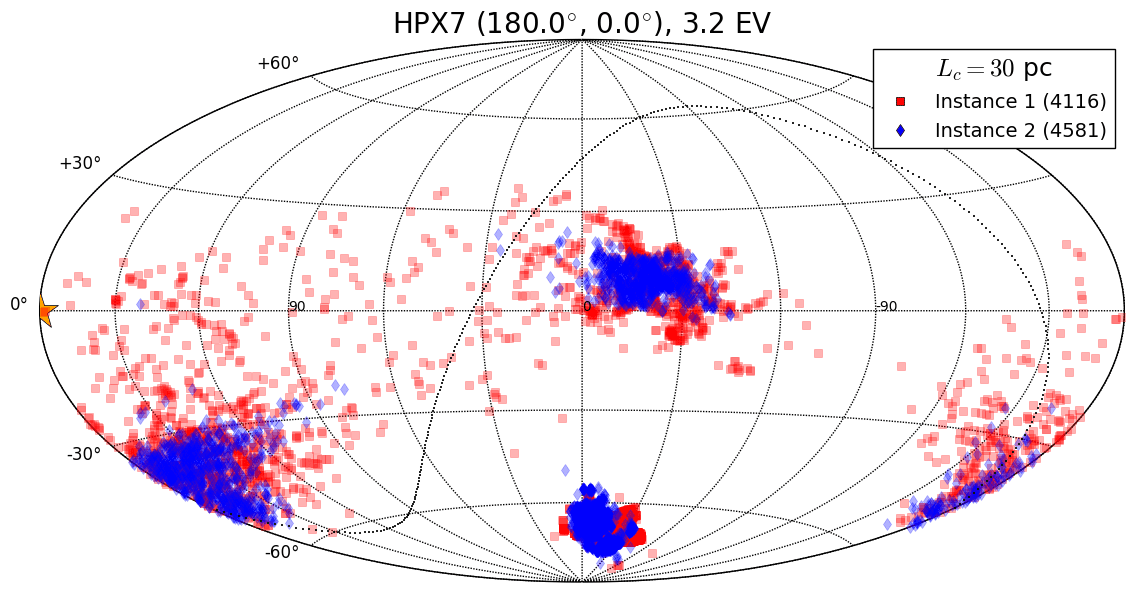}
\end{minipage}
\begin{minipage}[b]{0.48 \textwidth}
\includegraphics[width=1. \textwidth]{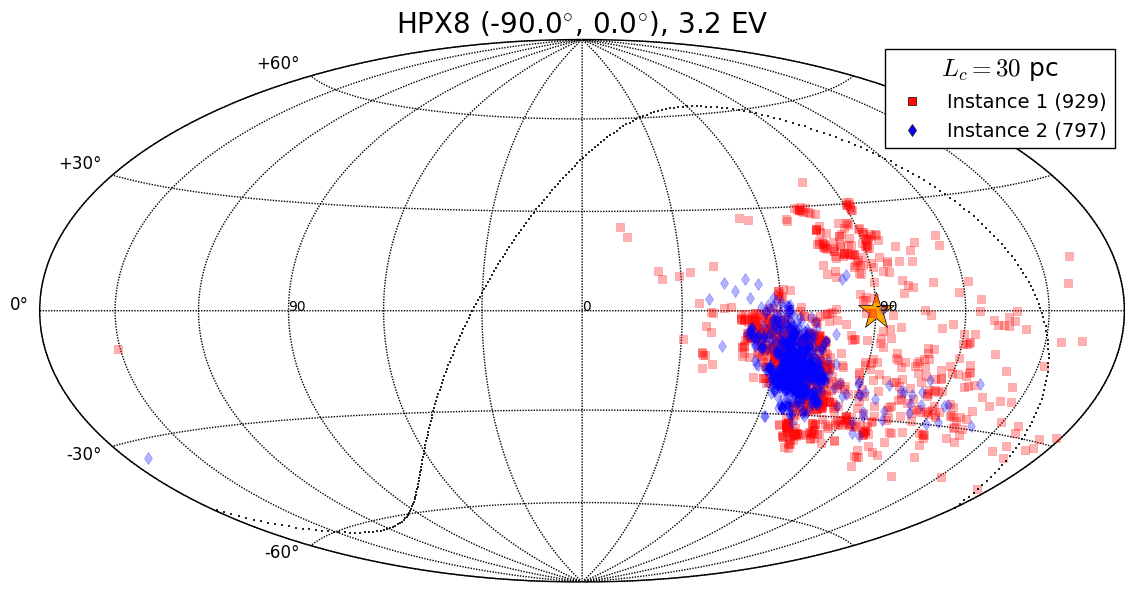}
\end{minipage}
\begin{minipage}[b]{0.48 \textwidth}
\includegraphics[width=1. \textwidth]{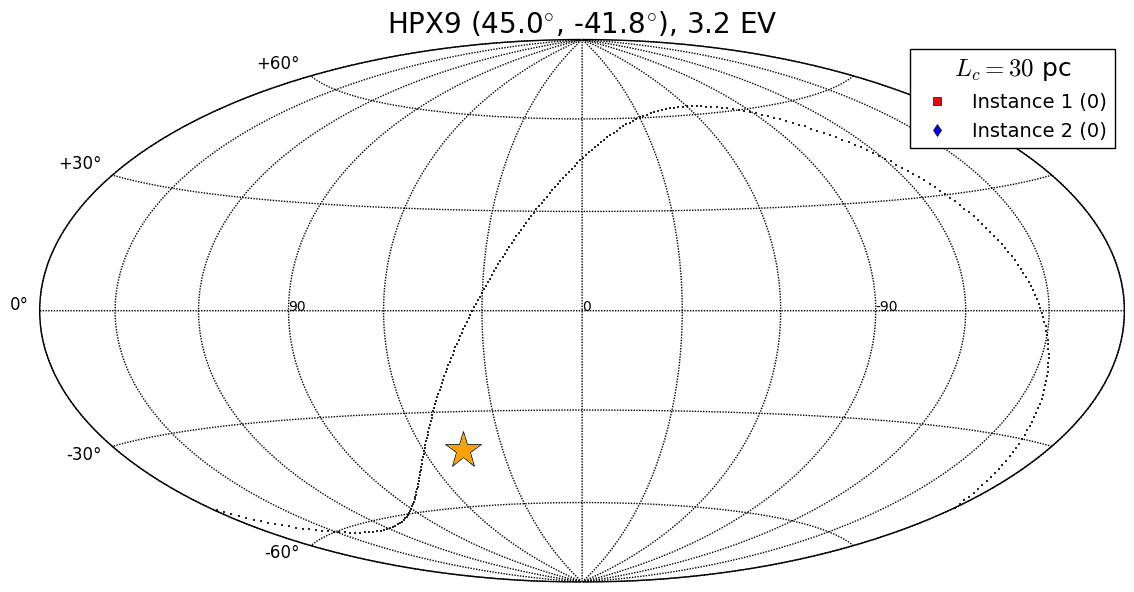}
\end{minipage}
\begin{minipage}[b]{0.48 \textwidth}
\includegraphics[width=1. \textwidth]{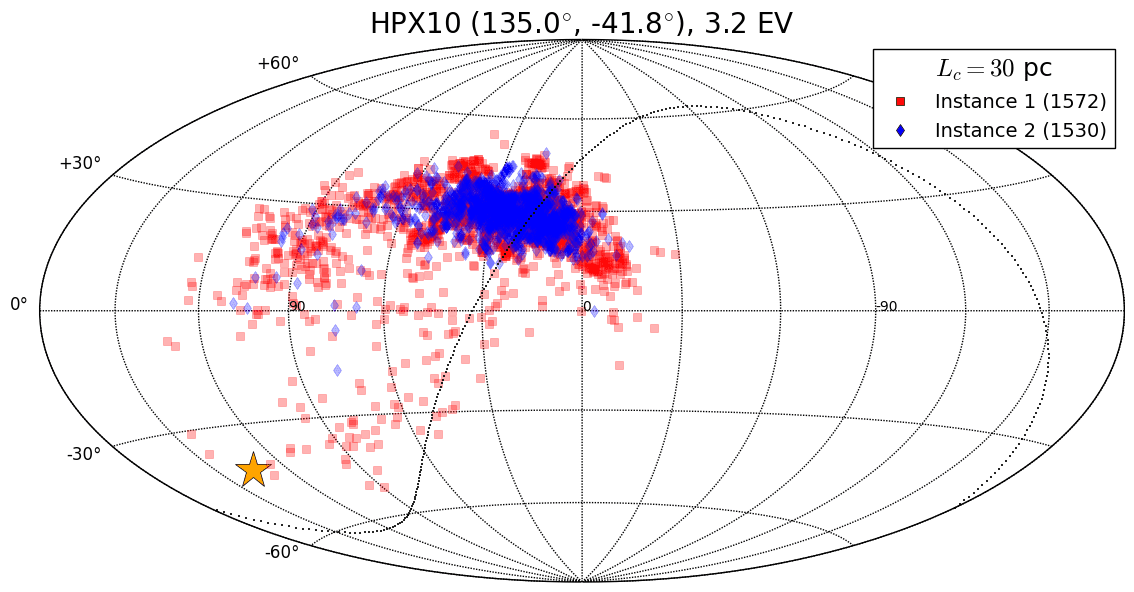}
\end{minipage}
\vspace{-0.3in}
\caption{Comparison of arrival direction distributions using different realizations of the $L_{coh} = 30$ pc random field, as in Fig. \ref{plt:hpx18o5_krf10_krf11}, but for log($R$ / V) = 18.5.} 
\label{plt:hpx18o5_krf10_krf11}
\vspace{-0.1in}
\end{figure}
\clearpage 

\pagebreak
\section{Tables of Arrival Direction Distributions from HEALPix Grid Sources}
\label{appdx:HEALPixTables}
\begin{table}
\tiny
\caption{Properties of arrival direction distributions from cosmic rays originating within 1\dg\ of the given HEALPix source directions.
Column 1 lists log$_{10}$ of the rigidity in V.
Column 2 is the magnification factor at that rigidity, i.e., the number of observed events from the source region in the angular window divided by the number which would be seen in the absence of magnetic lensing; a value of 0 means no events arrived from that source region.
In Col. 2, 3, 4 and 5, the first entry is for the 100 pc coherence length (KRF6) realization, followed by the result for the 30 pc realization (KRF10) in parentheses and the JF12 coherent-only case in curly brackets \{\}.
Dashes indicate no events in the distribution.
Columns 3 and 4 list the centroid of the distribution of arrival directions in Galactic coordinates. 
Column 5 gives the angular separation between the source direction and the centroid, and the median separation of the arrival directions around the centroid; i.e., $\sigma_{\Delta}$ measures the angular size of the arrival direction distribution.}
\begin{center}
 \hspace*{-1cm}\begin{tabular}[htpb]{ |c|c|c|c|c| }
  \hline
log$R$ & mag 	& $<{\ell}>$ ($^{\circ}$) & $<{b}>$ ($^{\circ}$) & $<{\Delta}>, \sigma_{\Delta}$  ($^{\circ}$) \\ \hline \hline
\multicolumn{5}{|c|}{North Pole } \\ \hline 
20.0  &  1.2 (0.96) \{0.98\}  &  -16 (-19) \{-20\}  &  87 (87) \{86\}  &  3.2, 0.8 (3.5, 0.7) \{3.5, 0.7\} \\ \hline
19.8  &  1.36 (0.79) \{0.96\}  &  -18 (-20) \{-21\}  &  85 (84) \{84\}  &  5.0, 0.8 (5.6, 0.6) \{5.6, 0.7\} \\ \hline
19.6  &  1.03 (0.65) \{0.94\}  &  -20 (-22) \{-21\}  &  82 (81) \{81\}  &  7.6, 0.7 (8.8, 0.6) \{8.8, 0.7\} \\ \hline
19.5  &  0.98 (0.95) \{0.92\}  &  -17 (-21) \{-21\}  &  81 (79) \{79\}  &  9.4, 0.7 (11.2, 0.9) \{11.1, 0.7\} \\ \hline
19.4  &  0.73 (0.73) \{0.9\}  &  -19 (-20) \{-21\}  &  78 (76) \{76\}  &  11.8, 0.7 (14.3, 0.6) \{14.1, 0.7\} \\ \hline
19.2  &  0.58 (0.58) \{0.86\}  &  -15 (-23) \{-22\}  &  67 (67) \{67\}  &  22.9, 0.5 (23.0, 0.8) \{22.9, 0.7\} \\ \hline
19.0  &  0.53 (0.64) \{0.8\}  &  -15 (-25) \{-23\}  &  44 (51) \{51\}  &  45.6, 6.9 (39.1, 1.3) \{38.9, 0.6\} \\ \hline
18.8  &  0.65 (0.03) \{0\}  &  -79 (-25) \{-\}  &  0 (17) \{-\}  &  89.7, 55.6 (72.6, 3.8) \{-, -\} \\ \hline
18.6  &  1.72 (0.01) \{0\}  &  -111 (-57) \{-\}  &  6 (-8) \{-\}  &  83.8, 14.2 (98.0, 46.0) \{-, -\} \\ \hline
18.5  &  1.72 (0.22) \{0.78\}  &  -107 (-49) \{-14\}  &  -17 (-10) \{1\}  &  107.3, 40.4 (100.1, 41.6) \{89.4, 2.4\} \\ \hline
18.4  &  2.5 (1.56) \{0.01\}  &  -90 (-89) \{-34\}  &  -20 (-34) \{3\}  &  109.9, 53.7 (124.3, 40.2) \{87.5, 5.4\} \\ \hline
18.3  &  2.5 (3.42) \{6.16\}  &  -50 (-89) \{-96\}  &  -19 (-27) \{-29\}  &  109.2, 63.1 (117.4, 35.6) \{118.7, 29.1\} \\ \hline
18.2  &  2.42 (4.25) \{6.62\}  &  13 (-98) \{-103\}  &  -5 (-7) \{-3\}  &  94.7, 79.5 (96.9, 23.9) \{93.2, 8.9\} \\ \hline
18.1  &  2.33 (1.66) \{0.43\}  &  23 (-79) \{-62\}  &  9 (-2) \{-33\}  &  81.3, 85.2 (92.1, 35.6) \{122.9, 15.1\} \\ \hline
18.0  &  1.31 (0.89) \{1.22\}  &  37 (-58) \{-79\}  &  2 (7) \{-5\}  &  87.6, 84.0 (83.1, 44.4) \{94.6, 18.8\} \\ \hline

\multicolumn{5}{|c|}{HPX 1:  $\ell = 45^{\circ}$,  $b = 41.8103^{\circ}$ }\\ \hline
20.0  &  1.24 (0.93) \{1.0\}  &  37 (38) \{38\}  &  41 (42) \{42\}  &  6.4, 0.8 (5.2, 0.7) \{5.2, 0.7\} \\ \hline
19.8  &  1.25 (1.06) \{0.99\}  &  32 (34) \{34\}  &  40 (42) \{42\}  &  10.0, 0.9 (8.3, 0.8) \{8.4, 0.7\} \\ \hline
19.6  &  1.99 (0.72) \{0.98\}  &  28 (26) \{26\}  &  39 (42) \{42\}  &  12.9, 1.1 (14.0, 0.7) \{13.8, 0.7\} \\ \hline
19.5  &  0.4 (1.45) \{0.98\}  &  22 (21) \{21\}  &  39 (41) \{41\}  &  17.5, 0.7 (18.1, 0.9) \{17.8, 0.7\} \\ \hline
19.4  &  0.2 (0.55) \{0.97\}  &  23 (15) \{14\}  &  33 (39) \{39\}  &  19.5, 5.4 (22.9, 0.9) \{23.3, 0.7\} \\ \hline
19.2  &  2.68 (0.64) \{0.71\}  &  -5 (-4) \{-5\}  &  33 (28) \{28\}  &  40.1, 2.7 (41.8, 1.7) \{42.3, 0.6\} \\ \hline
19.0  &  0.11 (0.15) \{0.28\}  &  35 (-18) \{-18\}  &  -4 (10) \{10\}  &  46.5, 42.3 (63.5, 3.7) \{63.6, 0.7\} \\ \hline
18.8  &  0.09 (0.01) \{0.0\}  &  44 (8) \{-31\}  &  -7 (-3) \{2\}  &  49.3, 28.2 (55.5, 20.5) \{77.9, 0.1\} \\ \hline
18.6  &  0.1 (0.01) \{0\}  &  56 (41) \{-\}  &  -18 (1) \{-\}  &  60.8, 35.9 (40.7, 15.8) \{-, -\} \\ \hline
18.5  &  0.16 (0.02) \{0\}  &  47 (34) \{-\}  &  -24 (-1) \{-\}  &  65.6, 36.6 (43.6, 23.0) \{-, -\} \\ \hline
18.4  &  0.18 (0.01) \{0\}  &  52 (32) \{-\}  &  -33 (0) \{-\}  &  75.2, 44.6 (43.5, 27.7) \{-, -\} \\ \hline
18.3  &  0.16 (0.05) \{0.03\}  &  56 (-18) \{-77\}  &  -21 (-7) \{-7\}  &  63.6, 49.6 (75.2, 60.9) \{118.5, 3.7\} \\ \hline
18.2  &  0.14 (0.13) \{0\}  &  52 (-19) \{-\}  &  -23 (-9) \{-\}  &  65.1, 51.5 (77.4, 59.5) \{-, -\} \\ \hline
18.1  &  0.11 (0.17) \{0.12\}  &  48 (-41) \{-68\}  &  -18 (-12) \{-1\}  &  59.8, 60.8 (95.3, 38.2) \{107.7, 14.8\} \\ \hline
18.0  &  0.1 (0.15) \{0.5\}  &  32 (-54) \{-98\}  &  -6 (-18) \{-14\}  &  49.8, 66.3 (108.5, 33.8) \{137.4, 6.2\} \\ \hline
\hspace*{-1cm} \end{tabular}
\label{tbl:dist_npol_hpx1}
\end{center}
\end{table}

\begin{table}
\tiny
\begin{center}
\hspace*{-2cm}\begin{tabular}[htpb]{ |c|c|c|c|c| }
    \hline 
log$R$ & mag 	& $<{\ell}>$ ($^{\circ}$) & $<{b}>$ ($^{\circ}$)  & $<{\Delta}>, \sigma_{\Delta}$  ($^{\circ}$) \\ \hline \hline
\multicolumn{5}{|c|}{HPX 2:  $\ell = 135^{\circ}$,  $b = 41.8103^{\circ}$ }\\ \hline
20.0  &  1.31 (0.88) \{1.05\}  &  132 (135) \{135\}  &  42 (42) \{42\}  &  2.1, 0.8 (0.5, 0.7) \{0.6, 0.7\} \\ \hline
19.8  &  1.18 (0.76) \{1.07\}  &  131 (134) \{134\}  &  43 (42) \{43\}  &  3.4, 0.8 (0.7, 0.6) \{0.9, 0.7\} \\ \hline
19.6  &  0.77 (0.71) \{1.12\}  &  128 (134) \{134\}  &  43 (43) \{43\}  &  5.4, 0.6 (1.5, 0.8) \{1.5, 0.7\} \\ \hline
19.5  &  0.54 (1.06) \{1.16\}  &  126 (133) \{134\}  &  43 (44) \{43\}  &  6.5, 0.5 (2.1, 0.8) \{2.0, 0.8\} \\ \hline
19.4  &  0.39 (1.23) \{1.21\}  &  125 (133) \{133\}  &  43 (44) \{44\}  &  7.8, 0.5 (2.5, 0.8) \{2.6, 0.8\} \\ \hline
19.2  &  0.07 (1.71) \{1.37\}  &  119 (131) \{132\}  &  42 (45) \{46\}  &  11.7, 2.2 (4.3, 1.3) \{4.5, 0.8\} \\ \hline
19.0  &  3.4 (1.45) \{1.76\}  &  146 (128) \{128\}  &  25 (49) \{49\}  &  19.0, 12.4 (8.4, 1.9) \{8.6, 0.9\} \\ \hline
18.8  &  6.64 (3.99) \{4.59\}  &  127 (101) \{113\}  &  27 (61) \{58\}  &  16.4, 23.6 (27.8, 8.1) \{21.5, 2.3\} \\ \hline
18.6  &  5.43 (2.37) \{1.74\}  &  132 (111) \{102\}  &  11 (-34) \{-46\}  &  30.8, 29.0 (79.2, 22.0) \{92.7, 7.4\} \\ \hline
18.5  &  3.81 (2.72) \{2.68\}  &  134 (103) \{127\}  &  -6 (-27) \{-27\}  &  48.0, 50.5 (74.6, 36.8) \{68.9, 7.4\} \\ \hline
18.4  &  4.92 (3.31) \{3.01\}  &  157 (99) \{106\}  &  -9 (-25) \{-28\}  &  54.8, 44.1 (74.9, 40.3) \{75.0, 27.0\} \\ \hline
18.3  &  3.24 (5.07) \{5.53\}  &  180 (111) \{77\}  &  -20 (-25) \{-24\}  &  74.8, 69.6 (70.3, 47.5) \{85.2, 53.7\} \\ \hline
18.2  &  4.19 (4.36) \{1.1\}  &  -167 (91) \{11\}  &  -12 (-29) \{7\}  &  75.6, 65.3 (81.3, 56.3) \{109.4, 46.7\} \\ \hline
18.1  &  7.75 (5.0) \{8.01\}  &  -162 (-154) \{41\}  &  8 (-26) \{-55\}  &  65.1, 49.7 (94.3, 70.9) \{125.1, 25.6\} \\ \hline
18.0  &  5.42 (3.53) \{2.06\}  &  152 (34) \{23\}  &  5 (1) \{21\}  &  39.4, 85.0 (97.8, 53.2) \{91.2, 34.2\} \\ \hline

\multicolumn{5}{|c|}{HPX 3:  $\ell =  -135^{\circ}$,  $b = 41.8103^{\circ}$ }\\ \hline
20.0  &  1.14 (1.06) \{1.01\}  &  -135 (-135) \{-135\}  &  41 (41) \{42\}  &  1.3, 0.8 (0.4, 0.7) \{0.4, 0.7\} \\ \hline
19.8  &  1.15 (1.17) \{1.01\}  &  -136 (-135) \{-135\}  &  40 (41) \{41\}  &  2.0, 0.8 (0.6, 0.8) \{0.6, 0.7\} \\ \hline
19.6  &  1.04 (1.22) \{1.02\}  &  -136 (-136) \{-136\}  &  39 (41) \{41\}  &  3.1, 0.7 (0.9, 0.8) \{0.9, 0.7\} \\ \hline
19.5  &  1.26 (1.16) \{1.03\}  &  -137 (-136) \{-136\}  &  38 (41) \{41\}  &  3.6, 0.8 (1.2, 0.8) \{1.1, 0.7\} \\ \hline
19.4  &  1.27 (1.09) \{1.04\}  &  -137 (-136) \{-136\}  &  38 (41) \{41\}  &  4.2, 0.9 (1.4, 0.9) \{1.4, 0.7\} \\ \hline
19.2  &  3.63 (0.76) \{1.06\}  &  -139 (-137) \{-137\}  &  38 (40) \{40\}  &  4.7, 4.9 (2.2, 1.0) \{2.3, 0.7\} \\ \hline
19.0  &  2.9 (0.76) \{1.11\}  &  -135 (-138) \{-138\}  &  22 (39) \{39\}  &  19.9, 11.5 (3.7, 2.1) \{3.9, 0.7\} \\ \hline
18.8  &  2.25 (1.65) \{1.2\}  &  -141 (-141) \{-141\}  &  7 (35) \{37\}  &  34.9, 19.3 (8.0, 3.5) \{6.5, 0.8\} \\ \hline
18.6  &  2.53 (3.48) \{2.07\}  &  -134 (-142) \{-140\}  &  6 (7) \{11\}  &  36.3, 25.5 (35.4, 28.5) \{30.9, 22.8\} \\ \hline
18.5  &  3.73 (4.55) \{4.54\}  &  -159 (-136) \{-136\}  &  23 (1) \{-1\}  &  27.1, 53.4 (41.0, 33.1) \{43.2, 32.4\} \\ \hline
18.4  &  3.93 (4.89) \{5.33\}  &  -36 (-130) \{-136\}  &  11 (4) \{5\}  &  89.7, 79.3 (37.7, 37.4) \{37.3, 23.6\} \\ \hline
18.3  &  6.24 (5.32) \{3.62\}  &  60 (-93) \{-38\}  &  32 (4) \{-4\}  &  104.4, 57.5 (53.0, 64.0) \{98.0, 30.4\} \\ \hline
18.2  &  5.77 (3.82) \{1.81\}  &  79 (-109) \{-60\}  &  34 (30) \{53\}  &  98.1, 44.0 (24.4, 47.3) \{49.8, 8.0\} \\ \hline
18.1  &  1.71 (3.24) \{2.04\}  &  93 (-118) \{-97\}  &  16 (33) \{60\}  &  107.4, 75.4 (16.2, 44.3) \{29.5, 9.3\} \\ \hline
18.0  &  2.47 (2.89) \{2.25\}  &  67 (-123) \{-138\}  &  4 (23) \{51\}  &  130.2, 83.6 (21.8, 54.8) \{9.4, 12.2\} \\ \hline
  \hspace*{-2cm}\end{tabular}
\label{tbl:dist_hpx2_hpx3}
\end{center}
\end{table}

\begin{table}
\tiny
\begin{center}
  \hspace*{-1cm}\begin{tabular}[htpb]{ |c|c|c|c|c| }
    \hline 
log$R$ & mag 	& $<{\ell}>$ ($^{\circ}$) & $<{b}>$ ($^{\circ}$)  & $<{\Delta}>, \sigma_{\Delta}$  ($^{\circ}$) \\ \hline \hline
\multicolumn{5}{|c|}{HPX 4:  $\ell = -45^{\circ}$,  $b = 41.8103^{\circ}$ }\\ \hline
20.0  &  1.14 (0.91) \{0.94\}  &  -45 (-45) \{-45\}  &  39 (39) \{39\}  &  2.5, 0.8 (2.4, 0.7) \{2.4, 0.7\} \\ \hline
19.8  &  0.47 (1.26) \{0.9\}  &  -44 (-44) \{-44\}  &  38 (38) \{38\}  &  3.8, 0.5 (3.7, 0.8) \{3.7, 0.7\} \\ \hline
19.6  &  0.5 (0.73) \{0.85\}  &  -42 (-44) \{-44\}  &  37 (36) \{36\}  &  5.8, 0.8 (5.9, 0.7) \{5.8, 0.7\} \\ \hline
19.5  &  0.61 (0.66) \{0.81\}  &  -41 (-44) \{-44\}  &  35 (34) \{35\}  &  7.2, 0.7 (7.4, 0.7) \{7.3, 0.6\} \\ \hline
19.4  &  0.38 (0.68) \{0.77\}  &  -38 (-43) \{-43\}  &  34 (33) \{33\}  &  10.0, 1.9 (9.2, 0.8) \{9.1, 0.6\} \\ \hline
19.2  &  0.04 (0.46) \{0.66\}  &  -75 (-42) \{-43\}  &  -1 (28) \{28\}  &  51.2, 5.7 (14.1, 1.4) \{14.0, 0.6\} \\ \hline
19.0  &  0.56 (0.51) \{0.49\}  &  -41 (-43) \{-43\}  &  7 (20) \{21\}  &  35.1, 13.8 (21.4, 2.3) \{21.0, 0.5\} \\ \hline
18.8  &  1.44 (0.67) \{1.62\}  &  -29 (-47) \{-41\}  &  16 (8) \{9\}  &  29.4, 8.5 (33.5, 7.2) \{33.4, 2.5\} \\ \hline
18.6  &  0.48 (0.57) \{1.12\}  &  -64 (-80) \{-80\}  &  6 (-2) \{-4\}  &  39.6, 28.7 (54.4, 11.8) \{55.6, 5.5\} \\ \hline
18.5  &  0.4 (0.58) \{1.15\}  &  -71 (-84) \{-85\}  &  -7 (-3) \{-3\}  &  54.0, 34.0 (56.8, 14.3) \{58.0, 8.2\} \\ \hline
18.4  &  0.32 (0.61) \{0.95\}  &  -72 (-88) \{-88\}  &  -16 (-1) \{-1\}  &  63.0, 42.9 (57.9, 14.2) \{57.8, 9.2\} \\ \hline
18.3  &  0.26 (0.54) \{1.03\}  &  -60 (-89) \{-99\}  &  -11 (-1) \{2\}  &  54.2, 43.0 (57.9, 16.7) \{62.6, 5.8\} \\ \hline
18.2  &  0.23 (0.43) \{0.6\}  &  -47 (-78) \{-83\}  &  -9 (-3) \{13\}  &  51.0, 49.8 (53.6, 26.4) \{43.5, 1.3\} \\ \hline
18.1  &  0.2 (0.48) \{0.37\}  &  -53 (-71) \{-75\}  &  -10 (-2) \{9\}  &  51.9, 68.2 (49.5, 27.9) \{41.8, 9.3\} \\ \hline
18.0  &  0.17 (0.51) \{0.41\}  &  -34 (-65) \{-79\}  &  -10 (-4) \{-2\}  &  53.1, 74.8 (49.5, 37.8) \{53.7, 26.5\} \\ \hline

\multicolumn{5}{|c|}{HPX 5:  Galactic Center, $\ell = 0^{\circ}$,  $b = 0^{\circ}$ }\\ \hline
20.0  &  1.86 (1.17) \{1.13\}  &  -10 (-9) \{-10\}  &  -4 (-1) \{-1\}  &  10.6, 3.1 (9.1, 2.6) \{9.9, 2.6\} \\ \hline
19.8  &  1.21 (0.67) \{0.7\}  &  -8 (-13) \{-14\}  &  3 (-2) \{-2\}  &  8.9, 3.5 (13.6, 2.9) \{13.8, 2.7\} \\ \hline
19.6  &  0.33 (0.37) \{0.34\}  &  -11 (-18) \{-19\}  &  -1 (-2) \{-3\}  &  11.5, 7.3 (18.3, 2.6) \{18.8, 2.5\} \\ \hline
19.5  &  0.34 (0.25) \{0.23\}  &  -25 (-21) \{-21\}  &  -1 (-3) \{-3\}  &  24.9, 10.1 (21.5, 2.2) \{21.5, 2.0\} \\ \hline
19.4  &  0.43 (0.18) \{0.16\}  &  -25 (-24) \{-24\}  &  0 (-3) \{-3\}  &  24.7, 10.9 (24.0, 2.1) \{24.3, 1.4\} \\ \hline
19.2  &  0.18 (0.08) \{0.07\}  &  -17 (-30) \{-31\}  &  2 (-2) \{-2\}  &  17.1, 7.6 (30.4, 2.9) \{30.6, 0.5\} \\ \hline
19.0  &  0.07 (0.02) \{0.02\}  &  -25 (-37) \{-38\}  &  -1 (0) \{0\}  &  24.8, 16.1 (36.9, 4.0) \{38.0, 0.4\} \\ \hline
18.8  &  0.01 (0.0) \{0.0\}  &  -1 (-42) \{-46\}  &  -2 (3) \{2\}  &  2.1, 31.1 (42.0, 4.2) \{46.3, 0.1\} \\ \hline
18.6  &  0.01 (0.0) \{0.0\}  &  25 (-53) \{-57\}  &  -8 (5) \{3\}  &  26.1, 49.5 (53.0, 2.8) \{57.0, 1.0\} \\ \hline
18.5  &  0.01 (0) \{0\}  &  -11 (-) \{-\}  &  -8 (-) \{-\}  &  13.1, 61.7 (-, -) \{-, -\} \\ \hline
18.4  &  0.01 (0.0) \{0.01\}  &  -8 (-52) \{0\}  &  -7 (8) \{0\}  &  10.8, 52.3 (52.4, 3.1) \{0.0, 0.5\} \\ \hline
18.3  &  0.01 (0.01) \{0\}  &  3 (-69) \{-\}  &  -24 (-6) \{-\}  &  24.4, 69.1 (69.0, 11.3) \{-, -\} \\ \hline
18.2  &  0.03 (0.02) \{0\}  &  -41 (-61) \{-\}  &  -24 (-8) \{-\}  &  46.3, 48.3 (61.8, 18.6) \{-, -\} \\ \hline
18.1  &  0.06 (0.04) \{0.01\}  &  -37 (-54) \{-72\}  &  -24 (-13) \{8\}  &  43.6, 60.7 (54.8, 23.0) \{72.3, 0.5\} \\ \hline
18.0  &  0.12 (0.09) \{0.13\}  &  -19 (-53) \{-45\}  &  -24 (-16) \{-1\}  &  30.5, 65.0 (54.4, 32.6) \{44.6, 2.1\} \\ \hline
 \hspace*{-1cm} \end{tabular}
\label{tbl:dist_hpx4_hpx5}
\end{center}
\end{table}

\begin{table}
\tiny
\begin{center}
  \hspace*{-1cm}\begin{tabular}[htpb]{ |c|c|c|c|c| }
    \hline 
log$R$ & mag 	& $<{\ell}>$ ($^{\circ}$) & $<{b}>$ ($^{\circ}$)  & $<{\Delta}>, \sigma_{\Delta}$  ($^{\circ}$) \\ \hline \hline
\multicolumn{5}{|c|}{HPX 6:  $\ell = 90^{\circ}$,  $b = 0^{\circ}$ }\\ \hline
20.0  &  0.29 (2.21) \{2.32\}  &  84 (89) \{89\}  &  -2 (-3) \{-3\}  &  6.8, 1.5 (3.3, 2.4) \{3.0, 2.6\} \\ \hline
19.8  &  1.38 (2.41) \{2.72\}  &  86 (88) \{88\}  &  -6 (-5) \{-4\}  &  7.6, 3.9 (5.3, 2.9) \{4.5, 3.8\} \\ \hline
19.6  &  0.86 (1.75) \{1.65\}  &  88 (87) \{87\}  &  -10 (-7) \{-8\}  &  10.6, 3.8 (7.7, 4.7) \{8.5, 4.2\} \\ \hline
19.5  &  1.42 (1.81) \{1.49\}  &  86 (86) \{86\}  &  -13 (-10) \{-10\}  &  14.0, 3.8 (10.5, 4.7) \{11.0, 4.3\} \\ \hline
19.4  &  1.54 (1.7) \{1.41\}  &  86 (84) \{84\}  &  -16 (-12) \{-13\}  &  16.6, 4.1 (13.7, 5.5) \{14.1, 5.0\} \\ \hline
19.2  &  0.98 (1.49) \{1.32\}  &  68 (77) \{76\}  &  -15 (-18) \{-19\}  &  26.4, 11.4 (22.1, 8.0) \{23.7, 6.1\} \\ \hline
19.0  &  0.95 (1.22) \{0.65\}  &  55 (65) \{53\}  &  4 (-3) \{-23\}  &  34.8, 28.2 (25.4, 20.6) \{43.1, 9.5\} \\ \hline
18.8  &  1.28 (2.55) \{1.21\}  &  60 (53) \{43\}  &  2 (12) \{8\}  &  29.8, 20.8 (39.0, 21.1) \{47.5, 31.5\} \\ \hline
18.6  &  0.53 (0.95) \{1.36\}  &  77 (53) \{50\}  &  -3 (4) \{9\}  &  13.2, 23.9 (36.9, 20.8) \{40.4, 15.6\} \\ \hline
18.5  &  0.37 (0.81) \{0\}  &  84 (56) \{-\}  &  -4 (4) \{-\}  &  7.4, 35.9 (34.2, 19.9) \{-, -\} \\ \hline
18.4  &  0.42 (0.74) \{0.01\}  &  76 (61) \{90\}  &  -22 (0) \{0\}  &  25.7, 51.1 (28.9, 21.6) \{0.0, 0.5\} \\ \hline
18.3  &  0.5 (0.82) \{0\}  &  68 (67) \{-\}  &  -25 (-10) \{-\}  &  32.9, 51.8 (25.3, 26.6) \{-, -\} \\ \hline
18.2  &  0.54 (0.73) \{0\}  &  62 (57) \{-\}  &  -24 (-14) \{-\}  &  36.0, 56.3 (35.9, 34.5) \{-, -\} \\ \hline
18.1  &  0.61 (0.47) \{0\}  &  40 (56) \{-\}  &  -11 (-1) \{-\}  &  50.7, 72.2 (34.0, 36.3) \{-, -\} \\ \hline
18.0  &  0.62 (0.37) \{0\}  &  38 (49) \{-\}  &  -5 (-2) \{-\}  &  52.4, 79.0 (40.8, 40.3) \{-, -\} \\ \hline

\multicolumn{5}{|c|}{HPX 7:  Galactic anti-center, $\ell = 180^{\circ}$,  $b = 0^{\circ}$ }\\ \hline
20.0  &  2.1 (1.02) \{0.9\}  &  179 (180) \{180\}  &  0 (-1) \{0\}  &  0.9, 3.8 (0.7, 1.0) \{0.5, 0.7\} \\ \hline
19.8  &  1.89 (0.82) \{0.75\}  &  -180 (180) \{180\}  &  -2 (-1) \{-1\}  &  2.5, 4.4 (1.3, 1.3) \{1.0, 0.6\} \\ \hline
19.6  &  0.79 (2.4) \{2.44\}  &  -180 (180) \{179\}  &  -4 (-9) \{-9\}  &  4.5, 4.0 (8.8, 2.8) \{8.7, 2.3\} \\ \hline
19.5  &  1.41 (2.76) \{2.81\}  &  179 (179) \{179\}  &  -10 (-11) \{-11\}  &  10.1, 3.3 (10.6, 3.1) \{10.9, 2.8\} \\ \hline
19.4  &  0.77 (2.41) \{1.96\}  &  179 (179) \{179\}  &  -13 (-13) \{-13\}  &  13.1, 2.9 (13.1, 4.4) \{12.7, 4.8\} \\ \hline
19.2  &  1.36 (1.48) \{1.53\}  &  176 (178) \{179\}  &  -21 (-19) \{-19\}  &  21.4, 10.2 (18.9, 7.1) \{19.3, 6.7\} \\ \hline
19.0  &  1.17 (0.79) \{1.12\}  &  168 (177) \{178\}  &  -25 (-27) \{-31\}  &  27.7, 15.0 (27.0, 11.3) \{31.1, 7.8\} \\ \hline
18.8  &  0.66 (1.37) \{0.86\}  &  157 (174) \{176\}  &  -24 (-51) \{-50\}  &  32.7, 20.7 (51.3, 6.6) \{50.1, 7.0\} \\ \hline
18.6  &  0.54 (0.91) \{0.8\}  &  144 (172) \{166\}  &  -23 (-73) \{-78\}  &  42.1, 43.1 (72.7, 13.6) \{78.5, 7.9\} \\ \hline
18.5  &  2.35 (1.19) \{1.17\}  &  23 (-8) \{-12\}  &  -23 (-52) \{-55\}  &  147.6, 59.2 (127.9, 18.1) \{124.2, 13.9\} \\ \hline
18.4  &  2.05 (0.54) \{0.56\}  &  32 (-9) \{-174\}  &  2 (-14) \{-22\}  &  148.1, 46.1 (163.6, 39.6) \{23.2, 35.2\} \\ \hline
18.3  &  1.72 (0.96) \{0.9\}  &  43 (4) \{-4\}  &  0 (28) \{39\}  &  136.8, 52.2 (152.1, 18.4) \{141.3, 3.6\} \\ \hline
18.2  &  1.62 (1.15) \{0.89\}  &  10 (11) \{10\}  &  2 (43) \{50\}  &  169.8, 66.5 (135.7, 19.8) \{129.1, 5.4\} \\ \hline
18.1  &  1.72 (1.66) \{1.34\}  &  -12 (41) \{37\}  &  -2 (42) \{55\}  &  167.8, 82.7 (124.5, 35.9) \{117.1, 14.1\} \\ \hline
18.0  &  2.23 (1.59) \{2.09\}  &  -94 (63) \{69\}  &  -1 (14) \{28\}  &  85.9, 87.6 (116.1, 63.4) \{108.8, 42.1\} \\ \hline
 \hspace*{-1cm} \end{tabular}
\label{tbl:dist_hpx6_hpx7}
\end{center}
\end{table}
\pagebreak
\begin{table}
\tiny
\begin{center}
\hspace*{-1cm}  \begin{tabular}[htpb]{ |c|c|c|c|c| }
    \hline 
log$R$ & mag 	& $<{\ell}>$ ($^{\circ}$) & $<{b}>$ ($^{\circ}$)  & $<{\Delta}>, \sigma_{\Delta}$  ($^{\circ}$) \\ \hline \hline
\multicolumn{5}{|c|}{HPX 8:  $\ell = -90^{\circ}$,  $b = 0^{\circ}$ }\\ \hline
20.0  &  2.19 (1.97) \{2.39\}  &  -90 (-91) \{-91\}  &  -2 (-2) \{-2\}  &  2.0, 1.6 (2.3, 1.2) \{2.1, 1.2\} \\ \hline
19.8  &  2.49 (2.4) \{3.02\}  &  -89 (-91) \{-91\}  &  -2 (-3) \{-3\}  &  2.3, 2.1 (3.5, 1.7) \{3.2, 1.6\} \\ \hline
19.6  &  1.71 (1.55) \{1.97\}  &  -88 (-90) \{-91\}  &  -3 (-5) \{-5\}  &  3.8, 2.6 (5.0, 2.9) \{4.6, 2.9\} \\ \hline
19.5  &  1.62 (1.64) \{1.68\}  &  -89 (-90) \{-91\}  &  -2 (-7) \{-6\}  &  2.7, 4.3 (7.2, 3.3) \{5.7, 3.6\} \\ \hline
19.4  &  0.76 (1.21) \{1.33\}  &  -89 (-90) \{-91\}  &  -2 (-8) \{-7\}  &  2.5, 4.1 (8.3, 3.9) \{7.3, 4.3\} \\ \hline
19.2  &  0.53 (0.72) \{0.47\}  &  -89 (-89) \{-88\}  &  -3 (-11) \{-16\}  &  2.8, 5.3 (11.5, 5.1) \{15.7, 0.5\} \\ \hline
19.0  &  0.18 (0.58) \{0.39\}  &  -84 (-85) \{-85\}  &  -2 (-18) \{-20\}  &  6.7, 18.1 (18.2, 3.7) \{20.6, 0.4\} \\ \hline
18.8  &  0.5 (0.35) \{0.28\}  &  -71 (-80) \{-80\}  &  -2 (-22) \{-23\}  &  19.1, 16.6 (24.1, 3.3) \{25.2, 0.4\} \\ \hline
18.6  &  0.51 (0.19) \{0.19\}  &  -71 (-73) \{-72\}  &  -8 (-20) \{-22\}  &  20.4, 21.5 (26.6, 5.7) \{28.4, 0.3\} \\ \hline
18.5  &  0.35 (0.21) \{0.16\}  &  -73 (-68) \{-67\}  &  0 (-16) \{-18\}  &  16.9, 27.6 (27.1, 7.1) \{29.1, 0.4\} \\ \hline
18.4  &  0.34 (0.3) \{0.2\}  &  -67 (-62) \{-64\}  &  -12 (-5) \{-8\}  &  25.9, 40.1 (28.8, 9.4) \{27.4, 1.7\} \\ \hline
18.3  &  0.33 (0.11) \{0.03\}  &  -57 (-64) \{-51\}  &  -13 (-7) \{9\}  &  35.4, 52.9 (26.5, 22.6) \{39.8, 0.2\} \\ \hline
18.2  &  0.44 (0.22) \{0.09\}  &  -57 (-55) \{-54\}  &  -14 (-4) \{15\}  &  35.9, 64.8 (34.8, 22.8) \{38.8, 0.8\} \\ \hline
18.1  &  0.47 (0.23) \{0.16\}  &  -45 (-51) \{-52\}  &  -22 (7) \{21\}  &  48.8, 67.4 (39.4, 29.4) \{42.9, 2.8\} \\ \hline
18.0  &  0.48 (0.23) \{0.15\}  &  -21 (-57) \{-47\}  &  -16 (4) \{20\}  &  69.9, 75.1 (33.7, 40.7) \{46.6, 15.3\} \\ \hline

\multicolumn{5}{|c|}{HPX 9:  $\ell = 45^{\circ}$,  $b = -41.8103^{\circ}$ }\\ \hline
20.0  &  0.72 (0.92) \{0.97\}  &  39 (38) \{38\}  &  -43 (-42) \{-42\}  &  4.9, 0.6 (5.0, 0.7) \{5.0, 0.7\} \\ \hline
19.8  &  0.46 (0.96) \{0.96\}  &  35 (34) \{34\}  &  -43 (-43) \{-43\}  &  7.1, 0.5 (8.0, 0.7) \{8.0, 0.7\} \\ \hline
19.6  &  1.5 (1.23) \{0.95\}  &  30 (28) \{27\}  &  -41 (-42) \{-42\}  &  11.3, 1.1 (12.9, 0.8) \{13.0, 0.7\} \\ \hline
19.5  &  1.63 (1.16) \{0.95\}  &  22 (22) \{23\}  &  -39 (-41) \{-41\}  &  17.5, 1.2 (16.9, 0.8) \{16.7, 0.7\} \\ \hline
19.4  &  0.62 (0.58) \{0.95\}  &  18 (16) \{16\}  &  -37 (-40) \{-40\}  &  21.5, 1.3 (21.8, 0.7) \{21.7, 0.7\} \\ \hline
19.2  &  0.82 (0.82) \{0.94\}  &  7 (-2) \{-2\}  &  -26 (-30) \{-30\}  &  34.9, 8.6 (39.1, 1.1) \{39.3, 0.7\} \\ \hline
19.0  &  0.12 (0.15) \{0.14\}  &  -19 (-19) \{-18\}  &  -15 (-10) \{-10\}  &  61.0, 6.0 (63.7, 2.6) \{63.2, 0.3\} \\ \hline
18.8  &  0.02 (0.01) \{0.01\}  &  -1 (-33) \{-33\}  &  -1 (1) \{1\}  &  57.9, 40.7 (81.8, 4.2) \{81.8, 0.1\} \\ \hline
18.6  &  0.01 (0) \{0\}  &  43 (-) \{-\}  &  -6 (-) \{-\}  &  36.2, 40.5 (-, -) \{-, -\} \\ \hline
18.5  &  0.01 (0) \{0\}  &  46 (-) \{-\}  &  -7 (-) \{-\}  &  34.8, 44.3 (-, -) \{-, -\} \\ \hline
18.4  &  0.01 (0) \{0\}  &  42 (-) \{-\}  &  -24 (-) \{-\}  &  17.6, 56.4 (-, -) \{-, -\} \\ \hline
18.3  &  0.03 (0) \{0\}  &  88 (-) \{-\}  &  -12 (-) \{-\}  &  47.5, 51.6 (-, -) \{-, -\} \\ \hline
18.2  &  0.03 (0.0) \{0\}  &  67 (4) \{-\}  &  -22 (-13) \{-\}  &  27.0, 59.5 (45.8, 56.9) \{-, -\} \\ \hline
18.1  &  0.04 (0.01) \{0\}  &  73 (-22) \{-\}  &  -12 (-15) \{-\}  &  38.4, 72.2 (63.4, 40.9) \{-, -\} \\ \hline
18.0  &  0.05 (0.02) \{0\}  &  32 (-33) \{-\}  &  -11 (-11) \{-\}  &  32.9, 72.1 (74.0, 39.7) \{-, -\} \\ \hline
 \hspace*{-1cm} \end{tabular}
\label{tbl:dist_hpx8_hpx9}
\end{center}
\end{table}
\pagebreak
\begin{table}
\tiny
\begin{center}
  \hspace*{-2cm}\begin{tabular}[htpb]{ |c|c|c|c|c| }
    \hline 
log$R$ & mag 	& $<{\ell}>$ ($^{\circ}$) & $<{b}>$ ($^{\circ}$)  & $<{\Delta}>, \sigma_{\Delta}$  ($^{\circ}$) \\ \hline \hline
\multicolumn{5}{|c|}{HPX 10:  $\ell = 135^{\circ}$,  $b = -41.8103^{\circ}$ }\\ \hline
20.0  &  0.91 (0.98) \{0.99\}  &  133 (133) \{133\}  &  -45 (-45) \{-45\}  &  3.8, 0.7 (4.0, 0.7) \{3.9, 0.7\} \\ \hline
19.8  &  0.98 (0.92) \{0.98\}  &  133 (131) \{131\}  &  -48 (-47) \{-47\}  &  6.1, 0.7 (6.1, 0.7) \{6.2, 0.7\} \\ \hline
19.6  &  0.44 (1.08) \{0.97\}  &  130 (129) \{129\}  &  -51 (-51) \{-51\}  &  10.1, 0.5 (9.7, 0.7) \{9.8, 0.7\} \\ \hline
19.5  &  1.07 (0.6) \{0.96\}  &  129 (127) \{127\}  &  -54 (-53) \{-53\}  &  12.5, 0.7 (12.2, 0.6) \{12.3, 0.7\} \\ \hline
19.4  &  0.22 (1.08) \{0.94\}  &  127 (124) \{124\}  &  -56 (-55) \{-55\}  &  15.4, 0.4 (15.3, 0.8) \{15.5, 0.7\} \\ \hline
19.2  &  0.12 (0.72) \{0.93\}  &  126 (112) \{112\}  &  -64 (-62) \{-63\}  &  22.3, 0.2 (24.8, 0.9) \{24.7, 0.7\} \\ \hline
19.0  &  0.5 (0.21) \{0.27\}  &  118 (86) \{83\}  &  -72 (-71) \{-71\}  &  31.2, 0.7 (38.1, 1.4) \{38.6, 0.4\} \\ \hline
18.8  &  0.93 (0.73) \{0.7\}  &  25 (24) \{26\}  &  -66 (-66) \{-66\}  &  59.5, 12.5 (60.0, 1.8) \{59.2, 0.6\} \\ \hline
18.6  &  0.35 (0.06) \{0\}  &  18 (2) \{-\}  &  -16 (17) \{-\}  &  97.9, 25.1 (133.1, 5.5) \{-, -\} \\ \hline
18.5  &  0.22 (0.4) \{0.65\}  &  61 (25) \{16\}  &  -1 (29) \{27\}  &  77.1, 50.1 (122.7, 10.7) \{128.8, 0.9\} \\ \hline
18.4  &  0.22 (0.85) \{0.59\}  &  66 (58) \{43\}  &  -7 (26) \{38\}  &  69.8, 55.7 (98.0, 22.5) \{115.6, 2.3\} \\ \hline
18.3  &  0.21 (0.62) \{0.4\}  &  83 (66) \{70\}  &  -18 (21) \{37\}  &  50.1, 59.9 (89.5, 28.1) \{98.5, 4.2\} \\ \hline
18.2  &  0.23 (0.35) \{0.19\}  &  64 (60) \{86\}  &  -20 (14) \{22\}  &  62.4, 64.9 (89.0, 38.4) \{77.9, 12.2\} \\ \hline
18.1  &  0.31 (0.26) \{0.3\}  &  30 (55) \{44\}  &  -14 (3) \{10\}  &  91.9, 77.1 (84.5, 48.9) \{97.2, 13.2\} \\ \hline
18.0  &  0.36 (0.19) \{0.05\}  &  -27 (47) \{65\}  &  -9 (-6) \{-5\}  &  126.8, 85.7 (84.7, 57.3) \{71.6, 5.9\} \\ \hline

\multicolumn{5}{|c|}{HPX 11:  $\ell = -135^{\circ}$,  $b = -41.8103^{\circ}$ }\\ \hline
20.0  &  0.83 (0.81) \{0.96\}  &  -134 (-133) \{-133\}  &  -45 (-44) \{-44\}  &  3.1, 0.6 (2.8, 0.6) \{2.9, 0.7\} \\ \hline
19.8  &  0.46 (0.59) \{0.93\}  &  -133 (-132) \{-132\}  &  -47 (-45) \{-46\}  &  5.0, 0.5 (4.1, 0.5) \{4.5, 0.7\} \\ \hline
19.6  &  0.73 (1.0) \{0.89\}  &  -130 (-131) \{-130\}  &  -49 (-48) \{-48\}  &  7.8, 0.7 (6.6, 0.9) \{7.0, 0.7\} \\ \hline
19.5  &  0.38 (0.95) \{0.86\}  &  -127 (-129) \{-129\}  &  -50 (-49) \{-49\}  &  9.7, 0.5 (8.6, 0.7) \{8.7, 0.7\} \\ \hline
19.4  &  1.33 (0.78) \{0.82\}  &  -123 (-126) \{-127\}  &  -46 (-51) \{-51\}  &  9.5, 5.0 (10.6, 0.9) \{10.8, 0.6\} \\ \hline
19.2  &  0.34 (0.74) \{0.7\}  &  -111 (-120) \{-120\}  &  -55 (-55) \{-55\}  &  20.4, 0.4 (16.5, 0.8) \{16.3, 0.6\} \\ \hline
19.0  &  0.79 (0.41) \{0.57\}  &  -90 (-106) \{-107\}  &  -63 (-59) \{-59\}  &  33.7, 2.0 (24.9, 1.2) \{24.4, 0.5\} \\ \hline
18.8  &  1.58 (0.45) \{0.46\}  &  -64 (-81) \{-83\}  &  -59 (-58) \{-58\}  &  45.9, 7.7 (36.9, 1.8) \{35.9, 0.5\} \\ \hline
18.6  &  1.33 (0.57) \{0.54\}  &  -35 (-54) \{-52\}  &  -40 (-41) \{-41\}  &  70.7, 13.6 (58.4, 4.5) \{59.3, 0.5\} \\ \hline
18.5  &  0.58 (0.85) \{0.69\}  &  -24 (-43) \{-44\}  &  -24 (-15) \{-19\}  &  88.8, 29.6 (81.6, 8.3) \{78.1, 0.7\} \\ \hline
18.4  &  0.32 (0.0) \{0\}  &  -62 (-97) \{-\}  &  -26 (-24) \{-\}  &  60.9, 68.0 (36.6, 24.7) \{-, -\} \\ \hline
18.3  &  0.25 (0.02) \{0\}  &  -126 (-60) \{-\}  &  -28 (-10) \{-\}  &  15.1, 76.2 (72.8, 47.1) \{-, -\} \\ \hline
18.2  &  0.25 (0.06) \{0\}  &  -8 (-24) \{-\}  &  -21 (14) \{-\}  &  100.5, 75.2 (114.8, 29.0) \{-, -\} \\ \hline
18.1  &  0.28 (0.15) \{0.01\}  &  -8 (-6) \{-18\}  &  -16 (19) \{10\}  &  104.5, 80.7 (131.5, 43.5) \{116.5, 6.6\} \\ \hline
18.0  &  0.32 (0.2) \{0.34\}  &  -50 (9) \{-11\}  &  -13 (8) \{38\}  &  77.7, 82.8 (134.0, 64.2) \{137.5, 26.1\} \\ \hline
  \hspace*{-2cm}\end{tabular}
\label{tbl:dist_hpx10_hpx11}
\end{center}
\end{table}
\pagebreak
\begin{table}
\tiny
\begin{center}
 \hspace*{-1cm} \begin{tabular}[htpb]{ |c|c|c|c|c| }
    \hline 
log$R$ & mag 	& $<{\ell}>$ ($^{\circ}$) & $<{b}>$ ($^{\circ}$)  & $<{\Delta}>, \sigma_{\Delta}$  ($^{\circ}$) \\ \hline \hline
\multicolumn{5}{|c|}{HPX 12:  $\ell = -45^{\circ}$,  $b = -41.8103^{\circ}$ }\\ \hline
20.0  &  0.63 (0.82) \{0.91\}  &  -44 (-45) \{-45\}  &  -39 (-39) \{-39\}  &  3.1, 0.6 (2.3, 0.6) \{2.3, 0.7\} \\ \hline
19.8  &  0.29 (0.62) \{0.86\}  &  -43 (-45) \{-45\}  &  -37 (-38) \{-38\}  &  4.8, 0.4 (3.8, 0.6) \{3.7, 0.7\} \\ \hline
19.6  &  0.09 (0.69) \{0.79\}  &  -43 (-45) \{-45\}  &  -35 (-36) \{-36\}  &  7.1, 0.2 (6.0, 0.7) \{5.8, 0.6\} \\ \hline
19.5  &  0.19 (1.02) \{0.75\}  &  -44 (-45) \{-45\}  &  -33 (-34) \{-35\}  &  8.5, 0.7 (7.5, 0.8) \{7.3, 0.6\} \\ \hline
19.4  &  0.42 (0.53) \{0.69\}  &  -42 (-45) \{-45\}  &  -30 (-32) \{-33\}  &  12.4, 1.2 (9.6, 1.1) \{9.2, 0.6\} \\ \hline
19.2  &  0.0 (0.56) \{0.57\}  &  -43 (-45) \{-44\}  &  -22 (-28) \{-27\}  &  20.2, 0.1 (14.0, 1.1) \{14.4, 0.5\} \\ \hline
19.0  &  0.14 (0.4) \{0.44\}  &  -54 (-45) \{-44\}  &  -19 (-19) \{-19\}  &  24.4, 1.9 (22.9, 2.1) \{22.9, 0.5\} \\ \hline
18.8  &  0.04 (0.19) \{0.23\}  &  -37 (-45) \{-45\}  &  -4 (-5) \{-5\}  &  38.0, 19.1 (36.5, 4.0) \{37.0, 0.3\} \\ \hline
18.6  &  0.01 (0.01) \{0.01\}  &  -56 (-47) \{-50\}  &  -7 (4) \{3\}  &  36.5, 27.4 (46.0, 4.7) \{45.1, 0.1\} \\ \hline
18.5  &  0.02 (0.0) \{0.0\}  &  -79 (-51) \{-55\}  &  -9 (6) \{4\}  &  44.5, 44.6 (48.5, 9.5) \{47.1, 0.1\} \\ \hline
18.4  &  0.03 (0) \{0\}  &  -62 (-) \{-\}  &  -21 (-) \{-\}  &  25.1, 37.2 (-, -) \{-, -\} \\ \hline
18.3  &  0.03 (0.0) \{0\}  &  -65 (-54) \{-\}  &  -20 (10) \{-\}  &  27.3, 50.6 (52.8, 5.4) \{-, -\} \\ \hline
18.2  &  0.04 (0.01) \{0\}  &  -51 (-58) \{-\}  &  -18 (-10) \{-\}  &  23.9, 66.6 (33.7, 24.8) \{-, -\} \\ \hline
18.1  &  0.06 (0.02) \{0.01\}  &  -39 (-52) \{-53\}  &  -26 (-17) \{14\}  &  16.5, 61.4 (25.9, 33.5) \{56.2, 3.2\} \\ \hline
18.0  &  0.07 (0.03) \{0.07\}  &  -5 (-48) \{-56\}  &  -22 (-23) \{15\}  &  39.3, 69.1 (19.3, 39.7) \{57.8, 13.6\} \\ \hline

\multicolumn{5}{|c|}{South Pole } \\ \hline 
20.0  &  0.78 (0.93) \{0.99\}  &  -26 (-23) \{-24\}  &  -87 (-87) \{-87\}  &  3.2, 0.6 (3.4, 0.7) \{3.3, 0.7\} \\ \hline
19.8  &  0.7 (1.17) \{0.97\}  &  -26 (-24) \{-24\}  &  -85 (-85) \{-85\}  &  5.1, 0.6 (5.3, 0.8) \{5.2, 0.7\} \\ \hline
19.6  &  0.51 (1.4) \{0.95\}  &  -26 (-25) \{-24\}  &  -82 (-82) \{-82\}  &  8.0, 0.5 (8.5, 0.8) \{8.3, 0.7\} \\ \hline
19.5  &  0.92 (0.67) \{0.94\}  &  -26 (-25) \{-24\}  &  -80 (-80) \{-80\}  &  10.3, 0.7 (10.5, 0.6) \{10.5, 0.7\} \\ \hline
19.4  &  0.57 (1.29) \{0.92\}  &  -26 (-25) \{-24\}  &  -77 (-77) \{-77\}  &  12.9, 0.5 (13.1, 0.8) \{13.3, 0.7\} \\ \hline
19.2  &  1.4 (1.28) \{0.88\}  &  -35 (-27) \{-25\}  &  -69 (-69) \{-69\}  &  21.2, 1.2 (20.8, 1.0) \{21.3, 0.7\} \\ \hline
19.0  &  0.03 (1.35) \{0.83\}  &  -34 (-24) \{-26\}  &  -55 (-55) \{-55\}  &  34.8, 2.7 (35.1, 1.4) \{34.7, 0.6\} \\ \hline
18.8  &  1.01 (1.17) \{0.92\}  &  -25 (-29) \{-29\}  &  -28 (-26) \{-27\}  &  62.5, 8.9 (64.1, 3.8) \{62.8, 0.9\} \\ \hline
18.6  &  0.0 (0) \{0\}  &  -78 (-) \{-\}  &  -19 (-) \{-\}  &  71.1, 41.1 (-, -) \{-, -\} \\ \hline
18.5  &  0.01 (0) \{0\}  &  -77 (-) \{-\}  &  -12 (-) \{-\}  &  78.1, 48.4 (-, -) \{-, -\} \\ \hline
18.4  &  0.01 (0) \{0\}  &  -105 (-) \{-\}  &  -21 (-) \{-\}  &  69.5, 68.2 (-, -) \{-, -\} \\ \hline
18.3  &  0.04 (0) \{0\}  &  45 (-) \{-\}  &  -13 (-) \{-\}  &  77.0, 67.9 (-, -) \{-, -\} \\ \hline
18.2  &  0.09 (0) \{0\}  &  -21 (-) \{-\}  &  -13 (-) \{-\}  &  76.7, 77.6 (-, -) \{-, -\} \\ \hline
18.1  &  0.15 (0.0) \{0\}  &  4 (-56) \{-\}  &  -13 (4) \{-\}  &  76.6, 83.7 (94.2, 19.0) \{-, -\} \\ \hline
18.0  &  0.21 (0.01) \{0\}  &  -29 (-43) \{-\}  &  -12 (-22) \{-\}  &  77.6, 85.3 (67.7, 43.6) \{-, -\} \\ \hline
 \hspace*{-1cm} \end{tabular}
\label{tbl:dist_hpx12_spol}
\end{center}
\end{table}
\vspace{-1.5cm}
\clearpage
\pagebreak

\section{Tables of Arrival Direction Distributions from Selected Sources}
\label{appdx:RadioTables}
\begin{table}
\tiny
\caption{Properties of arrival direction distributions from cosmic rays originating within 1$^{\circ}$ of the given source directions.  Source name and direction are given in horizontal bands. 
Column 1 lists log10 of the rigidity in V.
Column 2 is the magnification factor at that rigidity, i.e., the number of observed events from the source region in the angular window divided by the number which would be seen in the absence of magnetic lensing; a value of 0 means no events arrived from that source region.
In Col. 2, 3, 4 and 5, the first entry is for the 100 pc coherence length (KRF6) realization, followed by the result for the 30 pc realization (KRF10) in parentheses and the JF12 coherent-only case in curly brackets \{\}.
Columns 3 and 4 list the centroid of the distribution of arrival directions in Galactic coordinates.
Column 5 gives the angular separation between the source direction and the centroid, and the median separation of the arrival directions around the centroid; i.e., $\sigma_{\Delta}$ measures the angular size of the arrival direction distribution.}
\begin{center}
  \hspace*{-1cm}\begin{tabular}[htpb]{ |c|c|c|c|c| }
    \hline 
log$R$ & mag 	& $<{\ell}>$ ($^{\circ}$) & $<{b}>$ ($^{\circ}$)  & $<{\Delta}>, \sigma_{\Delta}$  ($^{\circ}$) \\ \hline \hline
\multicolumn{5}{|c|}{UGC 1841:  $\ell = 140.25^{\circ}$,  $b = -16.77^{\circ}$ }\\ \hline
20.0  &  0.96 (0.92) \{0.95\}  &  138 (139) \{139\}  &  -20 (-21) \{-21\}  &  3.9, 0.7 (4.4, 0.7) \{4.3, 0.7\} \\ \hline
19.8  &  0.88 (0.89) \{0.93\}  &  137 (139) \{139\}  &  -22 (-23) \{-23\}  &  6.0, 0.7 (6.7, 0.7) \{6.7, 0.7\} \\ \hline
19.6  &  0.46 (1.05) \{0.9\}  &  133 (138) \{137\}  &  -23 (-27) \{-27\}  &  9.5, 0.8 (10.5, 0.9) \{10.4, 0.7\} \\ \hline
19.5  &  0.97 (0.86) \{0.89\}  &  131 (137) \{136\}  &  -24 (-29) \{-29\}  &  11.5, 0.9 (12.9, 0.8) \{12.9, 0.7\} \\ \hline
19.4  &  1.92 (1.06) \{0.88\}  &  130 (135) \{135\}  &  -27 (-32) \{-32\}  &  14.0, 1.5 (16.2, 0.9) \{16.0, 0.7\} \\ \hline
19.2  &  1.94 (1.89) \{0.86\}  &  135 (131) \{130\}  &  -43 (-40) \{-40\}  &  26.3, 2.3 (24.8, 1.2) \{24.8, 0.7\} \\ \hline
19.0  &  0.35 (0.73) \{0.86\}  &  123 (120) \{119\}  &  -23 (-49) \{-51\}  &  17.2, 30.4 (35.8, 2.7) \{38.5, 0.7\} \\ \hline
18.8  &  8.58 (2.28) \{1.29\}  &  61 (87) \{85\}  &  -52 (-55) \{-57\}  &  70.1, 6.2 (55.5, 10.5) \{57.3, 7.9\} \\ \hline
18.6  &  0.79 (1.26) \{1.64\}  &  51 (16) \{17\}  &  -14 (-25) \{-24\}  &  85.7, 57.3 (111.4, 38.0) \{110.9, 40.0\} \\ \hline
18.5  &  0.63 (0.66) \{0.79\}  &  70 (63) \{85\}  &  -6 (-9) \{-18\}  &  68.9, 57.9 (75.8, 57.5) \{52.6, 36.0\} \\ \hline
18.4  &  0.53 (0.65) \{0.42\}  &  71 (63) \{19\}  &  -8 (-5) \{18\}  &  68.1, 57.6 (76.4, 47.7) \{124.6, 17.2\} \\ \hline
18.3  &  0.47 (0.56) \{0.73\}  &  90 (66) \{34\}  &  -20 (12) \{33\}  &  47.6, 61.9 (79.2, 35.9) \{112.7, 22.8\} \\ \hline
18.2  &  0.53 (0.37) \{0.38\}  &  71 (58) \{83\}  &  -20 (13) \{30\}  &  65.1, 76.1 (86.5, 39.7) \{72.6, 9.9\} \\ \hline
18.1  &  0.68 (0.32) \{0.97\}  &  21 (56) \{91\}  &  -10 (-8) \{22\}  &  113.8, 82.5 (82.6, 52.3) \{62.1, 7.1\} \\ \hline
18.0  &  0.8 (0.27) \{0.41\}  &  19 (57) \{65\}  &  -6 (-16) \{-11\}  &  117.8, 84.5 (79.2, 55.8) \{72.3, 7.6\} \\ \hline

\multicolumn{5}{|c|}{NGC1128:  $\ell = 170.26^{\circ}$,  $b = -44.93^{\circ}$ }\\ \hline
20.0  &  1.57 (0.99) \{0.98\}  &  170 (170) \{170\}  &  -48 (-48) \{-48\}  &  3.4, 0.9 (3.6, 0.7) \{3.6, 0.7\} \\ \hline
19.8  &  1.12 (0.99) \{0.97\}  &  171 (170) \{169\}  &  -51 (-50) \{-51\}  &  5.9, 0.7 (5.6, 0.7) \{5.6, 0.7\} \\ \hline
19.6  &  1.19 (1.09) \{0.95\}  &  169 (169) \{169\}  &  -54 (-54) \{-54\}  &  8.8, 0.8 (8.9, 0.8) \{8.8, 0.7\} \\ \hline
19.5  &  1.59 (0.58) \{0.93\}  &  165 (168) \{169\}  &  -55 (-56) \{-56\}  &  10.7, 1.6 (11.2, 0.6) \{11.1, 0.7\} \\ \hline
19.4  &  3.8 (0.7) \{0.91\}  &  156 (168) \{168\}  &  -57 (-59) \{-59\}  &  15.3, 1.4 (13.9, 0.6) \{13.9, 0.7\} \\ \hline
19.2  &  1.3 (0.52) \{0.83\}  &  148 (168) \{167\}  &  -64 (-66) \{-66\}  &  23.1, 1.2 (21.6, 0.6) \{21.6, 0.6\} \\ \hline
19.0  &  1.81 (0.42) \{0.78\}  &  153 (163) \{166\}  &  -78 (-78) \{-78\}  &  34.1, 3.5 (33.6, 1.1) \{33.5, 0.6\} \\ \hline
18.8  &  0.11 (0.86) \{0.65\}  &  -26 (-27) \{-27\}  &  -66 (-83) \{-83\}  &  68.7, 15.2 (52.0, 1.3) \{51.7, 0.6\} \\ \hline
18.6  &  0.64 (0.6) \{0.75\}  &  -3 (-24) \{-23\}  &  -27 (-48) \{-49\}  &  107.8, 19.5 (86.0, 3.9) \{85.3, 1.4\} \\ \hline
18.5  &  0.37 (0.03) \{0\}  &  20 (-10) \{-\}  &  -17 (14) \{-\}  &  112.2, 46.3 (148.7, 6.4) \{-, -\} \\ \hline
18.4  &  0.26 (0.15) \{0.33\}  &  83 (2) \{-4\}  &  -16 (26) \{25\}  &  76.9, 68.8 (159.3, 9.4) \{159.1, 0.5\} \\ \hline
18.3  &  0.25 (0.43) \{0.23\}  &  71 (27) \{12\}  &  -19 (32) \{39\}  &  82.8, 64.7 (149.1, 19.2) \{162.8, 1.1\} \\ \hline
18.2  &  0.27 (0.43) \{0.21\}  &  53 (44) \{29\}  &  -19 (32) \{46\}  &  94.2, 71.3 (137.3, 31.2) \{153.3, 4.4\} \\ \hline
18.1  &  0.35 (0.37) \{0.29\}  &  -5 (47) \{45\}  &  -15 (18) \{21\}  &  120.1, 80.1 (126.2, 46.5) \{129.8, 41.6\} \\ \hline
18.0  &  0.41 (0.26) \{0.39\}  &  -35 (47) \{58\}  &  -8 (1) \{-6\}  &  122.6, 83.7 (113.7, 63.4) \{100.9, 34.6\} \\ \hline

\hspace*{-1cm}\end{tabular}
\label{tbl:dist_UGC_NGC1128}
\end{center}
\end{table}
\vfill
\vspace{-1.5cm}
\begin{table}
\tiny
\begin{center}
  \hspace*{-1cm}\begin{tabular}[htpb]{ |c|c|c|c|c| }
    \hline 
log$R$ & mag 	& $<{\ell}>$ ($^{\circ}$) & $<{b}>$ ($^{\circ}$)  & $<{\Delta}>, \sigma_{\Delta}$  ($^{\circ}$) \\ \hline \hline
\multicolumn{5}{|c|}{NGC 4782:  $\ell = 304.14^{\circ}$,  $b = 50.29^{\circ}$ }\\ \hline
20.0  &  0.94 (1.06) \{0.96\}  &  -54 (-54) \{-54\}  &  47 (48) \{48\}  &  3.4, 0.7 (2.3, 0.7) \{2.3, 0.7\} \\ \hline
19.8  &  0.87 (0.94) \{0.95\}  &  -53 (-53) \{-53\}  &  46 (47) \{47\}  &  5.1, 0.7 (3.7, 0.7) \{3.7, 0.7\} \\ \hline
19.6  &  0.4 (1.21) \{0.91\}  &  -53 (-51) \{-51\}  &  43 (45) \{45\}  &  7.2, 0.5 (6.0, 0.9) \{5.9, 0.7\} \\ \hline
19.5  &  0.33 (0.84) \{0.89\}  &  -54 (-51) \{-50\}  &  42 (44) \{44\}  &  8.7, 0.4 (7.0, 0.7) \{7.4, 0.7\} \\ \hline
19.4  &  0.25 (1.1) \{0.86\}  &  -54 (-49) \{-49\}  &  40 (42) \{42\}  &  10.7, 0.5 (9.3, 1.1) \{9.3, 0.7\} \\ \hline
19.2  &  0.24 (1.01) \{0.78\}  &  -52 (-47) \{-47\}  &  37 (37) \{37\}  &  13.9, 0.9 (15.1, 1.2) \{14.6, 0.6\} \\ \hline
19.0  &  2.99 (0.45) \{0.63\}  &  -50 (-44) \{-44\}  &  13 (29) \{29\}  &  37.6, 4.8 (22.9, 2.2) \{22.9, 0.6\} \\ \hline
18.8  &  0.93 (0.48) \{0.36\}  &  -49 (-41) \{-42\}  &  9 (15) \{17\}  &  42.0, 32.7 (36.8, 5.3) \{34.8, 0.4\} \\ \hline
18.6  &  0.75 (0.52) \{0.06\}  &  -84 (-81) \{-61\}  &  -4 (-1) \{13\}  &  60.0, 40.8 (55.6, 18.6) \{37.7, 1.2\} \\ \hline
18.5  &  0.73 (0.88) \{1.4\}  &  -84 (-87) \{-77\}  &  -14 (-7) \{-12\}  &  68.7, 36.6 (62.8, 18.5) \{65.1, 3.7\} \\ \hline
18.4  &  0.66 (1.04) \{0.38\}  &  -76 (-92) \{-106\}  &  -14 (-1) \{-12\}  &  66.9, 40.2 (60.0, 19.3) \{75.9, 3.5\} \\ \hline
18.3  &  0.6 (0.88) \{1.01\}  &  -62 (-93) \{-105\}  &  -5 (2) \{7\}  &  55.6, 50.1 (57.6, 18.8) \{59.3, 7.7\} \\ \hline
18.2  &  0.57 (0.72) \{0.88\}  &  -72 (-82) \{-87\}  &  12 (0) \{19\}  &  40.7, 67.6 (55.4, 28.4) \{39.6, 2.6\} \\ \hline
18.1  &  0.5 (0.9) \{2.0\}  &  -62 (-75) \{-80\}  &  9 (-6) \{-18\}  &  41.6, 68.9 (58.5, 29.1) \{71.9, 18.2\} \\ \hline
18.0  &  0.3 (0.59) \{0.43\}  &  -18 (-72) \{-85\}  &  -8 (3) \{6\}  &  66.5, 79.0 (49.7, 39.2) \{50.3, 24.9\} \\ \hline

\hspace*{-1cm}\end{tabular}
\label{tbl:dist_NGC4782}
\end{center}
\end{table}
\vspace{-1.5cm}
\pagebreak

\begin{table}
\tiny
\begin{center}
  \hspace*{-1cm}\begin{tabular}[htpb]{ |c|c|c|c|c| }
    \hline 
log$R$ & mag 	& $<{\ell}>$ ($^{\circ}$) & $<{b}>$ ($^{\circ}$)  & $<{\Delta}>, \sigma_{\Delta}$  ($^{\circ}$) \\ \hline \hline
\multicolumn{5}{|c|}{Centaurus A:  $\ell = 309.52^{\circ}$,  $b = 19.42^{\circ}$ }\\ \hline
20.0  &  1.24 (0.89) \{0.97\}  &  -51 (-52) \{-52\}  &  18 (18) \{18\}  &  1.6, 0.8 (2.1, 0.7) \{2.0, 0.7\} \\ \hline
19.8  &  2.99 (0.41) \{0.74\}  &  -53 (-53) \{-53\}  &  15 (17) \{17\}  &  5.0, 1.9 (3.4, 0.6) \{3.2, 0.6\} \\ \hline
19.6  &  0.89 (0.71) \{0.64\}  &  -55 (-53) \{-54\}  &  11 (16) \{16\}  &  9.5, 3.4 (4.5, 0.7) \{4.7, 0.6\} \\ \hline
19.5  &  0.42 (0.76) \{0.58\}  &  -57 (-54) \{-54\}  &  7 (15) \{15\}  &  14.5, 7.3 (5.5, 1.3) \{5.6, 0.5\} \\ \hline
19.4  &  0.09 (0.71) \{0.52\}  &  -59 (-55) \{-55\}  &  1 (15) \{14\}  &  19.8, 4.8 (6.5, 1.3) \{6.7, 0.5\} \\ \hline
19.2  &  0.28 (0.56) \{0.39\}  &  -56 (-59) \{-57\}  &  1 (11) \{13\}  &  19.6, 15.8 (12.2, 3.6) \{9.2, 0.4\} \\ \hline
19.0  &  0.21 (0.33) \{0.32\}  &  -63 (-64) \{-61\}  &  -1 (8) \{10\}  &  23.6, 7.3 (17.3, 6.0) \{13.7, 1.6\} \\ \hline
18.8  &  0.44 (0.39) \{0.29\}  &  -55 (-67) \{-65\}  &  13 (6) \{8\}  &  7.3, 13.7 (21.6, 7.1) \{18.1, 2.8\} \\ \hline
18.6  &  0.17 (0.41) \{0.26\}  &  -60 (-75) \{-72\}  &  2 (6) \{9\}  &  19.4, 29.2 (27.2, 7.9) \{22.7, 0.7\} \\ \hline
18.5  &  0.19 (0.39) \{0.86\}  &  -57 (-78) \{-82\}  &  -3 (6) \{7\}  &  23.1, 32.0 (29.9, 9.0) \{32.9, 5.0\} \\ \hline
18.4  &  0.19 (0.31) \{0.06\}  &  -53 (-82) \{-64\}  &  -5 (5) \{6\}  &  24.9, 35.2 (33.6, 11.7) \{18.8, 0.5\} \\ \hline
18.3  &  0.18 (0.26) \{0.04\}  &  -52 (-78) \{-67\}  &  -9 (1) \{7\}  &  28.0, 43.4 (33.0, 17.1) \{20.3, 0.9\} \\ \hline
18.2  &  0.18 (0.35) \{0.06\}  &  -55 (-67) \{-66\}  &  -14 (-3) \{8\}  &  34.0, 52.2 (27.9, 23.3) \{18.9, 6.4\} \\ \hline
18.1  &  0.26 (0.47) \{0.71\}  &  -47 (-57) \{-45\}  &  -20 (-5) \{-8\}  &  40.1, 63.2 (25.7, 27.0) \{27.7, 9.2\} \\ \hline
18.0  &  0.32 (0.57) \{0.62\}  &  -31 (-52) \{-41\}  &  -24 (-11) \{6\}  &  47.0, 66.0 (30.3, 34.4) \{16.8, 20.1\} \\ \hline

\multicolumn{5}{|c|}{CGCG114-025:  $\ell = 46.97^{\circ}$,  $b = 10.54^{\circ}$ }\\ \hline
20.0  &  0.26 (1.04) \{0.95\}  &  41 (43) \{43\}  &  10 (14) \{14\}  &  5.9, 3.6 (5.2, 0.8) \{5.3, 0.7\} \\ \hline
19.8  &  0.53 (1.15) \{0.96\}  &  36 (40) \{40\}  &  1 (15) \{15\}  &  14.2, 8.5 (8.6, 0.9) \{8.5, 0.7\} \\ \hline
19.6  &  1.44 (2.09) \{1.96\}  &  33 (36) \{36\}  &  6 (6) \{5\}  &  14.6, 8.7 (11.7, 11.6) \{12.1, 12.0\} \\ \hline
19.5  &  2.19 (2.37) \{1.83\}  &  26 (32) \{31\}  &  1 (5) \{6\}  &  23.1, 12.8 (15.5, 12.8) \{15.9, 12.5\} \\ \hline
19.4  &  1.95 (2.91) \{3.36\}  &  27 (26) \{29\}  &  -4 (3) \{2\}  &  24.7, 12.3 (22.2, 14.3) \{20.1, 12.7\} \\ \hline
19.2  &  0.6 (1.01) \{0.5\}  &  18 (13) \{-12\}  &  -1 (0) \{0\}  &  30.9, 21.3 (35.4, 17.9) \{59.2, 4.4\} \\ \hline
19.0  &  0.2 (0.12) \{0.02\}  &  35 (26) \{-27\}  &  -5 (2) \{1\}  &  19.8, 23.7 (22.7, 10.5) \{74.2, 1.1\} \\ \hline
18.8  &  0.1 (0.03) \{0\}  &  58 (38) \{-\}  &  -6 (2) \{-\}  &  19.6, 24.1 (12.5, 12.1) \{-, -\} \\ \hline
18.6  &  0.09 (0.03) \{0.0\}  &  65 (44) \{-53\}  &  -8 (-1) \{4\}  &  25.7, 24.9 (11.9, 17.1) \{99.2, 0.1\} \\ \hline
18.5  &  0.12 (0.04) \{0\}  &  74 (40) \{-\}  &  -28 (-2) \{-\}  &  46.3, 40.1 (14.0, 19.5) \{-, -\} \\ \hline
18.4  &  0.12 (0.02) \{0\}  &  62 (45) \{-\}  &  -30 (3) \{-\}  &  42.9, 41.4 (7.5, 20.0) \{-, -\} \\ \hline
18.3  &  0.1 (0.02) \{0\}  &  61 (19) \{-\}  &  -28 (-1) \{-\}  &  41.2, 46.7 (29.8, 42.5) \{-, -\} \\ \hline
18.2  &  0.09 (0.04) \{0\}  &  52 (-19) \{-\}  &  -25 (-10) \{-\}  &  36.1, 51.5 (68.4, 54.9) \{-, -\} \\ \hline
18.1  &  0.11 (0.07) \{0\}  &  56 (-31) \{-\}  &  -15 (-10) \{-\}  &  27.2, 63.5 (80.7, 47.1) \{-, -\} \\ \hline
18.0  &  0.12 (0.18) \{0.0\}  &  8 (-39) \{-83\}  &  -6 (-7) \{3\}  &  42.1, 75.4 (87.3, 35.5) \{128.3, 2.9\} \\ \hline

\hspace*{-1cm}\end{tabular}
\label{tbl:dist_CenA_CGCG}
\end{center}
\end{table}
\vspace{-1.5cm}
\begin{table}
\tiny
\begin{center}
 \hspace*{-1cm} \begin{tabular}[htpb]{ |c|c|c|c|c| }
    \hline 
log$R$ & mag 	& $<{\ell}>$ ($^{\circ}$) & $<{b}>$ ($^{\circ}$)  & $<{\Delta}>, \sigma_{\Delta}$  ($^{\circ}$) \\ \hline \hline
\multicolumn{5}{|c|}{M82:  $\ell = 141.4^{\circ}$,  $b = 40.57^{\circ}$ }\\ \hline
20.0  &  1.68 (1.01) \{1.04\}  &  141 (141) \{141\}  &  40 (41) \{41\}  &  0.3, 0.9 (0.5, 0.7) \{0.4, 0.7\} \\ \hline
19.8  &  2.35 (1.0) \{1.06\}  &  141 (141) \{141\}  &  40 (41) \{41\}  &  0.4, 1.1 (0.8, 0.7) \{0.6, 0.7\} \\ \hline
19.6  &  3.96 (0.86) \{1.1\}  &  140 (140) \{141\}  &  41 (41) \{41\}  &  0.7, 1.4 (1.1, 0.7) \{1.1, 0.7\} \\ \hline
19.5  &  4.74 (0.83) \{1.13\}  &  140 (140) \{140\}  &  41 (42) \{42\}  &  1.0, 1.6 (1.5, 0.7) \{1.4, 0.8\} \\ \hline
19.4  &  5.24 (1.02) \{1.17\}  &  139 (139) \{140\}  &  41 (42) \{42\}  &  2.2, 1.8 (2.2, 0.9) \{1.8, 0.8\} \\ \hline
19.2  &  4.09 (1.56) \{1.29\}  &  140 (138) \{139\}  &  38 (43) \{43\}  &  2.5, 9.0 (3.3, 1.0) \{3.1, 0.8\} \\ \hline
19.0  &  4.61 (1.29) \{1.56\}  &  139 (136) \{137\}  &  34 (45) \{45\}  &  6.6, 10.9 (5.8, 1.2) \{5.7, 0.9\} \\ \hline
18.8  &  4.23 (2.19) \{2.41\}  &  119 (129) \{132\}  &  20 (50) \{51\}  &  28.3, 11.3 (12.6, 3.3) \{12.1, 1.3\} \\ \hline
18.6  &  7.12 (2.75) \{1.19\}  &  106 (124) \{115\}  &  29 (-20) \{-47\}  &  30.9, 43.1 (63.1, 29.9) \{91.2, 4.7\} \\ \hline
18.5  &  4.42 (3.83) \{4.4\}  &  130 (83) \{84\}  &  -10 (-30) \{-36\}  &  51.6, 54.0 (88.9, 51.5) \{93.0, 45.7\} \\ \hline
18.4  &  3.32 (3.13) \{2.55\}  &  145 (108) \{117\}  &  -14 (-24) \{-29\}  &  54.3, 58.7 (72.1, 39.5) \{73.2, 23.2\} \\ \hline
18.3  &  4.15 (3.95) \{4.4\}  &  166 (95) \{79\}  &  -11 (-22) \{-24\}  &  56.4, 56.5 (76.3, 49.7) \{86.4, 55.2\} \\ \hline
18.2  &  4.64 (6.39) \{6.25\}  &  173 (135) \{110\}  &  -10 (-23) \{-5\}  &  58.5, 61.7 (64.0, 33.2) \{53.9, 59.2\} \\ \hline
18.1  &  8.24 (4.46) \{2.18\}  &  -154 (58) \{-73\}  &  2 (-24) \{-26\}  &  69.6, 48.8 (100.5, 65.3) \{147.8, 63.7\} \\ \hline
18.0  &  5.85 (4.66) \{2.11\}  &  -177 (51) \{17\}  &  5 (-10) \{27\}  &  50.9, 85.9 (96.9, 74.0) \{95.3, 34.5\} \\ \hline

\multicolumn{5}{|c|}{M87:  $\ell = 283.8^{\circ}$,  $b = 74.49^{\circ}$ }\\ \hline
20.0  &  0.79 (0.96) \{0.96\}  &  -67 (-67) \{-67\}  &  73 (73) \{73\}  &  3.2, 0.6 (3.2, 0.7) \{3.0, 0.7\} \\ \hline
19.8  &  0.67 (0.91) \{0.94\}  &  -63 (-63) \{-63\}  &  71 (72) \{72\}  &  4.9, 0.6 (4.8, 0.7) \{4.8, 0.7\} \\ \hline
19.6  &  0.58 (0.96) \{0.9\}  &  -58 (-58) \{-57\}  &  70 (69) \{70\}  &  7.4, 0.5 (7.6, 0.7) \{7.7, 0.7\} \\ \hline
19.5  &  1.55 (0.88) \{0.88\}  &  -53 (-54) \{-53\}  &  68 (68) \{68\}  &  9.7, 0.9 (9.8, 0.7) \{9.7, 0.7\} \\ \hline
19.4  &  0.5 (0.75) \{0.86\}  &  -48 (-51) \{-50\}  &  66 (65) \{66\}  &  12.2, 0.5 (12.4, 0.8) \{12.2, 0.7\} \\ \hline
19.2  &  2.47 (0.43) \{0.79\}  &  -36 (-42) \{-43\}  &  55 (59) \{59\}  &  25.4, 3.2 (19.6, 1.0) \{19.7, 0.6\} \\ \hline
19.0  &  0.08 (0.66) \{0.67\}  &  -121 (-38) \{-36\}  &  -7 (46) \{47\}  &  86.5, 7.8 (32.6, 1.7) \{32.4, 0.6\} \\ \hline
18.8  &  0.77 (0.49) \{0.44\}  &  -82 (-32) \{-32\}  &  10 (21) \{24\}  &  64.4, 44.8 (58.6, 4.5) \{55.4, 0.5\} \\ \hline
18.6  &  1.3 (0.03) \{0\}  &  -128 (-114) \{-\}  &  -6 (-21) \{-\}  &  85.9, 35.1 (98.6, 10.0) \{-, -\} \\ \hline
18.5  &  1.28 (0.34) \{0.0\}  &  -109 (-119) \{-38\}  &  -14 (-22) \{5\}  &  90.5, 38.4 (99.9, 11.2) \{73.1, 1.6\} \\ \hline
18.4  &  1.81 (1.24) \{0.0\}  &  -101 (-111) \{-42\}  &  -13 (-25) \{6\}  &  89.0, 57.6 (101.9, 20.8) \{71.1, 5.7\} \\ \hline
18.3  &  2.25 (3.86) \{3.38\}  &  -104 (-102) \{-89\}  &  -4 (-8) \{-32\}  &  80.3, 73.8 (83.9, 28.2) \{106.8, 11.5\} \\ \hline
18.2  &  1.69 (2.94) \{2.99\}  &  -48 (-94) \{-100\}  &  6 (6) \{12\}  &  70.5, 83.4 (69.3, 25.6) \{63.7, 13.0\} \\ \hline
18.1  &  0.96 (1.07) \{0.7\}  &  -27 (-85) \{-92\}  &  0 (5) \{21\}  &  79.8, 76.9 (69.6, 35.8) \{53.7, 11.0\} \\ \hline
18.0  &  0.72 (0.92) \{0.99\}  &  1 (-76) \{-80\}  &  -1 (7) \{-19\}  &  87.0, 80.0 (67.6, 45.6) \{93.9, 45.8\} \\ \hline

\hspace*{-1cm}\end{tabular}
\label{tbl:dist_M82_M87}
\end{center}
\end{table}
\vspace{-1.5cm}
\pagebreak

\end{document}